\documentclass[fleqn,sort&compress,preprint,review,12pt,times]{elsarticle}

\usepackage{mathrsfs,amsmath,amssymb}
\usepackage{natbib}
\usepackage{txfonts}

\usepackage{longtable}
\usepackage{multirow}

\newcommand{\lc}{\left<}
\newcommand{\rc}{\right>}
\newcommand{\lr}{\left|}
\newcommand{\rl}{\right|}
\newcommand{\lb}{\left(}
\newcommand{\rb}{\right)}
\newcommand{\ls}{\left[}
\newcommand{\rs}{\right]}
\newcommand{\Lb}{\left\{}
\newcommand{\Rb}{\right\}}

\journal{Physics Reports}

\begin{document}

\begin{frontmatter}



\title{Hidden pseudospin and spin symmetries and their origins in atomic nuclei}


\author[PKU,RIKEN]{Haozhao Liang}
\ead{haozhao.liang@riken.jp}

\author[PKU,BUAA,SA]{Jie Meng\corref{cor1}}
\cortext[cor1]{Corresponding author}
\ead{mengj@pku.edu.cn}

\author[ITP,LZ]{Shan-Gui Zhou}
\ead{sgzhou@itp.ac.cn}

\address[PKU]{State Key Laboratory of Nuclear Physics and Technology, School of Physics,
    Peking University, Beijing 100871, China}
\address[RIKEN]{RIKEN Nishina Center, Wako 351-0198, Japan}
\address[BUAA]{School of Physics and Nuclear Energy Engineering, Beihang University,
    Beijing 100191, China}
\address[SA]{Department of Physics, University of Stellenbosch, Stellenbosch, South Africa}
\address[ITP]{State Key Laboratory of Theoretical Physics, Institute of Theoretical Physics,
    Chinese Academy of Sciences, Beijing 100190, China}
\address[LZ]{Center of Theoretical Nuclear Physics,
    National Laboratory of Heavy Ion Accelerator, Lanzhou 730000, China}

\begin{abstract}
Symmetry plays a fundamental role in physics.
The quasi-degeneracy between single-particle orbitals $(n, l, j = l + 1/2)$ and $(n-1, l + 2, j = l + 3/2)$ indicates a hidden symmetry in atomic nuclei, the so-called pseudospin symmetry (PSS).
Since the introduction of the concept of PSS in atomic nuclei, there have been comprehensive efforts to understand its origin.
Both splittings of spin doublets and pseudospin doublets play critical roles in the evolution of magic numbers in exotic nuclei discovered by modern spectroscopic studies with radioactive ion beam facilities.
Since the PSS was recognized as a relativistic symmetry in 1990s, many special features, including the spin symmetry (SS) for anti-nucleon, and many new concepts have been introduced.
In the present Review, we focus on the recent progress on the PSS and SS in various systems and potentials, including extensions of the PSS study from stable to exotic nuclei, from non-confining to confining potentials, from local to non-local potentials, from central to tensor potentials, from bound to resonant states, from nucleon to anti-nucleon spectra, from nucleon to hyperon spectra, and from spherical to deformed nuclei. Open issues in this field are also discussed in detail, including the perturbative nature, the supersymmetric representation with similarity renormalization group, and the puzzle of intruder states.
\end{abstract}


\end{frontmatter}


\tableofcontents

\section{Introduction}\label{Sect:1}


The establishment of nuclear shell model is one of the most important milestones in nuclear physics.
Similar to that of electrons orbiting in an atom, protons and neutrons in a nucleus form shell structures.
The corresponding so-called magic numbers are found to be $2$, $8$, $20$, $28$, $50$, and $82$ for both protons and neutrons as well as $126$ for neutrons in stable nuclei \cite{Haxel1949_PR075-1766,Mayer1949_PR075-1969}.
The abundance of a nucleus with the magic numbers of proton and/or neutron is normally more than its neighboring nuclei.
The magic numbers manifest themselves as a sudden jump in the plots of the two-nucleon separation energies \cite{Wang2012_ChinPhysC36-1603}, the $\alpha$-decay half-lives, neutron-capture cross sections, and also the isotope shifts as functions of nucleon number \cite{Hughes1953}.
They also appear as peaks in the abundance pattern in the solar systems in astrophysics.

In order to understand these magic numbers, starting from some simple models such as the square well or harmonic oscillator (HO) potential with analytical solutions, nuclear physicists tried to solve the corresponding Schr\"odinger equations.
In 1949, independently, Haxel, Jensen, and Suess \cite{Haxel1949_PR075-1766} and Mayer \cite{Mayer1949_PR075-1969} introduced the spin-orbit (SO) potential which largely splits the states with high orbital angular momentum.
In combination with the usual mean-field harmonic oscillator, square well, or Woods-Saxon potentials, the strong spin-orbit potential, although added by hand, excellently reproduces all traditional magic numbers in nuclear physics.
Apart from the magic numbers, it also provides wonderful descriptions for nuclear ground-state and some low-lying excited-state properties.
Therefore, the substantial spin symmetry (SS) breaking between the spin doublets $(n, l, j = l \pm 1/2)$ is one of the most important concepts in nuclear structure.

The success of the nuclear shell model or spin-orbit potential is unprecedented.
For light nuclei ($A \lesssim 25$) the rotational features of nuclear spectra can be understood in a many-particle spherical shell-model framework in terms of the SU(3) coupling scheme of Elliott and Harvey \cite{Elliott1958PRSA245-128, Elliott1958PRSA245-562, Elliott1963PRSA272-557}.
By introducing the deformation-dependent oscillator length, Nilsson \textit{et al.} \cite{Nilsson1955_DMFM29-16,Nilsson1969_NPA131-1} extended the shell model to the deformed cases and built the foundation for describing not only the deformed nuclei but also nuclear rotation phenomena.
Even nowadays, searching for new magic numbers and investigating shell structure evolution for unstable nuclei is still one of the key topics for the radioactive ion beam facilities worldwide \cite{Sorlin2008_PPNP61-602}.


After the successful reproduction of the magic numbers, the shell model with strong spin-orbit potential became the strongest candidate of the standard nuclear model and almost the entire nuclear physics community was exploring and enjoying the new physics brought in by this model.
Here we mentioned ``almost'' because there were a few groups who were examining the nuclear shell model in a different way.

\begin{figure}[tbhp]
\begin{center}
  \includegraphics[width=8cm]{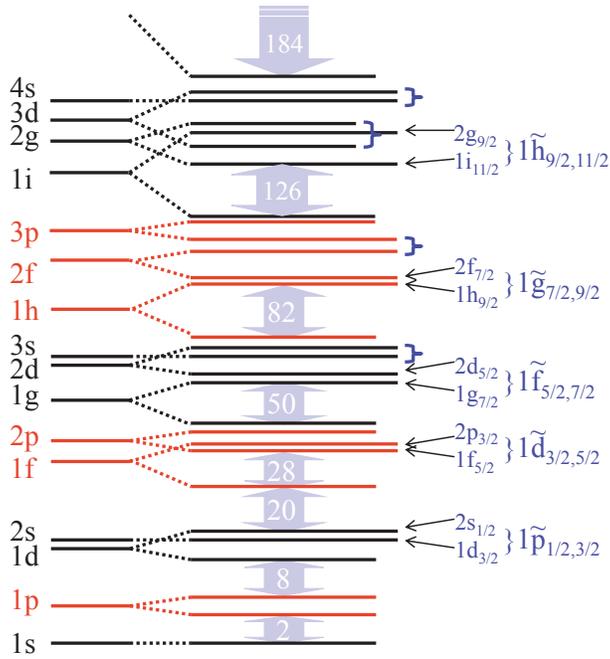}
\end{center}
\caption{(Color online) Schematic nuclear single-particle spectrum.
On one hand, the strong spin-orbit splittings between the spin doublets $(n, l, j = l \pm 1/2)$ lead to the traditional magic numbers.
On the other hand, pairs of single-particle states in braces, $(n, l, j = l + 1/2)$ and $(n-1, l + 2, j = l + 3/2)$, are quasi-degenerate.
They are defined as the pseudospin doublets $(\tilde{n}=n-1, \tilde{l}=l+1, j=\tilde{l}\pm1/2)$, and the pseudospin symmetry is introduced for such near degeneracy.
\label{Fig:1.PSS}}
\end{figure}

These groups were not satisfied with simply reproducing the magic numbers or the splittings between the spin doublets.
By examining the single-particle spectra, Hecht and Adler \cite{Hecht1969_NPA137-129} and Arima, Harvey, and Shimizu \cite{Arima1969_PLB30-517} found the near degeneracy between two single-particle states with quantum numbers $(n, l, j = l + 1/2)$ and $(n-1, l + 2, j = l + 3/2)$.
They introduced the so-called pseudospin symmetry (PSS) and defined the pseudospin doublets as $(\tilde{n}=n-1, \tilde{l}=l+1, j=\tilde{l}\pm1/2)$ to explain this near degeneracy.
This is illustrated in Fig.~\ref{Fig:1.PSS}.

The pseudospin symmetry remains an important physical concept in axially deformed \cite{Ratna-Raju1973_NPA202-433,Draayer1984_AP156-41,Troltenier1994_NPA576-351,Troltenier1995_NPA586-53} and even triaxially deformed \cite{Blokhin1997_NPA612-163,Beuschel1997_NPA619-119} nuclei.
Based on this concept, a simple but useful pseudo-SU(3) model was proposed, and this model was generalized to be the (pseudo-)symplectic model \cite{Rosensteel1976AP96-1, Rosensteel1979AP123-36, Rowe1980AP126-198, Rosensteel1980AP126-343, Troltenier1994_NPA576-351, Troltenier1995_NPA586-53}, (also see Ref.~\cite{Rowe1985RPP48-1419} and references therein).
The pseudospin symmetries have been also extensively used in the odd-mass nuclei in the context of the interacting Boson-Fermion model \cite{Iachello1981AP136-19, Bijker1984AP156-110, Bijker1985AP161-360, Bijker1988AP187-148}.

Although the concept of pseudospin symmetry was based on the observation of empirical single-particle spectra, it remained to be a purely theoretical subject for nearly $20$ years before the discovery of the nuclear superdeformed identical rotation bands \cite{Byrski1990_PRL64-1650}.
From then on, a lot of phenomena in nuclear structure have been successfully interpreted directly or implicitly by the pseudospin symmetry, including nuclear superdeformed configurations \cite{Dudek1987_PRL59-1405,Bahri1992_PRL68-2133}, identical bands \cite{Nazarewicz1990_PRL64-1654,Nazarewicz1990_NPA512-61,Zeng1991_PRC44-R1745}, quantized alignment \cite{Stephens1990_PRL65-301}, and pseudospin partner bands \cite{Xu2008_PRC78-064301,Hua2009_PRC80-034303}.
The pseudospin symmetry may also manifest itself in magnetic moments and transitions \cite{Troltenier1994_NPA567-591,Ginocchio1999_PRC59-2487,Neumann-Cosel2000_PRC62-014308} and $\gamma$-vibrational states in nuclei \cite{Jolos2012_PRC86-044320} as well as in nucleon-nucleus and nucleon-nucleon scatterings \cite{Ginocchio1999_PRL82-4599,Leeb2000_PRC62-024602,Ginocchio2002_PRC65-054002,Leeb2004_PRC69-054608}.
In addition, the relevance of pseudospin symmetry in the structure of halo nuclei \cite{Long2010_PRC81-031302R} and superheavy nuclei \cite{Jolos2007_PAN70-812,Li2014_PLB732-169} was pointed out.

In the 21st century, it has been intensively shown that the traditional magic numbers can change in nuclei far away from the stability line \cite{Sorlin2008_PPNP61-602}.
It is found that both splittings of spin and pseudospin doublets play critical roles in the shell structure evolution.
For example, the $N=28$ shell closure disappears due to the quenching of the spin-orbit splitting for the $\nu1f$ spin doublets \cite{Gaudefroy2006_PRL97-092501,Bastin2007_PRL99-022503,Tarpanov2008_PRC77-054316,Moreno-Torres2010_PRC81-064327}, and the $Z=64$ subshell closure is closely related to the restoration of pseudospin symmetry for the $\pi2\tilde p$ and $\pi1\tilde f$ pseudospin doublets \cite{Nagai1981_PRL47-1259,Long2007_PRC76-034314,Long2009_PLB680-428}.
Therefore, it will be quite interesting and challenging to understand shell closure and pseudospin symmetry on the same footing, in particular for superheavy and exotic nuclei near the limit of nucleus existence.


Since the recognition of pseudospin symmetry in atomic nuclei, there have been comprehensive efforts to understand its origin.
Apart from the formal relabelling of quantum numbers, various explicit transformations from the normal scheme to the pseudospin scheme have been proposed \cite{Bohr1982_PS26-267,Castanos1992_PLB277-238,Blokhin1995_PRL74-4149,Blokhin1996_JPA29-2039}.
In 1982, Bohr, Hamamoto, and Mottelson \cite{Bohr1982_PS26-267} discussed the pseudospin symmetry in rotating nuclear potentials, and found that such a symmetry is very helpful to understand qualitatively the feature of quasi-particle motion in rotating potentials.
Based on the single-particle Hamiltonian of the harmonic oscillator shell model, they tried to understand the origin of pseudospin symmetry in terms of the spin-orbit potential introduced by hand, and also the orbit-orbit term, which has been artificially added in the Nilsson model.
It turns out that the origin of pseudospin symmetry is connected with a special ratio between the strengths of the spin-orbit and orbit-orbit interactions.
This idea was followed by the groups at Louisiana State University, University of California, and National Autonomous University of Mexico (UNAM), and they tried to understand the spin-orbit and orbit-orbit potentials in order to explain the pseudospin symmetry \cite{Bahri1992_PRL68-2133, Castanos1992_PLB277-238, Blokhin1995_PRL74-4149, Blokhin1996_JPA29-2039}.


The relation between the pseudospin symmetry and the relativistic mean-field (RMF) theory \cite{Ring1996_PPNP37-193,Vretenar2005_PR409-101,Meng2006_PPNP57-470,Niksic2011_PPNP66-519,Meng2011_PP31-199,Meng2013_FPC8-55} was first noted in Ref.~\cite{Bahri1992_PRL68-2133}, where the relativistic mean-field theory was used to explain approximately such a special ratio between the strengths of the spin-orbit and orbit-orbit interactions.

In order to see the connection with the relativistic mean-field or the covariant density functional theory (CDFT), it will be quite illuminating to examine the Dirac equation governing the motion of nucleons.
The corresponding single-particle wave functions are given in the form of the Dirac spinor, which has both the upper and lower components.
For the spherical case, by looking into the Dirac spinor, it is interesting to note that the upper and lower components have the same total angular momentum $j$ but not the orbital angular momentum $l$.
Their orbital angular momenta $l$ differ by one unit.
In the normal labelling of the single-nucleon states, the $l$ of the dominant upper component is used.
Equivalently, one can also use the quantum number of the lower component to label the single-nucleon states.
In 1997, Ginocchio \cite{Ginocchio1997_PRL78-436} revealed that the pseudospin symmetry is essentially a relativistic symmetry of the Dirac Hamiltonian and the pseudo-orbital angular momentum $\tilde l$ is nothing but the orbital angular momentum of the lower component of the Dirac spinor.
He also showed that the pseudospin symmetry in nuclei is exactly conserved when the scalar potential $S(r)$ and the vector potential $V(r)$ have the same size but opposite sign, i.e., $\Sigma(r) \equiv S(r) + V (r) = 0$.
This discovery not only reveals the origin of pseudospin symmetry but also demonstrates an unexpected success of the relativistic mean-field theory.
It should be noted the pseudospin symmetry is a special case of the Bell-Reugg symmetries \cite{Bell1975_NPB98-151}, as pointed out in Section~2 of Ref.~\cite{Ginocchio2005_PR414-165}.

One can also go one step further to reduce the Dirac equation into the second-order differential equation for either the upper or lower component.
Then for the upper and lower components there will be the spin-orbit and pseudospin-orbit (PSO) potentials, respectively governing the relevant partner splittings.
If either the spin-orbit or pseudospin-orbit potential vanishes, it will lead to the corresponding spin or pseudospin symmetry.
The derivative for the sum of the scalar and vector potentials, i.e., $d\Sigma(r)/dr$, determines the pseudospin symmetry.
The pseudospin symmetry is exact under the condition $d\Sigma(r)/dr=0$ \cite{Meng1998_PRC58-R628}.
This condition means that the pseudospin symmetry becomes much better for exotic nuclei with a highly diffused potential \cite{Meng1999_PRC59-154}.
Approximately, the pseudospin symmetry is connected with the competition between the pseudo-centrifugal barrier (PCB) and the pseudospin-orbit potential.
However, in either limit, $\Sigma(r) = 0$ or $d\Sigma(r)/dr = 0$, there are no longer bound states, thus the pseudospin symmetry is always broken in realistic nuclei.
In this sense, the pseudospin symmetry is viewed as a dynamical symmetry \cite{Arima1999_RIKEN-AF-NP-276,Alberto2001_PRL86-5015,Alberto2002_PRC65-034307} or it is of the non-perturbative nature \cite{Marcos2001_PLB513-30,Lisboa2010_PRC81-064324,Ginocchio2011_JPCS267-012037,Castro2012_PRA86-032122}.

Following discussions for spherical nuclei, the study of pseudospin symmetry within the relativistic framework was quickly extended to deformed ones \cite{Lalazissis1998_PRC58-R45,Sugawara-Tanabe1998_PRC58-R3065}.
As the pseudospin symmetry is a relativistic symmetry, the wave functions of the pseudospin partners satisfy certain relations.
These relations have been tested in both spherical and deformed nuclei \cite{Lalazissis1998_PRC58-R45,Ginocchio1998_PRC57-1167,Ginocchio2002_PRC66-064312,Sugawara-Tanabe2002_PRC65-054313,
Ginocchio2004_PRC69-034303}.


Although the doubt on the connection between the pseudospin symmetry and the condition $\Sigma(r)=0$ or $d\Sigma(r)/dr = 0$ exists \cite{Marcos2007_EPJA34-429,Marcos2008_EPJA37-251,Desplanques2010_EPJA43-369}, following the pseudospin symmetry limit, a lot of discussions about the pseudospin symmetry in single-particle spectra have been made by exactly or approximately solving the Dirac equation with various potentials, for examples,
the one-dimensional Woods-Saxon potential \cite{Panella2010_JPA43-325302},
the two-dimensional Smorodinsky-Winternitz potential \cite{Zhang2009_PRA80-054102},
the spherical harmonic oscillator \cite{Chen2003_CPL20-358,Lisboa2004_PRC69-024319,Ginocchio2005_PRL95-252501,Guo2005_NPA757-411,deCastro2006_PRC73-054309,
Xu2007_HEPNP31-251,Zhang2008_PRA78-040101R,Akcay2009_PLA373-616,Akcay2009_IJMPC20-930},
anharmonic oscillator \cite{Zhang2009_APS58-61},
Coulomb \cite{Lisboa2003_PRC67-054305, Hamzavi2010_PLA374-4303, Castro2012_PRA86-032122, Barakat2013_FP43-1171, Eshghi2013_CPB22-030303},
Deng-Fan \cite{Ortakaya2014_CPB23-030306},
diatomic molecular \cite{Jia2007_PS75-388, Akcay2013_FBS54-1839},
Eckart \cite{Jia2006_JPA39-7737,Soylu2008_JPG41-065308},
Hellmann \cite{Hamzavi2013_CanJP91-411},
Hulth\'en \cite{Guo2003_CPL20-602, Ikhdair2011_AMC217-9019, Aydogdu2013_ChinPhysB22-010302, Hamzavi2013_CPB22-080302},
Manning-Rosen \cite{Wei2008_PLA373-49,Chen2009_PS79-055002,Wei2010_PLB686-288},
Mie-type \cite{Aydogdu2010_AoP325-373,Hamzavi2010_FBS48-171,Ikot2014_CPC38-013101},
Morse \cite{Berkdemir2006_NPA770-32, Bayrak2007_JPA40-11119, Qiang2007_JPG40-1677, Aydogdu2011_PLB703-379, Ikhdair2011_JMP52-052303, Ortakaya2013_AP338-250},
P\"oschl-Teller \cite{Jia2007_EPJA34-41,Jia2009_PLA373-1621,Wei2009_EPL87-40004,Wei2010_EPJA43-185,Candemir2012_IJMPE21-1250060,
Hamzavi2012_IJMPE21-1250097, Ikhdair2012_PS86-045002, Hamzavi2013_AdvHEP2013-196986, Falaye2013_ChinPhysB22-060305, Ikot2013_FBS54-2053},
Rosen-Morse \cite{Oyewumi2010_EPJA45-311, Wei2010_EPJA46-207, Chen2011_ChinPhysB20-062101, Aguda2013_CanJP91-689},
Tietz-Hua \cite{Ikhdair2012_FBS53-473},
Woods-Saxon \cite{Guo2005_PLA338-90,Xu2006_NPA768-161,Aydogdu2010_EPJA43-73,Chen2012_PRC85-054312,Maghsoodi2012_PS85-055007},
and Yukawa \cite{Aydogdu2011_PS84-025005, Maghsoodi2012_PS86-015005, Ikhdair2013_PS87-035002, Ikot2013_FBS54-2027}
potentials,
as well as the deformed harmonic oscillator \cite{Ginocchio2004_PRC69-034318,Guo2006_PLA353-378,Zhou2008_ChinPhysB17-380,Zhang2009_APS58-712,Zhang2009_PS80-065018,
Setare2010_MPLA25-549},
anharmonic oscillator \cite{Zhang2012_AP327-841,Hamzavi2014_AP341-153},
Hartmann \cite{Alhaidari2006_PLA349-87,Guo2007_IJMPA22-4825},
Hylleraas \cite{Hassanabadi2012_ChinPhysB21-120302},
Kratzer \cite{Berkdemir2008_JPA41-045302},
Makarov \cite{Zhou2009_CTP52-813},
Manning-Rosen \cite{Asgarifar2013_PS87-025703},
and ring-shaped \cite{Zhang2009_CEJP7-768,Zhang2009_IJTP48-2625,Zhang2011_JMP52-053518,Oyewumi2014-IJMPE23-1450005} potentials.
Self-consistently, the pseudospin symmetry in spherical \cite{Gambhir1998_EPJA3-255,Ginocchio1998_PRC57-1167,Marcos2000_PRC62-054309,Ginocchio2001_PRL87-072502,
Borycki2003_PRC68-014304,Marcos2001_PLB513-30,Chen2003_HEPNP27-324,Lopez-Quelle2003_NPA727-269,Marcos2003_EPJA17-173,
Alberto2005_PRC71-034313,Marcos2005_EPJA26-253,Long2006_PLB639-242,Xu2007_HEPNP31-168,Guo2010_EPJA45-179,
Long2010_PRC81-031302R} and deformed \cite{Lalazissis1998_PRC58-R45,Sugawara-Tanabe1998_PRC58-R3065,Sugawara-Tanabe2000_PRC62-054307,
Sugawara-Tanabe2002_PRC65-054313,Ginocchio2004_PRC69-034303,Sugawara-Tanabe2005_RMP55-277,Sun2012_EPJA48-18} nuclei have been investigated within relativistic mean-field and relativistic Hartree-Fock (RHF) \cite{Bouyssy1985_PRL55-1731,Bouyssy1987_PRC36-380,Long2006_PLB640-150,Long2010_PRC81-024308} theories.
One of interesting topics is the tensor effects on the pseudospin symmetry or spin symmetry, which has been investigated in some of the above-mentioned works \cite{Lisboa2004_PRC69-024319, Alberto2005_PRC71-034313, deCastro2006_PRC73-054309,
Long2007_PRC76-034314, Long2010_PRC81-031302R} and also in Refs.~\cite{Castro2012_PRC86-052201,Wang2013_JPG40-045105}.


For the many-body problems in quantum mechanics, the basis expansion is one of the standard methods, e.g., the solutions of the Schr\"odinger equation with a harmonic oscillator potential as a widely used basis.
For the Dirac equation, there exist not only the positive-energy states in the Fermi sea but also the negative-energy states in the Dirac sea, where the negative-energy states correspond to the anti-particle states.
When the solutions of the Dirac equation are used as a complete basis, e.g., the Dirac Woods-Saxon basis \cite{Zhou2003_PRC68-034323}, the states with both positive and negative energies must be included \cite{Zhou2003_PRC68-034323,Schunck2008_PRC77-011301R,Schunck2008_PRC78-064305,Long2010_PRC81-024308,Zhou2010_PRC82-011301R,
Chen2012_PRC85-067301,Li2012_PRC85-024312,Li2012_CPL29-042101,Li2012_AIPCP1491-208}.

When Zhou, Meng, and Ring developed the relativistic mean-field theory in a Dirac Woods-Saxon basis \cite{Zhou2003_PRC68-034323}, they examined carefully the negative-energy states in the Dirac sea and found that the pseudospin symmetry of those negative-energy states, or equivalently, the spin symmetry in the anti-nucleon spectra is very well conserved \cite{Zhou2003_PRL91-262501}.
Furthermore, they have shown that the spin symmetry in the anti-nucleon spectra is much better developed than the pseudospin symmetry in normal nuclear single-particle spectra.
It should be noted that, by applying the charge conjugate transformation, the spin symmetry for anti-nucleon states have been formally conjectured in Ref.~\cite{Ginocchio1999_PR315-231}.
The spin symmetry in the anti-nucleon spectra was also tested by investigating relations between Dirac wave functions of spin doublets with the relativistic mean-field theory \cite{He2006_EPJA28-265}.
Later, this spin symmetry was studied with the relativistic Hartree-Fock theory and the contribution from the Fock term was analyzed \cite{Liang2010_EPJA44-119}.
It has been pointed out in Ref.~\cite{Zhou2003_PRL91-262501} that an open problem related to the experimental study of the spin symmetry in the anti-nucleon spectra is the polarization effect caused by the annihilation of anti-nucleons in a normal nucleus.
Detailed calculations of the anti-baryon annihilation rates in the nuclear environment showed that the in-medium annihilation rates are strongly suppressed by a significant reduction of the reaction $Q$ values, leading to relatively long-lived anti-baryon-nucleus systems \cite{Mishustin2005_PRC71-035201}.
Recently, the spin symmetry in the anti-$\Lambda$ spectra of hypernuclei was studied quantitatively \cite{Song2009_CPL26-122102,Song2010_ChinPhysC34-1425,Song2011_CPL28-092101}, which may be free from the problem of annihilation.
This kind of study would be of great interests for possible experimental tests.


In recent years, there has been an increasing interest in the exploration of continuum and resonant states, especially in the studies of exotic nuclei with unusual $N/Z$ ratios.
In exotic nuclei, the neutron (or proton) Fermi surface is close to the particle continuum; thus, the contribution of the continuum is important \cite{Zhou2010_PRC82-011301R,Chen2012_PRC85-067301,Li2012_PRC85-024312,Li2012_CPL29-042101,Li2012_AIPCP1491-208,Dobaczewski1984_NPA422-103,Dobaczewski1996_PRC53-2809,
Meng1996_PRL77-3963,Meng1998_PRL80-460,Meng1998_NPA635-3,Meng2002_PRC65-041302R,Meng2011_SciChinaPMA54S1-119,
Poschl1997_PRL79-3841,Zhang2011_PRC83-054301,Zhang2012_PRC86-054318,Pei2008_PRC78-064306,Pei2011_PRC84-024311,
Pei2013_PRC87-051302R,He2011_SciChinaPMA54S1-32,Lin2011_SciChinaPMA54S1-73,Lu2011_SciChinaPMA54S1-136}.
Many methods or models developed for the studies of resonances \cite{Kukulin1989} have been adopted to locate the position and to calculate the width of a nuclear resonant state, e.g., the analytical continuation in coupling constant (ACCC) method \cite{Yang2001_CPL18-196,Zhang2004_PRC70-034308,Zhang2007_EPJA32-43,Zhang2009_IJMPE18-1761,Zhang2012_PRC86-032802,
Zhang2012_EPJA48-40}, the real stabilization method \cite{Zhang2007_APS56-3839,Zhang2008_PRC77-014312,Zhou2009_JPB42-245001,Mei2009_ChinPC33S1-101,Zhang2009_ChinPC33-187,
Zhang2010_MPLA25-727}, the complex scaling method (CSM) \cite{Guo2010_PRC82-034318,Guo2010_IJMPE19-1357,Guo2010_CPC181-550,Liu2012_PRC86-054312}, the coupled channels method \cite{Hagino2004_NPA735-55,Li2010_PRC81-034311,Li2010_SCG53-773}, and some others \cite{Fedorov2009_FBS45-191,Fernandez2012_AppMathComp218-5961}.

The study of symmetries in resonant states is certainly interesting, e.g., the pseudospin symmetry \cite{Guo2005_PRC72-054319,Guo2006_PRC74-024320,Liu2013_PRA87-052122,Zhang2006_HEPNP30S2-97,Zhang2007_CPL24-1199} and spin symmetry  \cite{Xu2012_IJMPEE21-1250096} in single-particle resonant states.
Recently, Lu, Zhao, and Zhou \cite{Lu2012_PRL109-072501} gave a rigorous verification of the pseudospin symmetry in single-particle resonant states.
They have shown that the pseudospin symmetry in single-particle resonant states in nuclei is exactly conserved under the same condition for the pseudospin symmetry in bound states, i.e., $\Sigma(r) = 0$ or $d\Sigma(r)/dr = 0$ \cite{Lu2012_PRL109-072501,Lu2013_AIPCP1533-63,Lu2013_PRC88-024323}.
The exact conservation and breaking mechanism of the pseudospin symmetry in single-particle resonances for spherical square-well potentials have been investigated, in which the pseudospin symmetry-breaking part can be separated from other parts in the Jost functions.
By examining zeros of Jost functions corresponding to the lower components of radial Dirac wave functions, general properties of pseudospin symmetry splittings of the energies and widths are examined.
As noted in Ref.~\cite{Lu2012_PRL109-072501}, it is straightforward to extend the study of the pseudospin symmetry in resonant states in the Fermi sea to that in the negative-energy states in the Dirac sea or spin symmetry in anti-particle continuum spectra.


Works are also in progress for understanding the origin of pseudospin symmetry and its breaking mechanism in a perturbative way.
On one hand, the perturbation theory was used in Refs.~\cite{Liang2011_PRC83-041301R,Li2011_ChinPhysC35-825} to investigate the symmetries of the Dirac Hamiltonian and their breaking in realistic nuclei.
This provides a clear way for investigating the perturbative nature of pseudospin symmetry.
An illuminating example is that the energy splittings of the pseudospin doublets can be regarded as a result of perturbation of the Hamiltonian with a relativistic harmonic oscillator (RHO) potential, where the pseudospin doublets are degenerate \cite{Liang2011_PRC83-041301R}.


On the other hand, supersymmetric (SUSY) quantum mechanics \cite{Cooper1995_PR251-267,Cooper2001} was used to investigate the symmetries of the Dirac Hamiltonian
\cite{Leviatan2004_PRL92-202501,Typel2008_NPA806-156,Leviatan2009_PRL103-042502}, also see Refs.~\cite{Hall2010_IJMPE19-1923,Zarrinkamar2010_AP325-2522,Alhaidari2011_PLB699-309,Zarrinkamar2011_IJMPA26-1011,
Zarrinkamar2011_PS83-015009,Maghsoodi2012_PS85-055007}.
In particular, by employing both exact and broken supersymmetries, the phenomenon that all states with $\tilde l > 0$ have their own pseudospin partners except for the so-called intruder states can be interpreted naturally within a unified scheme.
A pseudospin symmetry-breaking potential without a singularity can also be obtained with the supersymmetric technique \cite{Typel2008_NPA806-156}, in contrast to the singularities appearing in the reduction of the Dirac equation to a Schr\"odinger-like equation for the lower component of the Dirac spinor.
However, by reducing the Dirac equation to a Schr\"odinger-like equation for the upper component, the corresponding effective Hamiltonian is not Hermitian, since the upper component wave functions alone, as the solutions of the Schr\"odinger-like equation, are not orthogonal to each other.
In order to fulfill the orthonormality, an additional differential relation between the lower and upper components must be taken into account.
By doing so, effectively, the upper components alone are orthogonal with respect to a different metric \cite{Typel2008_NPA806-156}.
Such fact that the corresponding effective Hamiltonian is not Hermitian prevents us from being able to perform quantitative perturbation calculations.


Recent works by Guo and coauthors \cite{Guo2012_PRC85-021302R,Li2013_PRC87-044311,Guo2014_PRL112-062502} bridged the perturbation calculations and the supersymmetric descriptions by using the similarity renormalization group (SRG) \cite{Glazek1993_PRD48-5863, Glazek1994_PRD49-4214, Wegner1994_AP506-77, Bylev1998_PLB428-329, Wegner2001_PR348-77} to transform the Dirac Hamiltonian into a diagonal form.
The effective Hamiltonian expanded in a $1/M$ series in the Schr\"odinger-like equation is Hermitian.
This makes the perturbation calculations possible.
Therefore, one can understand the origin of pseudospin symmetry and its breaking by combining supersymmetric quantum mechanics, perturbation theory, and the similarity renormalization group \cite{Liang2013_PRC87-014334,Shen2013_PRC88-024311}.


Another open issue in the study of pseudospin symmetry is the special status of nodeless intruder states which do not have their own pseudospin partners.
The nodal structure of radial Dirac wave functions of pseudospin doublets was studied in Ref.~\cite{Leviatan2001_PLB518-214}, which was helpful particularly for understanding the reason why between a pair of pseudospin doublets the pseudospin-down state with $j_<=\tilde l-1/2$ has one more radial node than the pseudospin-up state with $j_>=\tilde l+1/2$.
However, in this case, there exist no bound states in the Fermi sea at the pseudospin symmetry limit.
In contrast, as pointed out in Ref.~\cite{Chen2003_CPL20-358}, there can exist bound states in the Fermi sea at the pseudospin symmetry limit if the potential $\Delta(r)\equiv V(r)-S(r)$ is confining.
Quite recently, the nodal structure of radial Dirac wave functions for this case was demonstrated in an analytical way by Alberto, de Castro, and Malheiro \cite{Alberto2013_PRC87-031301R}.
It is interesting to note that in such a case all states with $\tilde l>0$ have their own pseudospin partners, but instead some states with $l>0$ lose their own spin partners.


In this Review, we will mainly focus on the progress in the studies of pseudospin symmetry and spin symmetry hidden  in atomic nuclei and the related topics in the past decade.
Section~\ref{Sect:3} will be devoted to highlighting the progress from several different aspects, and Section~\ref{Sect:4} will be devoted to discussing the selected open issues.
Some of the topics covered in a former review \cite{Ginocchio2005_PR414-165} will not be repeated here, such as the Bell-Reugg symmetries, the pseudospin symmetry in the magnetic dipole, electric quadrupole, and Gamow-Teller transitions, the pseudospin symmetry in the nucleon-nucleon and nucleon-nucleus scattering, as well as the spin symmetry in hadrons.

The paper will be organized as follows.
The typical Dirac equation widely used in nuclear physics will be presented together with its Schr\"odinger-like equations in Section~\ref{Sect:2}.
In the same Section, different analytical solutions for Dirac equation at the pseudospin symmetry limit and the pseudospin symmetry breaking in realistic nuclei will be reviewed briefly.
Recent progress on the pseudospin symmetry, ranging from stable to exotic nuclei, non-confining to confining potentials, local to non-local potentials, central to tensor potentials, bound to resonant states, nucleon to anti-nucleon spectra, nucleon to hyperon spectra, and spherical to deformed nuclei, will be presented in Section~\ref{Sect:3}.
Section~\ref{Sect:4} will be devoted to discussing the open issues in this field, including the perturbative nature, puzzle of intruder states, and supersymmetric representation for pseudospin symmetry.
Finally, summary and perspectives will be given in Section~\ref{Sect:5}.

\section{General Features}\label{Sect:2}

\subsection{Dirac and Schr\"odinger-like equations}\label{Sect:2.1}

\subsubsection{Dirac equations}

In the relativistic framework, the motion of nucleons is described by the Dirac equation.
The corresponding eigenfunction equation for nucleons reads
\begin{equation}\label{Eq:2.1.eigeneq}
  H\psi(\mathbf{r})=\epsilon\psi(\mathbf{r})\,,
\end{equation}
where $\epsilon$ is the single-particle energy including the rest mass of nucleon $M$.
Originating from the minimal coupling of the scalar and vector mesons to the nucleons in the covariant density functional theory \cite{Ring1996_PPNP37-193,Vretenar2005_PR409-101,Meng2006_PPNP57-470,Niksic2011_PPNP66-519,Meng2011_PP31-199,Meng2013_FPC8-55},
the single-particle Dirac Hamiltonian $H$ is written as
\begin{equation}\label{Eq:2.1.HDirac}
  H = \boldsymbol{\alpha}\cdot\mathbf{p} + \beta[M+S(\mathbf{r})]+V(\mathbf{r})\,.
\end{equation}
In this expression, $\mathbf{\alpha}$ and $\beta$ are the Dirac matrices, while $S(\mathbf{r})$ and $V(\mathbf{r})$ are the scalar and vector potentials, respectively.
In addition, we set $\hbar=c=1$ in this paper.

From mathematical point of view, the conclusions concerning the symmetry limits discussed below remain valid if either the scalar or the vector potential is modified by an arbitrary constant, i.e.,
\begin{equation}\label{Eq:2.1.modifySV}
  S(\mathbf{r}) \rightarrow S(\mathbf{r})+c_S\,,\qquad V(\mathbf{r}) \rightarrow V(\mathbf{r})+c_V\,,
\end{equation}
because one can simply adjust the mass and energy by the same constant so that the Dirac equation remains unchanged \cite{Ginocchio2005_PR414-165}:
\begin{equation}\label{Eq:2.1.modifyME}
  M \rightarrow M - c_S\,,\qquad \epsilon \rightarrow \epsilon + c_V\,.
\end{equation}

When the spherical symmetry is imposed, the single-particle eigenstates are specified by a set of quantum numbers $\alpha=(a, m_a)=(n_a, l_a, j_a, m_a)$, and the single-particle wave functions can be factorized as
\begin{equation}\label{Eq:2.1.spwfR1}
    \psi_\alpha(\mathbf{r}) =
    \frac{1}{r}
    \lb \begin{array}{c}
        iG_a(r) \\ F_a(r)\hat{\boldsymbol{\sigma}}\cdot\hat{\mathbf{r}}
    \end{array} \rb \mathscr Y^{l_a}_{j_am_a}(\hat{\mathbf{r}})\,,
\end{equation}
with the spherical harmonics spinor $\mathscr Y^{l_a}_{j_am_a}(\hat{\mathbf{r}})$ for the angular and spin parts \cite{Varshalovich1988}.
The corresponding normalization condition reads
\begin{equation}\label{Eq:2.1.Normalization}
  \int \psi^\dag_\alpha(\mathbf{r})\psi_\alpha(\mathbf{r})d^3\mathbf{r}
  =\int \ls G_a^2(r)+F_a^2(r)\rs dr=1\,.
\end{equation}
Note that in this paper, to label the single-particle eigenstates we use the main quantum number $n$ equal to the number of the internal nodes plus one for the dominant component of the Dirac spinor. Namely, the single-particle spectra start from the $n=1$ states.

For the lower component of the Dirac spinor (\ref{Eq:2.1.spwfR1}), one has
\begin{equation}
    \hat{\boldsymbol{\sigma}}\cdot\hat{\mathbf{r}}\mathscr Y^{l_a}_{j_am_a}(\hat{\mathbf{r}})
    = -\mathscr Y^{\tilde l_a}_{j_am_a}(\hat{\mathbf{r}})\,,
\end{equation}
with
\begin{equation}\label{Eq:2.1.tildel}
  \tilde l = 2j - l\,.
\end{equation}
Thus, the single-particle wave functions can also be expressed as
\begin{equation}\label{Eq:2.1.spwfR}
    \psi_\alpha(\mathbf{r}) =
    \frac{1}{r}
    \lb \begin{array}{c}
        iG_a(r) \mathscr Y^{l_a}_{j_am_a}(\hat{\mathbf{r}})
        \\ -F_a(r)\mathscr Y^{\tilde l_a}_{j_am_a}(\hat{\mathbf{r}})
    \end{array} \rb\,.
\end{equation}
In such a way, the pseudo-orbital angular momentum $\tilde{l}$ is found to be the orbital angular momentum of the lower component of the Dirac spinor \cite{Ginocchio1997_PRL78-436}.

The corresponding radial Dirac equation reads
\begin{equation}\label{Eq:2.1.DiraceqR}
    \lb\begin{array}{cc}
    M+\Sigma(r) & -\displaystyle\frac{d}{dr}+\displaystyle\frac{\kappa_a}{r} \\
        \displaystyle\frac{d}{dr}+\displaystyle\frac{\kappa_a}{r} & -M+\Delta(r)
    \end{array}\rb
    \lb\begin{array}{c}
        G_a(r) \\ F_a(r)
    \end{array}\rb
    =\epsilon_a
    \lb\begin{array}{c}
        G_a(r) \\ F_a(r)
    \end{array}\rb\,,
\end{equation}
where $\Sigma(r)=S(r)+V(r)$ and $\Delta(r)=V(r)-S(r)$ denote the combinations of the scalar and vector potentials, and $\kappa$ is a good quantum number defined as $\kappa=\mp(j+1/2)$ for the $j=l\pm1/2$ orbitals.

The SS and PSS of the Dirac Hamiltonian in Eq.~(\ref{Eq:2.1.HDirac}) or (\ref{Eq:2.1.DiraceqR}) can be studied by the Bell-Reugg condition \cite{Bell1975_NPB98-151}, see also Section~2 of Ref.~\cite{Ginocchio2005_PR414-165}.
Alternatively, these symmetries can be investigated by reducing the Dirac equation to the Schr\"odinger-like equations.

\subsubsection{Schr\"odinger-like equations}

Focusing on the spherical case, one can derive the Schr\"odinger-like equation for the upper component $G(r)$ of the Dirac spinor by substituting
\begin{equation}\label{Eq:2.1.FG}
  F(r) = \frac{1}{M-\Delta(r)+\epsilon}\lb\frac{d}{dr}+\frac{\kappa}{r}\rb G(r)
\end{equation}
in Eq.~(\ref{Eq:2.1.DiraceqR}), and obtain
\begin{equation}\label{Eq:2.1.SchrG}
    \Lb -\frac{1}{M_+}\frac{d^2}{dr^2}
  +\frac{1}{M_+^2}\frac{dM_+}{dr}\frac{d}{dr}
  +\ls (M+\Sigma)+\frac{1}{M_+}\frac{\kappa(\kappa+1)}{r^2}
  +\frac{1}{M_+^2}\frac{dM_+}{dr}\frac{\kappa}{r}
  \rs \Rb G = \epsilon G\,,
\end{equation}
with the energy-dependent effective mass $M_+(r)=M-\Delta(r)+\epsilon$.
For brevity we omit the subscripts if there is no confusion.
Similarly, one can derive the Schr\"odinger-like equation for the lower component $F(r)$ by using
\begin{equation}\label{Eq:2.1.GF}
  G(r) = \frac{1}{-M-\Sigma(r)+\epsilon}\lb-\frac{d}{dr}+\frac{\kappa}{r}\rb F(r)\,,
\end{equation}
and obtain
\begin{equation}\label{Eq:2.1.SchrF}
    \Lb -\frac{1}{M_-}\frac{d^2}{dr^2}
    +\frac{1}{M_-^2}\frac{dM_-}{dr}\frac{d}{dr}
    +\ls (-M+\Delta)+\frac{1}{M_-}\frac{\kappa(\kappa-1)}{r^2}
    -\frac{1}{M_-^2}\frac{dM_-}{dr}\frac{\kappa}{r}
    \rs \Rb F= \epsilon F\,,
\end{equation}
with the energy-dependent effective mass $M_-(r)=-M-\Sigma(r)+\epsilon$.
In Refs.~\cite{Zhang2010_IJMPE19-55,Zhang2009_ChinPC33S1-113,Zhang2009_CPL26-092401,Li2011_SciChinaPMA54-231}, it has been shown that each of these two Schr\"odinger-like equations, together with its charge conjugated one, are fully equivalent to Eq.~(\ref{Eq:2.1.DiraceqR}).

For Eq.~(\ref{Eq:2.1.SchrG}), in analogy with the Schr\"{o}dinger equations, $\Sigma(r)$ is the central potential in which particles move; the term proportional to $l(l+1)=\kappa(\kappa+1)$ corresponds to the centrifugal barrier (CB); and the last term corresponds to the SO potential, which leads to the substantial SO splittings in single-particle spectra.
Namely,
\begin{equation}\label{Eq:2.1.VCBandVSO}
  V_{\rm CB}(r)=\frac{1}{M_+(r)}\frac{\kappa(\kappa+1)}{r^2}
  \qquad\mbox{and}\qquad
  V_{\rm SO}(r)=\frac{1}{M_+^2(r)}\frac{dM_+(r)}{dr}\frac{\kappa}{r}\,.
\end{equation}
It is well known that there is no SO splitting if the $V_{\rm SO}$ vanishes.
In other words,
\begin{equation}\label{Eq:2.1.SSlimit}
  -\frac{dM_+(r)}{dr}=\frac{d\Delta(r)}{dr}=0
\end{equation}
is the SS limit.

If one uses the Schr\"odinger-like equation (\ref{Eq:2.1.SchrF}) for the lower component instead, although usually $\Delta(r)$ does not stand for the potential in which particles move, all terms except one, $-(1/M_-^2)(dM_-/dr)(\kappa/r)$, are identical for the pseudospin doublets $a$ and $b$ with $\kappa_a(\kappa_a-1)=\kappa_b(\kappa_b-1)$, i.e., $\kappa_a + \kappa_b = 1$.
As pointed out in Refs.~\cite{Meng1998_PRC58-R628,Sugawara-Tanabe1998_PRC58-R3065}, if this term vanishes, i.e.,
\begin{equation}\label{Eq:2.1.PSSlimit}
  -\frac{dM_-(r)}{dr}=\frac{d\Sigma(r)}{dr}=0\,,
\end{equation}
each pair of pseudospin doublets will be degenerate and the PSS will be exactly conserved.
This is called the PSS limit, which is more general and includes the limit $\Sigma(r)=0$ discussed in Ref.~\cite{Ginocchio1997_PRL78-436}.
From the physical point of view, $\Sigma(r)=0$ is never fulfilled in realistic nuclei as in which there exist no bound states for nucleons \cite{Leviatan2001_PLB518-214}, but $d\Sigma(r)/dr\sim0$ can be approximately satisfied in exotic nuclei with highly diffuse potentials \cite{Meng1999_PRC59-154}.

Analogically, such a term is regarded as the PSO potential, while the term proportional to $\tilde l(\tilde l+1)=\kappa(\kappa-1)$ is regarded as the PCB, i.e.,
\begin{equation}\label{Eq:2.1.VPSOandVPCB}
  V_{\rm PCB}(r) = \frac{1}{M_-(r)}\frac{\kappa(\kappa-1)}{r^2}
  \qquad\mbox{and}\qquad
  V_{\rm PSO}(r) = -\frac{1}{M_-^2(r)}\frac{dM_-(r)}{dr}\frac{\kappa}{r}\,.
\end{equation}

\subsection{Analytical solutions at PSS limit}\label{Sect:2.2}

Within the pseudospin symmetry limit shown in Eq.~(\ref{Eq:2.1.PSSlimit}), the potential $\Sigma(r)$ is simply a constant $\Sigma_0$, and then Eq.~(\ref{Eq:2.1.SchrF}) for the lower component of the Dirac spinor can be reduced to
\begin{equation}\label{Eq:2.2.SchrFPSS}
    \ls \frac{d^2}{dr^2} - \frac{\kappa(\kappa-1)}{r^2} + (\epsilon-M-\Sigma_0)(\epsilon+M-\Delta(r)) \rs F(r) = 0\,.
\end{equation}

During the past decade, it is a very active field to investigate the exact or approximate analytical solutions of this equation within certain special forms of potential $\Delta(r)$, by using the Nikiforov-Uvarov (NU) method \cite{Nikiforov1988}, the supersymmetric quantum mechanics \cite{Cooper1995_PR251-267}, the asymptotic iteration method \cite{Ciftci2003_JPA36-11807}, the exact quantization rule \cite{Ma2005_EPL69-685}, and so on.
For the spherical case, extensive investigations have been made for
the spherical harmonic oscillator \cite{Chen2003_CPL20-358,Lisboa2004_PRC69-024319,Ginocchio2005_PRL95-252501,deCastro2006_PRC73-054309,Akcay2009_PLA373-616,
Akcay2009_IJMPC20-930},
anharmonic oscillator \cite{Zhang2009_APS58-61},
Coulomb \cite{Hamzavi2010_PLA374-4303, Castro2012_PRA86-032122, Barakat2013_FP43-1171, Eshghi2013_CPB22-030303},
Deng-Fan \cite{Ortakaya2014_CPB23-030306},
diatomic molecular \cite{Jia2007_PS75-388,  Akcay2013_FBS54-1839},
Eckart \cite{Jia2006_JPA39-7737,Soylu2008_JPG41-065308},
Hellmann \cite{Hamzavi2013_CanJP91-411},
Hulth\'en \cite{Guo2003_CPL20-602,Ikhdair2011_AMC217-9019,Aydogdu2013_ChinPhysB22-010302, Hamzavi2013_CPB22-080302},
Manning-Rosen \cite{Wei2008_PLA373-49,Chen2009_PS79-055002,Wei2010_PLB686-288},
Mie-type \cite{Aydogdu2010_AoP325-373,Hamzavi2010_FBS48-171, Ikot2014_CPC38-013101},
Morse \cite{Berkdemir2006_NPA770-32,Bayrak2007_JPA40-11119,Qiang2007_JPG40-1677,Aydogdu2011_PLB703-379,Ikhdair2011_JMP52-052303, Ortakaya2013_AP338-250},
P\"oschl-Teller \cite{Jia2007_EPJA34-41,Jia2009_PLA373-1621,Wei2009_EPL87-40004,Wei2010_EPJA43-185,Candemir2012_IJMPE21-1250060,
Hamzavi2012_IJMPE21-1250097, Ikhdair2012_PS86-045002, Hamzavi2013_AdvHEP2013-196986,Falaye2013_ChinPhysB22-060305, Ikot2013_FBS54-2053},
Rosen-Morse \cite{Oyewumi2010_EPJA45-311,Wei2010_EPJA46-207,Chen2011_ChinPhysB20-062101, Aguda2013_CanJP91-689},
Tietz-Hua \cite{Ikhdair2012_FBS53-473},
Woods-Saxon \cite{Guo2005_PLA338-90,Aydogdu2010_EPJA43-73,Maghsoodi2012_PS85-055007},
and Yukawa \cite{Aydogdu2011_PS84-025005, Maghsoodi2012_PS86-015005, Ikhdair2013_PS87-035002, Ikot2013_FBS54-2027}
potentials, etc.
Note that some of these potentials are good approximations to model the atom-atom or nucleon-nucleon interactions, but not nuclear mean-field potentials.

In this Section, we will take the relativistic harmonic oscillator and relativistic Morse potentials as examples, and introduce the corresponding analytical solutions of equation (\ref{Eq:2.2.SchrFPSS}) at the pseudospin symmetry limit.
In the second part, the Nikiforov-Uvarov method \cite{Nikiforov1988} and the Pekeris approximation \cite{Pekeris1934_PR045-98} for the non-vanishing (pseudo-)centrifugal barrier will be discussed as well.

\subsubsection{Relativistic harmonic oscillator potential}

In analogy with the Schr\"odinger equations with the harmonic oscillator potentials, the Dirac equations with the RHO potentials have received extensive attention in different fields of mathematics, physics, and chemistry.
These equations have analytical solutions in many cases, for example, the corresponding equations at the PSS limit \cite{Chen2003_CPL20-358,Lisboa2004_PRC69-024319,Ginocchio2005_PRL95-252501,deCastro2006_PRC73-054309,Akcay2009_PLA373-616,
Akcay2009_IJMPC20-930}.

By taking one of the simplest cases as an example, i.e.,
\begin{equation}\label{Eq:2.2.HO1}
  \Delta(r) = \frac{1}{2}M\omega^2r^2 \qquad\mbox{and}\qquad \Sigma_0 = 0\,,
\end{equation}
the Schr\"odinger-like equation~(\ref{Eq:2.2.SchrFPSS}) for the lower component of the Dirac spinor at the PSS limit becomes \cite{Chen2003_CPL20-358,Lisboa2004_PRC69-024319}
\begin{equation}\label{Eq:2.2.HO0}
    \ls \frac{d^2}{dr^2} - \frac{\tilde l(\tilde l+1)}{r^2} + (\epsilon-M)(\epsilon+M-\frac{1}{2}M\omega^2r^2) \rs F(r) = 0\,,
\end{equation}
as $\kappa(\kappa-1) = \tilde l(\tilde l+1)$.
Note that some notations here are changed from the original papers for the self-consistency through the present Review, and similar changes have been done for the whole paper.

By further introducing \cite{Lisboa2004_PRC69-024319}
\begin{equation}\label{Eq:2.2.HO2}
  \tilde y = \sqrt{\frac{M(\epsilon-M)}{2}}\omega r^2 = \tilde{a}^2r^2
  \qquad\mbox{and}\qquad
  \tilde \lambda = -\frac{\epsilon^2-M^2}{\tilde{a}^2}\,,
\end{equation}
the above equation is rewritten as
\begin{equation}\label{Eq:2.2.HO3}
  \Lb 4\tilde y\frac{d^2}{d\tilde y^2} + 2\frac{d}{d\tilde y} - \frac{\tilde l(\tilde l+1)}{\tilde y} - \tilde y - \tilde\lambda\Rb
  F(\tilde y) = 0\,.
\end{equation}
An asymptotic analysis suggests searching for the solutions of the type of
\begin{equation}\label{Eq:2.2.HO4}
  F_\kappa(\tilde y) = Be^{-\tilde y/2} \tilde y^{(\tilde l+1)/2} w(\tilde y)\,,
\end{equation}
where $w(\tilde y)$ is a function to be determined and $B$ is a normalization factor.
Inserting this expression into Eq.~(\ref{Eq:2.2.HO3}), the equation for $w(\tilde y)$ reads
\begin{equation}\label{Eq:2.2.HO5}
  \ls \tilde y\frac{d^2}{d\tilde y^2} + \lb \tilde l + \frac{3}{2} - \tilde y\rb \frac{d}{d\tilde y} - \frac{1}{2}\lb\tilde l+\frac{3}{2}+\frac{\tilde\lambda}{2}\rb\rs w(\tilde y) = 0\,.
\end{equation}
The solutions of this equation, which guarantee that $\lim_{\tilde y\rightarrow \infty}F_\kappa(\tilde y)=0$, are the generalized Laguerre polynomials of degree $\tilde n$, $L^{\tilde p}_{\tilde n}(\tilde y)$, where
\begin{equation}\label{Eq:2.2.HO6}
  \tilde n = - \frac{1}{2}\lb\tilde l+\frac{3}{2}+\frac{\tilde\lambda}{2}\rb
  \qquad\mbox{and}\qquad
  \tilde p = \tilde l+\frac{1}{2}\,.
\end{equation}
Finally, one can get the eigenenergies \cite{Chen2003_CPL20-358,Lisboa2004_PRC69-024319}
\begin{equation}\label{Eq:2.2.HO7}
    (\epsilon_{\tilde n\kappa}+M)\sqrt{\frac{\epsilon_{\tilde n\kappa}-M}{2M}}
    = \omega\lb 2\tilde n+\tilde l+\frac{3}{2}\rb\,,
\end{equation}
which are discrete since $\tilde n$ is an integer equal to or greater than zero.
The corresponding eigenfunctions for the lower component of the Dirac spinor read
\begin{equation}\label{Eq:2.2.HO8}
  F_{\tilde n\kappa}(r) = B e^{-\tilde{a}^2 r^2/2} (\tilde a r)^{\tilde l+1} L^{\tilde l+1/2}_{\tilde n}(\tilde{a}^2 r^2)\,.
\end{equation}
Here $\tilde n$ is the number of the internal nodes of $F(r)$, denoted as $n_F$ in the following Sections.

The solutions of the Dirac equation with the RHO potential and their special features will be re-visited in Section~\ref{Sect:3.1} for the PSS in confining potentials, Section~\ref{Sect:3.3} for the PSS in tensor potentials, Section~\ref{Sect:3.5} for the SS in anti-nucleon spectra, Section~\ref{Sect:4.1} for the perturbative nature of PSS, and Section~\ref{Sect:4.2} for the puzzle of the intruder states.

\subsubsection{Relativistic Morse potential}\label{Sect:2.2.2}

Another widely discussed potential is the relativistic Morse potential \cite{Morse1929_PR034-57--64},
\begin{equation}\label{Eq:2.2.NUex1}
  \Delta(r) = D\ls e^{-2a(r-r_0)}-2e^{-a(r-r_0)}\rs\,,
\end{equation}
for atomic systems.
In this expression, $D>0$ is the dissociation energy, $r_0$ is the equilibrium internuclear distance, and $a>0$ is a parameter controlling the width of potential well.
In Refs.~\cite{Berkdemir2006_NPA770-32,Bayrak2007_JPA40-11119,Qiang2007_JPG40-1677,Aydogdu2011_PLB703-379}, the analytical solutions of the relativistic Morse potential at the PSS limit were investigated by using the NU method \cite{Nikiforov1988}, the asymptotic iteration method \cite{Ciftci2003_JPA36-11807}, the exact quantization rule \cite{Ma2005_EPL69-685}, and the confluent hypergeometric functions, respectively.

In the following, we will briefly introduce one of the widely used methods, the Nikiforov-Uvarov method \cite{Nikiforov1988}, and the Pekeris approximation \cite{Pekeris1934_PR045-98} for the non-vanishing (pseudo-)centrifugal barrier, then discuss the solutions of the relativistic Morse potential \cite{Berkdemir2006_NPA770-32,Bayrak2007_JPA40-11119,Qiang2007_JPG40-1677,Aydogdu2011_PLB703-379}.

The Nikiforov-Uvarov method \cite{Nikiforov1988} is based on solving the hypergeometric-type second-order differential equations by means of the special orthogonal functions.
Detailed derivations about the NU method and specific examples for standard Schr\"odinger equations with the harmonic oscillator, Coulomb, Kratzer, Morse, and Hulth\'en potentials can be found in Ref.~\cite{Berkdemir2012}.

The main equation which is closely associated with the NU method reads
\begin{equation}\label{Eq:2.2.NU1}
  \psi''(s) + \frac{\tilde\tau(s)}{\sigma(s)}\psi'(s) + \frac{\tilde\sigma(s)}{\sigma^2(s)}\psi(s)=0\,,
\end{equation}
where $\sigma(s)$ and $\tilde\sigma(s)$ are polynomials at most second-degree and $\tilde\tau(s)$ is a first-degree polynomial.
By letting $\psi(s)=\phi(s)y(s)$ and
\begin{equation}\label{Eq:2.2.NU2}
  \frac{\phi'(s)}{\phi(s)}=\frac{\pi(s)}{\sigma(s)}\,,
\end{equation}
Eq.~(\ref{Eq:2.2.NU1}) can be finally reduced into an equation of hypergeometric type,
\begin{equation}\label{Eq:2.2.NU3}
  \sigma(s)y''(s)+\tau(s)y'(s)+\lambda y(s)=0\,,
\end{equation}
with $\tau(s)=\tilde\tau(s)+2\pi(s)$.
Here both $\tau(s)$ and $\pi(s)$ are polynomials of degree at most one and $\lambda$ is a constant.

The polynomial $\pi(s)$ satisfies a quadratic equation,
\begin{equation}\label{Eq:2.2.NU4}
  \pi^2(s) + [\tilde\tau(s)-\sigma'(s)]\pi(s) + [\tilde\sigma(s)-k\sigma(s)]=0\,,
\end{equation}
with
\begin{equation}\label{Eq:2.2.NU5}
  k=\lambda-\pi'(s)\,.
\end{equation}
The corresponding solutions read
\begin{equation}\label{Eq:2.2.NU6}
  \pi(s)=\frac{\sigma'(s)-\tilde\tau(s)}{2}\pm
    \sqrt{\lb\frac{\sigma'(s)-\tilde\tau(s)}{2}\rb^2-\tilde\sigma(s)+k\sigma(s)}\,.
\end{equation}
Since $\pi(s)$ is a polynomial of degree at most one, the expression under the square root has to be the square of a polynomial.
In such a way, the constant $k$ can be determined.
In addition, the derivative of $\tau(s)$ thus obtained must be negative for bound states.
This is the main essential condition for any choice of particular solutions.

To generalize the solutions of Eq.~(\ref{Eq:2.2.NU3}), it is shown that the $n$th-order derivative of $y(s)$, $v_n(s)\equiv d^n y(s)/ds^n$, is also a hypergeometric-type function, which satisfies
\begin{equation}\label{Eq:2.2.NU7}
  \sigma(s)v''_n(s)+\tau_n(s)v'_n(s)+\mu_n v_n(s)=0\,,
\end{equation}
with
\begin{equation}\label{Eq:2.2.NU8}
  \tau_n(s) = \tau(s) + n\sigma'(s) \quad \mbox{and} \quad \mu_n = \lambda+n\tau'(s)+\frac{n(n-1)}{2}\sigma''(s)\,.
\end{equation}

When $\mu_n=0$, Eq.~(\ref{Eq:2.2.NU7}) has a particular solution of the form $y(s) = y_n(s)$, which is a polynomial of degree $n$,
and Eq.~(\ref{Eq:2.2.NU8}) becomes
\begin{equation}\label{Eq:2.2.NU9}
    \lambda_n = -n\tau'(s) - \frac{n(n-1)}{2}\sigma''(s) \quad \mbox{with} \quad n = 0,1,2,\ldots
\end{equation}
The whole set of eigenvalues for the second-order differential equation (\ref{Eq:2.2.NU1}) can be obtained by comparing $\lambda$ in Eq.~(\ref{Eq:2.2.NU5}) and $\lambda_n$ in Eq.~(\ref{Eq:2.2.NU9}), i.e., $\lambda=\lambda_n$.

Finally, for the corresponding eigenfunctions $\psi(s)=\phi(s)y_n(s)$, $\phi(s)$ is obtained by solving Eq.~(\ref{Eq:2.2.NU2}), and the polynomial solutions $y_n(s)$ are given by the Rodrigues relation,
\begin{equation}\label{Eq:2.2.NU10}
  y_n(s) = \frac{B_n}{\rho(s)}\frac{d^n}{ds^n}\ls\sigma^n(s)\rho(s)\rs\,,
\end{equation}
where $B_n$ is a normalization constant and the weight function $\rho(s)$ must satisfy the following condition:
\begin{equation}\label{Eq:2.2.NU11}
  [\sigma(s)\rho(s)]'=\tau(s)\rho(s)\,.
\end{equation}

For the relativistic Morse potential shown in Eq.~(\ref{Eq:2.2.NUex1}), by assuming new variables $\alpha=ar_0$ and $x=(r-r_0)/r_0$, Eq.~(\ref{Eq:2.2.SchrFPSS}) becomes
\begin{equation}\label{Eq:2.2.NUex2}
  \ls \frac{d^2}{dx^2} - \frac{\kappa(\kappa-1)}{(1+x)^2} + r_0^2(M+\Sigma_0-\epsilon)D(e^{-2\alpha x}-2e^{-\alpha x})
   + r_0^2(\epsilon-M-\Sigma_0)(\epsilon+M)\rs F(x) = 0\,.
\end{equation}

First of all, this equation can be solved exactly by the NU method if the pseudo-centrifugal barrier is absent, i.e., for the $p_{1/2}$ orbitals ($\kappa=1$).
In this case, by making a new change of independent variable $s=e^{-\alpha x}$, Eq.~(\ref{Eq:2.2.NUex2}) is rewritten as
\begin{equation}\label{Eq:2.2.NUex3}
  \frac{d^2F(s)}{ds^2} + \frac{1}{s}\frac{dF(s)}{ds}
    + \frac{1}{s^2}\ls \varepsilon_3 s^2 - \varepsilon_2 s + \varepsilon_1\rs F(s) = 0\,,
\end{equation}
with $\varepsilon_1=r_0^2(\epsilon-M-\Sigma_0)(\epsilon+M)/\alpha^2$, $\varepsilon_2 = 2r_0^2D(M+\Sigma_0-\epsilon)/\alpha^2$, and $\varepsilon_3 = r_0^2D(M+\Sigma_0-\epsilon)/\alpha^2$.
With reference to Eq.~(\ref{Eq:2.2.NU1}), it indicates
\begin{equation}\label{Eq:2.2.NUex4}
  \tilde\tau(s)=1\,,\quad \sigma(s)=s\,,\quad \tilde\sigma(s) = \varepsilon_3 s^2 - \varepsilon_2 s + \varepsilon_1\,.
\end{equation}
Then, one has $\pi(s)=\pm\sqrt{-\varepsilon_3s^2+(k+\varepsilon_2)s-\varepsilon_1}$, and $k_\pm=-\varepsilon_2\pm2\sqrt{\varepsilon_1\varepsilon_3}$ to make the expression under the square root be the square of a polynomial of the first degree.
In such a case, there are four possible forms of $\pi(s)$:
\begin{align}\label{Eq:2.2.NUex5}
  \pi(s) &= \pm i(\sqrt{\varepsilon_3}s+\sqrt{\varepsilon_1}) \quad \mbox{for}\quad k_-=-\varepsilon_2-2\sqrt{\varepsilon_1\varepsilon_3}\,,\nonumber\\
  \pi(s) &= \pm i(\sqrt{\varepsilon_3}s-\sqrt{\varepsilon_1}) \quad \mbox{for}\quad k_+=-\varepsilon_2+2\sqrt{\varepsilon_1\varepsilon_3}\,.
\end{align}
One of these four possible forms must be chosen to obtain the bound state solutions.
The most suitable form can be $\pi(s) = - i(\sqrt{\varepsilon_3}s+\sqrt{\varepsilon_1})$ for $k_-$, as the derivative of $\tau(s)$ thus obtained, $\tau'(s)=-2i\sqrt{\varepsilon_3}$, is negative \cite{Berkdemir2006_NPA770-32}.

Finally, a particular solution of Eq.~(\ref{Eq:2.2.NUex3}) which is a polynomial of degree $n$ can be calculated from Eq.~(\ref{Eq:2.2.NU9}),
\begin{equation}\label{Eq:2.2.NUex6}
  \lambda=\lambda_n \quad \Rightarrow \quad
  -\varepsilon_2 - 2\sqrt{\varepsilon_1\varepsilon_3}-i\sqrt{\varepsilon_3}=2ni\sqrt{\varepsilon_3}\,.
\end{equation}
Namely, the eigenvalues of Eq.~(\ref{Eq:2.2.NUex2}) for the $p_{1/2}$ orbitals can be obtained by
\begin{equation}\label{Eq:2.2.NUex7}
  \lb 2\sqrt{D(M+\Sigma_0-\epsilon_n)}+2\sqrt{(\epsilon_n+M)(\epsilon_n-M-\Sigma_0)}\rb^2+(1+2n)^2a^2=0\,.
\end{equation}

For the non-vanishing pseudo-centrifugal barrier, the Pekeris approximation \cite{Pekeris1934_PR045-98} is widely used to expand the PCB about $r=r_0$ in a series of powers of $x=(r-r_0)/r_0$,
\begin{equation}\label{Eq:2.2.NUex8}
  V_{\rm PCB}(r) = \frac{\kappa(\kappa-1)}{r^2} = \frac{\gamma}{(1+x)^2} = \gamma(1-2x+3x^2-\cdots)\,,
\end{equation}
with $\gamma=\kappa(\kappa-1)/r_0^2$.
Here one should approximate this potential by the exponential forms in the Morse potential, then the approximate PCB reads
\begin{align}\label{Eq:2.2.NUex9}
  \tilde V_{\rm PCB}(r) &= \gamma(D_0+D_1e^{-\alpha x}+D_2e^{-2\alpha x})\nonumber\\
  &= \gamma\ls D_0 + D_1\lb 1-\alpha x+\frac{\alpha^2 x^2}{2} - \cdots\rb
    + D_2\lb 1-2\alpha x+\frac{4\alpha^2 x^2}{2} - \cdots\rb\rs\,.
\end{align}
Comparing with these two expressions, the coefficients are
\begin{equation}\label{Eq:2.2.NUex10}
  D_0 = 1-\frac{3}{\alpha}+\frac{3}{\alpha^2}\,,\quad
  D_1 = \frac{4}{\alpha}-\frac{6}{\alpha^2}\,,\quad
  D_2 = -\frac{1}{\alpha}+\frac{3}{\alpha^2}\,.
\end{equation}
In such a way, the original PCB is approximated as
\begin{equation}\label{Eq:2.2.NUex11}
  \frac{1}{s^2}\ls-\frac{r_0^2\gamma}{\alpha^2}\lb D_0+D_1s+D_2s^2\rb\rs\,,
\end{equation}
acting on $F(s)$ in Eq.~(\ref{Eq:2.2.NUex3}).
All procedures for solving Eq.~(\ref{Eq:2.2.NUex3}) remain, but simply substituting $\varepsilon_1 \rightarrow \varepsilon_1 - r_0^2\gamma D_0/\alpha^2$, $\varepsilon_2 \rightarrow \varepsilon_2 + r_0^2\gamma D_1/\alpha^2$, and $\varepsilon_3 \rightarrow \varepsilon_3 - r_0^2\gamma D_2/\alpha^2$.
The final solutions of eigenvalues can be obtained by
\begin{equation}\label{Eq:2.2.NUex12}
  \lb\frac{-2DM_-+\gamma D_1}{\sqrt{-D M_- -\gamma D_2}}+2\sqrt{(\epsilon_{n\kappa}+M) M_- -\gamma D_0}\rb^2+(1+2n)^2a^2=0\,,
\end{equation}
and here $M_- = -M-\Sigma_0+\epsilon_{n\kappa}$.
The readers are referred to Ref.~\cite{Aydogdu2011_PLB703-379} for the corresponding eigenfunctions, which can be expressed in terms of confluent hypergeometric function $_1F_1$.

The same results were obtained in Refs.~\cite{Bayrak2007_JPA40-11119} and \cite{Qiang2007_JPG40-1677} by using the asymptotic iteration method \cite{Ciftci2003_JPA36-11807} and the exact quantization rule \cite{Ma2005_EPL69-685} together with the Pekeris approximation, respectively.
Note that the validity of the Pekeris approximation deserves more careful examinations.

As a common example shown in Refs.~\cite{Berkdemir2006_NPA770-32,Bayrak2007_JPA40-11119,Qiang2007_JPG40-1677,Aydogdu2011_PLB703-379},
the parameters of the relativistic Morse potential in Eq.~(\ref{Eq:2.2.NUex1}) are taken as $D=5.0$~fm$^{-1}$, $r_0=2.40873$~fm, and $a=0.988879$~fm$^{-1}$ with the mass $M=10.0$~fm$^{-1}$.
The corresponding coefficients in Eq.~(\ref{Eq:2.2.NUex10}) deduced with the Pekeris approximation read $D_0=0.26928$, $D_1=0.62178$, and $D_2=0.10893$.
For different choices of the constant $\Sigma_0$, one can obtain different numerical solutions for the eigenvalues of Eq.~(\ref{Eq:2.2.NUex12}).

\begin{table}
\begin{center}
\caption{The bound-state eigenenergies $\epsilon_{n\kappa}$ of the Dirac particle in the pseudospin-symmetry Morse potential with the Pekeris approximation and $\Sigma_0=0$.
Energy units are in fm$^{-1}$.
The data are taken from Ref.~\cite{Bayrak2007_JPA40-11119}.
\label{Tab:2.2.Morse}}
\begin{tabular}{@{}ccccc@{}} \hline
$n$ & ($l,j$) & $\epsilon_{n\kappa}$ &  ($l,j$) & $\epsilon_{n\kappa}$ \\ \hline
1 & $s_{1/2}$ & $9.993 5101$ & $d_{3/2}$ & $9.993 5101$ \\
1 & $p_{3/2}$ & $9.983 8165$ & $f_{5/2}$ & $9.983 8165$ \\
1 & $d_{5/2}$ & $9.973 7712$ & $g_{7/2}$ & $9.973 7712$ \\
1 & $f_{7/2}$ & $9.965 6754$ & $h_{9/2}$ & $9.965 6754$ \\
2 & $s_{1/2}$ & $9.992 9544$ & $d_{3/2}$ & $9.992 9544$ \\
2 & $p_{3/2}$ & $9.980 7043$ & $f_{5/2}$ & $9.980 7043$ \\
2 & $d_{5/2}$ & $9.965 2868$ & $g_{7/2}$ & $9.965 2868$ \\
2 & $f_{7/2}$ & $9.948 4873$ & $h_{9/2}$ & $9.948 4873$ \\ \hline
\end{tabular}
\end{center}
\end{table}

\begin{figure}[tbhp]
\begin{center}
  \includegraphics[width=6cm]{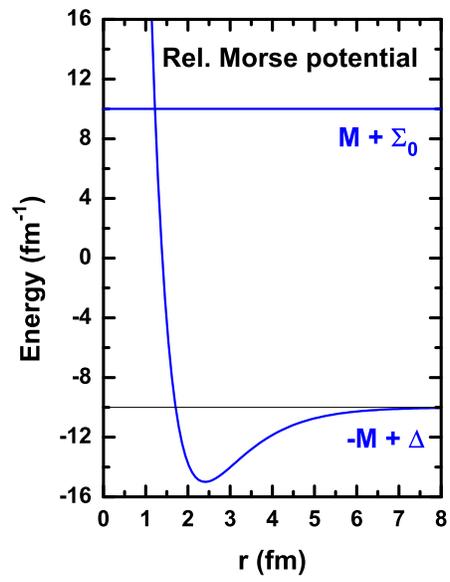}
\end{center}
\caption{(Color online) Relativistic Morse potential with $D=5.0$~fm$^{-1}$, $r_0=2.40873$~fm, $a=0.988879$~fm$^{-1}$, and $M=10.0$~fm$^{-1}$ corresponding to the PSS limit $\Sigma(r)=\Sigma_0=0$.
\label{Fig:2.2.Morse}}
\end{figure}

It is instructive to choose $\Sigma_0=0$ for the comparison with the results shown in the coming Sections.
In Table~\ref{Tab:2.2.Morse}, the bound-state eigenenergies $\epsilon_{n\kappa}$ of the Dirac particle in the PSS Morse potential are shown for several $n$ and $\kappa$ states.
It is confirmed that the single-particle energies of pseudospin doublets are exactly degenerate at the PSS limit.
However, it is found that the single-particle spectra are bound from the top, and for each given $\kappa$ the single-particle energies decrease when the radial quantum number $n$ increases.
This indicates these bound states are indeed of the characteristics of the states belonging to the Dirac sea, but not those in the Fermi sea, even though the eigenvalues are positive and close to the threshold of the continuum states in the Fermi sea.
It can be seen in a clearer way by showing the corresponding single-particle potentials explicitly in Fig.~\ref{Fig:2.2.Morse}.
Due to the particular shape of the Morse potential, the potential $\Delta(r)-M$ extends to the positive-energy region and forms a pocket for the bound states, which belong to the Dirac sea.

Furthermore, when $\Sigma_0$ is chosen as $-M$ instead of $0$, the general pattern of the single-particle spectra does not change \cite{Berkdemir2006_NPA770-32,Bayrak2007_JPA40-11119,Qiang2007_JPG40-1677,Aydogdu2011_PLB703-379}, since it is a trivial modification by a constant as shown in Eqs.~(\ref{Eq:2.1.modifySV}) and (\ref{Eq:2.1.modifyME}).
In this case, it is clear that all solutions of Eq.~(\ref{Eq:2.2.NUex12}) are of negative energy and there are no bound states with positive energy.

Therefore, more precisely, the phenomenon discussed in Section~\ref{Sect:2.2.2} corresponds to the spin symmetry of the anti-particle spectra, which will be discussed in more detail in Section~\ref{Sect:3.5}.

\subsection{PSS breaking in realistic nuclei}\label{Sect:2.3}

For the isolated atomic nuclei, both the scalar $S(r)$  and vector $V(r)$ potentials in the single-particle Dirac Hamiltonian $H$ in Eq.~(\ref{Eq:2.1.HDirac}) vanish at large distance from the center, i.e., $\lim_{r\rightarrow\infty}\Sigma(r) = \lim_{r\rightarrow\infty}\Delta(r) = 0$.
This kind of potentials will be specified as the non-confining potentials in Section~\ref{Sect:3.1}.
Within these potentials, it can be proven that there are no single-particle bound states in the Fermi sea at the pseudospin symmetry limit shown in Eq.~(\ref{Eq:2.1.PSSlimit}) \cite{Leviatan2001_PLB518-214}.
In other words, the pseudospin symmetry must be broken in realistic nuclei.

Therefore, much more meaningful and important tasks are to investigate to which extent the pseudospin symmetry is approximately conserved in realistic nuclei and what the symmetry-breaking mechanism is.
This approximate pseudospin symmetry can be examined by the quasi-degenerate single-particle energies, as well as by the relation between the pseudospin-orbit potential and pseudo-centrifugal barrier pointed out by Meng \textit{et al.} \cite{Meng1998_PRC58-R628,Meng1999_PRC59-154} and the relations between the single-particle wave functions pointed out by Ginocchio and Madland \cite{Ginocchio1998_PRC57-1167,Ginocchio2002_PRC66-064312}.

As the pseudospin symmetry is shown to be a relativistic symmetry \cite{Ginocchio1997_PRL78-436}, let us start with the introduction of the covariant density functional theory \cite{Ring1996_PPNP37-193,Vretenar2005_PR409-101,Meng2006_PPNP57-470,Niksic2011_PPNP66-519,Meng2011_PP31-199,Meng2013_FPC8-55}, which is one of the most appropriate microscopic and self-consistent approaches for studying the properties of pseudospin symmetry in realistic nuclei.

\subsubsection{Covariant density functional theory}

The CDFT \cite{Ring1996_PPNP37-193,Vretenar2005_PR409-101,Meng2006_PPNP57-470,Niksic2011_PPNP66-519,Meng2011_PP31-199,Meng2013_FPC8-55} can be traced back to the successful RMF models introduced by Walecka and Serot \cite{Serot1986_ANP16-1}.
The most popular RMF models are based on the finite-range meson-exchange representation, in which the nucleus is described as a system of Dirac nucleons that interact with each other via the exchange of mesons and photons.
The nucleons and mesons are described by the Dirac and Klein-Gordon equations, respectively.
Together with the electromagnetic field, the isoscalar-scalar $\sigma$ meson, the isoscalar-vector $\omega$ meson, and the isovector-vector $\rho$ meson build the minimal set of meson fields that is necessary for a description of bulk and single-particle nuclear properties.
Moreover, a quantitative treatment of nuclear matter and finite nuclei needs a medium dependence of effective mean-field interactions, which can be introduced either by including nonlinear meson self-interaction terms in the Lagrangian \cite{Boguta1977_NPA292-413, Sharma1993_PLB312-377,Sugahara1994_NPA579-557,Lalazissis1997_PRC55-540,Long2004_PRC69-034319} or by assuming
explicit density dependence for the meson-nucleon couplings \cite{Typel1999_NPA656-331,Niksic2002_PRC66-024306,Long2004_PRC69-034319,Lalazissis2005_PRC71-024312}.

The Lagrangian density of the RMF theory with nonlinear meson self-interactions \cite{Sharma1993_PLB312-377,Sugahara1994_NPA579-557,Lalazissis1997_PRC55-540,Long2004_PRC69-034319} can be written by using the conventions in Ref.~\cite{Meng2006_PPNP57-470} as
\begin{align}\label{Eq:2.3.Lagrangian}
    \mathscr{L}=&\bar\psi\left[i\gamma^\mu\partial_\mu-M-g_\sigma \sigma-g_\omega \gamma^\mu \omega_\mu
        -g_\rho \gamma^\mu \vec\tau \cdot \vec \rho_\mu
        -e\gamma^\mu\frac{1-\tau_3}{2}A_\mu\right]\psi\nonumber\\
    &+\frac{1}{2}\partial^\mu\sigma\partial_\mu\sigma-U_\sigma(\sigma)
        -\frac{1}{4}\Omega^{\mu\nu}\Omega_{\mu\nu}+U_\omega(\omega_\mu)
        -\frac{1}{4}\vec{R}^{\mu\nu}\cdot\vec{R}_{\mu\nu}+U_\rho(\vec \rho_\mu)-\frac{1}{4}F^{\mu\nu}F_{\mu\nu}\,,
\end{align}
where $M$ and $m_i$ ($g_i$) ($i = \sigma,\omega,\rho$) are the masses (coupling constants) of the nucleon and mesons, respectively, and
\begin{equation}\label{Eq:2.3.fieldT}
    \Omega^{\mu\nu} = \partial^\mu\omega^\nu-\partial^\nu\omega^\mu\,,\quad
    \vec{R}^{\mu\nu}= \partial^\mu\vec\rho^\nu-\partial^\nu\vec\rho^\mu\,,\quad
    F^{\mu\nu}= \partial^\mu A^\nu-\partial^\nu A^\mu\,,
\end{equation}
are the field tensors of the vector mesons and electromagnetic field.
We adopt the arrows to indicate vectors in isospin space and bold type for the space vectors.
Greek indices $\mu$ and $\nu$ run over $0, 1, 2, 3$ or $t, x, y, z$, while Roman indices $i,j$, etc. denote the spatial components.
The nonlinear self-coupling terms $U_\sigma(\sigma)$, $U_\omega(\omega_\mu)$, and $U_\rho(\vec \rho_\mu)$ for the $\sigma$, $\omega$, and $\rho$ mesons in the Lagrangian density (\ref{Eq:2.3.Lagrangian}) respectively have the following forms:
\begin{align}\label{Eq:2.3.nonlinear}
    U_\sigma(\sigma) &= \frac{1}{2}m^2_\sigma\sigma^2+\frac{1}{3}g_2\sigma^3+\frac{1}{4}g_3\sigma^4\,,\nonumber\\
    U_\omega(\omega_\mu) &= \frac{1}{2}m^2_\omega\omega^\mu\omega_\mu+\frac{1}{4}c_3(\omega^\mu\omega_\mu)^2\,,\nonumber\\
    U_\rho(\vec \rho_\mu) &= \frac{1}{2}m^2_\rho\vec\rho^\mu\cdot\vec\rho_\mu+\frac{1}{4}d_3(\vec\rho^\mu\cdot\vec\rho_\mu)^2\,.
\end{align}

The system Hamiltonian density can be obtained via the general Legendre transformation,
\begin{equation}\label{Eq:2.3.H_den}
    \mathscr{H}=\frac{\partial \mathscr{L}}{\partial \dot\phi_i}\dot\phi_i-\mathscr{L}\,,
\end{equation}
where $\phi_i$ represent the nucleon-, meson-, and photon-field operators.
The ground-state trial wave function is taken as a Slater determinant,
\begin{equation}\label{Eq:2.3.GS}
  \lr\Phi_0\rc = \prod_{\alpha=1}^A c_i^\dag\lr-\rc\,,
\end{equation}
where $\lr-\rc$ is the physical vacuum and the single-particle states $\alpha$ are confined to those with positive energies in the Fermi sea, i.e., the no-sea approximation.
Combining these two expressions together, one has the energy density functional for the whole system,
\begin{equation}\label{Eq:2.3.EDF}
  E_{\rm RMF} = \lc\Phi_0\rl\mathscr{H}\lr\Phi_0\rc\,.
\end{equation}
In the RMF theory, only the direct contributions of the meson and Coulomb fields, i.e., the so-called Hartree terms, are taken into account.

For the systems with time-reversal symmetry, the space-like components of the vector fields vanish.
Furthermore, one can assume that the nucleon single-particle states do not mix isospin, i.e., the single-particle states are eigenstates of $\tau_3$, therefore only the third component of $\vec\rho_\mu$ survives.

By the variation principle, the Dirac equation for nucleons reads [cf. Eqs.~(\ref{Eq:2.1.eigeneq}) and (\ref{Eq:2.1.HDirac})]
\begin{equation}\label{Eq:2.3.Dirac}
  \ls\boldsymbol{\alpha}\cdot\mathbf{p} + \beta(M+S(\mathbf{r}))+V(\mathbf{r})\rs\psi_\alpha(\mathbf{r})=\epsilon_\alpha\psi_\alpha(\mathbf{r})\,,
\end{equation}
and the Klein-Gordon equations for mesons and photons read
\begin{align}\label{Eq:2.3.Klein-Gordon}
  -\mathbf{\nabla}^2\sigma+U'_\sigma(\sigma) &= -g_\sigma\rho_S\,,\nonumber\\
  -\mathbf{\nabla}^2\omega_0+U'_\omega(\omega_0) &= g_\omega\rho_V\,,\nonumber\\
  -\mathbf{\nabla}^2\rho^3_0+U'_\rho(\rho^3_0) &= g_\rho\rho^{(3)}_V\,,\nonumber\\
  -\mathbf{\nabla}^2A_0 &= e\rho_C\,.
\end{align}
The scalar and vector potentials in Eq.~(\ref{Eq:2.3.Dirac}) are, respectively,
\begin{align}\label{Eq:2.3.SandV}
  S(\mathbf{r}) &= g_\sigma\sigma(\mathbf{r})\,,\nonumber\\
  V(\mathbf{r}) &= g_\omega\omega_0(\mathbf{r})+g_\rho\tau_3\rho^3_0(\mathbf{r})+\frac{1-\tau_3}{2}eA_0(\mathbf{r})\,.
\end{align}
The scalar density $\rho_S$, the baryonic density $\rho_V$, the isovector density $\rho^{(3)}_V$, and the charge density $\rho_C$ in the Klein-Gordon equations (\ref{Eq:2.3.Klein-Gordon}) are, respectively,
\begin{align}\label{Eq:2.3.densities}
  \rho_S(\mathbf{r}) &= \sum_{\alpha=1}^A \bar\psi_\alpha(\mathbf{r})\psi_\alpha(\mathbf{r})\,,\nonumber\\
  \rho_V(\mathbf{r}) &= \sum_{\alpha=1}^A \psi^\dag_\alpha(\mathbf{r})\psi_\alpha(\mathbf{r})\,,\nonumber\\
  \rho^{(3)}_V(\mathbf{r}) &= \sum_{\alpha=1}^A \psi^\dag_\alpha(\mathbf{r})\tau_3\psi_\alpha(\mathbf{r})\,,\nonumber\\
  \rho_C(\mathbf{r}) &= \sum_{p=1}^Z \psi^\dag_p(\mathbf{r})\psi_p(\mathbf{r})\,.
\end{align}

In the density-dependent RMF approach \cite{Typel1999_NPA656-331,Niksic2002_PRC66-024306,Long2004_PRC69-034319,Lalazissis2005_PRC71-024312}, the nonlinear meson self-couplings in the Lagrangian density are replaced by the density
dependence of the coupling strengths $g_\sigma(\rho)$, $g_\omega(\rho)$, and $g_\rho(\rho)$, and an additional term, i.e., the rearrangement term, will appear in the Dirac equation (\ref{Eq:2.3.Dirac}).

More recently, this framework has been re-interpreted by the relativistic Kohn-Sham density functional theory, and the functionals have been developed based on the zero-range point-coupling interaction \cite{Nikolaus1992_PRC46-1757,Burvenich2002_PRC65-044308,Niksic2008_PRC78-034318,Zhao2010_PRC82-054319,Zhao2011_PRL107-122501}, in which the meson exchange in each channel (isoscalar-scalar, isoscalar-vector, isovector-scalar, and isovector-vector) is replaced by the corresponding local four-point contact interaction between nucleons.
The point-coupling model has attracted more and more attention owing to the following advantages.
First, it avoids the possible physical constrains introduced by explicit usage of the Klein-Gordon equation to describe mean meson fields, especially the fictitious $\sigma$ meson.
Second, it is possible to study the role of naturalness \cite{Friar1996_PRC53-3085,Manohar1984_NPB234-189} in effective theories for nuclear-structure-related problems.
Third, it is relatively easy to include the Fock terms \cite{Liang2012_PRC86-021302}, and provides more opportunities to investigate its relationship to the non-relativistic approaches \cite{Sulaksono2003_AP308-354}.

In order to describe open-shell nuclei, the pairing correlation and the coupling to continuum must be taken into account properly, which are in particular crucial for the descriptions of drip line nuclei.
The extension of the RMF theory to take into account both bound states and (discretized) continuum via Bogoliubov transformation in a microscopic and self-consistent way has been done in Refs.~\cite{Meng1996_PRL77-3963,Meng1998_NPA635-3}, the so-called relativistic continuum Hartree-Bogoliubov (RCHB) theory.

\subsubsection{PSS in single-particle energies}

\begin{figure}[tbhp]
\begin{center}
  \includegraphics[width=6cm]{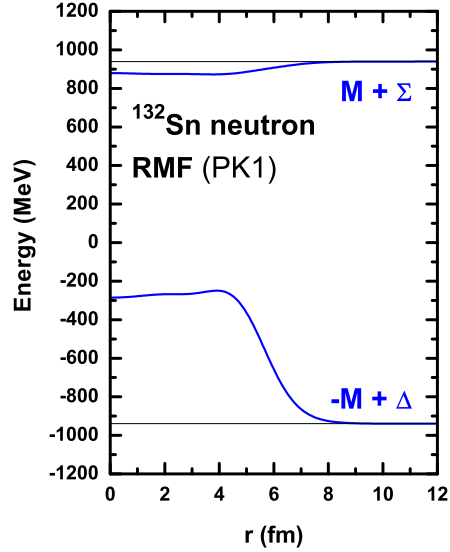}
\end{center}
\caption{(Color online) Self-consistent single-particle potentials for neutrons in $^{132}$Sn calculated by the RMF theory with the effective interaction PK1 \cite{Long2004_PRC69-034319}.
\label{Fig:2.3.132potl}}
\end{figure}

First of all, let us start with the neutron-rich doubly magic nucleus $^{132}$Sn as an example, which shows that the self-consistent CDFT or RMF theory can nicely reproduce its ground-state properties including the single-particle spectra \cite{Liang2011_PRC83-011302R}.

In Fig.~\ref{Fig:2.3.132potl}, the potentials $\Sigma(r)$ and $\Delta(r)$ for neutrons calculated by the RMF theory with the effective interaction PK1 \cite{Long2004_PRC69-034319} are shown.
The depths of potentials are $\Sigma(r)\sim70$~MeV and $\Delta(r)\sim700$~MeV, respectively.

\begin{figure}[tbhp]
\begin{center}
  \includegraphics[width=8cm]{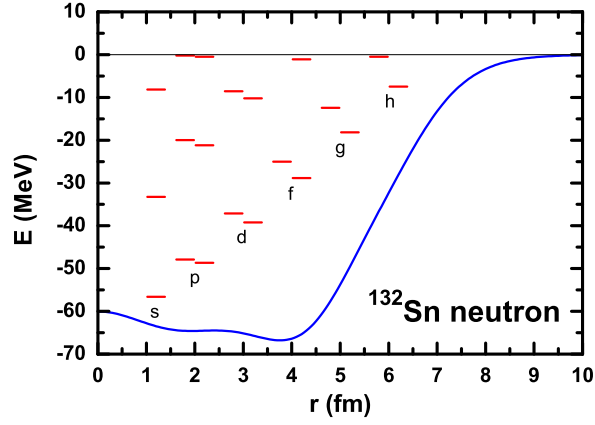}
\end{center}
\caption{(Color online) Neutron single-particle spectrum in $^{132}$Sn calculated by the RMF theory with PK1.
For each pair of spin doublets, the left levels are those with $j_<=l-1/2$ and the right ones with $j_>=l+1/2$.
Potential $\Sigma(r)$ is shown as the solid line.
\label{Fig:2.3.132spectra}}
\end{figure}

\begin{figure}[tbhp]
\begin{center}
  \includegraphics[width=6cm]{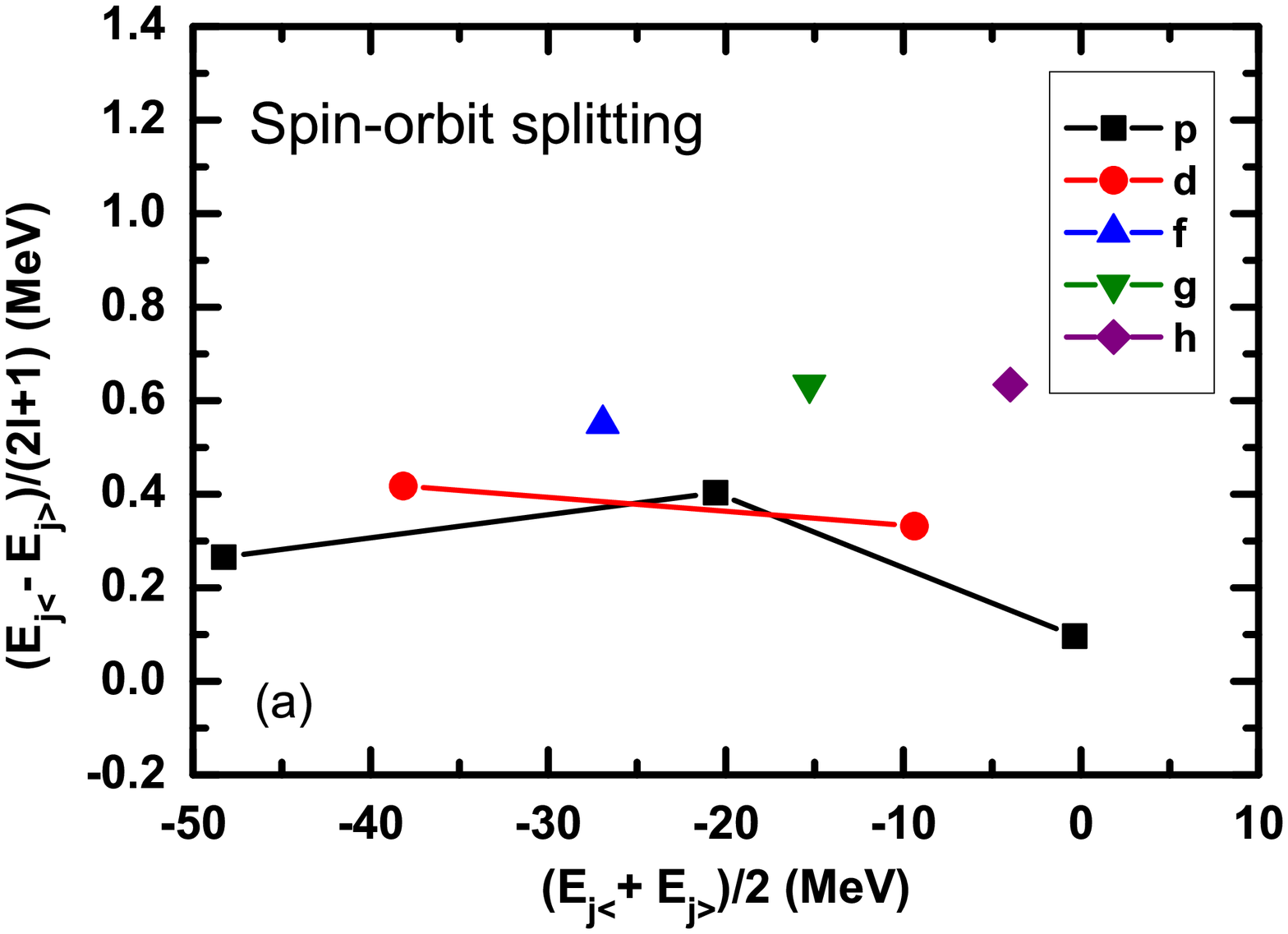}
  \includegraphics[width=6cm]{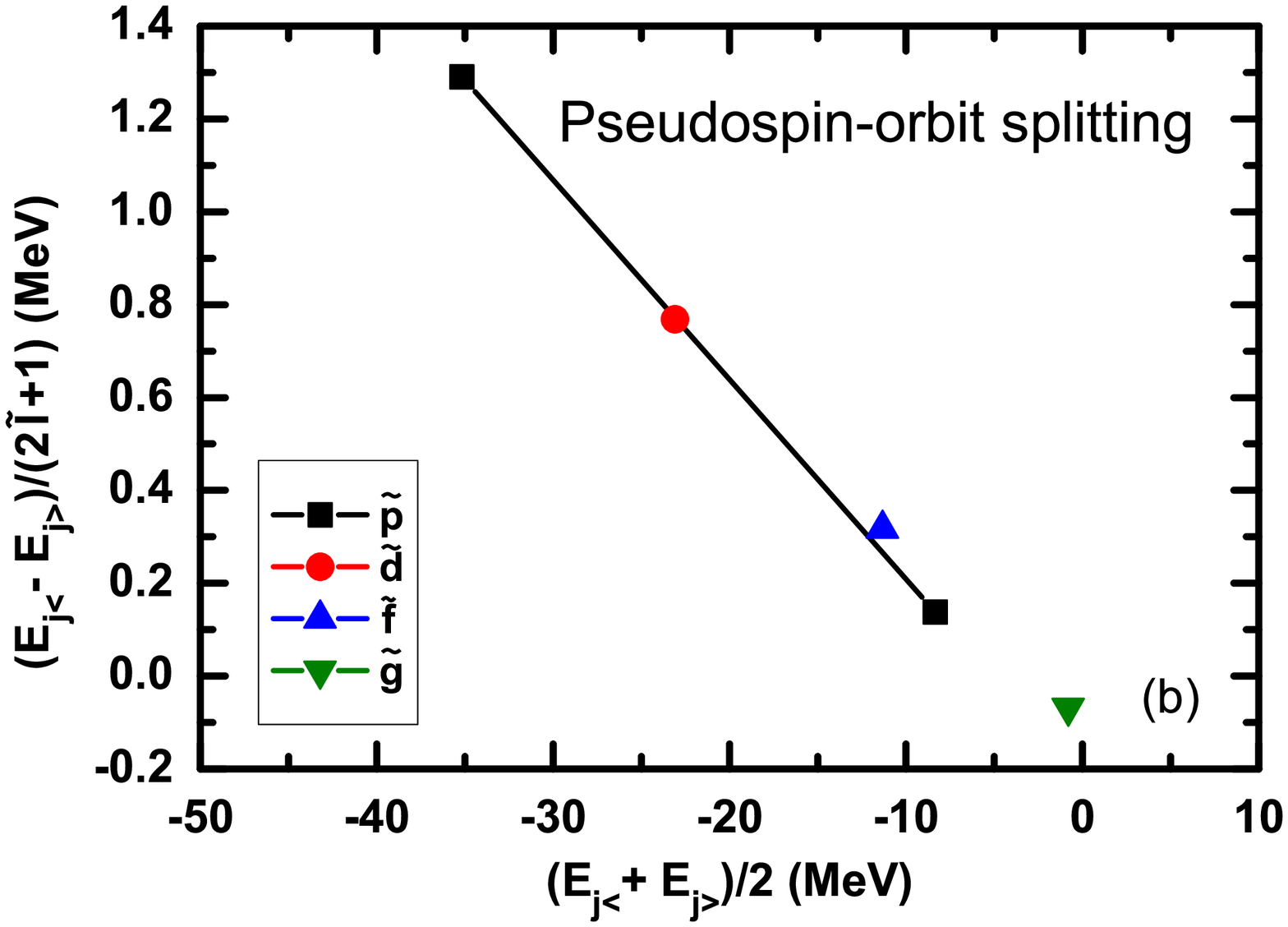}
\end{center}
\caption{(Color online) Reduced SO splittings $(E_{j_<} - E_{j_>})/(2l+1)$ (the left panel) and reduced PSO splittings $(E_{j_<} - E_{j_>})/(2\tilde l+1)$ (the right panel) versus their average single-particle energies $E_{\rm av}=(E_{j_<} + E_{j_>})/2$ in neutron single-particle spectrum of $^{132}$Sn calculated by the RMF theory with PK1.
\label{Fig:2.3.132split}}
\end{figure}

The single-particle energies of the neutron bound states thus obtained are shown in Fig.~\ref{Fig:2.3.132spectra}, where $E=\epsilon-M$ excluding the rest mass of nucleon.
In order to show the SO and PSO splittings and to see their energy dependence more clearly, the reduced SO splittings $\Delta E_{\rm SO}=(E_{j_<} - E_{j_>})/(2l+1)$ and the reduced PSO splittings $\Delta E_{\rm PSO}=(E_{j_<} - E_{j_>})/(2\tilde l+1)$ versus their average single-particle energies $E_{\rm av}=(E_{j_<} + E_{j_>})/2$ are plotted in the left and right panels of Fig.~\ref{Fig:2.3.132split}, respectively.
In this paper, $j_<$ ($j_>$) denotes the states with $j=l-1/2$ ($j=l+1/2$) for the spin doublets and the states with $j=\tilde l-1/2$ ($j=\tilde l+1/2$) for the pseudospin doublets.

A dramatic energy dependence can be seen in the reduced PSO splittings, whereas the reduced SO splittings are less energy dependent.
While the reduced PSO splitting for the $1\tilde p$ pseudospin doublets is $1.291$~MeV, that for the $2\tilde p$ doublets is $0.138$~MeV, roughly smaller than the former one by a factor of 10.
Thus, the PSS becomes better near the Fermi surface.
This is in agreement with the experimental observation.

Around the Fermi surface, $E\approx-8$~MeV in this case, $\Delta E_{\rm PSO}=0.138$~MeV for the $2\tilde p$ pseudospin doublets, compared to $\Delta E_{\rm SO}=0.332$~MeV for the $2d$ spin doublets.
Note that these two cases in the comparison share a common state $2d_{3/2}$.
Further approaching the single-particle threshold, on one hand, the reduced SO splitting for the $1h$ doublets is almost the same as those for the $1g$ and $1f$ doublets below the Fermi surface; on the other hand, the PSO splittings become smaller and even reversed, e.g., $\Delta E_{\rm PSO}=-0.068$~MeV for the $1\tilde g$ doublets.

To understand why the energy splitting between the pseudospin partners changes with the single-particle energies, the PSO potential $V_{\rm PSO}$ and the pseudo-centrifugal barrier $V_{\rm PCB}$ in Eq.~(\ref{Eq:2.1.VPSOandVPCB}) should be examined carefully \cite{Meng1998_PRC58-R628,Meng1999_PRC59-154}.
Their contribution to the single-particle energy $E$ can be evaluated by the integrals with the lower component of the Dirac spinor, i.e.,
\begin{equation}\label{Eq:2.3.Econtribution}
  E_i = \frac{\int F^*(r)V_i(r)F(r)dr}{\int F^*(r)F(r)dr}\,.
\end{equation}
It was pointed out in Refs.~\cite{Meng1998_PRC58-R628,Meng1999_PRC59-154} that
\begin{equation}\label{Eq:2.3.PSOvsPCB}
  \left|V_{\rm PSO}(r)\right| \ll \left|V_{\rm PCB}(r)\right|
\end{equation}
is the condition under which the PSS is conserved approximately.
Unfortunately, it is difficult to plot and compare these potentials, as both of them have a singularity at $r_0$ where $M_-(r)=0$, i.e., $E=\Sigma(r)|_{r=r_0}$.
As one is only interested in the relative magnitude of the PCB and the PSO potential, the effective PCB,
\begin{equation}\label{Eq:2.3.effPCB}
  V_{\rm PCB}^{\rm eff}(r) = M_-\kappa(\kappa-1)/r^2\,,
\end{equation}
and the effective PSO potential,
\begin{equation}\label{Eq:2.3.effPSO}
  V_{\rm PSO}^{\rm eff}(r) = (-dM_-/dr)(\kappa/r)\,,
\end{equation}
are introduced for comparison \cite{Meng1998_PRC58-R628, Meng1999_PRC59-154}.
They correspond respectively to the PCB and the PSO potential multiplied by a common factor $M_-^2$.

\begin{figure}[tbhp]
\begin{center}
  \includegraphics[width=6cm]{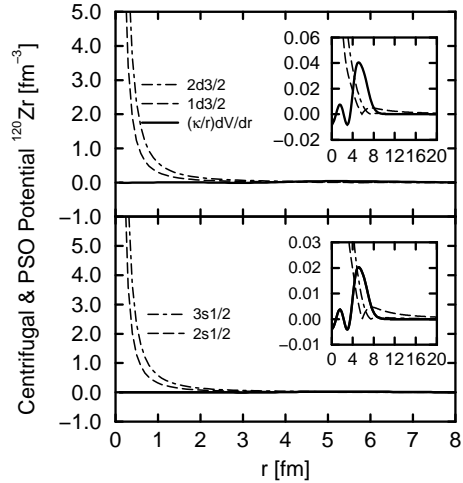}
\end{center}
\caption{Comparison of the effective PCB (dashed and dot-dashed lines) and the effective PSO potential (solid line) in arbitrary scale for $d_{3/2}$ (the upper panel) and $s_{1/2}$ (the lower panel) states in $^{120}$Zr.
The dashed lines are for $1d_{3/2}$ and $2s_{1/2}$ states, and the dot-dashed lines are for $2d_{3/2}$ and $3s_{1/2}$ states.
The inserted boxes show the behaviors of the effective PCB and the effective PSO potential near the nuclear surface.
Taken from Ref.~\cite{Meng1999_PRC59-154}.
\label{Fig:2.3.VPCBandVPSO}}
\end{figure}

The effective PSO potential in Eq.~(\ref{Eq:2.3.effPSO}) depends on the angular momentum and parity, but does not depend on the single-particle energy.
On the other hand, the effective PCB in Eq.~(\ref{Eq:2.3.effPCB}) depends on the energy.
They are given in Fig.~\ref{Fig:2.3.VPCBandVPSO} for the $s_{1/2}$ and $d_{3/2}$ orbitals of $^{120}$Zr in arbitrary scale, and their behavior near the nuclear surface are shown in the inserts.
Note that the solid lines in the figure are the effective PSO potentials, which are enlarged in the inserts.
It is found that the PSS is conserved much better for the less bound pseudospin partners, because the effective PCB is smaller for the more deeply bound states.

\begin{figure}[tbhp]
\begin{center}
  \includegraphics[angle=270,width=8cm]{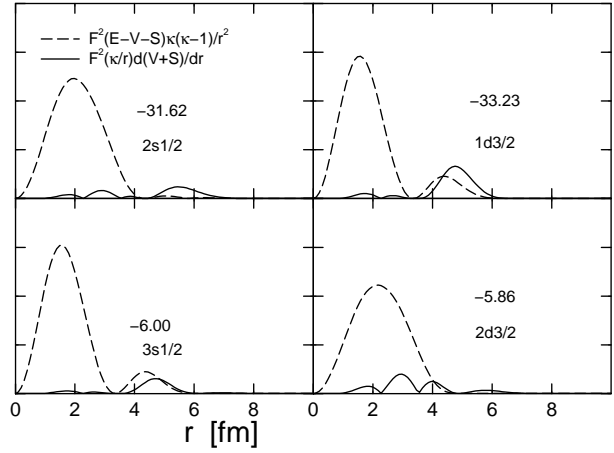}
\end{center}
\caption{Comparison of the effective PCB (dashed lines) and the effective PSO potentials (solid lines) multiplied by $F^2(r)$ in arbitrary scale for $s_{1/2}$ (left panels) and $d_{3/2}$ (right panels) states in $^{120}$Zr.
Taken from Ref.~\cite{Meng1998_PRC58-R628}.
\label{Fig:2.3.VFF}}
\end{figure}

To see clearly their contributions to the single-particle energy, the effective PCB and the effective PSO potential multiplied by the squares of the lower component wave function $F(r)$ are given in Fig.~\ref{Fig:2.3.VFF} for the $2s_{1/2}$, $3s_{1/2}$, $1d_{3/2}$, and $2d_{3/2}$ states of $^{120}$Zr in arbitrary scale.
It is clear that the contributions of the effective PCB are much bigger than those of the effective PSO potential, and generally they differ by two orders of magnitude.
In a semi-quantitative sense, this indicates the condition in Eq.~(\ref{Eq:2.3.PSOvsPCB}) is satisfied.

However, it was also pointed out in Ref.~\cite{Marcos2000_PRC62-054309} that, from a more quantitative point of view, one should directly compare the PSO potential and PCB, instead of using the effective ones, because the common factor $M_-^2$ multiplied depends on $r$.
As a result, the magnitude of the PSO potential is drastically modified around $r_0$, and the inequality $\left|M^2_-(r)V_{\rm PSO}(r)\right| \ll \left|M^2_-(r)V_{\rm PCB}(r)\right|$ differ from $\left|V_{\rm PSO}(r)\right| \ll \left|V_{\rm PCB}(r)\right|$ shown in Eq.~(\ref{Eq:2.3.PSOvsPCB}).
Furthermore, although $V_{\rm PSO}(r)$ and $V_{\rm PCB}(r)$ have a singularity, it has been proven that the principal values of the
integrals, $P\int F^*(r)V_i(r)F(r)dr$, are still finite due to the nodal structure of $F(r)$ \cite{Marcos2000_PRC62-054309,Alberto2002_PRC65-034307}.
This makes a direct comparison possible.
Several examples will be shown in Sections~\ref{Sect:3.2}, \ref{Sect:3.3}, \ref{Sect:3.6}, and \ref{Sect:4.1}.

\subsubsection{PSS in single-particle wave functions}

There are intensive discussions on the approximate energy degeneracy between the pseudospin doublets since the introduction of PSS in 1969, but less on their wave functions until the relativistic origin of PSS was revealed \cite{Ginocchio1998_PRC57-1167,Ginocchio2002_PRC66-064312}.

Within the PSS limit shown in Eq.~(\ref{Eq:2.1.PSSlimit}), the Schr\"odinger-like equation for the lower component $F(r)$ of the Dirac spinor is expressed as
\begin{equation}\label{Eq:2.3.SchrFPSS}
    \ls \frac{d^2}{dr^2} - \frac{\kappa(\kappa-1)}{r^2} + (\epsilon-M-\Sigma_0)(\epsilon+M-\Delta(r)) \rs F(r) = 0\,,
\end{equation}
as seen in Eq.~(\ref{Eq:2.2.SchrFPSS}).
It is clear that this equation is identical for the pseudospin doublets $a$ and $b$ with $\kappa_a(\kappa_a-1)=\kappa_b(\kappa_b-1)$, i.e., $\kappa_a + \kappa_b = 1$.
Therefore, the eigenfunctions $F_a(r)$ and $F_b(r)$ are exactly the same up to a normalization factor.

Moreover, there holds Eq.~(\ref{Eq:2.1.GF}),
\begin{equation}\label{Eq:2.3.GF}
  G(r) = \frac{1}{\epsilon-M-\Sigma_0}\lb-\frac{d}{dr}+\frac{\kappa}{r}\rb F(r)\,,
\end{equation}
and now $\Sigma_0$ is just a common constant.
Therefore, for the single-particle wave functions of the state $a$, the normalization condition in Eq.~(\ref{Eq:2.1.Normalization}) reads
\begin{equation}\label{Eq:2.3.Norm1}
  \int \ls G^2_a(r)+F^2_a(r)\rs dr
  =\int \Lb\ls\frac{1}{\epsilon_a-M-\Sigma_0}\lb-\frac{d}{dr}+\frac{\kappa_a}{r}\rb F_a(r)\rs^2+F^2_a(r)\Rb dr
  =1\,.
\end{equation}
Integrating by parts and using Eq.~(\ref{Eq:2.3.SchrFPSS}), one will end up with
\begin{equation}\label{Eq:2.3.Norm2}
  \int (2\epsilon_a-\Delta(r)-\Sigma_0) F^2_a(r)dr
  = \epsilon_a-M-\Sigma_0\,.
\end{equation}
In the same way, for its pseudospin partner $b$, one has
\begin{equation}\label{Eq:2.3.Norm3}
  \int (2\epsilon_b-\Delta(r)-\Sigma_0) F^2_b(r)dr
  = \epsilon_b-M-\Sigma_0\,.
\end{equation}
As $\epsilon_a=\epsilon_b$ at the PSS limit, the normalization factors for $F_a(r)$ and $F_b(r)$ are the same.
Therefore, for a pair of pseudospin doublets, their lower components of the Dirac spinor are identical (up to a phase) at the PSS limit \cite{Ginocchio1998_PRC57-1167},
\begin{equation}\label{Eq:2.3.FatPSS}
  F_a(r) = F_b(r)\,.
\end{equation}

As a step further, together with Eq.~(\ref{Eq:2.1.FG}), the first-order differential relation for their upper components of the Dirac spinor can be obtained \cite{Ginocchio2002_PRC66-064312},
\begin{equation}\label{Eq:2.3.GatPSS}
    \lb\frac{d}{dr}+\frac{\kappa_a}{r}\rb G_a(r) = \lb\frac{d}{dr}+\frac{\kappa_b}{r}\rb G_b(r)\,,
\end{equation}
or written as
\begin{equation}\label{Eq:2.3.GatPSS2}
    \lb\frac{d}{dr}-\frac{\tilde l}{r}\rb G_{j_<}(r) = \lb\frac{d}{dr}+\frac{\tilde l+1}{r}\rb G_{j_>}(r),
\end{equation}
with $j_<$ ($j_>$) labelling the $j=\tilde l-1/2$ ($j=\tilde l+1/2$) orbital.

The wave-function relations (\ref{Eq:2.3.FatPSS}) and (\ref{Eq:2.3.GatPSS}) between the pseudospin doublets can also be derived from the pseudospin SU(2) generator \cite{Ginocchio1998_PLB425-1}
\begin{equation}\label{Eq:2.3.PSSgenerator}
  \tilde{\mathbf{S}}
  = \lb\begin{array}{cc}
    \tilde{\mathbf{s}} & 0 \\ 0 & \mathbf{s}
    \end{array}\rb\,,
\end{equation}
with $\mathbf{s} = \boldsymbol{\sigma}/2$ and $\tilde{\mathbf{s}} = (\boldsymbol{\sigma}\cdot\hat{\mathbf{p}}) \mathbf{s} (\boldsymbol{\sigma}\cdot\hat{\mathbf{p}})$.
The details can be found in Ref.~\cite{Ginocchio2002_PRC66-064312}.

\begin{figure}[tbhp]
\begin{center}
  \includegraphics[width=4cm]{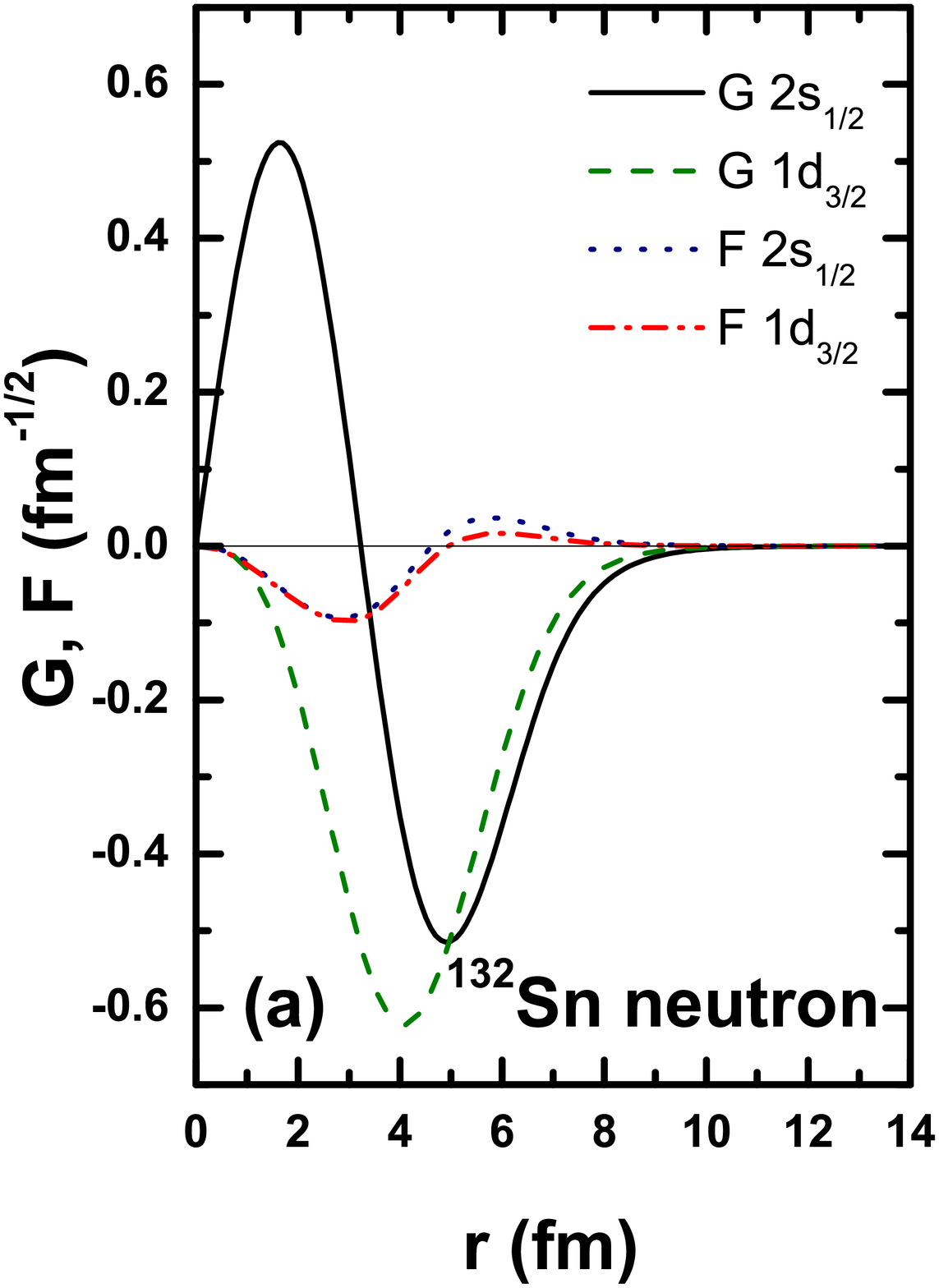}
  \includegraphics[width=4cm]{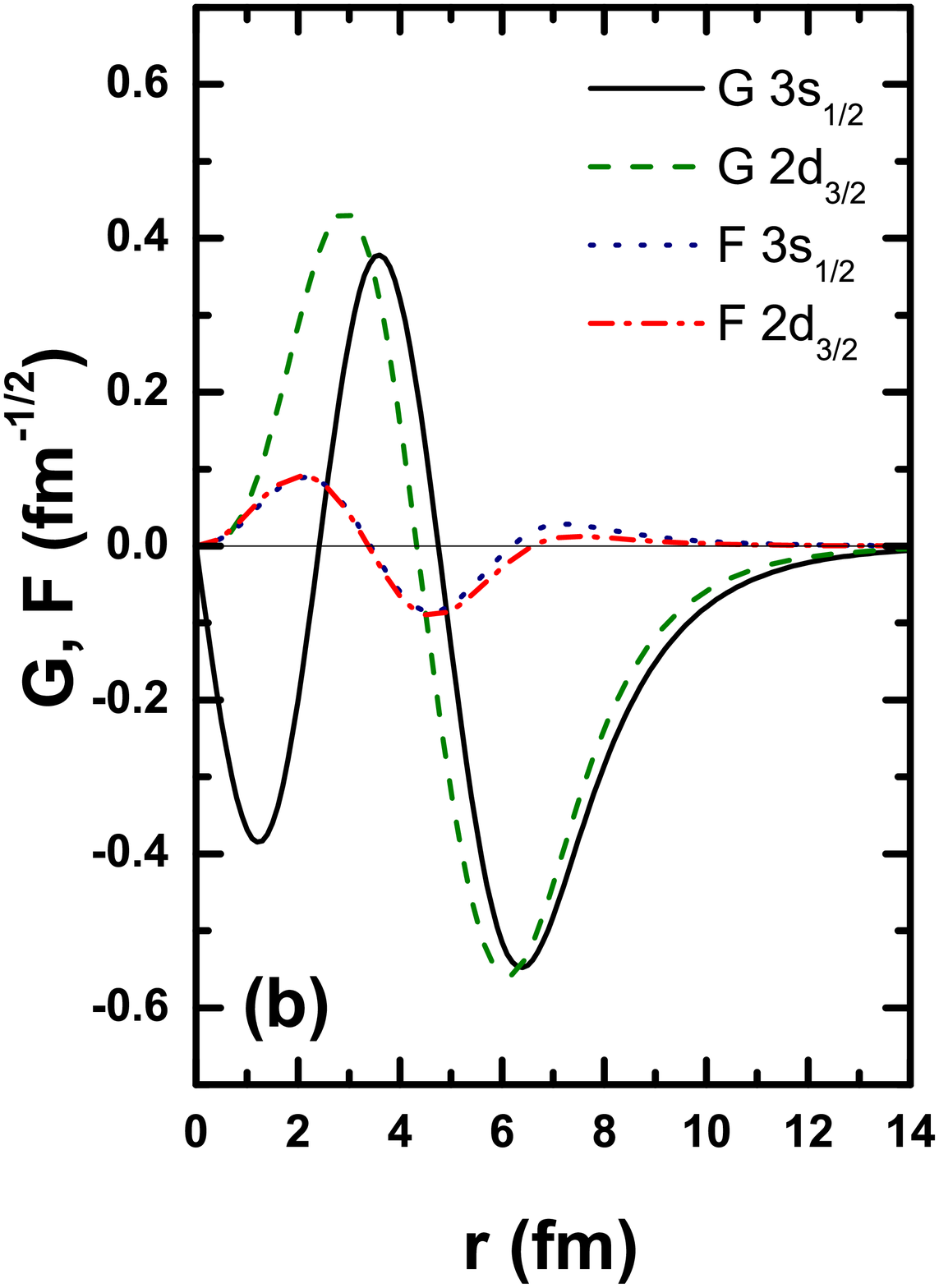}
  \includegraphics[width=4.03cm]{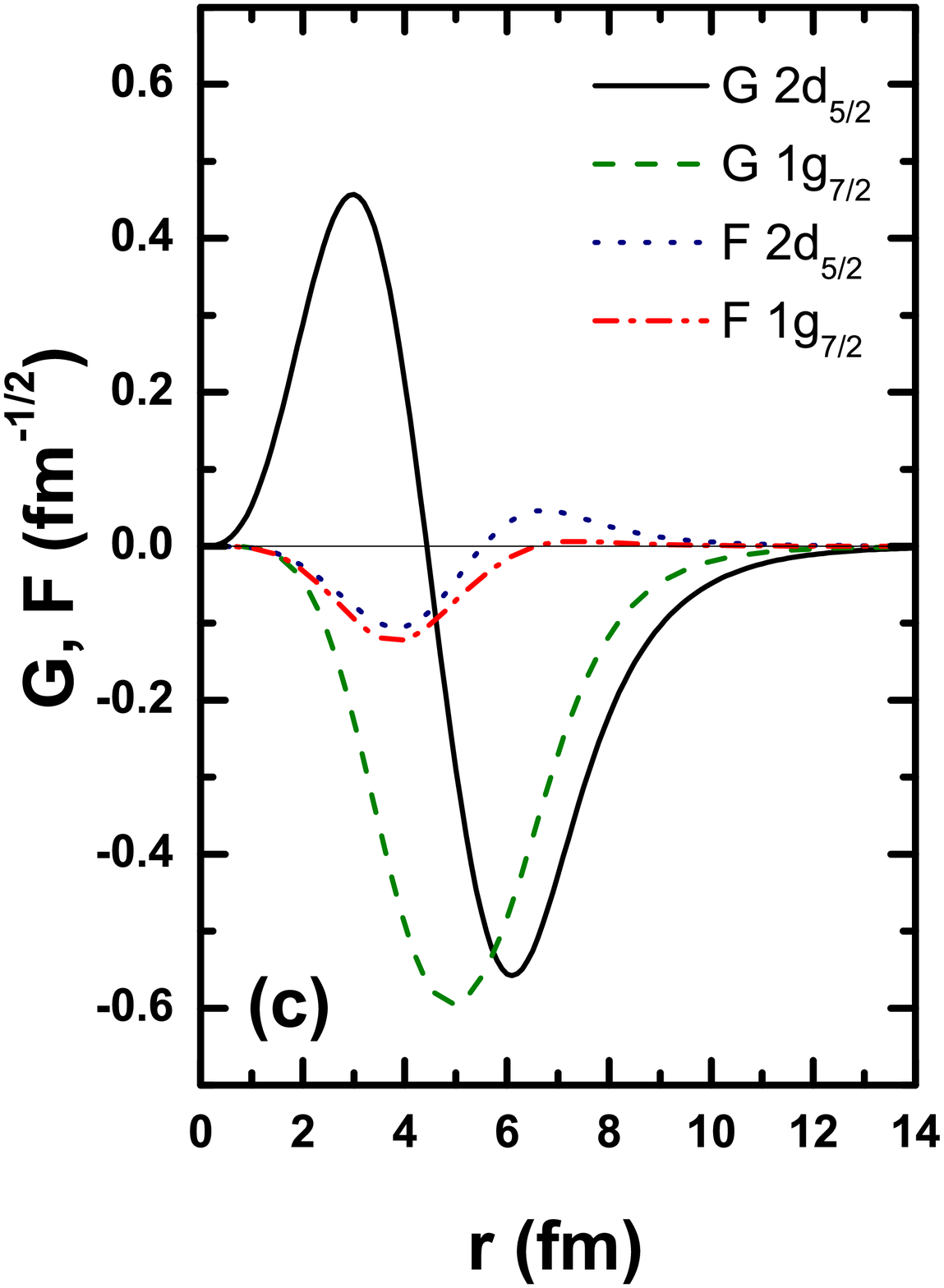}
\end{center}
\caption{(Color online) Neutron single-particle wave functions of the $1\tilde p$, $2\tilde p$, and $1\tilde f$ pseudospin doublets in $^{132}$Sn calculated by the RMF theory with the effective interaction PC-PK1 \cite{Zhao2010_PRC82-054319}.
\label{Fig:2.3.wf}}
\end{figure}

To test the relation shown in Eq.~(\ref{Eq:2.3.FatPSS}), the single-particle wave functions for the neutrons in $^{132}$Sn are calculated by the self-consistent point-coupling RMF theory with the effective interaction PC-PK1 \cite{Zhao2010_PRC82-054319}.
In panels (a), (b), and (c) of Fig.~\ref{Fig:2.3.wf} are shown the wave functions of the $1\tilde p$, $2\tilde p$, and $1\tilde f$ pseudospin doublets, respectively.
For each pair of pseudospin doublets, their upper components $G$ have different number of nodes and radial shape, however, their lower components $F$ are almost identical except on the nuclear surface.
By comparing these three panels, it is found that the relation in Eq.~(\ref{Eq:2.3.FatPSS}) is better satisfied for smaller $\tilde l$.

\begin{figure}[tbhp]
\begin{center}
  \includegraphics[width=4cm]{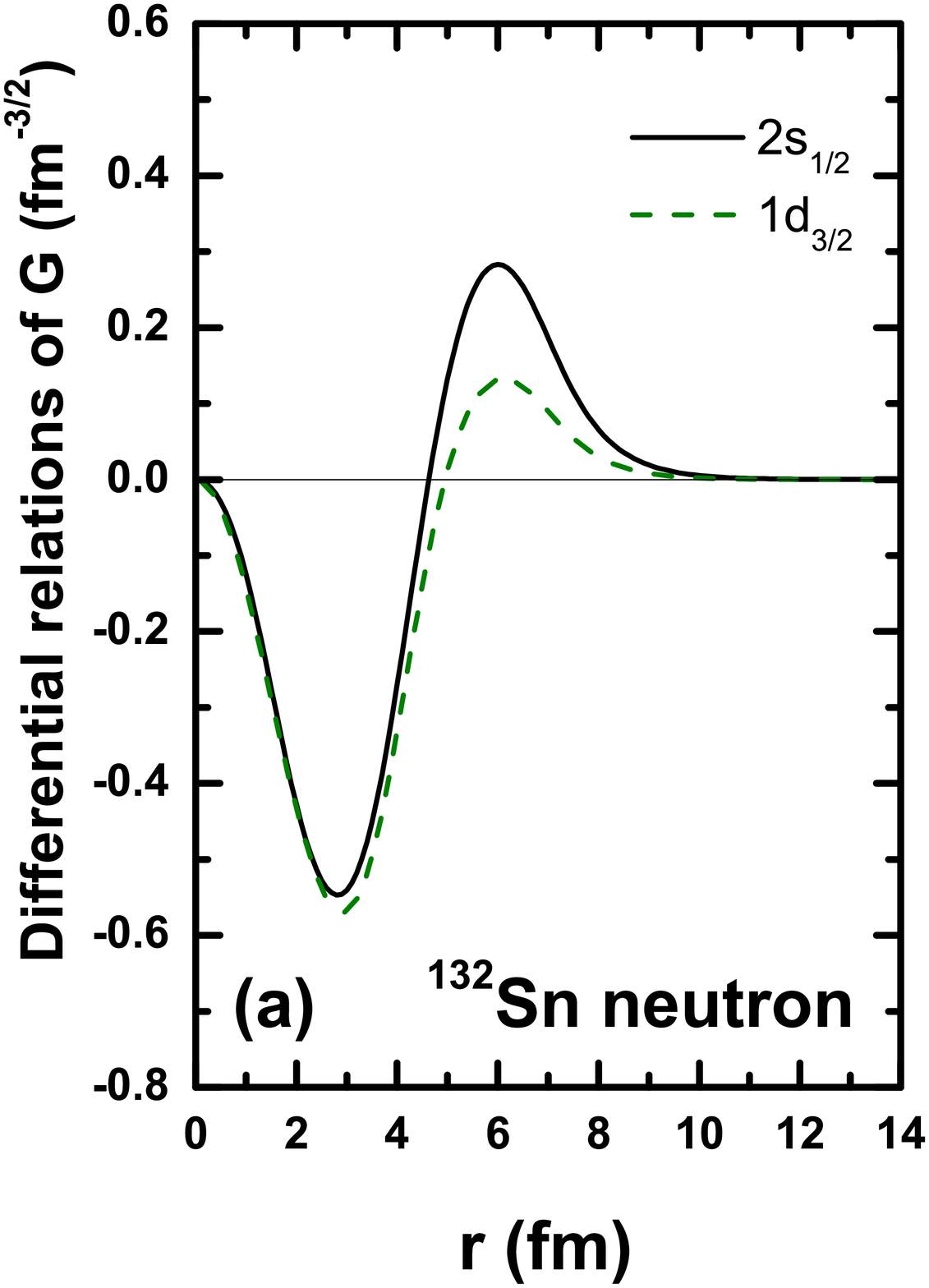}
  \includegraphics[width=4cm]{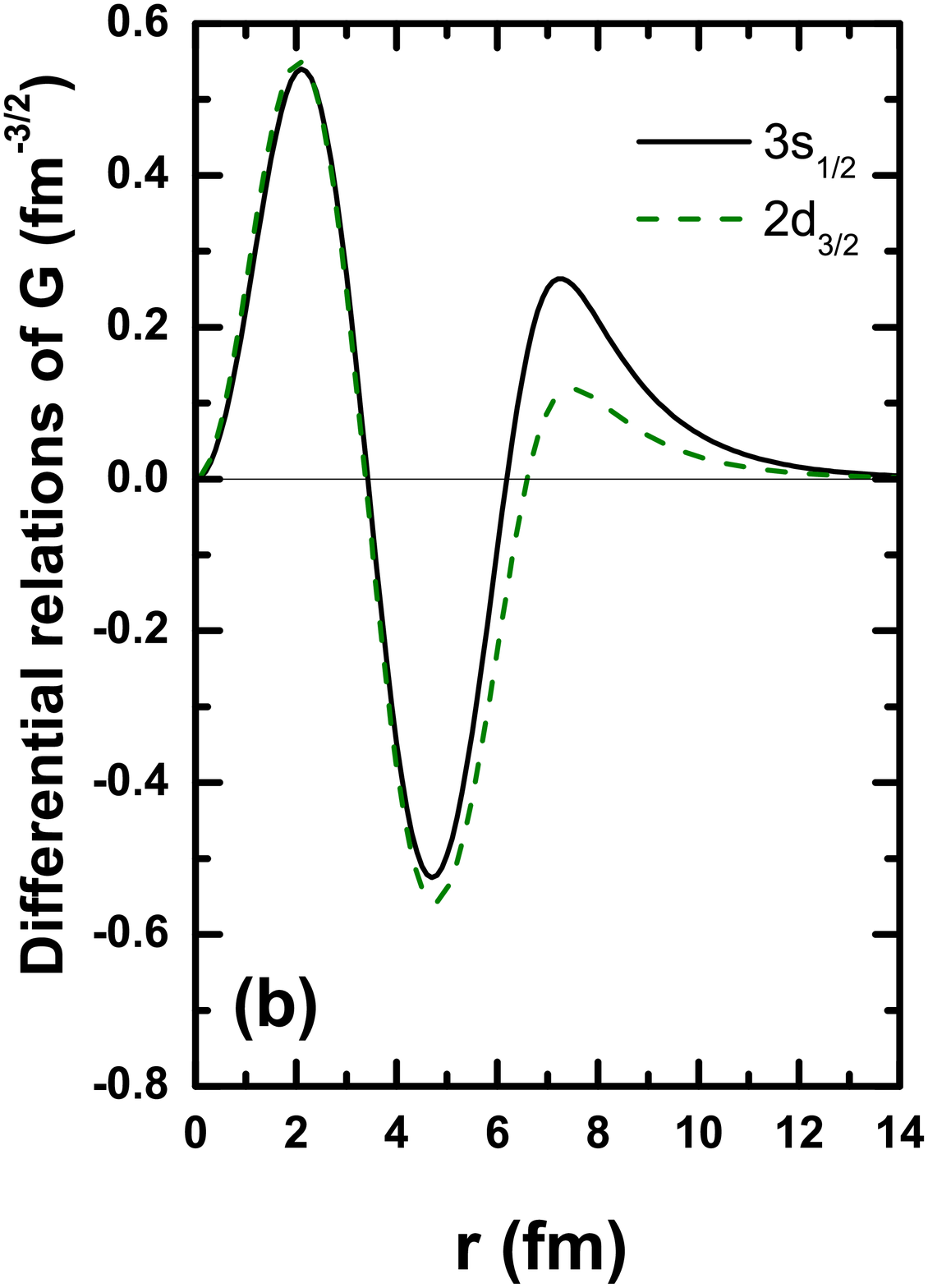}
  \includegraphics[width=4cm]{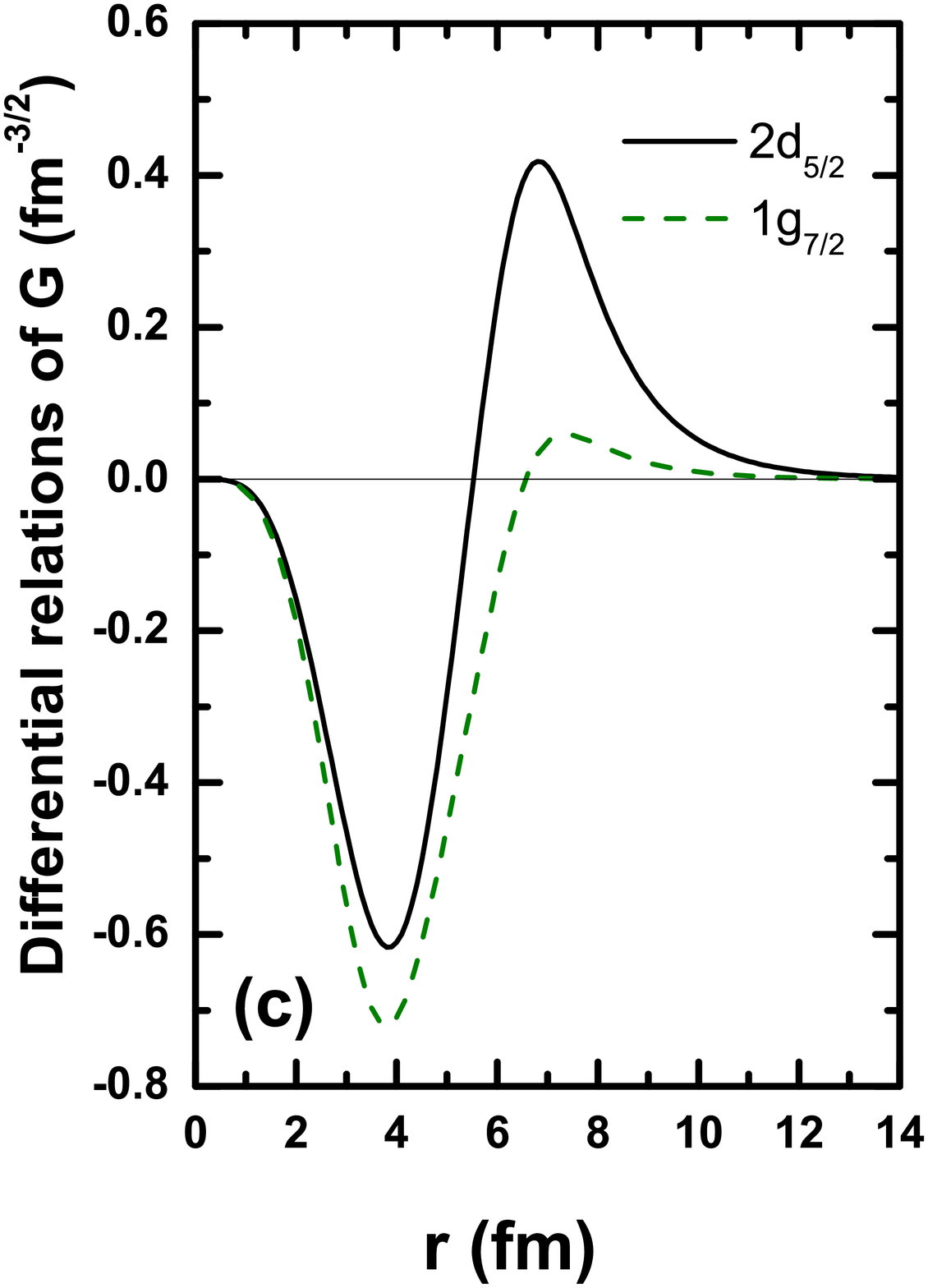}
\end{center}
\caption{(Color online) Differential relation of the upper components $G$ in Eq.~(\ref{Eq:2.3.GatPSS}) for the $1\tilde p$, $2\tilde p$, and $1\tilde f$ pseudospin doublets in $^{132}$Sn calculated by the RMF theory with PC-PK1.
\label{Fig:2.3.dwf}}
\end{figure}

To test the differential relation of the upper components shown in Eq.~(\ref{Eq:2.3.GatPSS}), the corresponding results obtained by using the first-order differential operators are plotted in Fig.~\ref{Fig:2.3.dwf}.
It is found that, with the $\kappa$-dependent first-order differential operators, one obtains a remarkable similarity in the differential wave functions except near the nuclear surface.
By comparing three panels, it is also found that the relation in Eq.~(\ref{Eq:2.3.GatPSS}) is better satisfied with smaller $\tilde l$.

In Section~\ref{Sect:2}, the general features for the Dirac equation and its corresponding Schr\"odinger-like equations were discussed.
The analytical solutions for Dirac equation at the pseudospin symmetry limit were shown by taking the relativistic harmonic oscillator and relativistic Morse potentials as examples.
The pseudospin symmetry and its breaking in realistic nuclei were discussed in a general framework of the covariant density functional theory.
The evaluations of the pseudospin symmetry in the single-particle energies and wave functions were reviewed.

\section{PSS and SS in Various Systems and Potentials}\label{Sect:3}

\subsection{From stable nuclei to exotic nuclei}\label{Sect:3.0}

The concept of pseudospin symmetry \cite{Hecht1969_NPA137-129,Arima1969_PLB30-517} was introduced originally based on the observation of the empirical single-particle spectra in stable nuclei.
Since then, intensive discussions of PSS were mainly concentrated around the nuclear $\beta$-stability valley.
During the past decades, more and more highly unstable nuclei with extreme $N/Z$ ratios have been accessible with the radioactive ion beam facilities.
The physics connected to the extreme neutron richness in these nuclei and the low density in the tails of their matter distributions have attracted much attention, and new exciting discoveries have been made by exploring hitherto inaccessible regions in the nuclear chart.
One of the examples is the investigations of PSS from stable to exotic nuclei \cite{Meng1998_PRC58-R628,Meng1999_PRC59-154}.

From the theoretical point of view, for open-shell exotic nuclei, the RCHB theory \cite{Meng2006_PPNP57-470} is able to take the pairing correlation and the coupling to continuum into account properly.
Furthermore, as pointed out in Ref.~\cite{Meng1998_PRL80-460}, in the RCHB theory, the particle levels for the bound states in the canonical basis are the same as those by solving the Dirac equation with the corresponding scalar and vector potentials.
Therefore, the Schr\"odinger-like equations (\ref{Eq:2.1.SchrG}) and (\ref{Eq:2.1.SchrF}) remain the same in the canonical basis even after the pairing interaction has been taken into account.

One observes from Eq.~(\ref{Eq:2.1.SchrF}) that formally the only term which breaks the PSS is the PSO potential $V_{\rm PSO}$ (\ref{Eq:2.1.VPSOandVPCB}), which is proportional to $d\Sigma(r)/dr$.
Therefore, it is expected that the PSS is better conserved when $|d\Sigma(r)/dr|$ becomes small \cite{Meng1998_PRC58-R628}.
This conjecture can be verified with the exotic nuclei, whose potentials can be much more diffuse than the stable ones \cite{Meng1999_PRC59-154}.

\begin{figure}[tbhp]
\begin{center}
  \includegraphics[width=5.8cm]{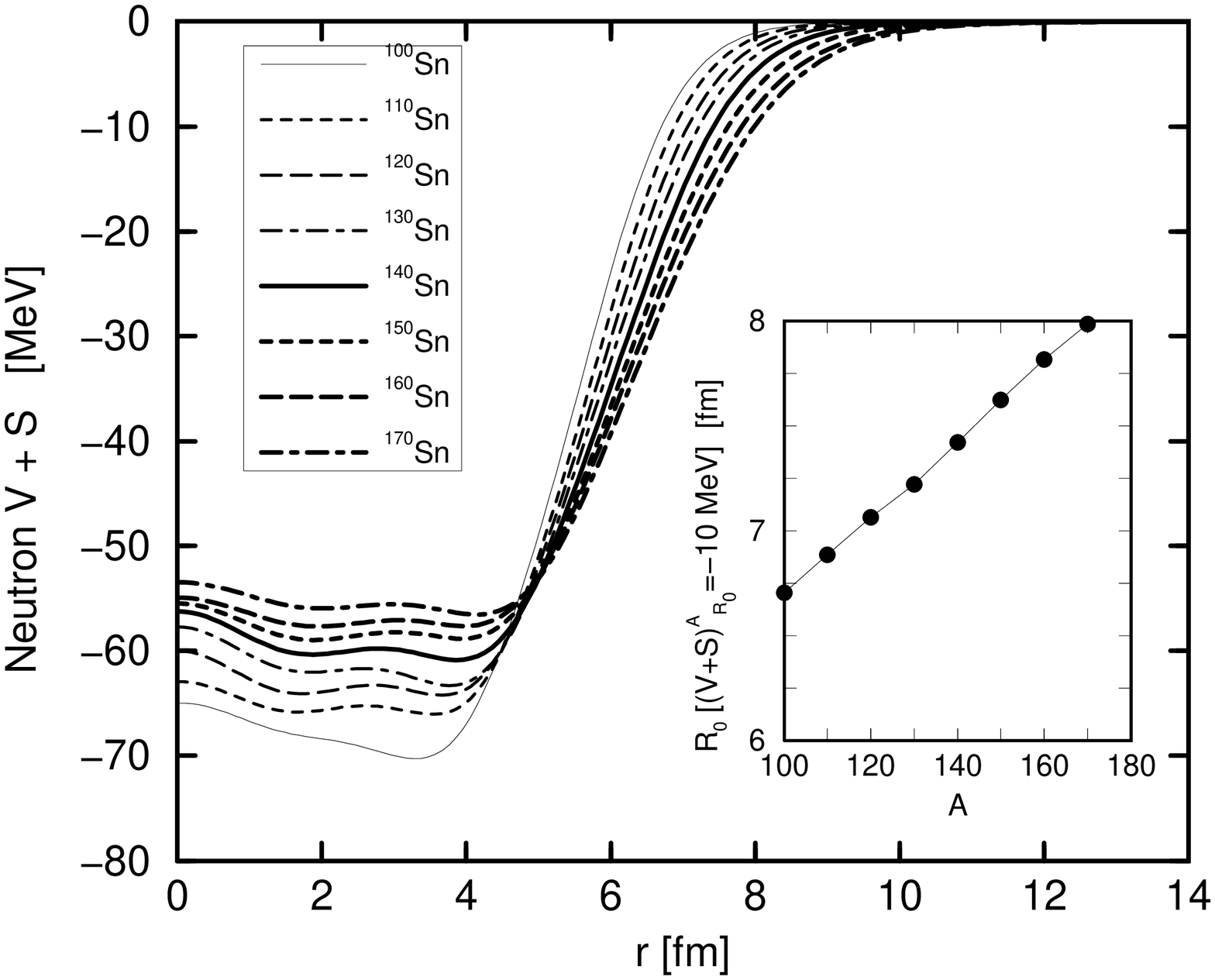}
  \includegraphics[width=6.6cm]{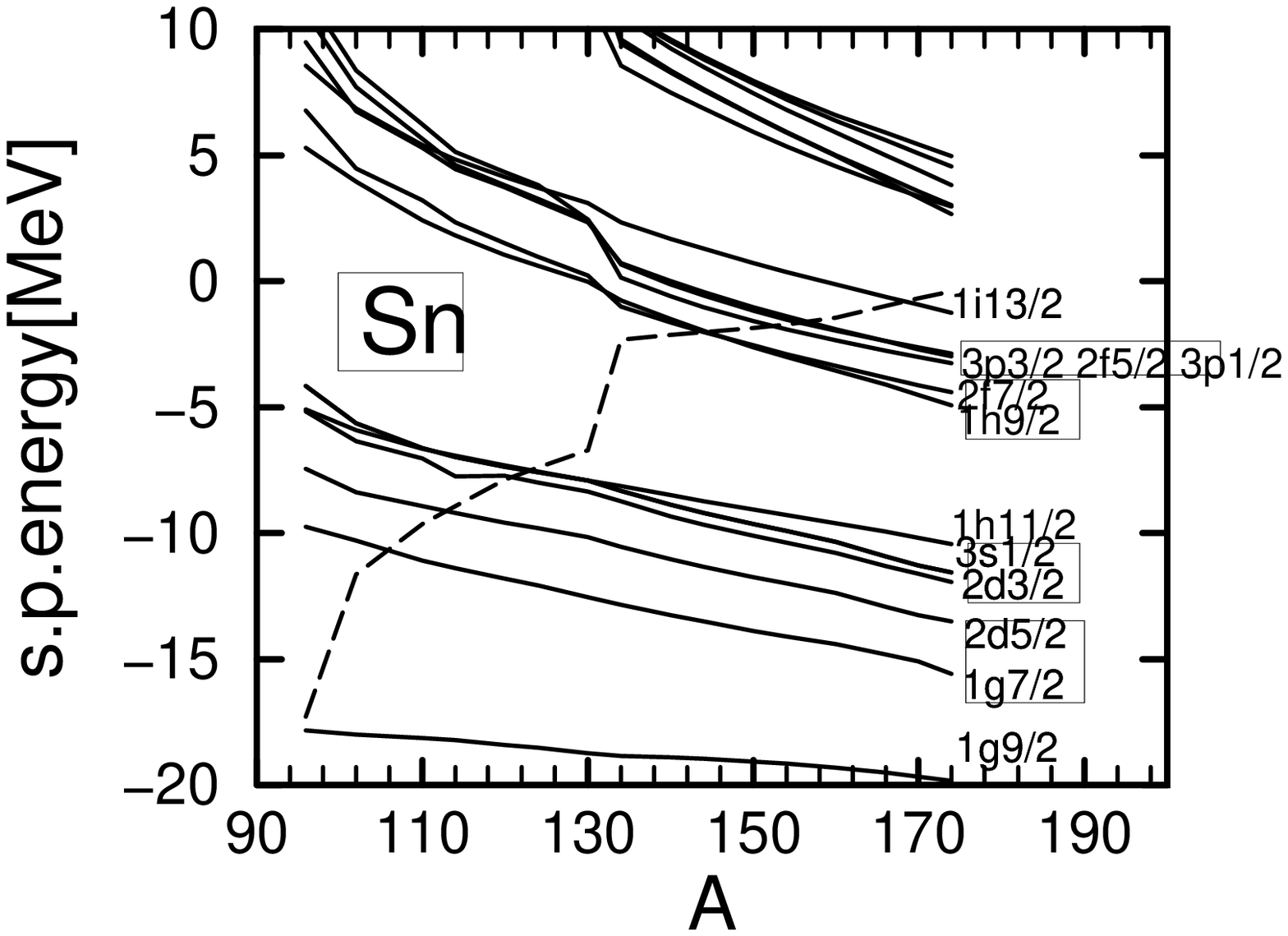}
\end{center}
\caption{Left panel: Neutron potentials $\Sigma(r)$ for Sn isotopes calculated by the RCHB theory \cite{Meng2006_PPNP57-470} with the effective interaction NLSH \cite{Sharma1993_PLB312-377}.
Taken from Ref.~\cite{Meng1999_NPA650-176}.
Right panel: The corresponding neutron single-particle energies in the canonical basis as a function of the mass number, where the pseudospin doublets are marked by boxes and the Fermi surface is shown by a dashed line.
Taken from Ref.~\cite{Meng1999_PRC59-154}.
\label{Fig:3.0.Sn}}
\end{figure}

In the left panel of Fig.~\ref{Fig:3.0.Sn}, the potentials $\Sigma(r)$ for neutrons in Sn isotopes calculated by the self-consistent RCHB theory with the effective interaction NLSH \cite{Sharma1993_PLB312-377} are shown.
One can see a gradual change in the diffuseness of $\Sigma(r)$ from the neutron-deficient nucleus $^{100}$Sn to the extremely neutron-rich nucleus $^{170}$Sn.
The corresponding evolution of the single-particle energies in the canonical basis can be seen in the right panel of Fig.~\ref{Fig:3.0.Sn}.

\begin{figure}[tbhp]
\begin{center}
  \includegraphics[angle=270,width=8cm]{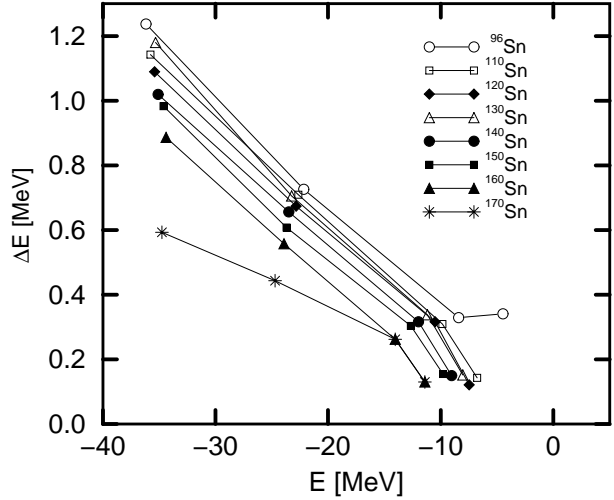}
\end{center}
\caption{Reduced PSO splittings for neutrons in Sn isotopes versus their average single-particle energies $[(\tilde l+1)E_{j_<} + \tilde l E_{j_>}]/(2\tilde l+1)$ calculated by RCHB theory with NLSH.
Taken from Ref.~\cite{Meng1999_PRC59-154}.
\label{Fig:3.0.dE}}
\end{figure}

The reduced PSO splittings $\Delta E_{\rm PSO}$ versus their average single-particle energies are plotted in Fig.~\ref{Fig:3.0.dE}.
It is seen that the PSO splittings in Sn isotopes have a monotonous decreasing behavior with increasing isospin.
In particular, for the $1\tilde p$ doublets, $\Delta E_{\rm PSO}$ in $^{170}$Sn is only half of that in $^{96}$Sn.
Furthermore, a monotonous decreasing behavior of $\Delta E_{\rm PSO}$ with increasing single-particle energies $E$ maintains from the proton drip line to the neutron drip line.

From these studies, the pseudospin symmetry remains a good approximation for both stable and exotic nuclei.
A better pseudospin symmetry can be expected for the orbitals near the threshold, in particular for nuclei near the particle drip line.

\subsection{From non-confining potentials to confining potentials}\label{Sect:3.1}

It is observed that the main quantum numbers $n$ of a pair of pseudospin doublets differ by one, e.g., ($2s_{1/2}$, $1d_{3/2}$), ($2p_{3/2}$, $1f_{5/2}$), etc.
This phenomenon motivated Leviatan and Ginocchio \cite{Leviatan2001_PLB518-214} for the analytical proof on the nodal structure of the Dirac spinor.
For the so-called non-confining potentials, which mean $S(r),V(r)\rightarrow0$ for $r\rightarrow\infty$, it is proven that the number of internal nodes of the upper $G$ and lower $F$ components of the Dirac spinor, $n_G$ and $n_F$, obeys \cite{Leviatan2001_PLB518-214}
\begin{equation}\label{Eq:3.1.GFnodes0}
  n_F = n_G
  \quad\mbox{for}\quad
  \kappa<0\,,\qquad
  n_F = n_G + 1
  \quad\mbox{for}\quad
  \kappa>0\,.
\end{equation}
It is also proven that there exist no bound states in the Fermi sea at the pseudospin symmetry limit (\ref{Eq:2.1.PSSlimit}) within the non-confining potentials.

In contrast to the non-confining potentials, Chen \textit{et al.} \cite{Chen2003_CPL20-358} showed there exist bound states in the Fermi
sea at the pseudospin symmetry limit (\ref{Eq:2.1.PSSlimit}) when the potential $\Delta(r)$ is confining.
A typical example is the relativistic harmonic oscillator potential \cite{Chen2003_CPL20-358,Lisboa2004_PRC69-024319,Ginocchio2005_PRL95-252501,deCastro2006_PRC73-054309}.
Recently, the corresponding nodal structure of the Dirac spinor within the confining potentials was derived analytically by Alberto, de Castro, and Malheiro \cite{Alberto2013_PRC87-031301R}.

In this Section, we will highlight the key steps of these analytical proofs.
We will then discuss the single-particle spectra and wave functions given by the relativistic harmonic oscillator potential, which conserves the pseudospin symmetry exactly.

\subsubsection{Nodal structure for non-confining potentials}\label{Sect:3.1.1}

First of all, let us focus on the so-called non-confining potentials, which generally occur in isolated atomic nuclei.
Their scalar and vector potentials satisfy $rS(r), rV(r) \rightarrow 0$ for $r\rightarrow0$, and $S(r),V(r)\rightarrow0$ for $r\rightarrow\infty$ \cite{Leviatan2001_PLB518-214}.
A typical example for such potentials calculated by the self-consistent RMF theory is shown in Fig.~\ref{Fig:2.3.132potl}.

From the radial Dirac equations (\ref{Eq:2.1.DiraceqR}), it is seen that the radial wave functions follow
\begin{align}\label{Eq:3.1.G2F2}
  \frac{d^2G(r)}{dr^2}&\sim -M_+(r)M_-(r)G(r) \sim(M^2-\epsilon^2)G(r)\,,\nonumber\\
  \frac{d^2F(r)}{dr^2}&\sim -M_+(r)M_-(r)F(r) \sim(M^2-\epsilon^2)F(r)\,,
\end{align}
at large $r$.
As for a bound state, the wave functions $G(r)$ and $F(r)$ should vanish exponentially, $G(r) \propto F(r) \propto e^{-\lambda r}$ with $\lambda=\sqrt{M^2-\epsilon^2}$.
Therefore, $-M<\epsilon<M$ is the condition for the bound states.

Focusing on the single-particle bound states in the Fermi sea, $\epsilon>0$, the effective mass $M_+(r)$ is always positive, and the effective mass $M_-(r)$ is positive at the origin and becomes negative at large $r$, changing its sign at $r_0$.
The asymptotic behaviors of their radial wave functions at $r\rightarrow0$ read
\begin{align}\label{Eq:3.1.GF0}
  &G_\kappa(r)\propto r^{-\kappa}\,,\quad
  F_\kappa(r)\propto r^{-\kappa+1}\,,\quad
  \lim_{r \rightarrow 0}\frac{F(r)}{G(r)} = -\frac{M_-(0)}{-2\kappa+1}r<0
  \quad\mbox{for}\quad\kappa<0\,,\nonumber\\
  &G_\kappa(r)\propto r^{\kappa+1}\,,\quad
  F_\kappa(r)\propto r^{\kappa}\,,\quad
  \lim_{r \rightarrow 0}\frac{G(r)}{F(r)} = \frac{M_+(0)}{2\kappa+1}r>0
  \quad\mbox{for}\quad\kappa>0\,.
\end{align}

In order to study the properties of the radial wave functions, it is helpful to introduce $\mathcal{G}_\kappa(r)=r^\kappa G_\kappa(r)$ and $\mathcal{F}_\kappa(r)=r^{-\kappa} F_\kappa(r)$, and then Eqs.~(\ref{Eq:2.1.DiraceqR}) can be simplified as  \cite{Leviatan2001_PLB518-214}
\begin{equation}\label{Eq:3.1.GF}
  \frac{d\mathcal{G}_\kappa(r)}{dr} = r^{2\kappa}M_+(r)\mathcal{F}_\kappa(r)
  \quad\mbox{and}\quad
  \frac{d\mathcal{F}_\kappa(r)}{dr} = -r^{-2\kappa}M_-(r)\mathcal{G}_\kappa(r)\,.
\end{equation}
In the open interval $(0,\infty)$, the nodes of $\mathcal{G}$ and $\mathcal{F}$ coincide with the nodes of $G$ and $F$, respectively.

Equations (\ref{Eq:3.1.GF}) lead to a number of observations \cite{Leviatan2001_PLB518-214}.
First, it is impossible for $\mathcal{G}$ and $\mathcal{F}$, or $G$ and $F$, to vanish simultaneously at the same point because, if they did, then all other higher-order derivatives would vanish at that point and hence the functions themselves would vanish everywhere.
Moreover, a node of $\mathcal{F}$ corresponds to an extremum of $\mathcal{G}$, and a node of $\mathcal{G}$ corresponds to an extremum of $\mathcal{F}$.
Since $M_-(r)$ changes sign at $r_0$, $\mathcal{F}$ can have an additional extremum at this point, which does not correspond to a node of $\mathcal{G}$.
It follows that the nodes of $\mathcal{F}$ and $\mathcal{G}$ alternate, i.e., between every pair of adjacent nodes of $\mathcal{F}$ ($\mathcal{G}$) there is one node of $\mathcal{G}$ ($\mathcal{F}$).

Furthermore, for bound states, both $\mathcal{G}(r)$ and $\mathcal{F}(r)$ vanish at $r=\infty$ and their extrema are concave towards the $r$-axis.
Therefore, the nodes of $\mathcal{G}$ and $\mathcal{F}$ can occur only where $M_+(r)M_-(r)>0$, i.e., both $M_+(r)>0$ and $M_-(r)>0$ in the present cases.
This is consistent with the non-relativistic case that the nodes of the radial wave function can occur only in the region of classically allowed motion, where the kinetic energy is positive.

One can now use the above results to obtain a relation between the radial nodes of $G(r)$ and $F(r)$, together with
\begin{equation}\label{Eq:3.1.GFprime}
  [\mathcal{G}_\kappa(r)\mathcal{F}_\kappa(r)]'
  = [G_\kappa(r)F_\kappa(r)]'
  = M_+(r)F_\kappa^2(r) - M_-(r)G_\kappa^2(r)\,.
\end{equation}
For the single-particle bound states in the Fermi sea within the non-confining potentials, $(GF)' = (\epsilon+M)F^2 - (\epsilon-M)G^2>0$ at large $r$.
At small $r$, $(GF)' = -M_-(0)G^2 < 0$ for $\kappa < 0$, while $(GF)' = M_+(0)F^2 > 0$ for $\kappa > 0$.
Since $GF$ vanishes at both $r = 0$ and $r = \infty$, one confirms that
\begin{equation}\label{Eq:3.1.GF0v2}
  \lim_{r\rightarrow0}G_\kappa(r)F_\kappa(r)<0
  \quad\mbox{for}\quad
  \kappa<0\,,\qquad
  \lim_{r\rightarrow0}G_\kappa(r)F_\kappa(r)>0
  \quad\mbox{for}\quad
  \kappa>0\,,
\end{equation}
and
\begin{equation}\label{Eq:3.1.GFinftyv2}
  \lim_{r\rightarrow\infty}G_\kappa(r)F_\kappa(r)<0\,.
\end{equation}
Furthermore, since both $M_+(r)$ and $M_-(r)$ are positive at nodes of $G$ and $F$, one has
\begin{equation}\label{Eq:3.1.GFatr1}
  \left.[G_\kappa(r)F_\kappa(r)]'\right|_{r=r_1} = -M_-(r_1)G^2(r_1) < 0\,,
\end{equation}
at $F(r_1)=0$, and
\begin{equation}\label{Eq:3.1.GFatr2}
  \left.[G_\kappa(r)F_\kappa(r)]'\right|_{r=r_2} = M_+(r_2)F^2(r_2) > 0\,,
\end{equation}
at $G(r_2)=0$.
This indicates $G_\kappa(r)F_\kappa(r)$ is a decreasing function at the nodes of $F$, and an increasing function at the nodes of $G$.

Combining all the properties discussed above, one can conclude that:
(i) For $\kappa>0$, $G_\kappa(r)F_\kappa(r)$ has an odd number of zeroes, and the first and the last zeroes belong to the nodes of $F$.
As a result, $F$ has one more node than $G$.
(ii) For $\kappa<0$, $G_\kappa(r)F_\kappa(r)$ has an even number of zeroes, and the first and the last zeroes belong to the nodes of $G$ and $F$, respectively.
As a result, $G$ and $F$ have the same number of nodes.
Note that this also includes the case that neither $G$ nor $F$ has internal nodes.
Therefore, the number of internal nodes of $G$ and $F$, $n_G$ and $n_F$, obeys \cite{Leviatan2001_PLB518-214}
\begin{equation}\label{Eq:3.1.GFnodes}
  n_F = n_G
  \quad\mbox{for}\quad
  \kappa<0\,,\qquad
  n_F = n_G + 1
  \quad\mbox{for}\quad
  \kappa>0\,.
\end{equation}

\begin{figure}[tbhp]
\begin{center}
  \includegraphics[width=4cm]{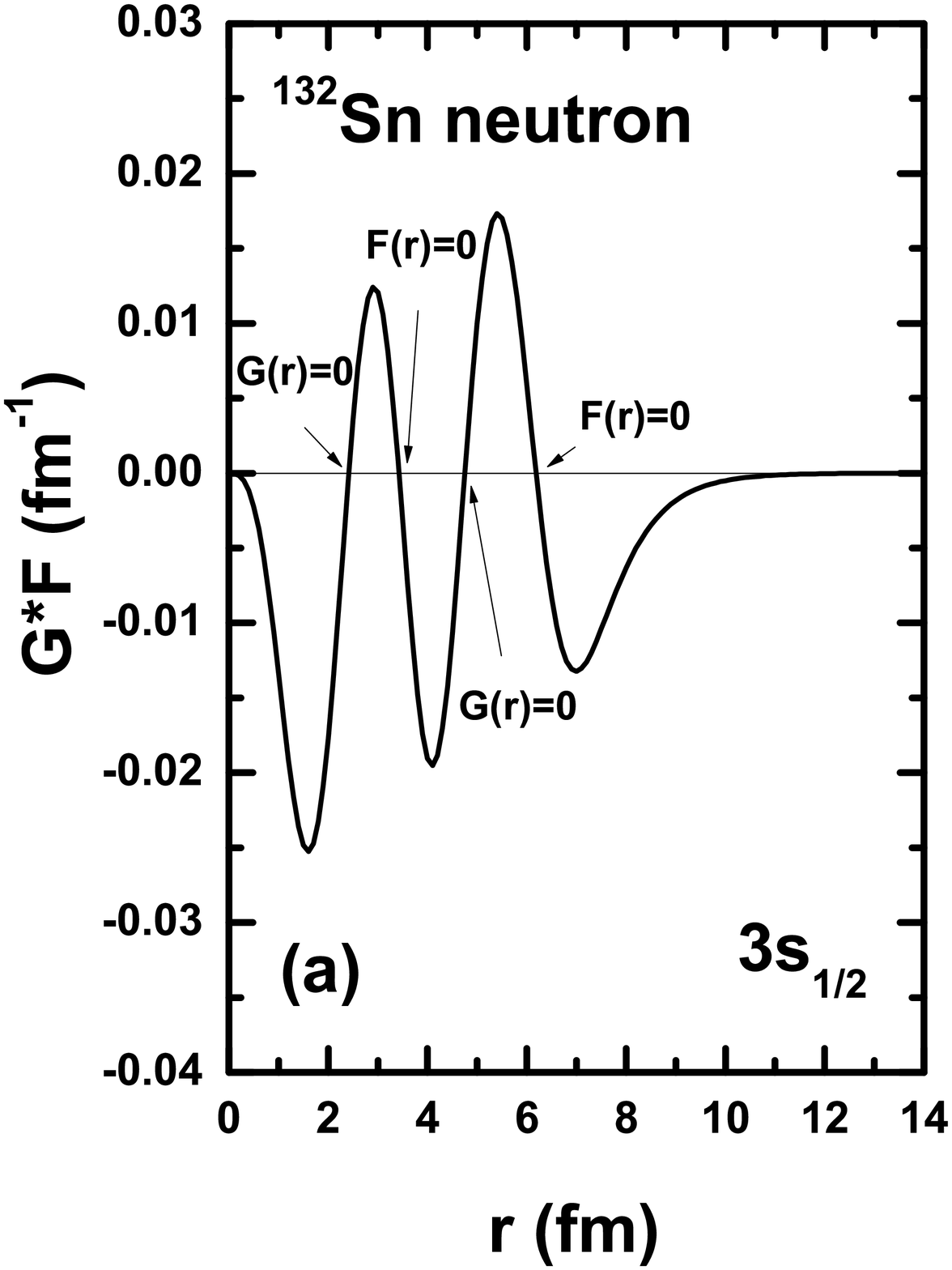}
  \includegraphics[width=4cm]{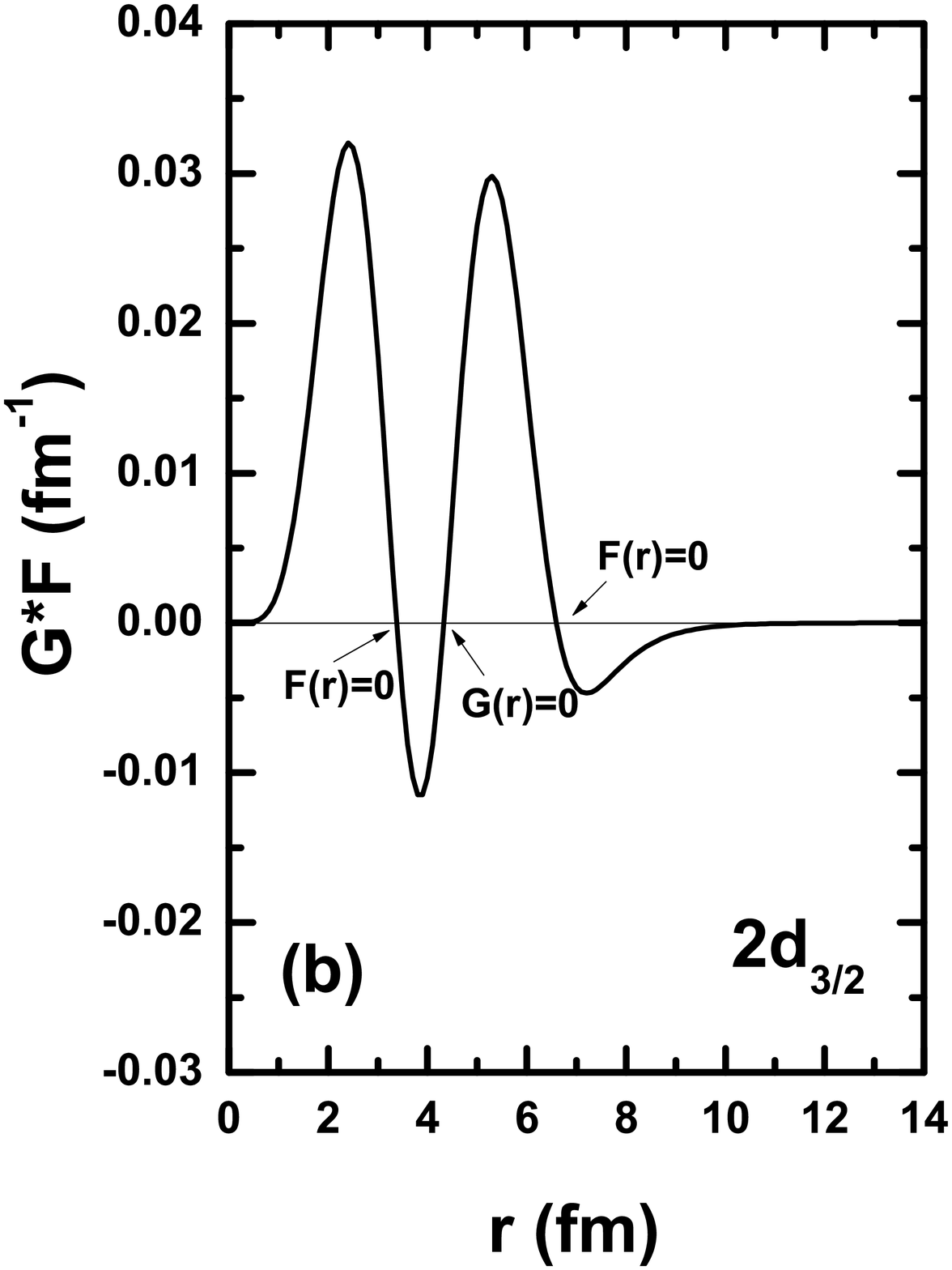}
  \includegraphics[width=4cm]{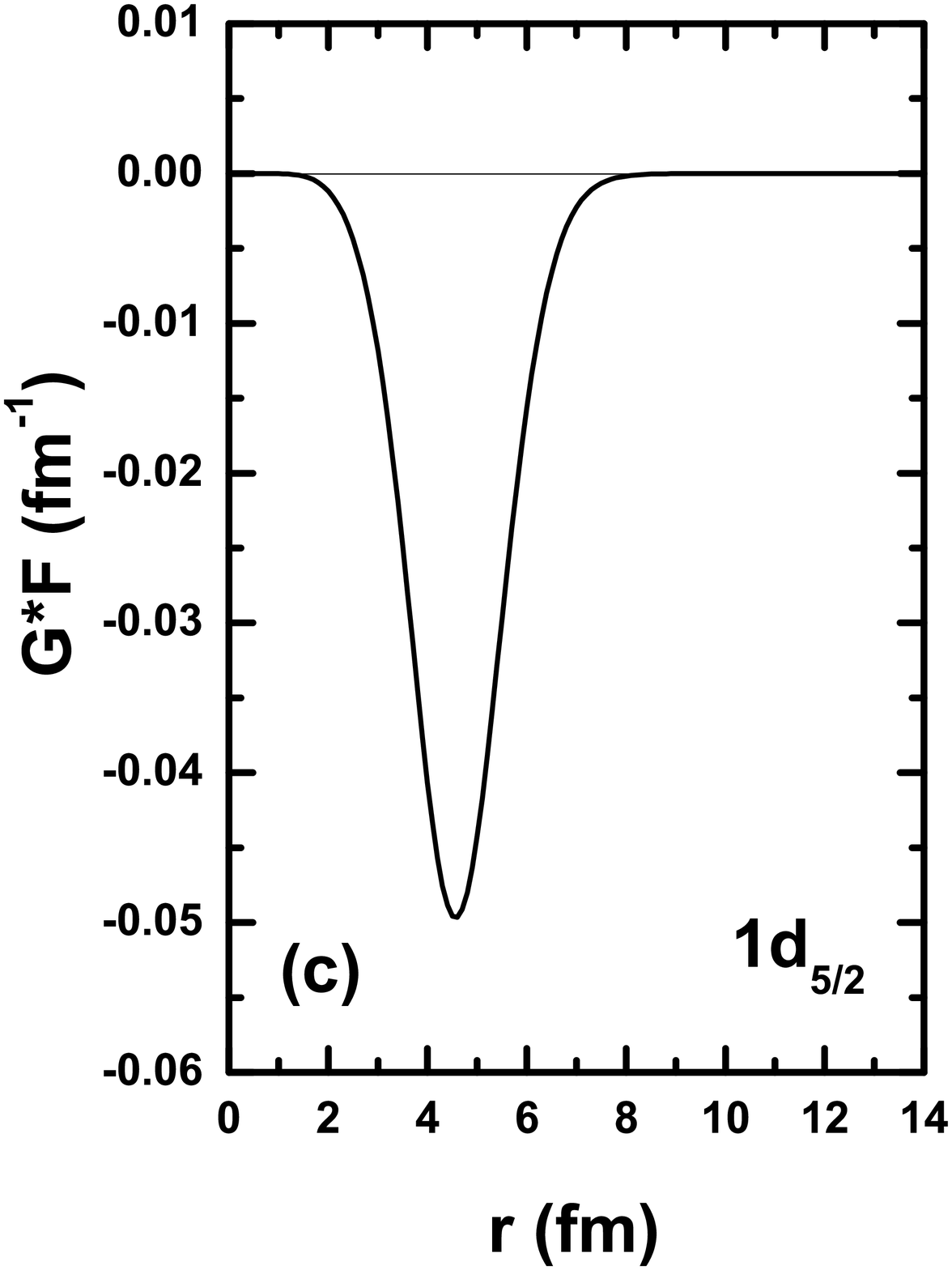}
\end{center}
\caption{Products of the upper and lower components $G(r)F(r)$ of the neutron $3s_{1/2}$, $2d_{3/2}$, and $1d_{5/2}$ states in $^{132}$Sn calculated by the RMF theory with the effective interaction PC-PK1 \cite{Zhao2010_PRC82-054319}.
\label{Fig:3.1.nodes}}
\end{figure}

In Fig.~\ref{Fig:3.1.nodes}, the products of the upper and lower components $G(r)F(r)$ are shown for the $3s_{1/2}$, $2d_{3/2}$, and $1d_{5/2}$ states.
From panels (a) and (b), one can confirm all the properties discussed above.
If there exist internal nodes, the last one must belong to $F$.
For the $\kappa>0$ states, $F$ has one node more than $G$, while for the $\kappa<0$ states, $F$ has the same number of nodes as $G$.
Meanwhile, $G(r)F(r)$ is a decreasing function at the nodes of $F$ and an increasing function at the nodes of $G$.

It should be noted that there is a kind of special case that $n_F = n_G=0$ for all the lowest $j=l+1/2$ orbitals, e.g., the $1d_{5/2}$ state shown in panel (c) of Fig.~\ref{Fig:3.1.nodes}.
These orbitals are special because they have no pseudospin partners.
We will come back to this point with the discussions on the puzzle of intruder states in Section~\ref{Sect:4.2}.

Before ending this part, it is important to point out that, within the non-confining potentials, there exist no bound states in the Fermi sea at the PSS limit \cite{Leviatan2001_PLB518-214}.
The reason is following:
On one hand, for the bound states, $M_-(r)$ must be negative at large $r$ because $M_+(r)$ is always positive.
The exact PSS limit (\ref{Eq:2.1.PSSlimit}) means $M_-$ is simply a constant, which means $M_-$ is a negative constant here.
On the other hand, from Eqs.~(\ref{Eq:3.1.GF0v2}) and (\ref{Eq:3.1.GFinftyv2}), we know that $G(r)F(r)$ goes to zero for both small and large $r$, but $G(r)F(r)$ cannot be identically zero for all $r$.
This means $[G(r)F(r)]'$ in Eq.~(\ref{Eq:3.1.GFprime}) must be negative for some range of $r$ and positive for some other range of $r$.
However, this is impossible if $M_+(r)$ is always positive and $M_-(r)$ is a negative constant.
This contradiction demonstrates that within the non-confining potentials there exist no bound states in the Fermi sea at the PSS limit.

\subsubsection{Nodal structure for confining potentials}

Chen \textit{et al.} pointed out in Ref.~\cite{Chen2003_CPL20-358} that there can exist bound states in the Fermi sea at the PSS limit (\ref{Eq:2.1.PSSlimit}), $d\Sigma(r)/dr=0$, if the potential $\Delta(r)$ satisfies
\begin{equation}\label{Eq:3.1.boundatPSS}
  \lim_{r\rightarrow\infty}\Delta(r) > M+\epsilon > 2M + \Sigma_0\,.
\end{equation}

As discussed in the previous part, for a bound state, its wave functions $G(r)$ and $F(r)$ should go to zero exponentially at large $r$.
From Eqs.~(\ref{Eq:3.1.G2F2}), one finds that this requirement can be fulfilled either by $M_+(r)>0$ and $M_-(r)<0$ or by $M_+(r)<0$ and $M_-(r)>0$ at large $r$.
The bound states in the non-confining potentials belong to the former case.
On the other hand, the exact PSS limit (\ref{Eq:2.1.PSSlimit}) indicates $M_-$ is simply a constant, in particular, it should be positive for all possible solutions belonging to the Fermi sea.
These solutions become the bound states as long as $M_+(r)<0$ at large $r$, which corresponds to the condition given in Eq.~(\ref{Eq:3.1.boundatPSS}).

\begin{figure}[tbhp]
\begin{center}
  \includegraphics[width=6cm]{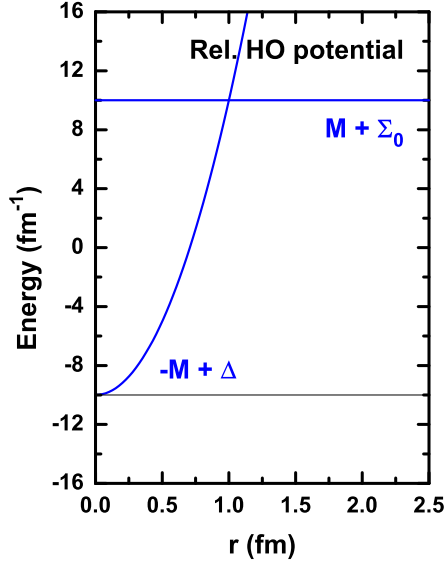}
\end{center}
\caption{(Color online) Relativistic harmonic oscillator potential for $V(r) = -S(r) = \frac{1}{4}M\omega^2 r^2$ with $M=10.0$~fm$^{-1}$ and $\omega=2.0$~fm$^{-1}$.
\label{Fig:3.1.HOpotl}}
\end{figure}

A typical example for such kind of potential is the relativistic harmonic oscillator potential \cite{Chen2003_CPL20-358,Lisboa2004_PRC69-024319,Ginocchio2005_PRL95-252501,deCastro2006_PRC73-054309},
\begin{equation}\label{Eq:3.1.HO}
  \Delta(r) = \frac{1}{2}M\omega^2r^2\,,
\end{equation}
as illustrated in Fig.~\ref{Fig:3.1.HOpotl}.
This is called the confining potential \cite{Alberto2013_PRC87-031301R} as $\Delta(r)\rightarrow\infty$ for $r\rightarrow\infty$, and it is different from the non-confining potential as shown in Fig.~\ref{Fig:2.3.132potl}.

Recently, the nodal structure of the wave functions obtained within the confining potential was demonstrated in an analytical way by Alberto, de Castro, and Malheiro \cite{Alberto2013_PRC87-031301R}.
Assuming the potential $\Delta(r)$ goes to infinity at large $r$ as a power law, $\lim_{r\rightarrow\infty}\Delta(r) = C r^a$, 
with $C>0$ and $a>0$, the wave functions $G(r)$ and $F(r)$ go to zero exponentially as $G(r) \propto F(r) \propto e^{-\lambda r^{a/2+1}}$.
In this case, $\lambda = 2\sqrt{C(\epsilon-M-\Sigma_0)}/(a+2)$ indicating that $\epsilon>M+\Sigma_0$ is the condition for the bound states.

For these single-particle bound states, the effective mass $M_-(r)$ is a positive constant, and the effective mass $M_+(r)$ is positive at the origin and becomes negative at large $r$, changing its sign at $r_0$.
Therefore, the asymptotic behaviors of their radial wave functions at small $r$ still have the properties shown in Eqs.~(\ref{Eq:3.1.GF0}).
In addition, Eqs.~(\ref{Eq:3.1.GFprime}), (\ref{Eq:3.1.GF0v2}), (\ref{Eq:3.1.GFatr1}), and (\ref{Eq:3.1.GFatr2}) also remain valid.
It follows that nodes of $G$ and $F$ can occur only where both $M_+(r)>0$ and $M_-(r) > 0$, as well as $G(r)F(r)$ is a decreasing function at the nodes of $F$ and an increasing function at the nodes of $G$.

Compared with the non-confining potentials, the difference happens in Eq.~(\ref{Eq:3.1.GFinftyv2}).
Within the confining potentials, $M_-(r)>0$ and $M_+(r)<0$ at large $r$, and thus the right hand side of Eq.~(\ref{Eq:3.1.GFprime}) is negative.
This indicates
\begin{equation}\label{Eq:3.1.GFinftycon2}
  \lim_{r\rightarrow\infty}G_\kappa(r)F_\kappa(r)>0\,,
\end{equation}
and thus if there exist internal nodes in the wave function the last one must belong to $G$.
Therefore, the number of internal nodes of $G$ and $F$, $n_G$ and $n_F$, obeys \cite{Alberto2013_PRC87-031301R}
\begin{equation}\label{Eq:3.1.GFnodescon}
  n_F = n_G - 1
  \quad\mbox{for}\quad
  \kappa<0\,,\qquad
  n_F = n_G
  \quad\mbox{for}\quad
  \kappa>0\,.
\end{equation}

\subsubsection{Exact PSS in confining potentials}

As pointed out above, the exact PSS can be fulfilled for the single-particle bound states in the Fermi sea within the confining potentials.
The first example was shown in Ref.~\cite{Chen2003_CPL20-358}.

\begin{figure}[tbhp]
\begin{center}
  \includegraphics[width=4cm]{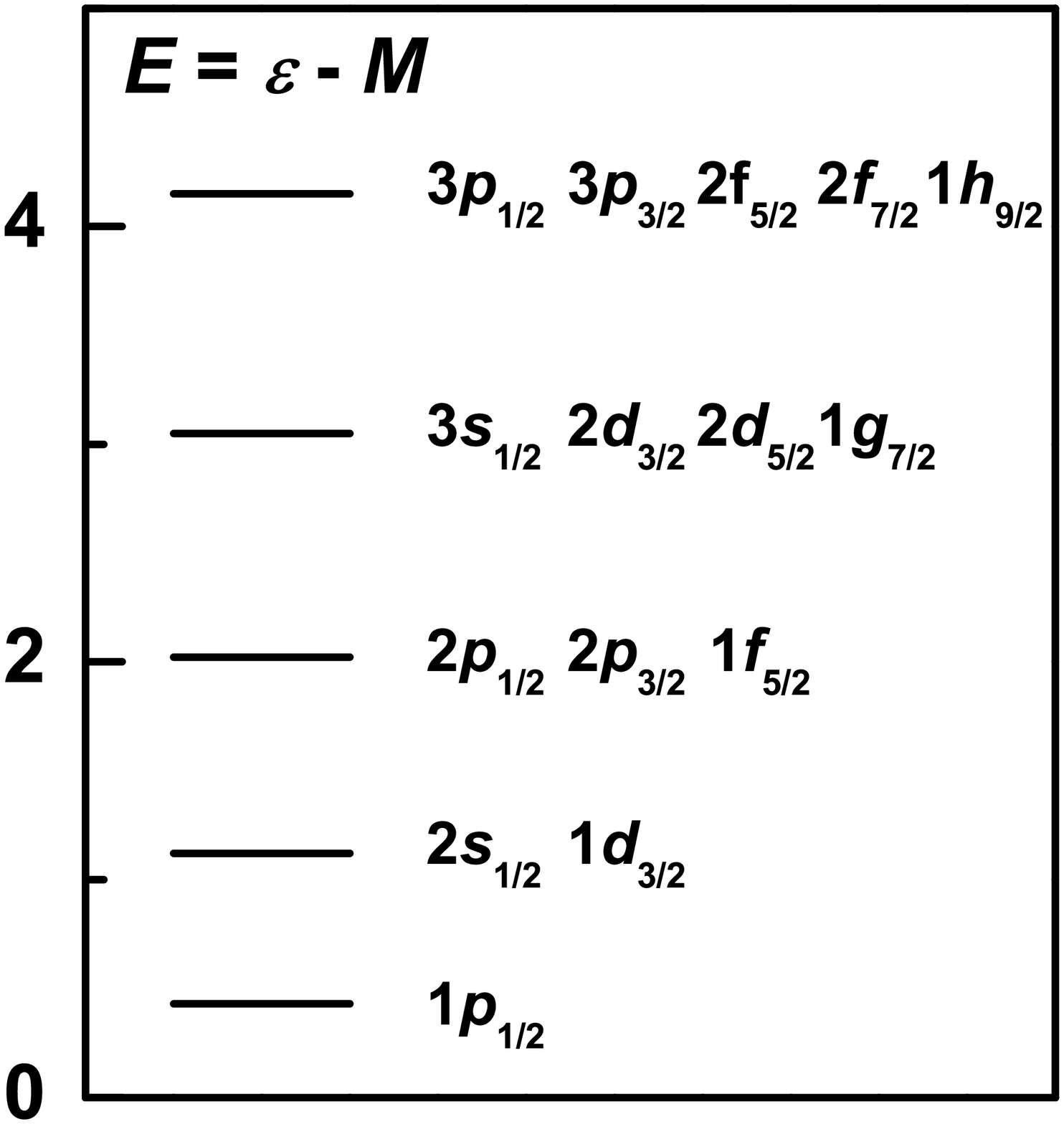}
  \includegraphics[width=4cm]{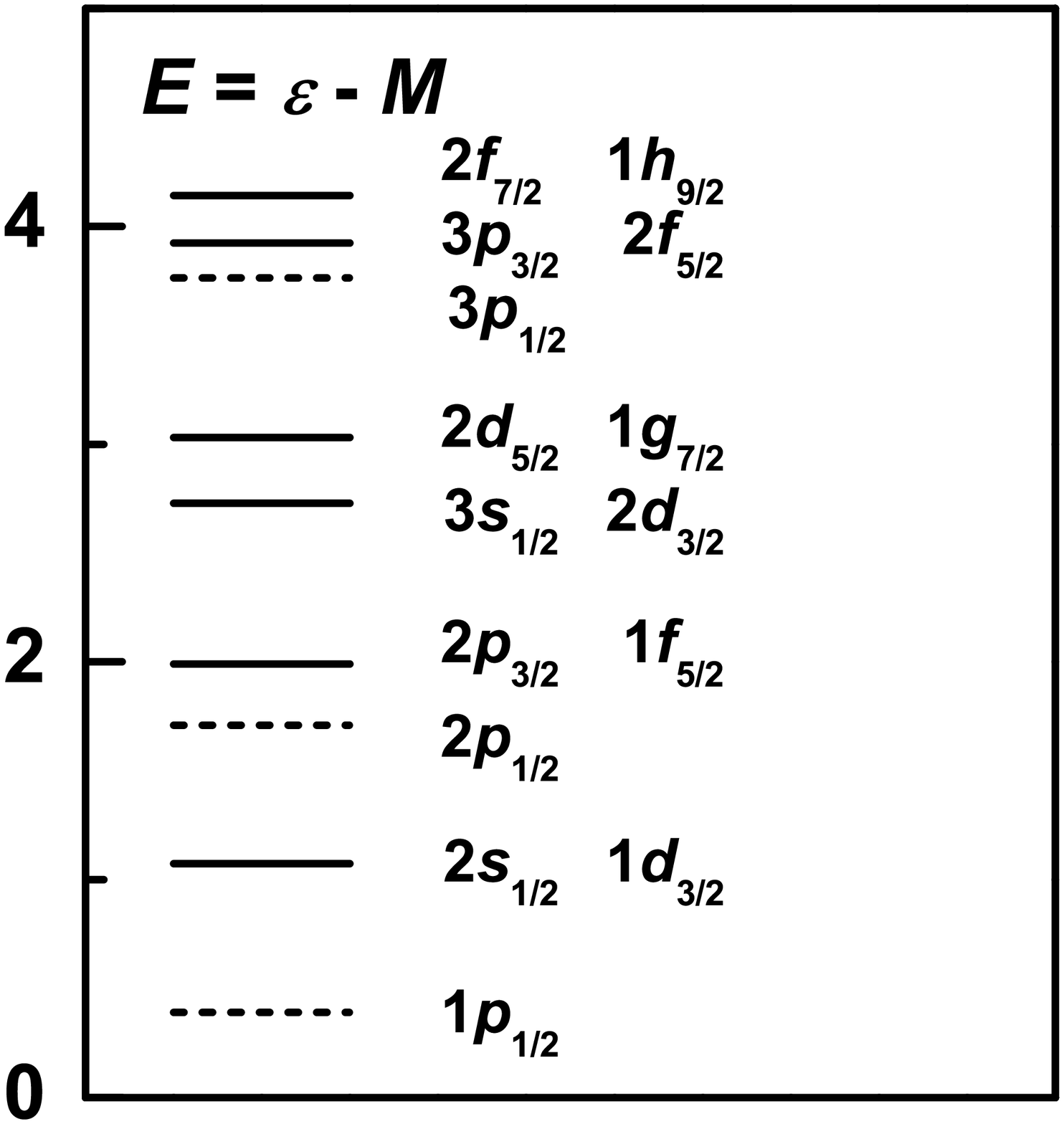}
\end{center}
\caption{Single-particle spectrum in units of fm$^{-1}$ for $V(r) = -S(r) = M\omega^2r^2/4$ (the left panel) and $V(r) = -S(r) = M\omega^2r^2/4 - V_0/[1+e^{(r-R)/a}]$ (the right panel) with $M=10.0$~fm$^{-1}$, $\omega=2.0$~fm$^{-1}$, $V_0 = 5.0$~fm$^{-1}$, $R = 0.3$~fm, and $a = 0.05$~fm.
The pseudospin singlets are shown with the dashed line, and the pseudospin doublets with the solid lines.
Taken from Ref.~\cite{Chen2003_CPL20-358} and modified to present notations.
\label{Fig:3.1.HOE}}
\end{figure}

In the left panel of Fig.~\ref{Fig:3.1.HOE}, the single-particle spectrum for $V(r) = -S(r) = M\omega^2r^2/4$ with $M=10.0$~fm$^{-1}$ and $\omega=2.0$~fm$^{-1}$ is shown.
Note that different from the figures in Ref.~\cite{Chen2003_CPL20-358}, here the main quantum number $n$ of each state equals the number of its internal nodes for the dominant component of the Dirac spinor plus one.
It can be seen that there are the exact pseudospin degeneracies for the pseudospin doublets, such as ($2s_{1/2}$, $1d_{3/2}$), ($2p_{3/2}$, $1f_{5/2}$), etc.
This spectrum has higher degeneracy than that of PSS due to the speciality of the harmonic oscillator potential.
To eliminate the extra degeneracies, an additional term of Woods-Saxon potential is introduced, $V(r) = -S(r) = M\omega^2r^2/4 - V_0/[1+e^{(r-R)/a}]$ with $V_0 = 5.0$~fm$^{-1}$, $R = 0.3$~fm, and $a = 0.05$~fm.
The single-particle spectrum thus obtained is shown in the right panel of Fig.~\ref{Fig:3.1.HOE}, where the pseudospin singlets are shown with the dashed lines, and the doublets with the solid lines.
By comparing these two panels, we see this new term reduces all redundant degeneracies of harmonic oscillator potential and keeps the pseudospin symmetry.

\begin{figure}[tbhp]
\begin{center}
  \includegraphics[width=8cm]{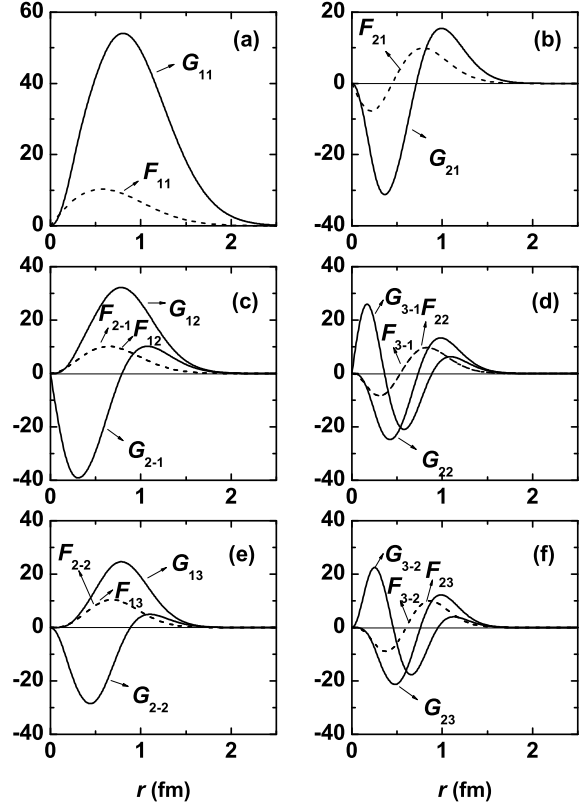}
\end{center}
\caption{Unnormalized radial wave functions for $V(r) = -S(r) = M\omega^2r^2/4 - V_0/[1+e^{(r-R)/a}]$ with $M=10.0$~fm$^{-1}$, $\omega=2.0$~fm$^{-1}$, $V_0 = 5.0$~fm$^{-1}$, $R = 0.3$~fm, and $a = 0.05$~fm.
Pseudospin singlet states: (a) $1p_{1/2}$ and (b) $2p_{1/2}$.
Pseudospin doublet states: (c) ($2s_{1/2}$, $1d_{3/2}$), (d) ($3s_{1/2}$, $2d_{3/2}$), (e) ($2p_{3/2}$, $1f_{5/2}$), and (f) ($3p_{3/2}$, $2f_{5/2}$).
Taken from Ref.~\cite{Chen2003_CPL20-358} and modified to present notations.
\label{Fig:3.1.HOWF}}
\end{figure}

The radial wave functions for the pseudospin singlets, $1p_{1/2}$ and $2p_{1/2}$, are shown in Fig.~\ref{Fig:3.1.HOWF}(a) and (b), where the dashed lines are for the lower components $F_{n\kappa}$ of the Dirac spinor and the solid lines for the upper components $G_{n\kappa}$.
Those for the pseudospin doublets, ($2s_{1/2}$, $1d_{3/2}$), ($3s_{1/2}$, $2d_{3/2}$), ($2p_{3/2}$, $1f_{5/2}$), and ($3p_{3/2}$, $2f_{5/2}$), are shown in Fig.~\ref{Fig:3.1.HOWF}(c)--(e).
It can be seen that for every pair of doublets, their lower components are identical as proven before Eq.~(\ref{Eq:2.3.FatPSS}).
Meanwhile, their upper components look very different since they obey the first-order differential relation shown in Eq.~(\ref{Eq:2.3.GatPSS}).

Furthermore, Fig.~\ref{Fig:3.1.HOWF} confirms that if there exist internal nodes in the radial wave functions, the last one belongs to the upper components $G$.
It is also seen that the upper and lower components are now in phase at large $r$, rather than out of phase for the cases of the non-confining potentials shown in Figs.~\ref{Fig:2.3.wf} and \ref{Fig:3.1.nodes}.

It is interesting and important to point out that within the confining potentials all of the $n=1$ eigenstates for the $j_> = l + 1/2$ ($\kappa<0$) orbitals disappear in the Fermi sea, such as $1s_{1/2}$, $1p_{3/2}$, $1d_{5/2}$, etc., as seen in Fig.~\ref{Fig:3.1.HOE}.
This is because the nodal structure (\ref{Eq:3.1.GFnodescon}) of the wave functions obtained within the confining potentials is different from the conventional nodal structure (\ref{Eq:3.1.GFnodes}) appearing in the isolated atomic nuclei.
For the $\kappa<0$ orbitals, it has been proven that $n_F = n_G - 1$, and since $n_F$ cannot be negative, all of such eigenstates have at least one internal node in their upper components of the Dirac spinor.
In other words, the single-particle spectra for $\kappa<0$ orbitals start from the main quantum number $n=2$.
As a result, there are no intruder states in the single-particle spectrum at all, and every $\tilde l >0$ state has its own pseudospin partner.

In summary, the nodal structure of the Dirac spinor can be derived analytically.
For the non-confining and confining potentials, the corresponding relations are shown in Eq.~(\ref{Eq:3.1.GFnodes}) and Eq.~(\ref{Eq:3.1.GFnodescon}), respectively.
This also explains the reason why at the pseudospin symmetry limit there are no bound states in the Fermi sea within the non-confining potentials, but there exist bound states within the confining potentials.
Meanwhile, it is interesting to see that all states with $\tilde l>0$ have their own pseudospin partners in the confining potentials, as shown in Fig.~\ref{Fig:3.1.HOE}, but this is not the case for the non-confining potentials appearing in the isolated atomic nuclei, as shown in Fig.~\ref{Fig:2.3.132spectra}.
We consider this as a puzzle of intruder states and will discuss in more detail in Section~\ref{Sect:4.2}.

\subsection{From local potentials to non-local potentials}\label{Sect:3.2}

So far, most of studies on the pseudospin symmetry mainly focus on the single-particle Hamiltonian with only local potentials.
L\'opez-Quelle \textit{et al.} \cite{Lopez-Quelle2003_NPA727-269} performed one of the first investigations of the pseudospin symmetry with non-local potentials in the framework of the relativistic Hartree-Fock theory \cite{Bouyssy1985_PRL55-1731,Bouyssy1987_PRC36-380}.
Recently, the relativistic Hartree-Fock theory achieved lots of success in describing nuclear ground-state and excited-state properties, by introducing the density-dependent meson-nucleon couplings by Long and coauthors \cite{Long2006_PLB640-150,Long2007_PRC76-034314,Long2008_EPL82-12001}.
Along this direction, more detailed investigations of the pseudospin symmetry in the relativistic Hartree-Fock theory were performed in Refs.~\cite{Long2006_PLB639-242,Long2010_PRC81-031302R}, as well as the spin symmetry in the anti-nucleon spectra in Ref.~\cite{Liang2010_EPJA44-119}.

In this Section, we will briefly introduce the relativistic Hartree-Fock theory, and then mainly focus on the special features of pseudospin symmetry due to the non-local potentials.
The spin symmetry in the anti-nucleon spectra will be discussed in Section~\ref{Sect:3.5}.

\subsubsection{Relativistic Hartree-Fock theory}

Similar to the RMF theory, the starting point of the RHF theory \cite{Bouyssy1985_PRL55-1731,Bouyssy1987_PRC36-380} is a Lagrangian density $\mathscr{L}$ (\ref{Eq:2.3.Lagrangian}), in which the nucleus is described as a system of Dirac nucleons that interact with each other via the exchange of mesons and photons.
With the general Legendre transformation (\ref{Eq:2.3.H_den}) as well as the mean-field and no-sea approximations (\ref{Eq:2.3.GS}), one has the energy density functional for the whole system,
\begin{equation}\label{Eq:3.2.EDF_RHF}
  E_{\rm RHF} = \lc\Phi_0\rl\mathscr{H}\lr\Phi_0\rc\,.
\end{equation}
However, different from the RMF theory, here both the direct and exchange contributions of the meson and Coulomb fields, the so-called Hartree and Fock terms, are included.

In such a way, the effective nucleon-nucleon tensor interactions can be naturally taken into account, which are practically absent at the Hartree level.
The corresponding parts in the Lagrangian density read \cite{Long2007_PRC76-034314,Long2008_EPL82-12001}
\begin{align}\label{Eq:3.2.Ltensor}
  \mathscr{L}^T_\rho &= \frac{f_\rho}{2M} \bar\psi \sigma^{\mu\nu} \partial_\nu \vec{\rho}_\mu \cdot \vec\tau\psi\,,\nonumber\\
  \mathscr{L}_\pi &= -\frac{f_\pi}{m_\pi} \bar\psi \gamma_5\gamma^\mu \partial_\mu \vec\pi \cdot \vec\tau \psi
    + \frac{1}{2} \partial_\mu \vec\pi \cdot \partial^\mu \vec\pi - \frac{1}{2} m^2_\pi \vec\pi \cdot \vec\pi\,,
\end{align}
corresponding to the $\rho$-$N$ tensor and $\pi$-$N$ pseudovector couplings.
These tensor interactions are crucial for improving the descriptions of the nuclear shell structures and their evolutions \cite{Long2007_PRC76-034314,Long2008_EPL82-12001,Tarpanov2008_PRC77-054316,Long2009_PLB680-428,Moreno-Torres2010_PRC81-064327}.

In recent studies, it has been also shown that the meson exchange terms play very important roles in the nucleon effective mass splitting \cite{Long2006_PLB640-150}, symmetry energies \cite{Sun2008_PRC78-065805}, halo structure \cite{Long2010_PRC81-031302R}, deformation \cite{Ebran2011_PRC83-064323}, and spin-isospin resonances \cite{Liang2008_PRL101-122502,Liang2012_PRC85-064302}.
Meanwhile, the Coulomb exchange term \cite{Gu2013_PRC87-041301} is crucial for understanding the isospin symmetry-breaking corrections to the superallowed $\beta$ decays \cite{Liang2009_PRC79-064316} in order to test the unitarity of the Cabibbo-Kobayashi-Maskawa matrix \cite{Towner2010_RPP73-046301}.

In the RHF theory, the eigenfunction equations for nucleons involve non-local potentials due to the finite-range meson and photon exchanges.
For the spherical case, the corresponding radial Dirac equations are expressed as the coupled integro-differential equations \cite{Bouyssy1987_PRC36-380},
\begin{equation}\label{Eq:3.2.DiracRHF}
    \lb\begin{array}{cc}
    M+\Sigma^D(r) & -\displaystyle\frac{d}{dr}+\displaystyle\frac{\kappa}{r} \\
        \displaystyle\frac{d}{dr}+\displaystyle\frac{\kappa}{r} & -M+\Delta^D(r)
    \end{array}\rb
    \lb\begin{array}{c}
        G(r) \\ F(r)
    \end{array}\rb
    +\lb\begin{array}{c}
        Y(r) \\ X(r)
    \end{array}\rb
    =\epsilon
    \lb\begin{array}{c}
        G(r) \\ F(r)
    \end{array}\rb\,,
\end{equation}
where $\Sigma^D$ and $\Delta^D$ contain the contributions from the Hartree terms as shown in Eq.~(\ref{Eq:2.1.DiraceqR}), while $X$ and $Y$ functions represent the results of the non-local Fock potentials acting on the wave functions $F$ and $G$, respectively.

\subsubsection{PSS in non-local potentials}

As shown in Refs.~\cite{Lopez-Quelle2003_NPA727-269,Long2006_PLB639-242,Long2010_PRC81-031302R}, in many cases the single-particle spectra obtained in the self-consistent RHF theory show a good PSS.
It was also found that the stability of neutron halo structures, e.g., in the drip-line Ce isotopes, is closely related to the PSS conservation of the single-proton spectrum.
In this PSS conservation the $\rho$-$N$ tensor coupling in Eq.~(\ref{Eq:3.2.Ltensor}) plays an essential role via the Fock terms \cite{Long2010_PRC81-031302R}.
However, in these cases, it is more sophisticated to trace the origin of the PSS and the exact PSS limit is no longer as simple as $d\Sigma^D(r)/dr=0$ (\ref{Eq:2.1.PSSlimit}) because of the non-local potentials \cite{Lopez-Quelle2003_NPA727-269}.

One possible way for tracing the origin of the PSS with non-local potentials was pointed out by Long \textit{et al.} \cite{Long2006_PLB639-242} by introducing the localized equivalent potentials,
\begin{align}\label{Eq:3.2.XYGF}
  X(r) &= \frac{G(r)X(r)}{G^2(r)+F^2(r)}G(r) + \frac{F(r)X(r)}{G^2(r)+F^2(r)}F(r)
    \equiv X_G(r)G(r) + X_F(r)F(r)\,,\nonumber\\
  Y(r) &= \frac{G(r)Y(r)}{G^2(r)+F^2(r)}G(r) + \frac{F(r)Y(r)}{G^2(r)+F^2(r)}F(r)
    \equiv Y_G(r)G(r) + Y_F(r)F(r)\,,
\end{align}
so that the radial Dirac equations (\ref{Eq:3.2.DiracRHF}) can be rewritten as
\begin{align}\label{Eq:3.2.Diracloc}
  \ls\frac{d}{dr} - \frac{\kappa}{r} - Y_F(r)\rs F(r) - \ls\Sigma(r)-E\rs G(r) &= 0\,,\nonumber\\
  \ls\frac{d}{dr} + \frac{\kappa}{r} + X_G(r)\rs G(r) + \ls\Delta(r)-E\rs F(r) &= 0\,,
\end{align}
where $E=\epsilon-M$, $\Sigma=\Sigma^D + Y_G$, and $\Delta = \Delta^D + X_F - 2M$.

The corresponding Schr\"odinger-like equation for the lower component $F(r)$ reads
\begin{equation}\label{Eq:3.2.SchrF}
  \frac{d^2}{dr^2}F + V_1\frac{d}{dr}F + (\mathcal{V}_{\rm PCB} + \mathcal{V}_{\rm PSO}+V_2)F = -(\Delta^D-E)(\Sigma^D-E)F\,,
\end{equation}
with
\begin{align}\label{Eq:3.2.Vs}
  V_1 &= (X_G-Y_F) - \frac{1}{\Sigma-E}\frac{d\Sigma}{dr}\,,\nonumber\\
  V_2 &= Y_F\frac{1}{\Sigma-E}\frac{d\Sigma}{dr} - X_GY_F - \frac{d}{dr}Y_F + Y_G(\Delta^D-E) + X_F(\Sigma-E)\,,\nonumber\\
  \mathcal{V}_{\rm PSO} &= \frac{\kappa}{r}\ls \frac{1}{\Sigma-E}\frac{d\Sigma}{dr} - (X_G + Y_F)\rs\,,\nonumber\\
  \mathcal{V}_{\rm PCB} &= \frac{\kappa(1-\kappa)}{r^2}\,,
\end{align}
where $\mathcal{V}_{\rm PCB}$ and $\mathcal{V}_{\rm PSO}$ correspond to the PCB and PSO potential, respectively. While the potential $V_2$ entirely originates from the Fock contributions, the Hartree and Fock contributions to $V_1$ and the PSO potential can be approximately separated as,
\begin{align}\label{Eq:3.2.V1VPSO}
  V^D_1 &= - \frac{1}{\Sigma-E}\frac{d\Sigma^D}{dr}\,,\qquad
  V^E_1 = (X_G-Y_F) - \frac{1}{\Sigma-E}\frac{dY_G}{dr}\,,\nonumber\\
  \mathcal{V}^D_{\rm PSO} &= \frac{\kappa}{r}\frac{1}{\Sigma-E}\frac{d\Sigma^D}{dr}\,,\qquad
  \mathcal{V}^E_{\rm PSO} = \frac{\kappa}{r}\ls \frac{1}{\Sigma-E}\frac{dY_G}{dr} - (X_G + Y_F)\rs\,,
\end{align}
denoted with superscripts ``$D$'' and ``$E$'', respectively.

Finally, to have a better understanding of the PSO splitting, especially the effects of non-local Fock terms, it will be more transparent to rewrite Eq.~(\ref{Eq:3.2.SchrF}) as \cite{Long2006_PLB639-242}
\begin{equation}\label{Eq:3.2.SchrF_HF}
  \frac{1}{\Delta^D-E}\frac{d^2}{dr^2}F + \frac{1}{\Delta^D-E}\ls \mathcal{V}_{\rm PCB} + \mathcal{V}^D + \mathcal{V}^E\rs F
  +\Sigma^D F = EF\,,
\end{equation}
where the operators $\mathcal{V}^D$ and $\mathcal{V}^E$ are respectively
\begin{equation}\label{Eq:3.2.VHF}
    \mathcal{V}^D = V_1^D\frac{d}{dr} + \mathcal{V}^D_{\rm PSO}\qquad\mbox{and}\qquad
    \mathcal{V}^E = V_1^E\frac{d}{dr} + \mathcal{V}^E_{\rm PSO} + V_2\,.
\end{equation}
From Eq.~(\ref{Eq:3.2.SchrF_HF}), one can estimate the contributions of the potentials $\mathcal{V}_{\rm PCB}$, $\mathcal{V}^D$, and $\mathcal{V}^E$ to the single-particle energy $E$.
For example, the PCB contribution can be evaluated by
\begin{equation}\label{Eq:3.2.Eeach}
  \left.\int_0^\infty\frac{\mathcal{V}_{\rm PCB}}{\Delta^D-E}F^2dr\right/\int_0^\infty F^2dr\,.
\end{equation}
Although there exists a singularity in $1/(\Delta^D-E)$, it has been proven that the principal values of these
integrals are finite due to the nodal structure of $F(r)$ \cite{Marcos2000_PRC62-054309,Alberto2002_PRC65-034307}.

\begin{table}
\begin{center}
\caption{Single-particle energies $E$ in $^{132}$Sn and the contributions from each term in the left-hand side of Eq.~(\ref{Eq:3.2.SchrF_HF}) given by the RHF theory with the effective interaction PKO1 \cite{Long2006_PLB640-150}, in comparison with those by the RMF theory with the effective interaction PKDD \cite{Long2004_PRC69-034319}.
All units are in MeV.
The data are taken from Ref.~\cite{Long2006_PLB639-242}.\label{Tab:3.2.Sn132}}
\begin{tabular}{@{}llrrrrrr@{}} \hline
     & State & $E$~~~ &  $F''$ & $\Sigma^D$ & $\mathcal{V}_{\rm PCB}$ & $\mathcal{V}^D$ & $\mathcal{V}^E$ \\ \hline
RHF  & $\nu 2s_{1/2}$ & $-31.41$ & $18.11$ & $-75.35$ &  $9.30$ & $-2.99$ & $19.51$ \\
PKO1 & $\nu 1d_{3/2}$ & $-34.90$ & $14.87$ & $-79.01$ &  $9.54$ &  $0.44$ & $19.26$ \\
     & $\nu 3s_{1/2}$ &  $-8.33$ & $34.25$ & $-72.00$ & $11.11$ &  $0.09$ & $18.22$ \\
     & $\nu 2d_{3/2}$ &  $-8.66$ & $31.93$ & $-73.96$ & $11.32$ &  $3.89$ & $18.17$ \\ \hline
RMF  & $\nu 2s_{1/2}$ & $-34.81$ & $21.86$ & $-64.65$ & $11.04$ & $-3.07$ & --- \\
PKDD & $\nu 1d_{3/2}$ & $-38.87$ & $18.17$ & $-68.08$ & $11.41$ & $-0.37$ & --- \\
     & $\nu 3s_{1/2}$ &  $-8.15$ & $40.13$ & $-61.97$ & $13.02$ &  $0.67$ & --- \\
     & $\nu 2d_{3/2}$ &  $-8.44$ & $37.65$ & $-63.75$ & $13.36$ &  $4.30$ & --- \\ \hline
\end{tabular}
\end{center}
\end{table}

\begin{figure}[tbhp]
\begin{center}
  \includegraphics[width=6cm]{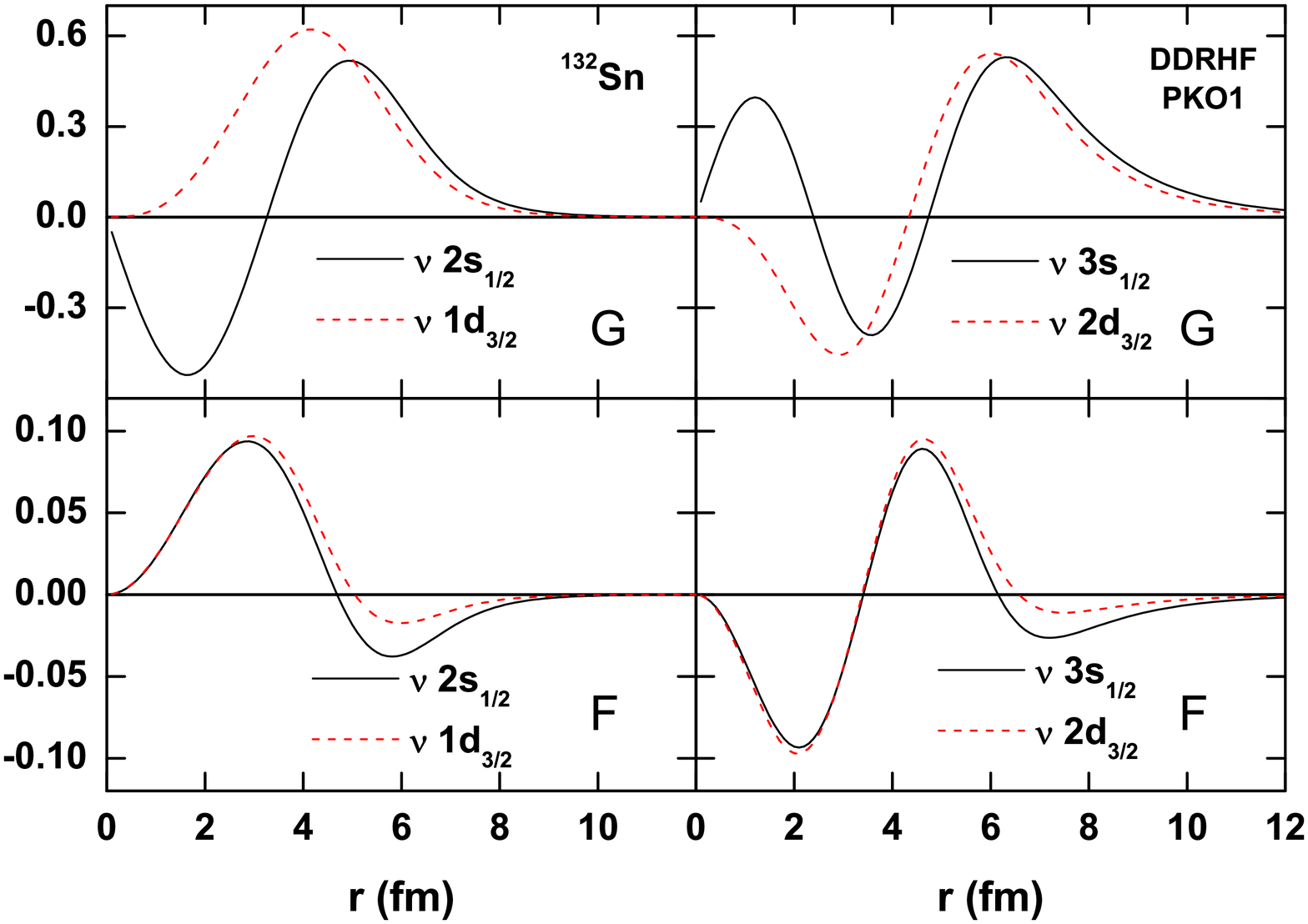}
  \includegraphics[width=6.15cm]{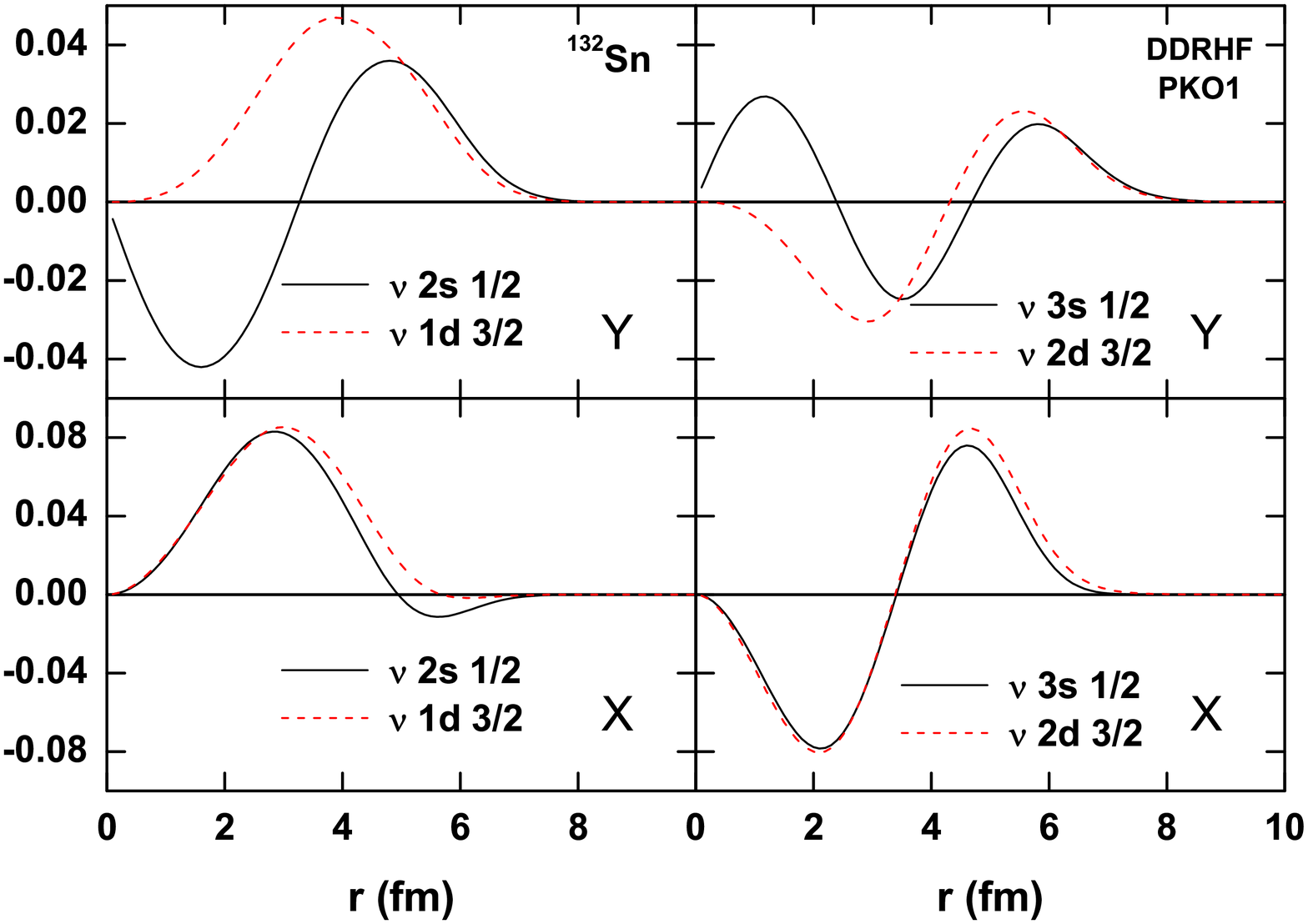}
\end{center}
\caption{(Color online) Radial wave functions $G$ and $F$ (left panels), and the non-local terms $X$ and $Y$ (right panels) for the $\nu1\tilde p$ and $\nu2\tilde p$ doublets in $^{132}$Sn given by the RHF theory with the effective interaction PKO1 \cite{Long2006_PLB640-150}.
Taken from Ref.~\cite{Long2006_PLB639-242}.
\label{Fig:3.2.GFXY}}
\end{figure}

In Table~\ref{Tab:3.2.Sn132}, the results calculated by the self-consistent RHF theory with the effective interaction PKO1 \cite{Long2006_PLB640-150} are shown for the neutron $1\tilde p$ and $2\tilde p$ pseudospin doublets in $^{132}$Sn.
For comparison, the corresponding results calculated by the RMF theory with the effective interaction PKDD \cite{Long2004_PRC69-034319} are shown in the lower part.
In addition, their radial wave functions $G(r)$ and $F(r)$ as well as the non-local terms $X(r)$ and $Y(r)$ given by the RHF theory are shown in Fig.~\ref{Fig:3.2.GFXY}.

First of all, the RHF and RMF theories share the same properties on the $F''$, $\Sigma^D$, $\mathcal{V}_{\rm PCB}$, and $\mathcal{V}^D$ terms.
The differences between the pseudospin partners in the PCB are very small, whereas the differences in the $F''$, $\Sigma^D$, and $\mathcal{V}^D$ terms are substantial individually.
However, these three terms cancel largely one another and the PSS is conserved very well.

\begin{figure}[tbhp]
\begin{center}
  \includegraphics[width=8cm]{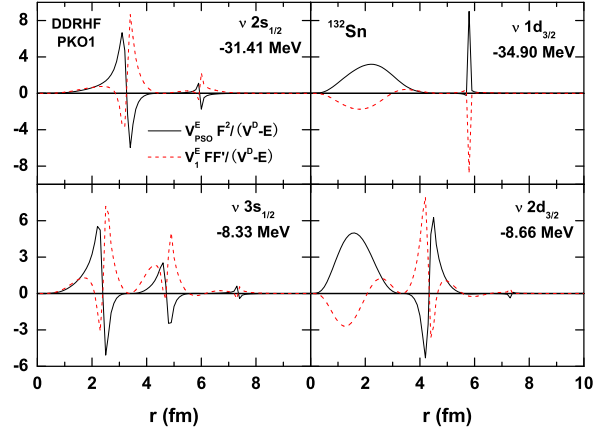}
\end{center}
\caption{(Color online) Functions $\mathcal{V}^E_{\rm PSO}F^2/(\Delta^D-E)$ and $V^E_1FF'/(\Delta^D-E)$ for the $\nu1\tilde p$ and $\nu2\tilde p$ doublets in $^{132}$Sn given by the Fock terms of the RHF theory with PKO1.
The singular points at $r\simeq6$~fm for $\nu1\tilde p$ and at $r\simeq7.5$~fm for $\nu2\tilde p$ are due to the denominator $(\Sigma-E)$ in the PSO potential, while the other local peaks are due to the nodes of the upper component $G$.
Taken from Ref.~\cite{Long2006_PLB639-242}.
\label{Fig:3.2.VE}}
\end{figure}

For the special features in the RHF theory, it is found that the gross differences in the Fock terms $\mathcal{V}^E$ are almost negligible.
The main reason is that there exist significant cancellations between $\mathcal{V}^E_{\rm PSO}$ and $V_1^E$ in Eq.~(\ref{Eq:3.2.VHF}), especially in the inner part of the nucleus.
The functions $\mathcal{V}^E_{\rm PSO}F^2/(\Delta^D-E)$ and $V^E_1FF'/(\Delta^D-E)$ for the neutron $1\tilde p$ and $2\tilde p$ doublets are illustrated in Fig.~\ref{Fig:3.2.VE}.
In other words, although the Fock terms bring substantial contributions to the PSO potential, these contributions are canceled by the other exchange term $V_1^E$, which stems mainly from the non-locality of the state-dependent exchange potentials.
Furthermore, it was pointed out that these cancellations are not accidental but because of the similar radial dependence between the Fock-related terms ($X$, $Y$) and the wave functions ($F$, $G$) \cite{Long2006_PLB639-242}, as shown in Fig.~\ref{Fig:3.2.GFXY}.

In summary, the Fock terms in the relativistic Hartree-Fock theory bring significant contributions to the pseudospin-orbit potential and make it comparable to the pseudo-centrifugal barrier.
However, these Fock terms in the pseudospin-orbit potential are counteracted by other exchange terms due to the non-locality of the exchange potentials.
The physical mechanism of these cancellations was discussed in relation to the similarity between the exchange potentials and the
Dirac wave functions.
Therefore, although the non-local potentials make the analysis more much complicated, they do not substantially violate the general features of pseudospin symmetry described by only using the local potentials.

\subsection{From central potentials to tensor potentials}\label{Sect:3.3}

The tensor component is a non-central contribution of the nucleon-nucleon interaction, which has been extensively discussed during the past decades.
\textit{Ab initio} calculations based on realistic nucleon-nucleon interactions have demonstrated the important role played by the bare tensor force in the description of the binding energy of nuclei \cite{Pudliner1997_PRC56-1720,Fabrocini1998_PRC57-1668}.
Its role has also been investigated for nuclear matter with realistic potentials \cite{Wiringa1988_PRC38-1010}.

With the rich spin-isospin structure of finite nuclei, it is reasonable to expect that some nuclear observables are sensitive to the tensor force.
For instance, the shell evolution and the modification of magic numbers far from stability are often discussed and interpreted in terms of tensor effects \cite{Sorlin2008_PPNP61-602}.
In the framework of the shell model, the tensor contribution in shell evolution has been explored and its effects have been underlined by Otsuka \textit{et al.} \cite{Otsuka2001_PRL87-082502,Otsuka2005_PRL95-232502,Otsuka2010_PRL104-012501}.
In the mean-field models, the tensor component has been also extensively discussed during the past years, including the non-relativistic Skyrme \cite{Brown2006_PRC74-061303R,Brink2007_PRC75-064311,Colo2007_PLB646-227,Lesinski2007_PRC76-014312,Grasso2013_PRC88-054328} and Gogny \cite{Otsuka2006_PRL97-162501,Anguiano2012_PRC86-054302} theories.
The readers are referred to Ref.~\cite{Sagawa2014_PPNP76-76} for a recent review.

As the covariant symmetry is conserved in the covariant density functional theory, the tensor interactions in this scheme usually indicate the interactions with Lorentz-type tensor couplings, i.e., those involving the Dirac matrix $\sigma^{\mu\nu}$ or $\gamma_5\gamma^\mu$.
The Lorentz-type tensor interactions lead to both central and non-central potentials, and the later ones correspond to the non-relativistic tensor components embedded in the covariant couplings.
The Lorentz-type tensor effects on the shell evolution have been discussed in the relativistic Hartree-Fock theory \cite{Long2006_PLB640-150,Long2007_PRC76-034314,Long2008_EPL82-12001}.
Furthermore, the tensor effects in the relativistic and non-relativistic theories were systematically compared in Refs.~\cite{Tarpanov2008_PRC77-054316,Moreno-Torres2010_PRC81-064327}.

In this Section, we will mainly focus on three different examples for illustrating the tensor effects on the pseudospin symmetry.
The first example is the relativistic harmonic oscillator with a linear tensor potential, which admits analytical solutions \cite{Lisboa2004_PRC69-024319}.
Self-consistently, the $\omega$-$N$ tensor coupling is one of the widely used ways to include the tensor potential for the covariant density functional theory in the Hartree level \cite{Furnstahl1998_NPA632-607}.
The corresponding effects on the pseudospin symmetry were investigated in Ref.~\cite{Alberto2005_PRC71-034313}.
Similarly, the $\omega$-$\bar\Lambda$ tensor effect on the spin symmetry in the single-$\bar\Lambda$ spectra will be discussed in Section~\ref{Sect:3.6}.
Finally, we will discuss the tensor effects in the relativistic Hartree-Fock theory \cite{Long2007_PRC76-034314}, which is considered as a more complete way to include the effective nucleon-nucleon tensor interactions in the scheme of covariant density functional theory.

\subsubsection{Linear tensor potential}

When the Lorentz-type tensor couplings of the vector mesons to nucleons are taken into account in the CDFT \cite{Furnstahl1998_NPA632-607}, the single-particle Dirac Hamiltonian $H$ in Eq.~(\ref{Eq:2.1.HDirac}) becomes
\begin{equation}\label{Eq:3.3.Hten1}
  H = \boldsymbol{\alpha}\cdot\mathbf{p} + \beta[M+S(\mathbf{r})]+V(\mathbf{r}) - i\beta\boldsymbol{\alpha}\cdot\mathbf{T}(\mathbf{r})\,,
\end{equation}
where $\mathbf{T}(\mathbf{r})$ is the tensor potential.
If the tensor potential is simply a function of the radial coordinate $r$, this Hamiltonian can be simplified as
\begin{equation}\label{Eq:3.3.Hten2}
  H = \boldsymbol{\alpha}\cdot\mathbf{p} + \beta[M+S(\mathbf{r})]+V(\mathbf{r}) - i\beta\boldsymbol{\alpha}\cdot\hat{\mathbf{r}}T(r)\,.
\end{equation}
When the spherical symmetry is adopted, the corresponding radial Dirac equation shown in Eq.~(\ref{Eq:2.1.DiraceqR}) becomes
\begin{equation}\label{Eq:3.3.Dirac}
    \lb\begin{array}{cc}
    M+\Sigma(r) & -\displaystyle\frac{d}{dr}+\displaystyle\frac{\kappa}{r}-T(r) \\
        \displaystyle\frac{d}{dr}+\displaystyle\frac{\kappa}{r}-T(r) & -M+\Delta(r)
    \end{array}\rb
    \lb\begin{array}{c}
        G(r) \\ F(r)
    \end{array}\rb
    =\epsilon
    \lb\begin{array}{c}
        G(r) \\ F(r)
    \end{array}\rb\,.
\end{equation}

As mentioned in Section~\ref{Sect:3.1}, there exist bound states in the Fermi sea at the PSS limit (\ref{Eq:2.1.PSSlimit}), $d\Sigma(r)/dr=0$, as long as the potential $\Delta(r)$ is confining.
An example shown in Figs.~\ref{Fig:3.1.HOE} and \ref{Fig:3.1.HOWF} corresponds to the case of relativistic harmonic oscillator (\ref{Eq:2.2.HO1}) with $\Sigma(r)=0$ and $\Delta(r)=M\omega_1 r^2/2$.
On top of this PSS limit, Lisboa \textit{et al.} \cite{Lisboa2004_PRC69-024319} investigated the effects of the tensor potential $T(r)$, which was assumed as a linear function of $r$,
\begin{equation}\label{Eq:3.3.Tlinear}
  T(r) = M\omega_2 r\,.
\end{equation}
The corresponding single-particle energies can be obtained analytically as \cite{Lisboa2004_PRC69-024319}
\begin{equation}\label{Eq:3.3.Elinear}
  \epsilon^2 -M^2 -(2\kappa+1)M\omega_2
  = \lb 2\tilde n + \tilde l + \frac{3}{2}\rb \sqrt{2M(\epsilon-M)\omega_1^2 + 4M^2\omega_2^2}\,.
\end{equation}
Here $\tilde n$ is the number of the internal nodes of $F(r)$, i.e., $n_F$.

\begin{figure}[tbhp]
\begin{center}
  \includegraphics[width=8cm]{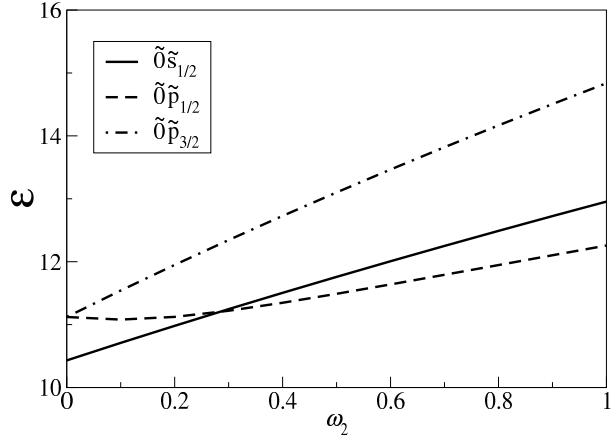}
\end{center}
\caption{Single-particle energies as a function of $\omega_2$ with $\Sigma(r)=0$, $\omega_1=2$, and $M=10$.
Here $\tilde n=n_F$ is the number of the internal nodes of $F(r)$.
Taken from Ref.~\cite{Lisboa2004_PRC69-024319}.
\label{Fig:3.3.linear}}
\end{figure}

In Fig.~\ref{Fig:3.3.linear}, the single-particle energies of $1\tilde s$ state and $1\tilde p$ pseudospin doublets are shown as a function of the tensor potential strength $\omega_2$.
It is shown that in general the single-particle energies increase with $\omega_2$, as $\omega^2_2$ appears in the right hand side of Eq.~(\ref{Eq:3.3.Elinear}).
More importantly, the pseudospin-up states with $j_>=\tilde l+1/2$ increase faster than the pseudospin-down states $j_<=\tilde l-1/2$.
This is because $\kappa>0$ ($\kappa<0$) for the $j_>$ ($j_<$) orbitals appearing in the left hand side of Eq.~(\ref{Eq:3.3.Elinear}) .

In this simple example, the SO splittings are not taken into account.
Therefore, the $j_>$ orbitals go higher in energy than the $j_<$ orbitals once the tensor potential is included.

\subsubsection{Tensor potential of $\omega$ meson}

For the CDFT in the Hartree level, i.e., the RMF theory, the $\omega$-$N$ tensor coupling is one of the widely used ways to include the tensor potential self-consistently \cite{Furnstahl1998_NPA632-607,Mao2003_PRC67-044318},
\begin{equation}\label{Eq:3.3.Lomega}
  \mathscr{L}^T_\omega = -\frac{f_\omega}{4M} \bar\psi \sigma^{\mu\nu}\Omega_{\mu\nu}\psi\,,
\end{equation}
where $f_\omega$ is the $\omega$-$N$ tensor coupling strength and $\Omega_{\mu\nu}$ is the field tensor defined in Eq.~(\ref{Eq:2.3.fieldT}).
In the spherical systems, the tensor potential thus obtained is
\begin{equation}\label{Eq:3.3.Tomega}
  T(r) = \frac{f_\omega}{2M}\omega'_0(r) \approx \frac{1}{2M}\frac{f_\omega}{g_\omega}V'(r)\,,
\end{equation}
which is approximately proportional to the derivative of the vector potential $V(r)$ if the contribution from the $\rho$ meson is negligible.
With this tensor potential the Schr\"odinger-like equation for the lower component $F(r)$ reads
\begin{equation}\label{Eq:3.3.SchrF}
  \Lb \frac{d^2}{dr^2} -\frac{\kappa(\kappa-1)}{r^2} -\frac{1}{M_-}\frac{dM_-}{dr} \ls \frac{d}{dr} - \frac{\kappa}{r} + T\rs  +2\kappa\frac{T}{r} + T' -T^2\Rb F(r) = -M_+M_-F(r)\,.
\end{equation}
In Ref.~\cite{Alberto2005_PRC71-034313}, this equation is further expressed as
\begin{align}\label{Eq:3.3.SchrF2}
  &p^2F(r) + \lb T^2 - T' - 2\frac{T}{r} - \frac{\Sigma'T}{E-\Sigma}\rb F(r) -\frac{\Sigma'}{E-\Sigma} \frac{dF(r)}{dr} +\lb -4T +2\frac{\Sigma'}{E-\Sigma}\rb\frac{\mathbf{L}\cdot \mathbf{S}}{r}F(r) \nonumber\\
  =& (E-\Sigma)(E+2M-\Delta)F(r)\,.
\end{align}
By dividing it by $2M^*=E+2M-\Delta$, the energy decomposition of the above equation reads \cite{Alberto2005_PRC71-034313}
\begin{equation}\label{Eq:3.3.Tdecom}
  \lc \frac{p^2}{2M^*}\rc +\lc V_T\rc +\lc V_{\Sigma'T}\rc + \lc V_{\rm Darwin}\rc + \lc V_{\rm PSO}\rc + \lc \Sigma\rc = E\,,
\end{equation}
where
\begin{equation}
  \lc V_i \rc \equiv \frac{\int F^* V_i F dr}{\int F^*Fdr}\,,
\end{equation}
and
\begin{align}\label{Eq:3.3.Vi}
  V_T &= \frac{1}{2M^*}\lb T^2 - T' - 2\frac{T}{r}\rb\,,\nonumber\\
  V_{\Sigma'T} &= -\frac{1}{2M^*}\frac{\Sigma'T}{E-\Sigma}\,,\nonumber\\
  V_{\rm Darwin}&= -\frac{1}{2M^*}\frac{\Sigma'}{E-\Sigma}\frac{d}{dr}\,,\nonumber\\
  V_{\rm PSO} &= \frac{1}{2M^*}\lb -2T +\frac{\Sigma'}{E-\Sigma}\rb \frac{\mathbf{L}\cdot \mathbf{S}}{r}\,,
\end{align}
which correspond to the tensor, Darwin, and PSO contributions, respectively.
For the terms with $E-\Sigma$ in the denominator, the integral is taken in the principal value sense.

\begin{table}
\begin{center}
\caption{Single-particle energies $E$ of $1\tilde p$ and $1\tilde{h}$ pseudospin doublets in $^{208}$Pb and the contributions from each term in the left-hand side of Eq.~(\ref{Eq:3.3.Tdecom}) calculated by the Woods-Saxon potentials without and with tensor potential.
The Woods-Saxon parameters are $R=7$~fm, $\Delta_0=650$~MeV, $\Sigma_0=-66$~MeV, and $a=0.6$~fm.
All units are in MeV.
The data are taken from Ref.~\cite{Alberto2005_PRC71-034313}.
\label{Tab:3.3.omega}}
\begin{tabular}{@{}clrrrrrrr@{}} \hline
$f_\omega/g_\omega$ & State & $\lc p^2/2M^*\rc$ & $\lc V_T \rc$ & $\lc V_{\Sigma'T} \rc$ & $\lc V_{\rm Darwin} \rc$ & $\lc V_{\rm PSO} \rc$ & $\lc \Sigma \rc$ & $E$\\ \hline
$0.0$ & $2s_{1/2}$  & $24.4396$ & $0.0000$ & $0.0000$ & $-3.9527$ & $-0.5852$ & $-61.4644$ & $-41.5627$ \\
      & $1d_{3/2}$ &  $21.1032$ & $0.0000$ & $0.0000$ & $-0.8106$ &  $0.0966$ & $-64.4159$ & $-44.0266$ \\
      & $\Delta E$ &  $ 3.3364$ & $0.0000$ & $0.0000$ & $-3.1421$ & $-0.6818$ &   $2.9515$ &   $2.4639$ \\ \hline
$1.3$ & $2s_{1/2}$  & $23.9037$ & $0.2351$ & $-0.0943$& $-3.6870$ & $-0.5632$ & $-61.9114$ & $-42.1170$ \\
      & $1d_{3/2}$ &  $20.7075$ & $0.1459$ & $-0.0712$& $-0.4170$ & $ 0.1559$ & $-64.6181$ & $-44.1170$ \\
      & $\Delta E$ &  $ 3.1962$ & $0.0892$ & $-0.0232$& $-3.2700$ & $-0.7191$ &   $2.7067$ &   $2.0000$ \\ \hline
$0.0$ & $2g_{9/2}$  & $55.7666$ & $0.0000$ & $0.0000$ &  $2.8295$ & $-6.4555$ & $-53.3816$ &  $-1.2410$ \\
      & $1i_{11/2}$ & $51.2033$ & $0.0000$ & $0.0000$ &  $3.1788$ &  $3.3530$ & $-61.0308$ &  $-3.2958$ \\
      & $\Delta E$ &  $ 4.5633$ & $0.0000$ & $0.0000$ & $-0.3493$ & $-9.8085$ &   $7.6492$ &   $2.0548$ \\ \hline
$1.3$ & $2g_{9/2}$  & $56.0706$ &  $0.5622$& $0.2855$ &  $2.6487$ & $-7.2110$ & $-54.4705$ &  $-2.1140$ \\
      & $1i_{11/2}$ & $51.6221$ &  $0.8997$& $0.1529$ &  $2.9268$ &  $4.1391$ & $-61.2232$ &  $-1.4826$ \\
      & $\Delta E$ &  $ 4.4485$ & $-0.3375$& $0.1326$ & $-0.2781$ &$-11.3501$ &   $6.7527$ &  $-0.6314$ \\ \hline
\end{tabular}
\end{center}
\end{table}

As an example, the single-particle energies of $1\tilde p$ and $1\tilde{h}$ pseudospin doublets in $^{208}$Pb and the contributions from each term in the left-hand side of Eq.~(\ref{Eq:3.3.Tdecom}) are calculated with Woods-Saxon potentials.
The results without and with tensor potential in Eq.~(\ref{Eq:3.3.Tomega}) are compared in Table~\ref{Tab:3.3.omega}.

First of all, because the tensor interaction is a higher order interaction in the Lagrangian scaled by $1/M$, the contributions of the $V_T$ and $V_{\Sigma'T}$ terms in energy are small compared to the others.
However, because the SO or PSO interaction is a term of the same order, the tensor effects on the SO and PSO splittings can be significant.

Focusing on the pseudospin-orbit splittings $\Delta E_{\rm PSO}$, for the deeply bound pseudospin doublets, the contribution of the PSO potential $V_{\rm PSO}$ almost does not change with $f_\omega$, whereas the contribution of terms such as $\lc p^2/2M^*\rc$ and $\lc \Sigma\rc$, which do not depend explicitly on the tensor potential $T$, is larger than that of $V_{\rm PSO}$.
This means that the main contribution to the change of $\Delta E_{\rm PSO}$ is mainly due to the change of the wave function induced by $T$.
Furthermore, the energy splitting results from a strong cancellation of the $\Delta\lc\Sigma\rc$, $\Delta\lc V_{\rm Darwin} \rc$, and $\Delta\lc p^2/2M^*\rc$ contributions \cite{Alberto2005_PRC71-034313}.

For the loosely bound pseudospin doublets, much of the previous analysis still holds, except for the fact that the PSO potential $V_{\rm PSO}$ is much stronger and also changes sensibly as $f_\omega$ changes, being responsible for most of the pseudospin splitting.
For the $1\tilde h$ doublets, the $\Delta E_{\rm PSO}$ even becomes negative when $f_\omega/g_\omega$ greater than $0.95$.
It is found that the tensor part of $V_{\rm PSO}$ in Eq.~(\ref{Eq:3.3.Vi}) is the dominant ingredient to understand this evolution \cite{Alberto2005_PRC71-034313}.
Because the tensor potential here is proportional to the derivative of the vector potential, such tensor effects are much more profound for the surface states than the deeply bound ones.

In short, the pseudospin-orbit splittings $\Delta E_{\rm PSO} = E_{j_<} - E_{j_>}$ decrease, i.e., the PSS is conserved better, when the $\omega$-$N$ tensor interaction is included.
Due to the surface-peaked shape of the tensor potential, the tensor effects are much more profound for the loosely bound states than the deeply bound states.

\subsubsection{Tensor potential of $\rho$ and $\pi$ mesons}

As shown in Section~\ref{Sect:3.2}, a more complete way to include the effective nucleon-nucleon tensor interactions in the CDFT scheme is the RHF theory \cite{Long2007_PRC76-034314}.
The $\rho$-$N$ tensor and $\pi$-$N$ pseudovector couplings shown in Eq.~(\ref{Eq:3.2.Ltensor}) are crucial for improving the descriptions of the nuclear shell structures and their evolutions \cite{Long2007_PRC76-034314,Long2008_EPL82-12001,Tarpanov2008_PRC77-054316,Long2009_PLB680-428,Moreno-Torres2010_PRC81-064327}.
In particular, these tensor interactions play important roles via the exchange terms.

\begin{figure}[tbhp]
\begin{center}
  \includegraphics[width=6cm]{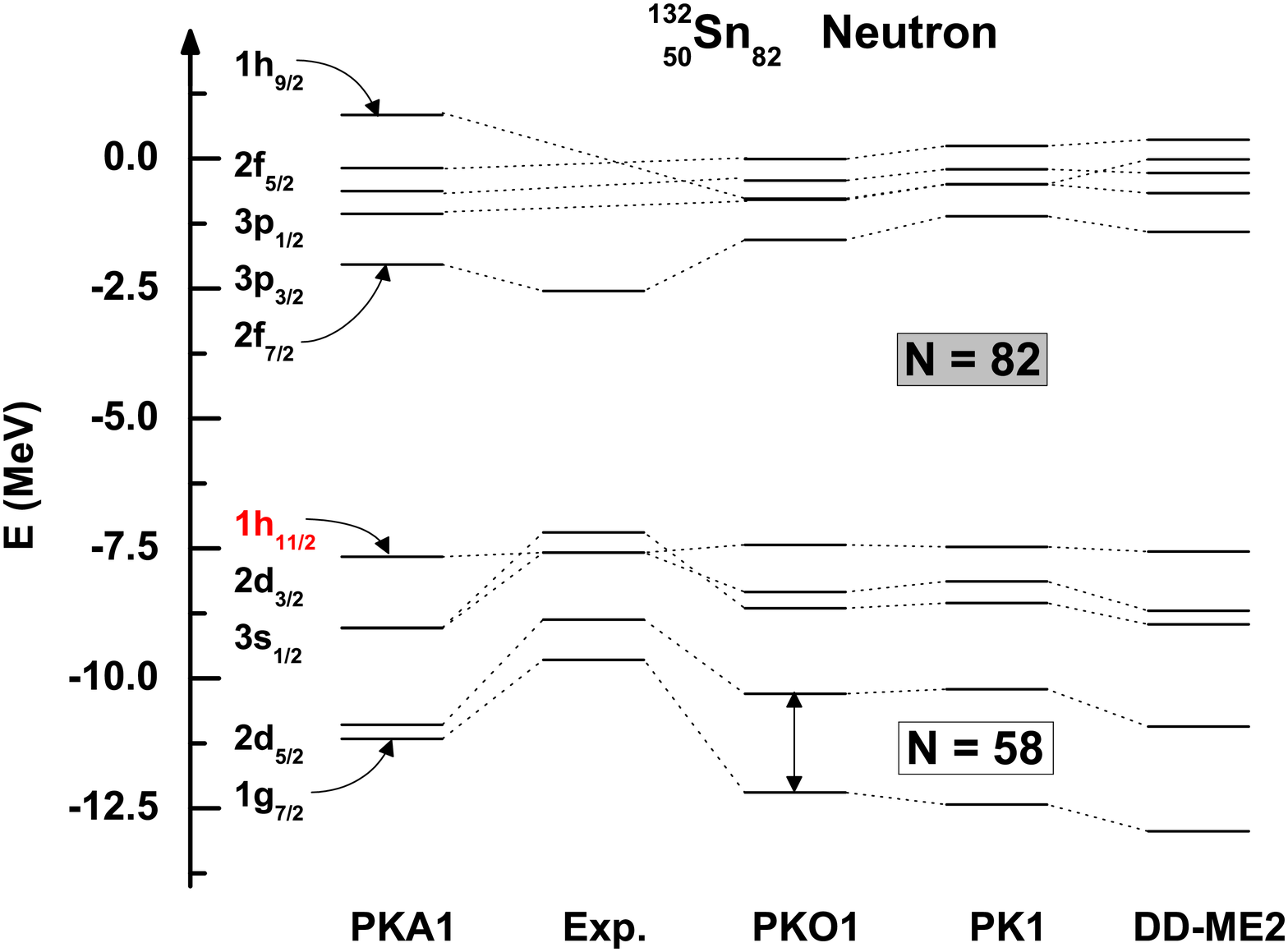}
  \includegraphics[width=6cm]{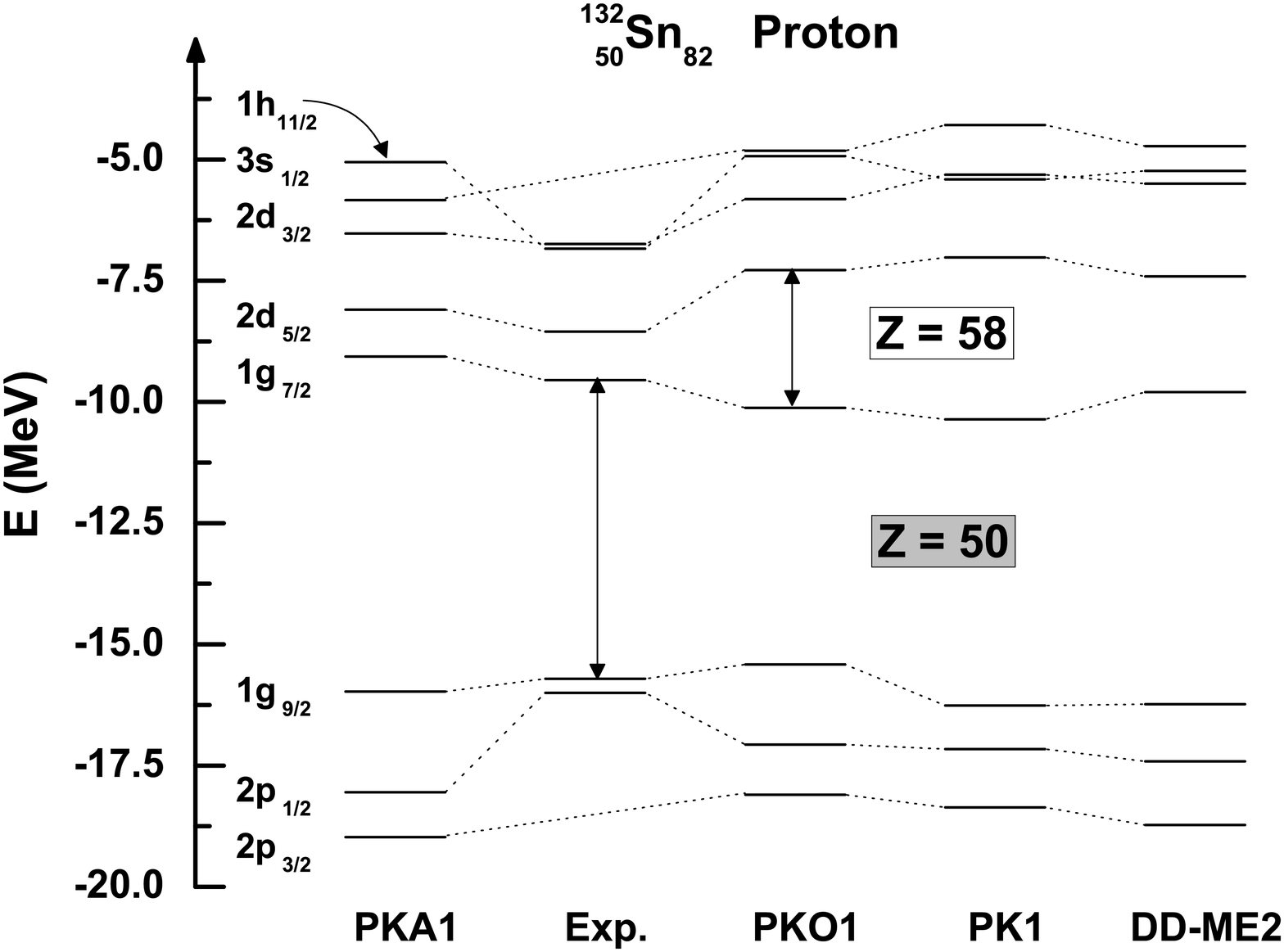}
\end{center}
\caption{Single-particle energies in $^{132}$Sn calculated by the RHF theory with the effective interactions PKA1 \cite{Long2007_PRC76-034314} and PKO1 \cite{Long2006_PLB640-150}, and by the RMF theory with the effective interactions PK1 \cite{Long2004_PRC69-034319} and DD-ME2 \cite{Lalazissis2005_PRC71-024312}, compared with the experimental data \cite{Oros1996PhD}.
Taken from Ref.~\cite{Long2007_PRC76-034314}.
\label{Fig:3.3.Sn132}}
\end{figure}

\begin{figure}[tbhp]
\begin{center}
  \includegraphics[width=6cm]{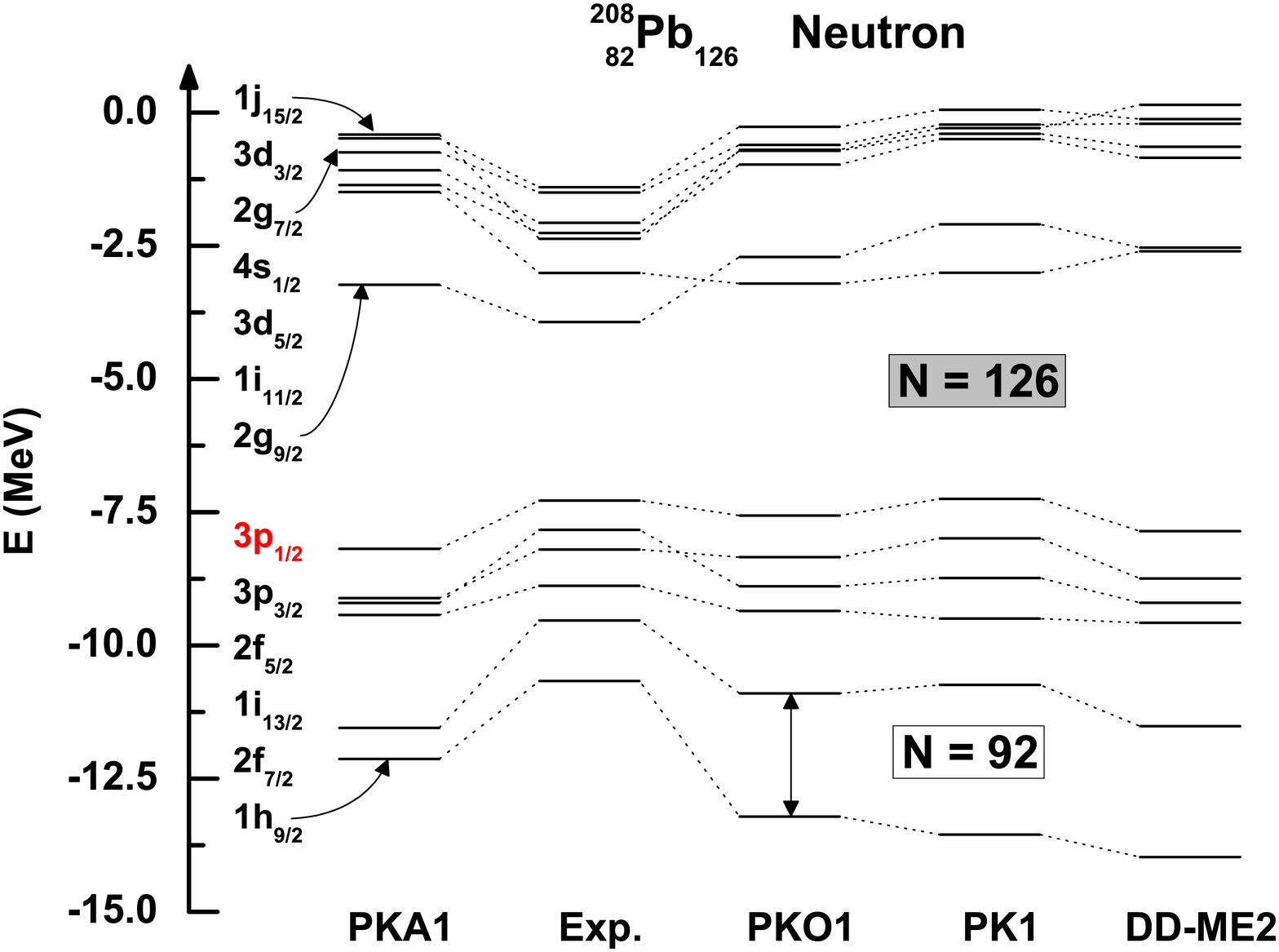}
  \includegraphics[width=6cm]{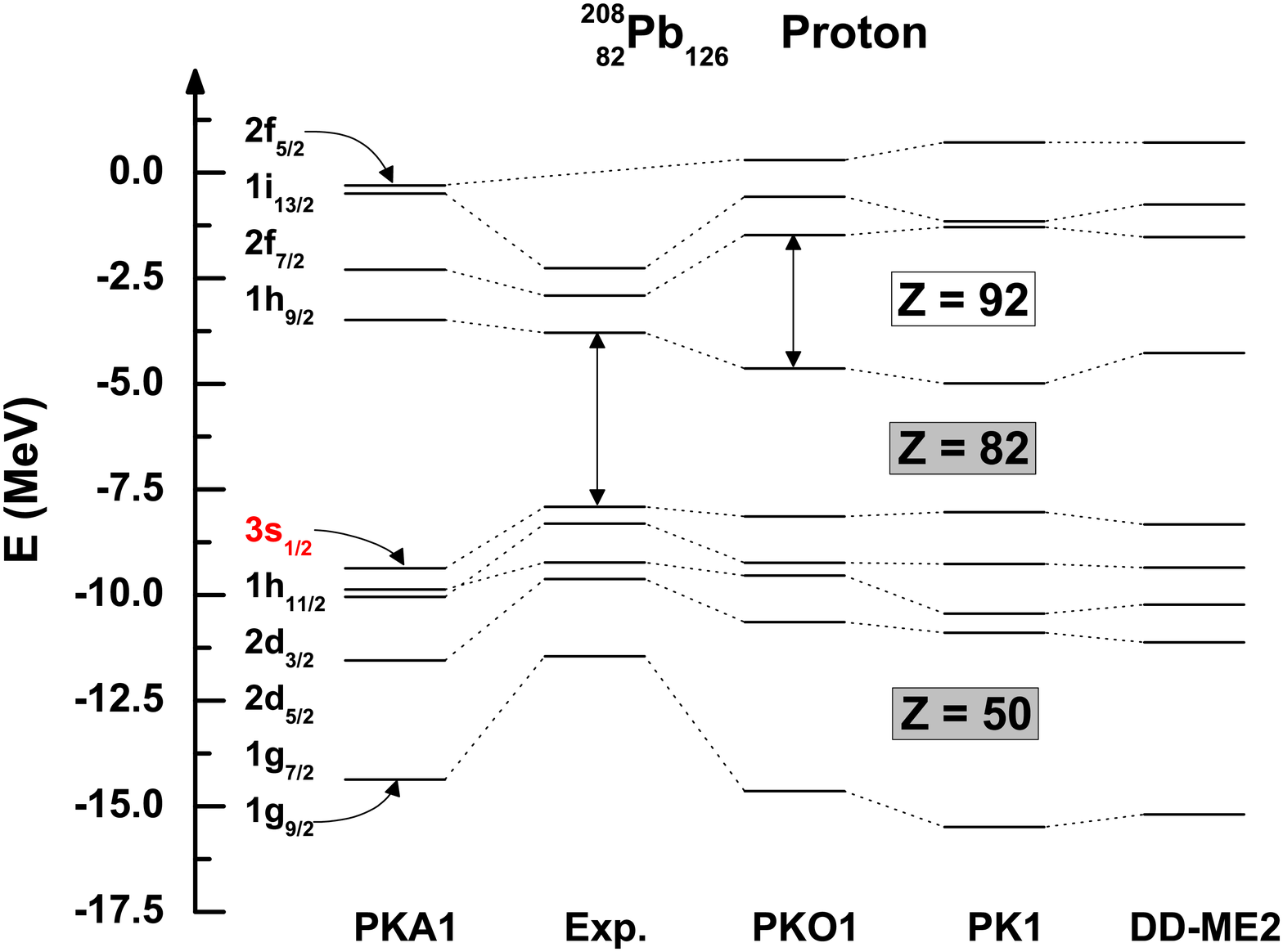}
\end{center}
\caption{Same as Fig.~\ref{Fig:3.3.Sn132}, but for the single-particle energies in $^{208}$Pb.
Taken from Ref.~\cite{Long2007_PRC76-034314}.
\label{Fig:3.3.Pb208}}
\end{figure}

In Figs.~\ref{Fig:3.3.Sn132} and \ref{Fig:3.3.Pb208}, the single-particle energies in $^{132}$Sn and $^{208}$Pb calculated by the self-consistent RHF theory with  the effective interactions PKA1 \cite{Long2007_PRC76-034314} and PKO1 \cite{Long2006_PLB640-150} effective interactions are compared with the corresponding results calculated by the RMF theory with the effective interactions PK1 \cite{Long2004_PRC69-034319} and DD-ME2 \cite{Lalazissis2005_PRC71-024312}.
Note that both $\rho$-$N$ and $\pi$-$N$ tensor couplings are included in PKA1, and only $\pi$-$N$ tensor coupling is included in PKO1, while the tensor effects are practically absent in the RMF models.

Compared with the experimental data \cite{Oros1996PhD}, the RMF results always show too large PSO splittings between the $2d_{5/2}$ and $1g_{7/2}$ states in $^{132}$Sn as well as the splittings between the $2f_{7/2}$ and $1h_{9/2}$ states in $^{208}$Pb.
This is the case not only for the neutron side but also for the proton side.
As a result, the overestimated sub-shell structures, e.g., $N$ or $Z =58$ and $N$ or $Z=92$, appear commonly in the RMF calculations without tensor interactions \cite{Geng2006_CPL23-1139}.

The artificial sub-shell closure is depressed to some extent by using the RHF theory with PKO1 effective interaction.
It is found the PSO splittings of the $1\tilde f$ and $1\tilde g$ doublets decrease compared to the RMF results.
However, because of the strong density dependence of $\pi$-$N$ coupling strength $f_\pi(\rho)$ in PKO1, the net tensor effects are mild, and the calculated PSO splittings are still substantially larger than the experimental ones.

The PSO splittings of the $1\tilde f$ and $1\tilde g$ doublets decrease further when both the $\pi$-$N$ and $\rho$-$N$ tensor couplings are included, which is the case of PKA1, and eventually the artificial sub-shells disappear.
Indeed, these tensor effects can be seen in all the pseudospin doublets shown in the figures by comparing the PKA1 results to the others.
This is also crucial for the descriptions of the ordering of the single-particle levels, e.g., the neutron states $2g_{9/2}$ and $1i_{11/2}$ in $^{208}$Pb.

In summary, three different models of tensor forces were discussed in this Section.
From a phenomenological linear tensor potential to more microscopic meson-nucleon tensor couplings, these models show consistent effects on the pseudospin symmetry, i.e., $\Delta E_{\rm PSO} = E_{j_<} - E_{j_>}$ decrease or become even negative when the strengths of tensor interactions increase.
This mechanism plays an important role in the nuclear shell structure and its evolutions.
For example, the overestimated sub-shell closure at $N$ or $Z= 58$ and $N$ or $Z= 92$ can be better reproduced.

\subsection{From bound states to resonant states}\label{Sect:3.4}

Weakly bound or unbound nuclei with unusual $N/Z$ ratios are
open quantum many-body systems in which the continuum
plays an important role \cite{Michel2009_JPG36-013101}.
In these nuclei,
the neutron (or proton) Fermi surface is close to the particle threshold,
thus the contribution of the continuum is crucial \cite{Dobaczewski1984_NPA422-103,
Dobaczewski1996_PRC53-2809,
Meng1996_PRL77-3963, Meng1998_PRL80-460, Meng1998_NPA635-3,
Meng2002_PRC65-041302R, Meng2006_PPNP57-470, Meng2011_SciChinaPMA54S1-119,
Poschl1997_PRL79-3841, Vretenar2005_PR409-101,
Zhou2010_PRC82-011301R, Li2012_PRC85-024312, Li2012_CPL29-042101, Li2012_AIPCP1491-208,
Zhang2011_PRC83-054301, Zhang2012_PRC86-054318,
Pei2008_PRC78-064306, Pei2011_PRC84-024311, Pei2013_PRC87-051302R,
He2011_SciChinaPMA54S1-32,
Lin2011_SciChinaPMA54S1-73,
Lu2011_SciChinaPMA54S1-136}.
Many approaches developed for resonances \cite{Kukulin1989},
e.g., the analytical continuation in coupling constant
method \cite{Yang2001_CPL18-196, Zhang2004_PRC70-034308,
 Zhang2007_EPJA32-43, Zhang2009_IJMPE18-1761, Zhang2012_PRC86-032802, Zhang2012_EPJA48-40},
the real stabilization method \cite{Zhang2008_PRC77-014312,
 Zhang2007_APS56-3839, Zhou2009_JPB42-245001, Mei2009_ChinPC33S1-101,
Zhang2009_ChinPC33-187, Zhang2010_MPLA25-727},
the complex scaling method \cite{Guo2010_PRC82-034318, Guo2010_IJMPE19-1357,
 Guo2010_CPC181-550, Liu2012_PRC86-054312},
the coupled-channel method \cite{Hagino2004_NPA735-55, Li2010_PRC81-034311, Li2010_SCG53-773},
and some others \cite{Fedorov2009_FBS45-191, Fernandez2012_AppMathComp218-5961},
have been used to
study nuclear single-particle resonant states.
Based on some of these methods, the pseudospin symmetry \cite{Guo2005_PRC72-054319, Guo2006_PRC74-024320, Liu2013_PRA87-052122,
Zhang2006_HEPNP30S2-97, Zhang2007_CPL24-1199} and
the spin symmetry \cite{Xu2012_IJMPEE21-1250096} in single-particle resonant states have been investigated.
The pseudospin symmetry and/or spin symmetry in nucleon-nucleus and nucleon-nucleon scatterings have
been also investigated \cite{Ginocchio1999_PRL82-4599, Ginocchio2002_PRC65-054002,
Leeb2000_PRC62-024602, Leeb2004_PRC69-054608}.

In Ref.~\cite{Lu2012_PRL109-072501}, by examining the asymptotic behaviors of the Dirac wave functions,
a rigorous verification of the pseudospin symmetry in single-particle
resonant states was given by Lu, Zhao, and Zhou.
It was shown that the pseudospin symmetry in single-particle resonant states in nuclei
is exactly conserved under the same condition for the pseudospin symmetry in bound states,
i.e., $\Sigma(r)=0$ or $d\Sigma(r)/dr=0$ given in Eq.~(\ref{Eq:2.1.PSSlimit}).
To understand more deeply the pseudospin symmetry in single-particle resonant states,
in Ref.~\cite{Lu2013_PRC88-024323}, the exact conservation and breaking mechanism
of the pseudospin symmetry in single-particle resonant states in square well potentials
were extensively studied.
A threshold effect in the energy splitting and an anomaly in the width splitting
of pseudospin partners were found when the depth of the potential varies
from zero to a finite value.

In this Section, we will
introduce the pseudospin symmetry in single particle resonant states.
For the pseudospin symmetry in nucleon-nucleus and nucleon-nucleon
scatterings \cite{Ginocchio1999_PRL82-4599,Leeb2000_PRC62-024602,
Ginocchio2002_PRC65-054002,Leeb2004_PRC69-054608}, see Ref.~\cite{Ginocchio2005_PR414-165}.

\subsubsection{PSS in single-particle resonant states with ACCC}

\begin{figure}[tbhp]
\begin{center}
  \includegraphics[width=5.5cm]{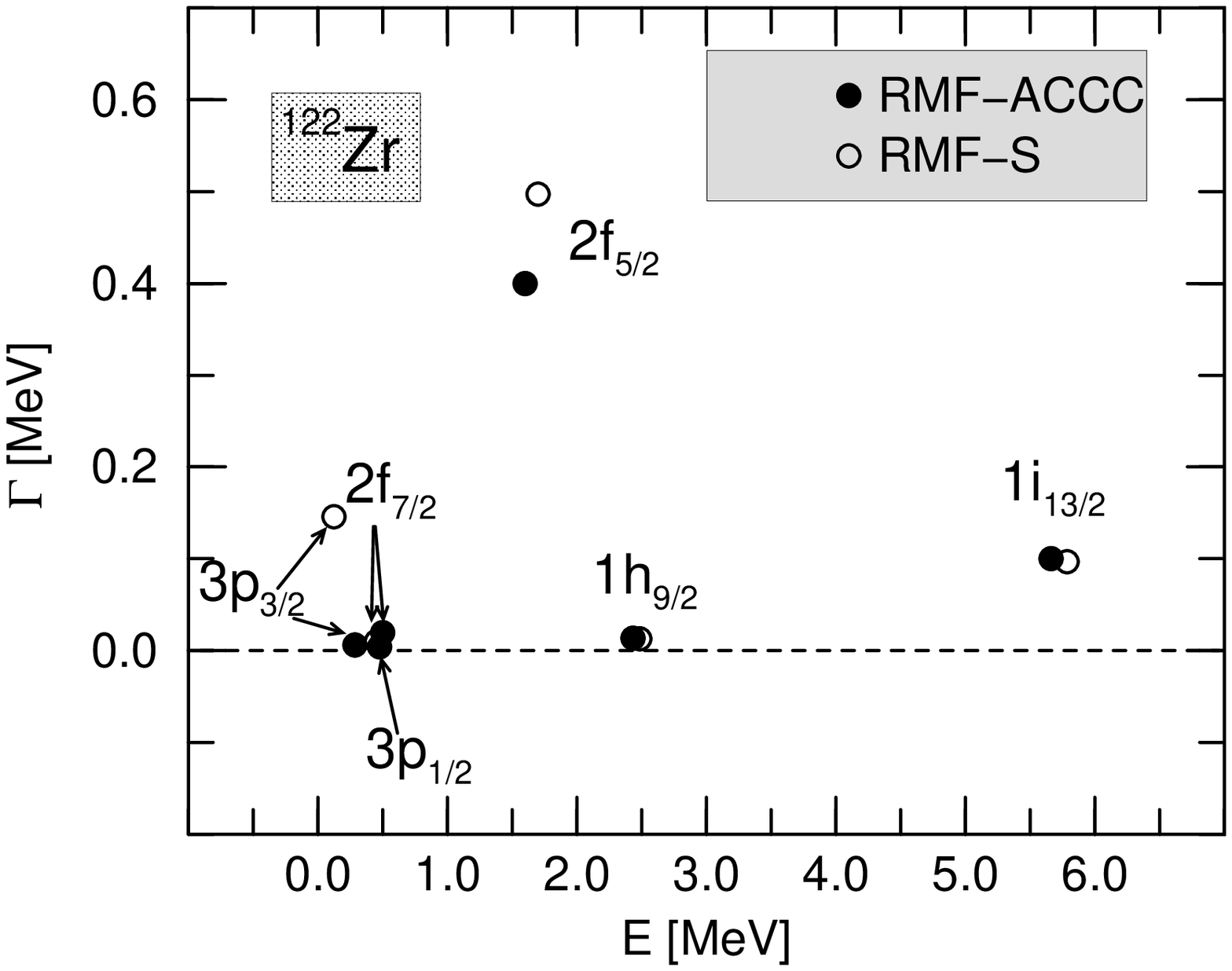}\hspace{2em}
  \includegraphics[width=6.2cm]{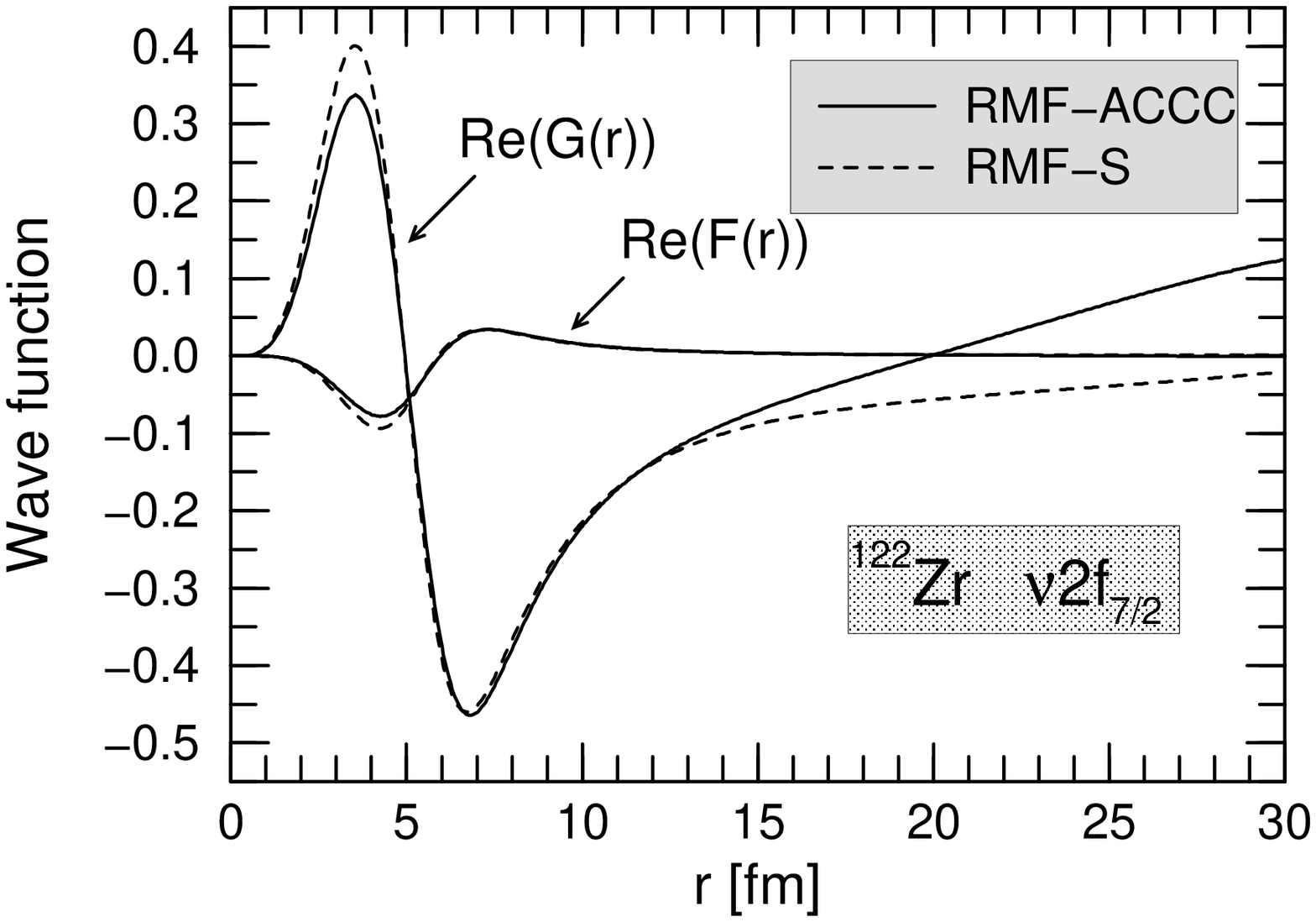}
\end{center}
\caption{
Left panel: Energies and widths for the neutron resonant states $3p_{3/2}$, $3p_{1/2}$, $2f_{7/2}$, $2f_{5/2}$, $1h_{9/2}$, and $1i_{13/2}$ in the potential of $^{122}$Zr.
Solid circles represent the results of the ACCC method, while open circles denote the results of the scattering method.
Right panel: Real parts of the upper and lower components of radial wave functions for the neutron resonant state $2f_{7/2}$.
Solid and dashed curves represent the results of the ACCC and the scattering method, respectively.
Taken from Ref.~\cite{Zhang2004_PRC70-034308}.
\label{fig:Zhang2004_PRC70-034308_Fig5}
}
\end{figure}

In Ref.~\cite{Yang2001_CPL18-196},
the ACCC method was combined with the covariant density functional theory
to study the unbound states in spherical nuclei.
The ACCC method \cite{Kukulin1977_JPA10-L33, Kukulin1979_SovJNP29-421}
is based on the following idea: An unbound state can be considered as
the continuation of a bound one when the strength of the attractive potential decreases.
That is, an unbound state will become bound if the coupling strength of attractive potential is increased
and becomes strong enough.
As a function of the strength,
the energy of a bound state is analytically continued
to the complex plane in order to get the width and energy of a resonant state.

In Ref.~\cite{Zhang2004_PRC70-034308},
the analyticity of the eigenvalues and eigenfunctions of the Dirac equation
with respect to the coupling constant was examined.
In the left panel of Fig.~\ref{fig:Zhang2004_PRC70-034308_Fig5}, the energies and widths for the neutron resonant
states $3p_{3/2}$, $3p_{1/2}$, $2f_{7/2}$, $2f_{5/2}$, $1h_{9/2}$, and $1i_{13/2}$ in the potential of $^{122}$Zr are shown in a
planar $E$-$\Gamma$ plot.
The results from the ACCC method agree satisfactorily with those
from the scattering calculation.
The wave functions for the neutron resonant state $2f_{7/2}$ are
exhibited in the right panel of Fig.~\ref{fig:Zhang2004_PRC70-034308_Fig5}.
The ACCC and the scattering method give
not only nearly the same resonance parameters, energies and widths, but also very similar wave functions.

In Ref.~\cite{Guo2005_PRC72-054319}, the PSS for the resonant states in $^{208}$Pb
was investigated by solving the Dirac equation with the Woods-Saxon scalar and
vector potentials, and the ACCC method was used to determine the resonance parameters.
The PSS breaking is shown in correlation with the nuclear mean field
shaped by the central depth $\Sigma_0$, the radius $R$, and the diffuseness $a$.
The energy-level crossings appear in several pseudospin partners of resonant states.
The width is found to be different for the pseudospin doublets
even when their energies are fully degenerate.
This was attributed to the difference in the centrifugal barrier
for the pseudospin partners \cite{Guo2005_PRC72-054319}.

The ACCC method, combined with the CDFT,
was used to study the $N$-dependence of PSS
in nuclear resonant states in Ref.~\cite{Guo2006_PRC74-024320}.
The energies and widths of
single-particle resonant states in Sn isotopes were investigated systematically.
An $N$-dependence of PSS is clearly shown in the resonant states
and is consistent with that observed in the bound states.
The splittings of energies and widths between pseudospin doublets are
correlated with the quantum numbers of the single-particle states,
as well as the nuclear mass number.
Although the $N$-dependence of the pseudospin splitting in energy is rather
complicated, the width splitting decreases almost monotonically with increasing mass number.

\begin{figure}[tbhp]
\begin{center}
  \includegraphics[width=5.8cm]{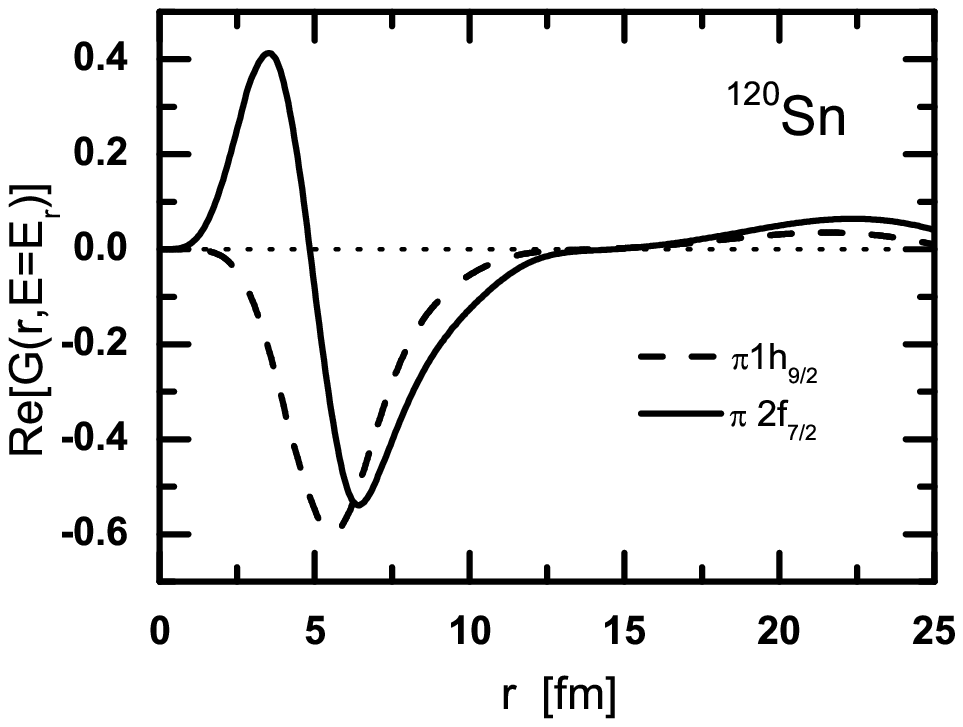}
  \includegraphics[width=6.1cm]{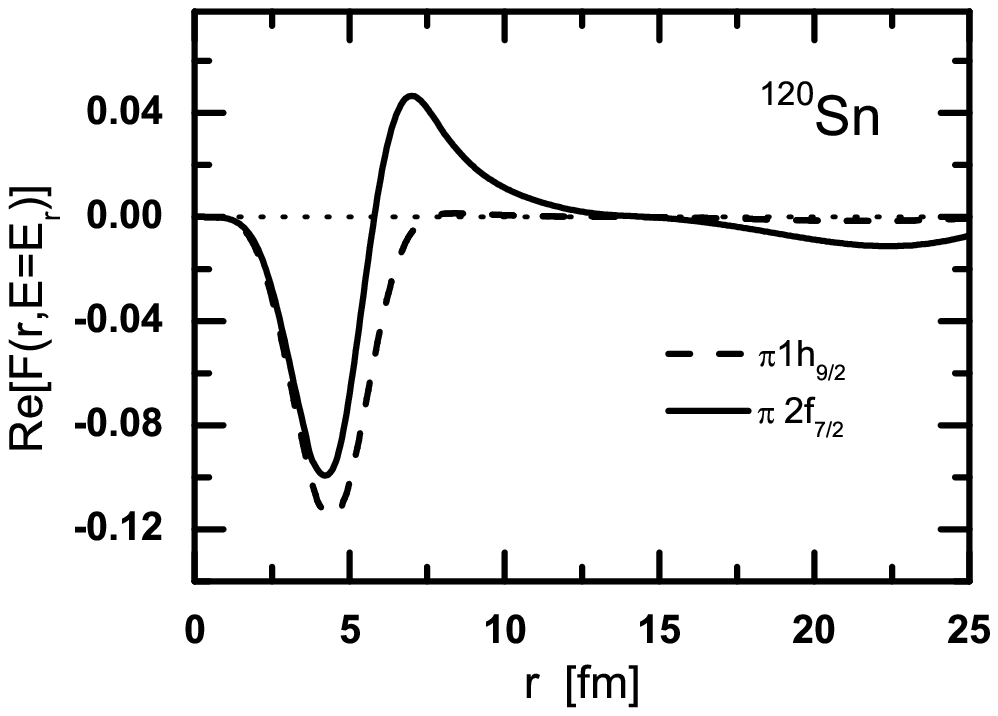}
\end{center}
\caption{The real part of the upper component $G(r)$ (the left panel)
and the lower component $F(r)$ (the right panel) for the proton
resonant state $1h_{9/2}$ (dashed line) compared with the proton
resonant state $2f_{7/2}$ (solid line) in $^{120}$Sn.
Taken from Ref.~\cite{Zhang2007_CPL24-1199}.
\label{fig:3.4.Zhang2007_CPL24-1199_Fig1-2}}
\end{figure}

In Refs.~\cite{Zhang2006_HEPNP30S2-97, Zhang2007_CPL24-1199},
the ACCC combined with CDFT was also used to discuss the PSS in the single-proton resonant states.
Not only the resonant energies and widths, but also
the wave functions for pseudospin partners in $^{120}$Sn were examined.
In Fig.~\ref{fig:3.4.Zhang2007_CPL24-1199_Fig1-2} are shown the real parts of the upper
and lower components of the radial wave functions for the $\pi1\tilde g$ pseudospin doublets in $^{120}$Sn.
It is seen that the lower
components agree very well in the region where nuclear potential
dominates, except for some disagreement on the surface, with the
number of radial nodes being the same.

\subsubsection{PSS in single-particle resonant states with CSM}

\begin{figure}[tbhp]
\begin{center}
  \includegraphics[width=12cm]{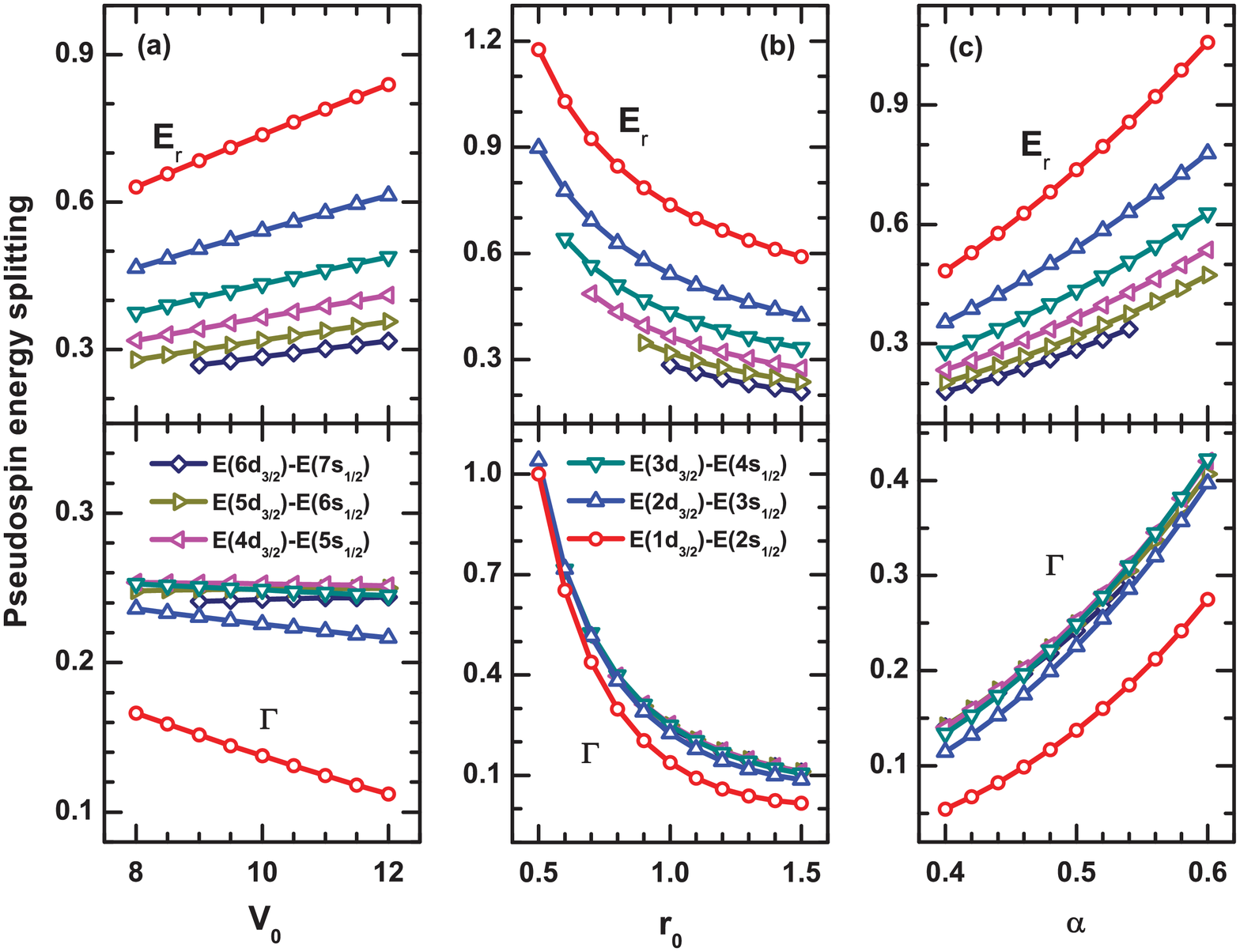}
\end{center}
\caption{(Color online)
Pseudospin splittings in energy and width as a function of
Morse potential parameters, (a) $V_0$,
(b) $r_0$, and (c) $\alpha$, for single-particle resonant states.
Taken from Ref.~\cite{Liu2013_PRA87-052122}.
\label{fig:3.4.Liu2013_PRA87-052122_Fig5}
}
\end{figure}

In Ref.~\cite{Liu2013_PRA87-052122}, the CSM was applied to study
the PSS of resonant states of a Dirac particle in a Morse potential,
\begin{equation}
V(r) = V_{0} e^{-\left( r-r_{0}\right) \alpha }
        \left( 2-e^{-\left( r-r_{0}\right) \alpha }\right)\,,
\end{equation}
By comparing the energies and widths of the pseudospin doublets, the PSS was examined and
the relationship between the PSS and the parameters of the Morse potential was studied.
In Fig.~\ref{fig:3.4.Liu2013_PRA87-052122_Fig5},
pseudospin splittings in energy and width for single-particle resonant states
are shown as a function of the Morse potential parameters $V_0$, $r_0$, and $\alpha$, respectively.
With $r_{0}$ and $\alpha $ fixed, the energy
splitting of a pair of pseudospin partners increases with increasing $V_{0}$.
However, the opposite is true for the width splitting.
This is because, with increasing $V_{0}$, the barrier becomes higher
and the potential well becomes deeper.
The dependence of the energy and width splittings
on $r_0$ or $\alpha$ were also made, and the details
can be found in Ref.~\cite{Liu2013_PRA87-052122}.

\subsubsection{PSS in single-particle resonant states with Jost functions}

Although the PSS in resonant states have been studied numerically, it
is still desirable to have a mathematical verification of the PSS in
the symmetry limit. In particular, there remains an open question about
the width splitting of the pseudospin doublets.
Recently, a rigorous justification of the PSS in single-particle resonant states
has been given by examining the asymptotic behavior of the nucleon
Dirac wave functions \cite{Lu2012_PRL109-072501}.

The first-order coupled equation~(\ref{Eq:2.1.DiraceqR})
can be rewritten as two decoupled second-order differential ones, Eqs.~(\ref{Eq:2.1.SchrG})
and (\ref{Eq:2.1.SchrF}).
In Refs.~\cite{Zhang2010_IJMPE19-55, Zhang2009_ChinPC33S1-113,
Zhang2009_CPL26-092401, Li2011_SciChinaPMA54-231},
it has been shown that each of these two Schr\"odinger-like equations,
together with its charge conjugated one, are fully equivalent to Eq.~(\ref{Eq:2.1.DiraceqR}).
The one for the small component is given in Eq.~(\ref{Eq:2.1.SchrF})
to which the PSS is directly connected.
Note that for bound states, there is always a singularity in $1/M_{-}(r)$
in Eq.~(\ref{Eq:2.1.SchrF}), whereas such a singularity does not exist for resonant states discussed here.

For the continuum in the Fermi sea, i.e., $\epsilon \geq M$, there exist two
independent solutions for Eq.~(\ref{Eq:2.1.SchrF}).
The physically acceptable solution is the one that vanishes at the origin.
As usual we define the regular solution $F(r)$ as the one that behaves like
$j_{\tilde{l}}(pr)$ function as $r \rightarrow 0$ \cite{Taylor1972},
\begin{equation}
  \lim_{r \rightarrow 0} F(r)/j_{\tilde{l}}(pr) = 1\,,
  \qquad p = \sqrt{\epsilon^2 - M^2}\,.
 \label{eq:regularsolution}
\end{equation}

At large $r$ the potentials for neutrons vanish and the radial wave functions oscillate.
Equation~(\ref{Eq:2.1.SchrF})
becomes a Ricatti-Bessel equation with angular momentum $\tilde{l}$, and
the solution can be written as a combination of the
Ricatti-Hankel functions,
\begin{equation}\label{Eq:3.4.Ricatti-Hankel}
 F(r)
 = \frac{i}{2} \left[
                       \mathcal{J}_{\kappa}^{F}(p)     h_{\tilde{l}}^{-}(pr)
                     - \mathcal{J}_{\kappa}^{F}(p)^{*} h_{\tilde{l}}^{+}(pr)
               \right]\,,
   \qquad r\rightarrow\infty\,,
\end{equation}
where $\mathcal{J}_{\kappa}^{F}(p)$ is the Jost function for the
small component and $h_{\tilde{l}}^{\pm}(pr)$ the Ricatti-Hankel functions.

\begin{figure}[tbhp]
\begin{center}
  \includegraphics[width=6cm]{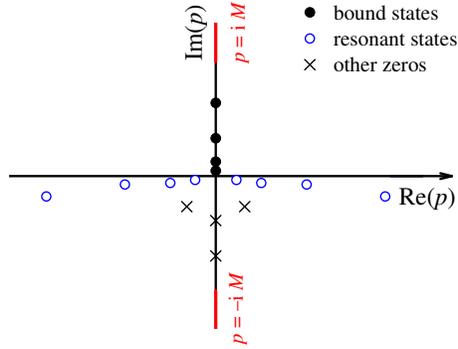}
\end{center}
\caption{(Color online)
Schematic picture of the zeros of the Jost function $\mathcal{J}_{\kappa}^{F}$ on the
complex momentum plane.
A cut is made on the imaginary axis, from $p=iM$ to infinity and back to $p=-iM$.
Taken from Ref.~\cite{Lu2012_PRL109-072501}.
}
~\label{fig:zeros_sche}
\end{figure}

In nuclei the vector $V(r)$ and scalar $S(r)$ potentials share some general properties,
e.g., they are analytical functions of $r$, vanish as
$r\rightarrow\infty$, and have no singularities.
Under such conditions, the Jost function is an analytic function of $p$,
and can be analytically continued to a large area in the complex $p$ plane.
Here the structure of the $p$ Riemann surface on which the Jost functions are defined
is more complex than the non-relativistic case.
For example, the square root in the relativistic energy-momentum relation
$\epsilon^2=p^{2}+M^{2}$ creates branching points at $p=\pm iM$, thus the
corresponding Riemann surface has at least two sheets.

In Fig.~\ref{fig:zeros_sche}, the zeros of the Jost function
$\mathcal{J}_{\kappa}^{F}(p)$ on the complex momentum plane
are schematically shown.
Only the first sheet with Re$(\epsilon)\geq0$ is presented in
Fig.~\ref{fig:zeros_sche} which contains positive-energy bound states and resonant states, while the other sheet with
Re$(\epsilon)\leq0$ can be used to investigate negative-energy ones.
These two sheets are connected by a cut on the imaginary axis,
from $p=iM$ to infinity and back to $p=-iM$.
Restricted to the first sheet and relatively small $|p|$, $\mathcal{J}_{\kappa}^{F}(p)$ is a
single-valued analytical function of $p$.
The zeros of $\mathcal{J}_{\kappa}^{F}(p)$ are denoted by the full circles for bound states,
open circles for resonant states, and crosses for other zeros, respectively.
The zeros on the positive imaginary axis of the $p$ plane represent
bound states of the original eigenvalue problem,
while the zeros on the lower $p$ plane and
near the real axis correspond to resonant states.
The resonance energy $E_{{\rm res}}$ and width $\Gamma_{{\rm res}}$
are determined by the relation
$E=E_{{\rm res}}-i{\Gamma_{{\rm res}}}/{2}=\sqrt{p^{2}+M^{2}}-M$.
By examining the zeros of the Jost function, one can study the bound
and resonant states on the same footing, and
many known properties of bound states can be generalized
to resonances straightforwardly.

At the PSS limit, Eq.~(\ref{Eq:2.1.SchrF}) is reduced to Eq.~(\ref{Eq:2.2.SchrFPSS}).
For the bound states, it is an eigenfunction equation that determines the eigenenergies $\epsilon$.
While for the continuum $\epsilon$ can be any value larger than or equal to $M$ and
one should focus on the wave functions and their asymptotic behavior.
For pseudospin doublets with different quantum numbers $\kappa_a$ and $\kappa_b$ with
$\kappa_b=-\kappa_a+1$, the small components satisfy the same equation because
they have the same pseudo-orbital angular momentum $\tilde{l}$ \cite{Ginocchio1997_PRL78-436}.
In particular, for continuum states, $F_{\kappa_a}(\epsilon,r)=
F_{\kappa_b}(\epsilon,r)$ for any energy $\epsilon$.
Because the definition of the Jost function $\mathcal{J}_{\kappa}^{F}(p)$ only
depends on the asymptotic behavior of the small component,
on the positive real axis,
$\mathcal{J}_{\kappa_a}^{F}(p)=\mathcal{J}_{\kappa_b}^{F}(p)$.
This equivalence can be generalized into the complex $p$ plane
due to the uniqueness of the analytical continuation.
Thus the zeros are the same for $\mathcal{J}_{\kappa_a}^{F}(p)$ and
$\mathcal{J}_{\kappa_b}^{F}(p)$: If there exists a resonant state with
energy $E_{{\rm res}}$ and width $\Gamma_{{\rm res}}$ and the quantum number $\kappa_a$,
there must be another one with the same energy and width and quantum number $\kappa_b$.
That is to say, the PSS in single-particle resonant states
in nuclei is exactly conserved when the attractive scalar and repulsive vector
potentials have the same magnitude but opposite sign.
If one focuses on the zeros of the Jost functions of pseudospin doublets
on the positive imaginary axis of the $p$ plane,
the well-known PSS for bound states can be investigated similarly.

In scattering theories, one can also determine resonance parameters
from the change of cross section or phase shift, which
give more insights into the resonant phenomena.
Using the asymptotic behavior of the Ricatti-Bessel functions,
one obtains from Eqs.~(\ref{Eq:2.2.SchrFPSS}) and (\ref{Eq:3.4.Ricatti-Hankel}),
\begin{equation}
            F_{\kappa}(r)
  \propto   \sin  \left(
                         pr
                       - \frac{\tilde{l}\pi}{2}
                       + \delta_{\kappa}^{F}(p)
                  \right)\,,\qquad
 r\rightarrow\infty\,,
 \label{eq:asymp}
\end{equation}
where the phase shift $\delta_{\kappa}^{F}(p)$ is related to the Jost function
through
\begin{equation}
 \mathcal{J}_{\kappa}^{F}(p)=|\mathcal{J}_{\kappa}^{F}(p)|e^{-i\delta_{\kappa}^{F}(p)}.
\end{equation}
Whenever $\delta_{\kappa}^{F}(p)=(n+1/2)\pi$, there is a resonant state
and its width is determined by the tangent of the phase shift function $\delta_{\kappa}^{F}(p)$.
In the PSS limit, the coincidence between $F_{\kappa_a}(r)$ and
$F_{\kappa_b}(r)$ means that
$\delta_{\kappa_a}^{F}(p)=\delta_{\kappa_b}^{F}(p)$ for any value of $p$.
Therefore, resonance parameters of pseudospin doublets are the same.

As it has been noted in Ref.~\cite{Lu2012_PRL109-072501},
it is straightforward to extend the study of PSS
in resonant states in the Fermi sea to that in the negative-energy states
in the Dirac sea or SS in anti-particle continuum spectra.
More investigations along this line should be made.

\subsubsection{PSS in single-particle resonant states in square-well potentials}

To extract the energies and widths of resonant states in realistic potentials
is relatively complex.
In particular, to study the PSS and examine the origin and
the symmetry breaking mechanism, it is better to start from analytically solvable models.
In Ref.~\cite{Lu2012_PRL109-072501}, the square-well potentials were
taken as examples to illustrate the conservation and breaking of the PSS
in the single-particle resonant states.
Although the diffuseness of realistic potentials cannot be included,
it is still a good starting point to study general properties of the PSS
for the resonant as well as bound states by using the square-well potentials,
because the PSS-breaking term in the Jost function is separated from the PSS-conserving term.

Spherical square-well potentials for $\Sigma(r)$ and $\Delta(r)$ read
\begin{equation}
 \Sigma(r) = \left\{ \begin{array}{c}
                     C\,,\qquad r<R\,,\\
                     0\,,\qquad r\geq R\,,
                     \end{array}\right.\\
 \Delta(r) = \left\{ \begin{array}{c}
                     D\,,\qquad r<R\,,\\
                     0\,,\qquad r\geq R\,,
                     \end{array}\right.
 \label{eq:S.W.}
\end{equation}
where $C$ and $D$ are depths and $R$ is the width.
The Jost function $\mathcal{J}_{\kappa}^{F}(p)$ is derived as \cite{Lu2012_PRL109-072501}
\begin{equation}
 \mathcal{J}_{\kappa}^{F}(p)
 =
 - \frac{p^{\tilde{l}}}{2ik^{\tilde{l}+1}}
   \left\{ j_{\tilde{l}}(kR) p h_{\tilde{l}}^{+\prime}(pR)
        -kj_{\tilde{l}}^{\prime}(kR) h_{\tilde{l}}^{+}(pR)
   - \frac{C}{\epsilon-M-C}
     \left[ kj_{\tilde{l}}^{\prime}(kR) - \frac{\kappa}{R} j_{\tilde{l}}(kR) \right]
     h_{\tilde{l}}^{+}(pR)
   \right\}\,,
 \label{eq:Jost_function}
\end{equation}
with $k = \sqrt{\left( \epsilon - C - M \right) \left( \epsilon - D + M \right)}$.
The PSS in both bound states and resonant states can be explained explicitly.
If $C = 0$,
the second term in $\mathcal{J}_{\kappa}^{F}(p)$ vanishes and
the first term only depends on the pseudo-orbital angular momentum $\tilde{l}$,
then the Jost functions with different $\kappa$ but the same $\tilde{l}$
are identical. The energies
and widths of resonant pseudospin partners are exactly the same.

\begin{figure}[tbhp]
\begin{center}
  \includegraphics[width=8cm]{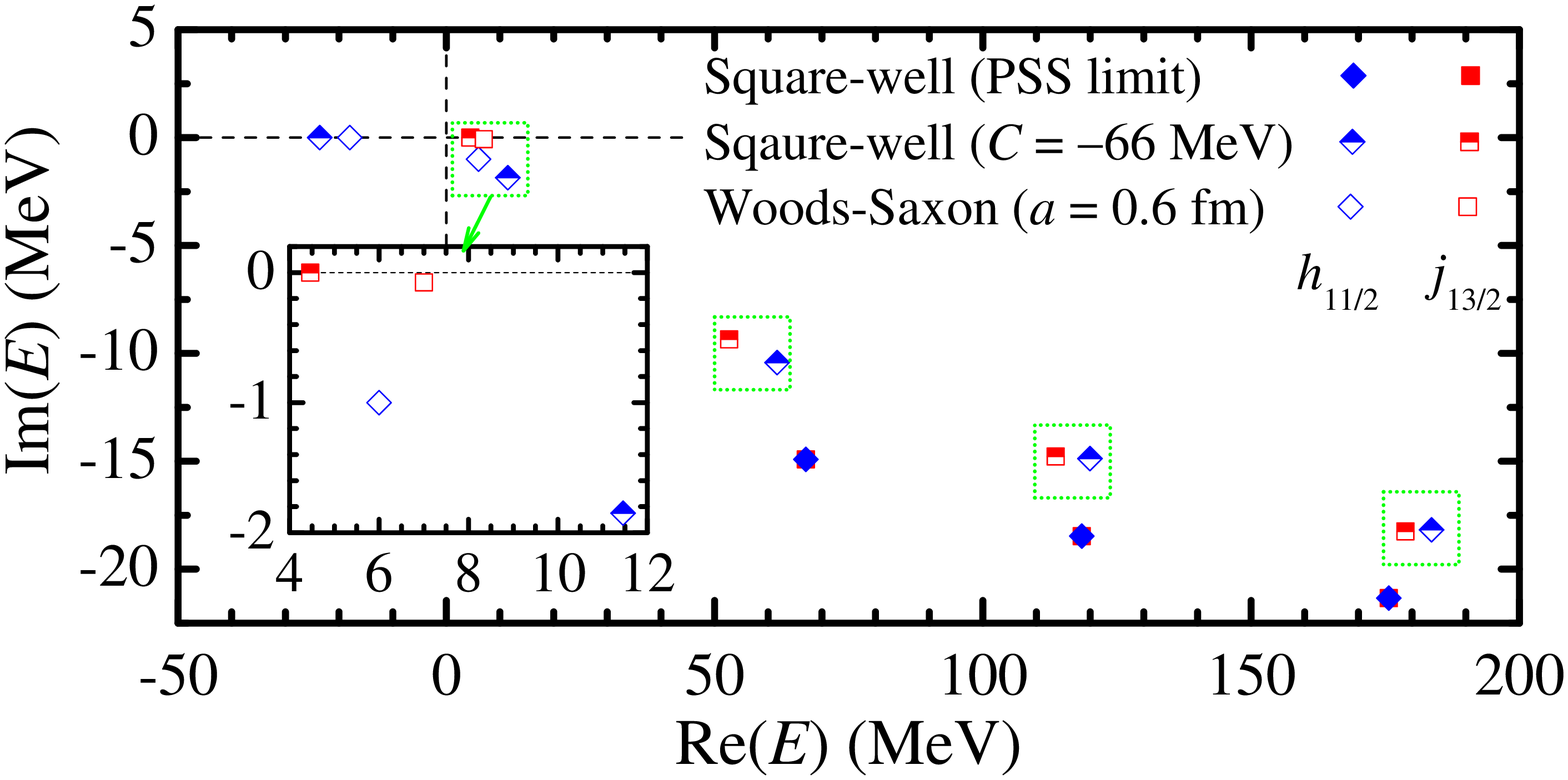}
\end{center}
\caption{(Color online)
The zeros of the Jost function $\mathcal{J}_{\kappa}^{F}$ on the complex energy plane
in square-well potentials (\ref{eq:S.W.}) with
$C=0$ (solid symbols) and $C=-66$~MeV (half-filled symbols)
for pseudospin partners $h_{11/2}$ (diamond) and $j_{13/2}$ (square).
The results with Woods-Saxon-like scalar and vector potentials are also
shown as open symbols.
Taken from Ref.~\cite{Lu2012_PRL109-072501}.
}
\label{fig:zeros}
\end{figure}

For the large component $G(r)$, there is a similar expression for the
asymptotic behavior,
\begin{equation}
 G(r) = \frac{i}{2}
        \left[ \mathcal{J}_{\kappa}^{G}(p)     h_{{l}}^{-}(pr) -
               \mathcal{J}_{\kappa}^{G}(p)^{*} h_{{l}}^{+}(pr)
        \right]\,,
 \qquad r\rightarrow\infty\,.
\end{equation}
At the origin,
\begin{equation}
 \lim_{r\rightarrow0} G(r) / j_{l}(pr) = 1\,,
 \qquad
 p=\sqrt{\epsilon^{2}-M^{2}}\,.
\end{equation}
The $\mathcal{J}_{\kappa}^{G}(p)$ is derived as
\begin{equation}
 \mathcal{J}_{\kappa}^{G}(p)=
 -\frac{p^{l}}{2ik^{l+1}}
  \left\{ j_{l}(kR) p h_{l}^{+\prime}(pR) - k j_{l}^{\prime}(kR) h_{l}^{+}(pR)
  -\frac{D}{\epsilon+M-D}
   \left[ k j_{l}^{\prime}(kR) + \frac{\kappa}{R} j_{l}(kR) \right] h_{l}^{+}(pR)
  \right\}\,.
\end{equation}
It looks similar to that of the small component $\mathcal{J}_{\kappa}^{F}(p)$,
with the exception that the potential parameter $C$ is substituted by
$D$ and the pseudo-orbital angular momentum $\tilde{l}$
is substituted by $l$.
In the case of $D\rightarrow0$, this form of Jost function can be used to
investigate the spin symmetry of single-particle levels.

Although the solutions of $\mathcal{J}_{\kappa}^{F}(p)=0$ cannot
be derived analytically, the secant method can be used for searching the roots,
because the Jost function is analytical near its zeros.
In Fig.~\ref{fig:zeros} are shown solutions in the complex energy plane
for PSS doublets with $\tilde{l}=6$, i.e., $h_{11/2}$ with $\kappa_a = -6$
and $j_{13/2}$ with $\kappa_b = 7$,
for square-well potentials with $D=650$~MeV and $R=7$~fm.
In the PSS limit, i.e., $C=0$, all the roots locate in the lower
half plane and there are no bound states.
In Fig.~\ref{fig:zeros} three pairs of pseudospin resonant partners
are shown by full diamonds and squares.
The conservation of the PSS for single-particle resonant states is clearly seen.
When $C=-66$~MeV, there is one bound state only for $h_{11/2}$.
Three pairs of pseudospin partners of resonant states are shown by
half-filled diamonds and squares.
One finds the breaking of the PSS both in the bound states and in the resonant states.
For pseudospin doublets with other values of $\tilde{l}$, similar
behaviors are observed concerning the exact conservation and the breaking of the PSS.
The resonances in Woods-Saxon potentials,
$W(r) = W_0 / (1+\exp[(r-R)/a])$ ($W=V$ or $S$) were also studied and
the potential parameters are the following:
the depths $V_0 - S_0 = 650$~MeV and $V_0 + S_0 = -66$~MeV, the diffuseness parameter
$a=0.6$~fm, and $R=7$~fm \cite{Guo2005_PRC72-054319}.
Resonance parameters are obtained with the real stabilization
method \cite{Zhang2008_PRC77-014312}.
The results are shown as open diamonds and squares for $h_{11/2}$ and $j_{13/2}$, respectively.
It is found that splittings of energy and width both become smaller
compared with the results with the square-well potentials.
The reason is that the derivative of $\Sigma(r)$ is smaller due to
a non-zero diffuseness parameter.

\begin{figure}[tbhp]
\begin{center}
  \includegraphics[width=6cm]{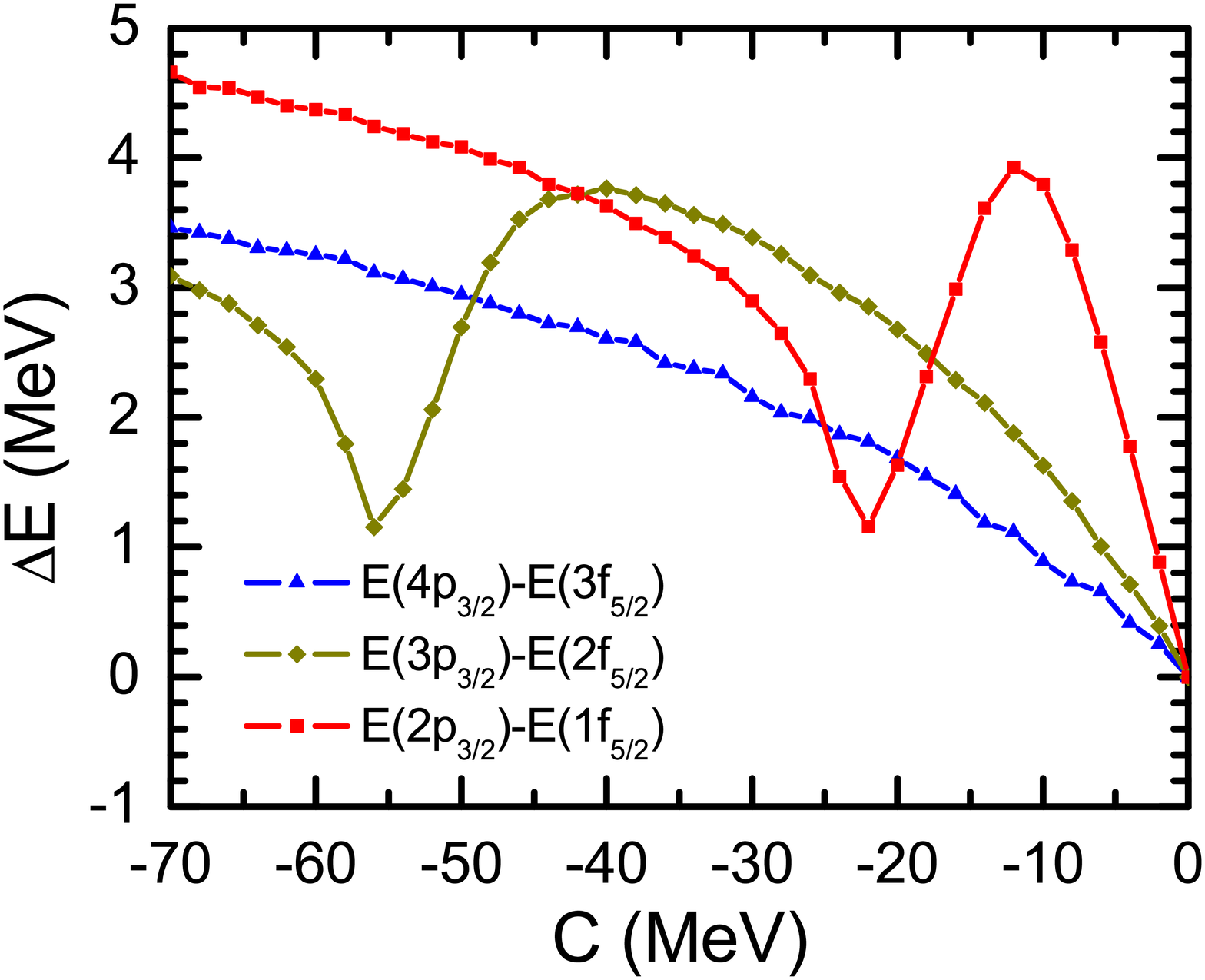}\hspace{1em}
  \includegraphics[width=6cm]{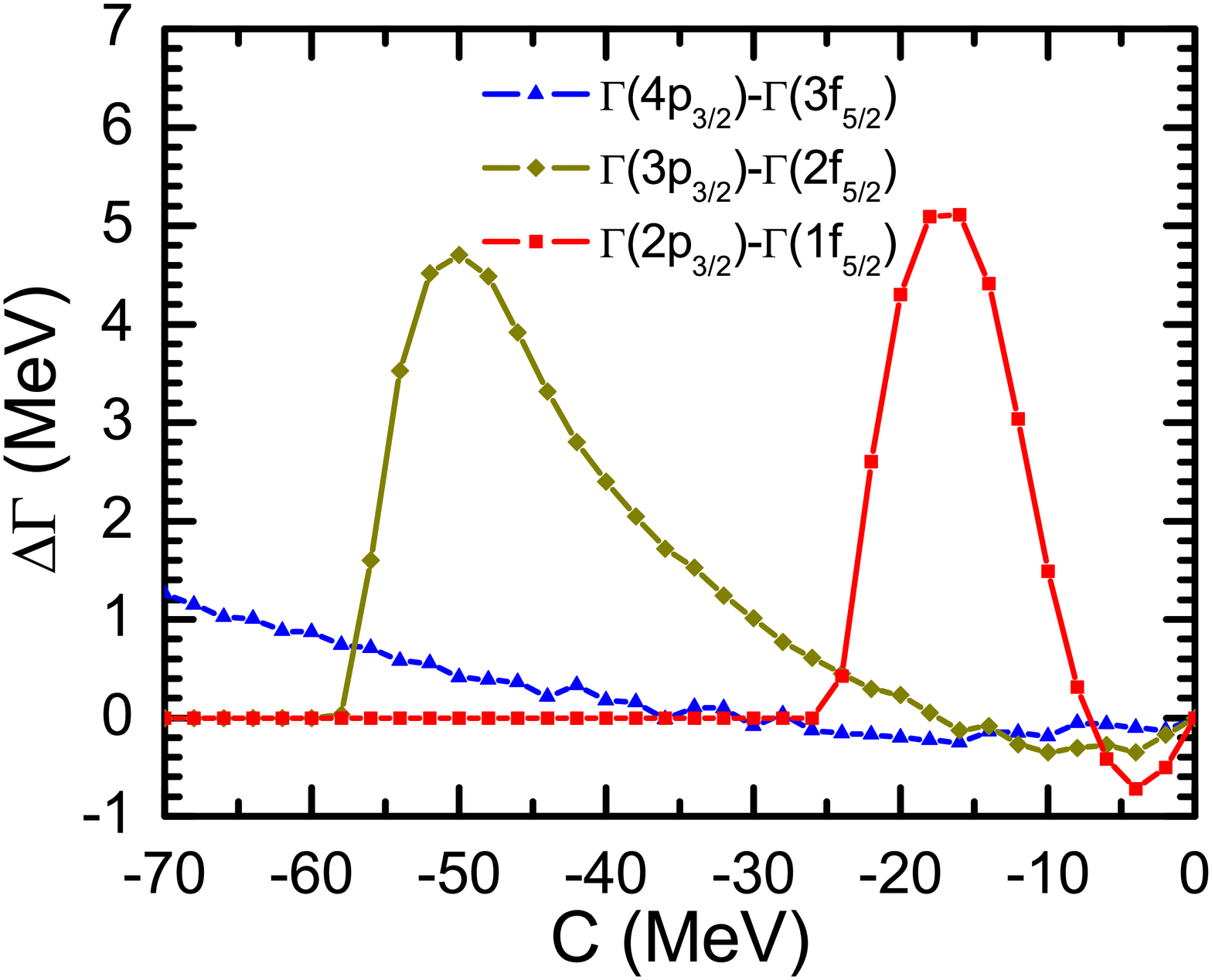}
\end{center}
\caption{(Color online)
Energy (the left panel) and width (the right panel)
splitting between the $\tilde d$ pseudospin doublets as a function of the potential depth $C$.
Taken from Ref.~\cite{Lu2013_PRC88-024323}.
}
\label{fig:Delta_EG}
\end{figure}

By examining the Jost function, one can trace continuously the PSS partners
from the PSS limit to the case with a finite potential depth.
This has been done in Ref.~\cite{Lu2013_PRC88-024323}
in which
were found a threshold effect in the energy splitting and
an anomaly in the width splitting of pseudospin partners
when $C$ varies from zero to a finite value.
As the depth of the single-particle potential becomes larger,
the PSS is broken more and
a threshold effect in the energy splitting appears which
can be seen in the left panel of Fig.~\ref{fig:Delta_EG}:
The energy splitting first increases then decreases until the pseudospin doublets encounter
the threshold where one of the levels becomes a bound state
and the splitting takes a minimum value;
When the potential becomes even deeper, the splitting increases again.
When the depth of the single-particle potential increases from zero,
there also appears an anomaly in
the width splitting of pseudospin partners which is shown in the right panel of
Fig.~\ref{fig:Delta_EG}: It first decreases from zero to a maximum value with
negative sign, then increases and becomes zero again;
after the inversion of the width splitting, the splitting increases
and reaches a positive maximum value, then it becomes smaller and eventually reaches zero.

In summary, the pseudospin and spin symmetries in single-particle resonant states in the Woods-Saxon
potentials and self-consistent relativistic mean-field potentials have been investigated
by using the analytical continuation in coupling constant method and the complex scaling method.
Not only the pseudospin-orbit splittings in energy and width,
but also the wave functions of pseudospin partners have been systematically examined.
A rigorous verification of the pseudospin symmetry in single-particle
resonant states was given by studying the Jost function of
the lower components of the Dirac wave functions.
In an extensive study of the pseudospin symmetry in single-particle resonant states in square well potentials,
a threshold effect in the energy splitting and an anomaly in the width splitting
of pseudospin partners were found when the depth of the single-particle potential varies
from zero to a finite value.
There are still many open problems concerning the pseudospin and spin symmetries in
single-particle resonant states,
as listed in Refs.~\cite{Lu2013_AIPCP1533-63,Lu2013_PRC88-024323}.

\subsection{From nucleon spectra to anti-nucleon spectra}\label{Sect:3.5}

In the spectrum of a Dirac Hamiltonian, one finds single-particle states with positive
energies in the Fermi sea as well as those with negative energies in the Dirac sea.
The latter are interpreted as anti-particles under a charge conjugation.
In the relativistic mean-field theory, the no-sea approximation is made:
Those negative-energy states are assumed to be empty \cite{Serot1986_ANP16-1}.
Therefore, less attention was paid to the negative-energy states.
However, if the eigenstates of the Dirac Hamiltonian are used as
basis states in a Hilbert space, the negative-energy states in the Dirac sea
must be included for the completeness.
When Zhou, Meng, and Ring developed the relativistic mean-field theory in a Dirac Woods-Saxon
basis \cite{Zhou2003_PRC68-034323}, they included and
examined in detail the negative-energy states
and found that the pseudospin symmetry of those negative-energy states in the Dirac sea,
or equivalently, the spin symmetry in the anti-nucleon spectra
is very well conserved \cite{Zhou2003_PRL91-262501}.
In Ref.~\cite{Zhou2003_PRL91-262501}, it was shown that the spin symmetry
in the anti-nucleon spectra is much better developed than the pseudospin symmetry
in normal nuclear single-particle spectra.
The spin symmetry in the anti-nucleon spectra of a nucleus was later tested by investigating
the relations between the Dirac wave functions of the spin doublets and
examining these relations within the relativistic mean-field theory \cite{He2006_EPJA28-265}.
Since then, the study of the spin symmetry has often been made in combination with that
of the pseudospin symmetry in various local potentials, see, e.g., Refs.~\cite{deCastro2006_PRC73-054309,
Lisboa2010_PRC81-064324}.
The spin symmetry in the Dirac negative-energy spectrum and its origin were also
investigated within the relativistic Hartree-Fock theory
in which the potentials are non-local \cite{Liang2010_EPJA44-119}.
Note that it was found that the equality of the vector and scalar
potentials in the Dirac Hamiltonian results in the spin symmetry in
Refs.~\cite{Smith1971_APNY65-352, Bell1975_NPB98-151},
where the authors suggested applications to meson spectra.
This symmetry was revealed to be valid for
mesons with one heavy quark \cite{Page2001_PRL86-204}.
It should also be noted that, by applying the charge conjugate transformation, the spin symmetry for the anti-nucleon states have been formally conjectured in Ref.~\cite{Ginocchio1999_PR315-231}.

In this Section, we will introduce the spin symmetry in the anti-nucleon spectra
in nuclei, the test of the spin symmetry by examining the wave functions,
and the spin symmetry in single-(anti-)particle spectra in various potentials.

\subsubsection{SS in single-anti-nucleon spectra}

\newcommand{\tN}{{\text{N}}}
\newcommand{\tA}{{\text{A}}}

In this Section, a nucleon state is explicitly labelled with ``N''
and an anti-nucleon state with ``A'' for convenience.
The Dirac equation for nucleons reads
\begin{equation}\label{eq:Dirac0}
 \left[ \boldsymbol{\alpha}\cdot \mathbf{p} + V_\tN(\mathbf{r}) + \beta (M+S_\tN(\mathbf{r}))
 \right] \psi_\tN(\mathbf{r},s) = \epsilon_\tN \psi_\tN(\mathbf{r})\,,
\end{equation}
where $V_\tN(\mathbf{r}) = V(\mathbf{r})$ and $S_\tN(\mathbf{r}) = S(\mathbf{r})$.
For a spherical system, the Dirac spinor
$\psi_\tN$ has the form (cf. Eq.~(\ref{Eq:2.1.spwfR}))
\begin{equation}\label{eq:SRHspinor}
 \psi_\tN (\mathbf{r}) = \frac{1}{r}
  \left(
   \begin{array}{c}
    i G_{n\kappa}(r)         \mathscr Y_{jm}^{l}(\hat{\mathbf{r}})        \\
    - F_{\tilde{n}\kappa}(r) \mathscr Y_{jm}^{\tilde{l}}(\hat{\mathbf{r}})
   \end{array}
  \right)\,,
  \qquad {j=l\pm \frac{1}{2}\,.}
\end{equation}

Charge conjugation leaves the scalar potential
$S_\tN(\mathbf{r})$ invariant, while it changes the sign of the
vector potential $V_\tN(\mathbf{r})$. That is, for
anti-nucleons (labelled by ``A''), $V_\tA(\mathbf{r}) = -
V_\tN(\mathbf{r}) = - V(\mathbf{r})$ and $S_\tA(\mathbf{r}) =
S_\tN(\mathbf{r}) = S(\mathbf{r})$. Charge conjugation of
Eq.~(\ref{eq:SRHspinor}) gives the Dirac spinor for an
anti-nucleon,
\begin{equation}
 \psi_\tA (\mathbf{r}) = \frac{1}{r}
  \left(
   \begin{array}{c}
    - F_{\tilde{n}\tilde\kappa}(r) \mathscr Y_{jm}^{\tilde{l}}(\hat{\mathbf{r}})\\
    i G_{n\tilde\kappa}(r)         \mathscr Y_{jm}^{l}(\hat{\mathbf{r}})        \\
   \end{array}
  \right)\,,
  \qquad {j=l\pm \frac{1}{2}\,,}
 \label{eq:SRHspinor2}
\end{equation}
with $\tilde\kappa = -\kappa$.

For particles there are positive- and negative-energy solutions; the same
is true for anti-particles.
For positive-energy states of the Dirac equations, the normal quantum numbers follow
the upper components which are dominant. A particle state is
labelled by $\{nl\kappa m\}$, while its pseudo-quantum
numbers are $\{\tilde{n}\tilde{l} \tilde{\kappa}m\}$.
Following Ref.~\cite{Leviatan2001_PLB518-214}, $\tilde{n}=n+1$ for $\kappa
>0$; $\tilde{n}=n$ for $\kappa <0$ [cf. Eq.~(\ref{Eq:3.1.GFnodes})].
An anti-particle state is labelled by $\{\tilde{n}\tilde{l}\tilde{\kappa}m\}$ and
its pseudo-quantum numbers are $\{nl\kappa m\}$. In analogy
to Ref.~\cite{Leviatan2001_PLB518-214}, the following relation holds for
anti-nucleon states,
\begin{equation}\label{eq:node2}
   n = \tilde{n}+1\quad \text{for}\quad \tilde{\kappa}>0\,;\qquad
   n = \tilde{n}\quad \text{for}\quad \tilde{\kappa}<0\,.
\end{equation}
With $\kappa(\kappa+1) = \tilde\kappa(\tilde\kappa-1) = l(l+1)$ and  $\kappa(\kappa-1)=\tilde\kappa(\tilde\kappa+1) = \tilde{l}(\tilde{l}+1)$
 in mind, one derives
the Schr\"{o}dinger-like equations for the upper and the lower
components [cf. Eqs.~(\ref{Eq:2.1.SchrG}) and (\ref{Eq:2.1.SchrF})],
\begin{equation}\label{eq:origin-of-symmetry1}
\Lb -\frac{1}{M_+}\frac{d^2}{dr^2}
  +\frac{1}{M_+^2}\frac{dM_+}{dr}\frac{d}{dr}
  +\ls (M+\Sigma)+\frac{1}{M_+}\frac{l(l+1)}{r^2}
  +\frac{1}{M_+^2}\frac{dM_+}{dr}\frac{\kappa}{r}
  \rs \Rb G(r)
 = \left\{
    \begin{array}{l}
     + \epsilon_\tN G(r)\,, \\
     - \epsilon_\tA G(r)\,,
    \end{array}
 \right.
\end{equation}
and
\begin{equation}\label{eq:origin-of-symmetry2}
    \Lb -\frac{1}{M_-}\frac{d^2}{dr^2}
    +\frac{1}{M_-^2}\frac{dM_-}{dr}\frac{d}{dr}
    +\ls (-M+\Delta)+\frac{1}{M_-}\frac{\tilde l(\tilde l+1)}{r^2}
    +\frac{1}{M_-^2}\frac{dM_-}{dr}\frac{\tilde\kappa}{r}
    \rs \Rb F(r)
 = \left\{
    \begin{array}{l}
     + \epsilon_\tN F(r)\,, \\
     - \epsilon_\tA F(r)\,,
    \end{array}
 \right.
\end{equation}
where
$M_+(r)=M-\Delta(r)+\epsilon$
and
$M_-(r)=-M-\Sigma(r)+\epsilon$ with
$\epsilon = +\epsilon_\tN$ for particle states or
$-\epsilon_\tA$ for anti-particle states. Both equations
are fully equivalent to the exact Dirac equation with the
full spectrum of particle and anti-particle states.

\begin{table}
\begin{center}
\caption{Relation between symmetries and external fields.
Taken from Ref.~\cite{Zhou2003_PRL91-262501}.
}
\label{tab:symmetries}
\begin{tabular}{@{}ccc@{}}
\hline
             & Particle             & Anti-particle        \\
\hline
 $d\Delta/dr=0$ & Spin symmetry        & Pseudospin symmetry \\
 $d\Sigma/dr=0$ & Pseudospin symmetry & Spin symmetry        \\
\hline
\end{tabular}
\end{center}
\end{table}

The relation between SS or PSS and the
external fields is given in Table~\ref{tab:symmetries}.
If $d\Delta/dr$ = 0,
there is an exact SS in the particle spectrum and exact
PSS in the anti-particle spectrum, because states
with the same $l$ but different $\kappa$ are degenerate in
Eq.~(\ref{eq:origin-of-symmetry1}), where $l$ is the orbital angular
momentum of particle states and pseudo-orbital angular momentum of
anti-particle states. When $d\Delta/dr\neq 0$, the symmetries are
broken, but if $d\Delta/dr$ is so small that the SO term is
much smaller than the centrifugal barrier, there will be approximate symmetries.

\begin{figure}[tbhp]
\begin{center}
\includegraphics[width=8cm]{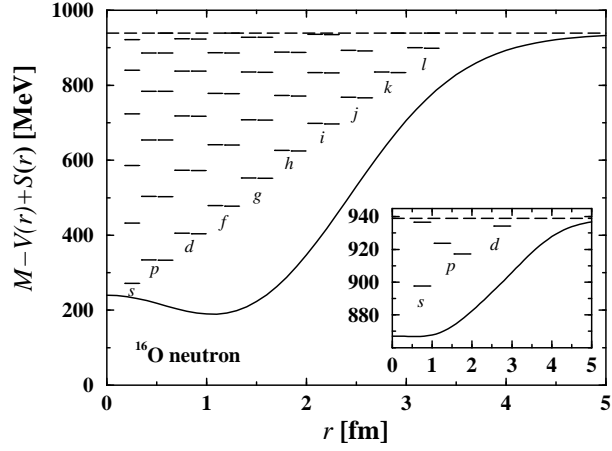}
\end{center}
\caption{\label{fig:spec-o16-neu}
Anti-neutron potential $M-V(r)+S(r)$ and spectrum of $^{16}$O.
For each pair of spin doublets, the left level is with $\tilde\kappa < 0$
and the right one with $\tilde\kappa > 0$.
The inset gives neutron potential $M+V(r)+S(r)$
and spectrum.
Taken from Ref.~\cite{Zhou2003_PRL91-262501} and modified to present notations.
}
\end{figure}

Similarly, when $d\Sigma/dr$ = 0 in
Eq.~(\ref{eq:origin-of-symmetry2}), there is an exact PSS in the particle spectra. On the
other hand, for anti-particle states, the SS is exactly conserved
because now $\tilde{l}$ is the orbital
angular momentum. If $d\Sigma/dr \neq 0$ but small, there are
approximate PSS in particle spectra and
approximate SS in anti-particle spectra. This implies
that the SS in the anti-particle spectra has the same
origin as the PSS in particle spectra.
However, as revealed in Ref.~\cite{Zhou2003_PRL91-262501},
there is an essential difference in the degree to which the symmetry is broken:
The factor $1/M_{-}^{2} = 1 /(\epsilon -\Sigma-M)^{2}$ is
much smaller for anti-nucleon states than that for nucleon states.
The bound anti-particle energies $\epsilon_\tA$ are in the region
between $M-\Delta(0) \lesssim \epsilon_\tA \lesssim M$, approximately $0.3$~GeV $ \lesssim
\epsilon_\tA \lesssim 1$~GeV. On the other hand the bound particle
states are in the region of $ M-|\Sigma(0)| \lesssim \epsilon_\tN
\lesssim M$, i.e., for realistic nuclei close to $1$~GeV.
Then $|M_{-}(\epsilon_\tA)| > 2|M-S(0)|$ and $|M_{-}(\epsilon_\tN)| < |\Sigma(0)|$.
Thus the factor in front of
the $\tilde{\kappa}$ term for anti-particle states is smaller than for
particle states by more than
a factor $(2|M-S(0)|/|\Sigma(0)|)^{2} \approx 400$.
The SS for anti-particle states is
therefore much less broken than the PSS for particle
states \cite{Zhou2003_PRL91-262501}, as shown in Fig.~\ref{fig:spec-o16-neu} by taken the neutron and anti-neutron spectra in $^{16}$O as examples.

\subsubsection{SS in single-anti-nucleon wave functions}

\begin{figure}[tbhp]
\begin{center}
\includegraphics[width=12cm]{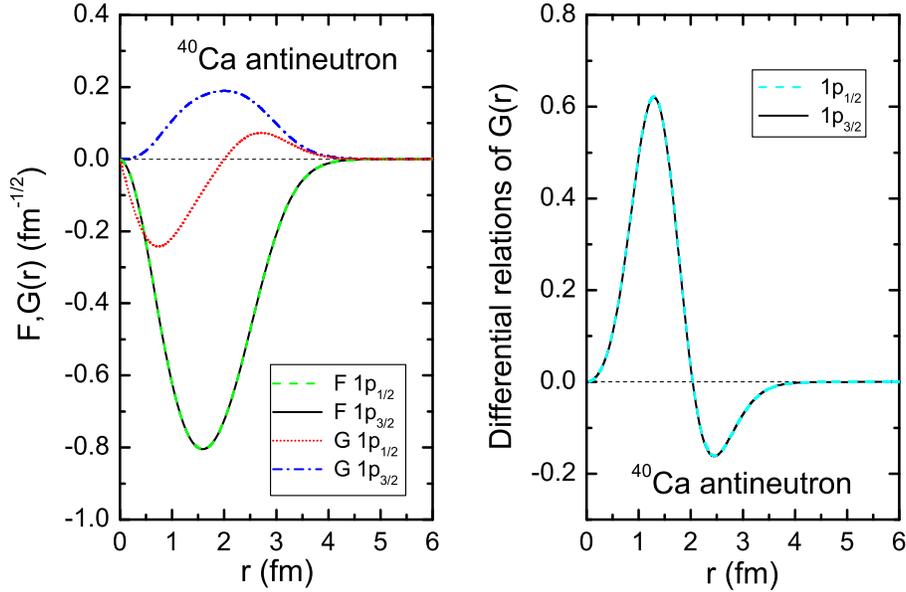}
\end{center}
\caption{(Color online) \label{Fig:Ca-40-0p}
Radial wave functions (the left panel) and
the differential relation (\ref{eq:differrelation}) of the
lower components (the right panel) for the anti-neutron spin
doublets $1p_{1/2}$ ($\varepsilon = 271.91$~MeV) and $1p_{3/2}$ ($\varepsilon = 271.55$~MeV)
in $^{40}$Ca.
Taken from Ref.~\cite{He2006_EPJA28-265} and modified to present notations.
}
\end{figure}

Since the spin-orbit term in Eq.~(\ref{eq:origin-of-symmetry2}) is so small,
for a pair of spin partners in the anti-nucleon spectrum,
the dominant components of radial wave functions should be very similar to each other
in nuclei
and be the same at the SS limit, i.e.,
\begin{equation}
 F_{j_<}(r)= F_{j_>}(r)
 \quad \mbox{if}\quad d\Sigma/dr = 0\,,
 \label{eq:Frelation}
\end{equation}
with $j_<$ ($j_>$) labelling the $j=\tilde l-1/2$ ($j=\tilde l+1/2$) orbital.
This relation has been tested in Refs.~\cite{Zhou2003_PRL91-262501, He2006_EPJA28-265}:
The dominant components $F(r)$ are nearly identical for the
two spin partners; on the other hand, their small components $G(r)$
show dramatic deviations from each other.
However, the small components should satisfy a certain relation
at the SS limit \cite{He2006_EPJA28-265},
\begin{equation}
  \left(\frac{d}{dr}+\frac{\tilde{l}+1}{r}\right)G_{j_>}(r)
 =\left(\frac{d}{dr}-\frac{\tilde{l}}  {r}\right)G_{j_<}(r)\,.
 \label{eq:differrelation}
\end{equation}
Here the radial quantum number $\tilde{n}$ is omitted for brevity.

In Ref.~\cite{He2006_EPJA28-265}, it was examined to what extent the relations given in
Eqs.~(\ref{eq:Frelation}) and (\ref{eq:differrelation}) are fulfilled
in nuclei.
The RMF calculations were performed for $^{40}$Ca, $^{90}$Zr, $^{124}$Sn, and $^{208}$Pb
with the effective interaction NL3 \cite{Lalazissis1997_PRC55-540}.
A good SS is found in both the anti-proton and anti-neutron spectra.
The radial wave functions for the anti-neutron $1p$ doublets
in $^{40}$Ca are shown in the left panel
of Fig.~\ref{Fig:Ca-40-0p}.
The energies of these two
states are $271.91$ and $271.55$~MeV, respectively. One can see that the
upper components $F(r)$ of the eigenfunctions for the spin doublets
are almost identical with each other due to the good SS.
But the lower component $G(r)$ of the wave function of an
anti-neutron state deviates dramatically from that of its spin
partner. In the right panel of Fig.~\ref{Fig:Ca-40-0p}, the
differential relation of the lower components given in
Eq.~(\ref{eq:differrelation}) is presented.
This differential relation is satisfied remarkably well, which gives a
further support to the SS in the anti-nucleon spectra in nuclei.

\subsubsection{SS in single-anti-nucleon spectra in local potentials}

In the past decade, there have been intensive investigations of the SS in
single-particle or anti-particle spectra in various local potentials.
For some special potentials,
exact or approximate analytical solutions can be obtained,
which makes it very convenient to discuss not only the PSS but also the SS.
The readers are referred to Section~\ref{Sect:2.2} for further details.

The HO potential is widely used and discussed in nuclear physics.
For the study of the PSS and/or the SS, the RHO potentials are of the most relevance.
Several examples have been illustrated in Sections~\ref{Sect:2.2}, \ref{Sect:3.1}, and \ref{Sect:3.3}.

In Ref.~\cite{Ginocchio2004_PRC69-034318}, the eigenfunctions and eigenenergies
for a Dirac Hamiltonian with equal scalar and vector HO
potentials for spherical, axially deformed, and triaxially deformed shapes are derived.
It has been shown that under the condition of equal scalar and vector potentials,
the spectrum has a SS.
In particular, for the spherical case, a higher symmetry
analogous to the SU(3) symmetry of the non-relativistic HO was discussed.

In Ref.~\cite{deCastro2006_PRC73-054309}, the generalized RHO Hamiltonian in $1+1$ dimensions was solved.
Both positive and negative quadratic potentials were considered and
the bound-state solutions for particles and anti-particles were discussed.
The main features of these bound states are the same as the ones of
the generalized three-dimensional RHO bound states.
The solutions found for zero pseudoscalar potential are related to the SS
and PSS of the Dirac equation in $3+1$ dimensions.
It has been shown how the charge conjugation and $\gamma^5$ chiral transformations relate
the several spectra obtained and that for massless particles
the SS- and PSS-related problems have the same spectrum
but different spinor solutions.

\begin{figure}[tbhp]
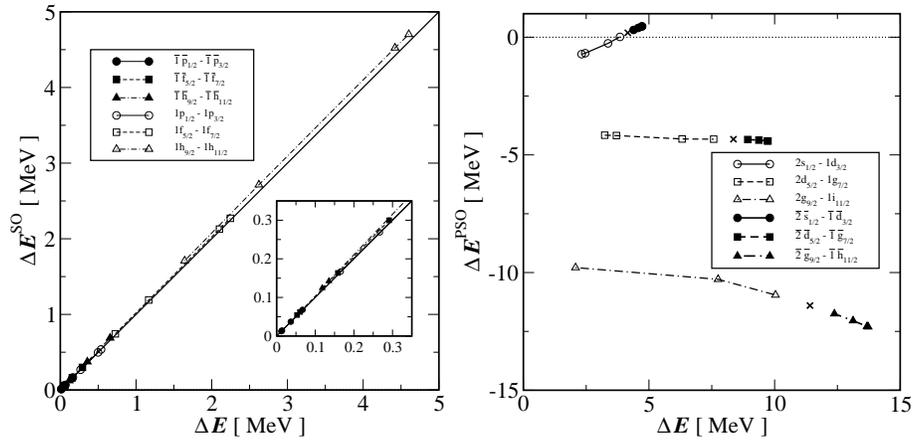

\begin{center}
\includegraphics[width=5.8cm]{figs/3.5.Lisboa2010_PRC81-064324_ws_cc_fig3.eps}
\includegraphics[width=6.1cm]{figs/3.5.Lisboa2010_PRC81-064324_ws_cc_fig4.eps}
\end{center}
\caption
{Left panel: Spin-orbit terms splittings for three neutrons and
anti-neutrons spin partners plotted against the respective energy splittings $\Delta E$.
The thicker solid line represents the values for which $\Delta E^{\rm SO}=\Delta E$.
Right panel: Pseudospin-orbit terms splittings for three neutrons and anti-neutrons
spin partners plotted against the respective energy splittings $\Delta E$.
The points labelled by `\textsf{x}' mark the point of vector potential $V_0=0$ for each pair of levels.
Taken from Ref.~\cite{Lisboa2010_PRC81-064324}.
}
\label{fig:spin_LS_DE}
\end{figure}

In Ref.~\cite{Lisboa2010_PRC81-064324}, the SS and PSS in
the spectra of nucleons and anti-nucleons in $^{208}$Pb are studied in
scalar and vector Woods-Saxon potentials with different depths.
Lisboa \textit{et al.} examined the SO and PSO couplings
for selected spin and pseudospin partners in both spectra.
To assess the perturbative nature of the spin and pseudospin
symmetries for particles and anti-particles,
the SO and PSO contributions are defined for the energy of the level
with quantum numbers $n\kappa$ (particles),
\begin{align}\label{E_spin_pspin-orbit}
  E^{\rm SO}_{n\kappa} &=
  - \left.{\displaystyle \int_0^\infty \frac{\Delta^{\prime}}{(\epsilon_{n\kappa}+2M-\Delta)^2}
                                      \frac{1+\kappa }{r}\, |G_{n\kappa}|^2\,r^2\,dr}\,\right/\,
         {\displaystyle \int_0^\infty |G_{n\kappa}|^2\,r^2\,dr}\,,
\nonumber\\
  E^{\rm PSO}_{n\kappa} &=
  - \left.{\displaystyle \int_0^\infty\frac{\Sigma ^{\prime}}
                                          {(\epsilon_{n\kappa}-\Sigma)(\epsilon_{n\kappa}+2M-\Delta)}
                                      \frac{1-\kappa }{r}\, |F_{n\kappa}|^2\,r^2\,dr}\,\right/\,
         {\displaystyle\int_0^\infty  |F_{n\kappa}|^2\,r^2\,dr}\,,
\end{align}
and with quantum numbers $\bar n\bar\kappa$ (anti-particles),
\begin{align}\label{E_spin_pspin-orbit_cc}
  E^{\rm SO}_{\bar{n}\bar\kappa} &=
  \left.{\displaystyle \int_0^\infty\frac{\Sigma^{\prime}}
                                        {(\epsilon_{\bar{n}\bar\kappa}+2M+\Sigma)^2}
                                   \frac{1+\bar\kappa }{r}\,
                                   |F_{\bar{n}\bar\kappa}|^2\,r^2\,dr}\,\right/\,
       {\displaystyle \int_0^\infty |F_{\bar{n}\bar\kappa}|^2\,r^2\,dr}\,,
\nonumber\\
  E^{\rm PSO}_{\bar{n}\bar\kappa} &=
  \left.\displaystyle \int_0^\infty\frac{\Delta^{\prime}}
                                        {(\epsilon_{\bar{n}\bar\kappa}+\Delta)(\epsilon_{\bar{n}\bar\kappa}+2M+\Sigma)}
                                   \frac{1-\bar\kappa }{r}\,
                                   |G_{\bar{n}\bar\kappa}|^2\,r^2\,dr \,\right/\,
       \int_0^\infty |G_{\bar{n}\bar\kappa}|^2\,r^2\,dr\,.
\end{align}

In the left panel of Fig.~\ref{fig:spin_LS_DE}
are shown the SO splittings for three spin partners for both neutrons
and anti-neutrons, i.e., the difference of the SO term
defined in Eq.~(\ref{E_spin_pspin-orbit}) for neutron spin partners and
those in Eq.~(\ref{E_spin_pspin-orbit_cc}) for anti-neutron spin partners.
In the right panel of Fig.~\ref{fig:spin_LS_DE}
are shown the PSO splittings for three pseudospin partners, again for both neutrons
and anti-neutrons.
From these figures one can clearly see the different behaviors of spin and
pseudospin in nuclei for both neutrons and anti-neutrons.
There is a correlation between the values the SO splittings and
the energy splittings for spin partners, the ratio $\Delta E^{\rm SO}/\Delta E$
being very close to $1$ for anti-neutrons.
This is a sign of the perturbative nature of SS in nuclei
for both neutrons and anti-neutrons.
The situation for the pseudospin partners is completely different.
There is no correlation between the PSO splittings $\Delta E^{\rm PSO}$ and the
energy splittings $\Delta E$, even for small values of $\Delta E$.
Thus it was concluded that the PSS in nuclei is not perturbative
for both neutrons and anti-neutrons \cite{Lisboa2010_PRC81-064324}.
The readers are referred to Section~\ref{Sect:4.1} for more discussions
of the non-perturbative behaviors of PSS.

\subsubsection{SS in single-anti-nucleon spectra in non-local potentials}

In the above investigations, the SS in the (anti-)particle spectrum
of local potentials has been discussed.
In fact, the PSS and its origin as well as the importance of
the Fock terms have also been investigated \cite{Lopez-Quelle2003_NPA727-269,Long2006_PLB639-242,Long2010_PRC81-031302R}.
Although the PSS was still found to be a good approximation
in the RHF theory, its mechanism becomes
rather complicated by the presence of the non-local potentials,
see Section~\ref{Sect:3.2} for details.

In Ref.~\cite{Liang2010_EPJA44-119},
the SS in the Dirac negative-energy spectrum and its
origin were investigated in non-local potentials
within the RHF theory.
Taking the nucleus $^{16}$O as an example, the SS in the
negative-energy spectrum was found to be a good approximation and the
dominant components of the Dirac wave functions for the spin
doublets are nearly identical. In comparison with the relativistic
Hartree approximation where the origin of SS lies in the
equality of the scalar and vector potentials, in RHF the
cancellation between the Hartree and Fock terms is responsible for
the better SS properties and determines the subtle
spin-orbit splitting.

In the RHF theory, the radial Dirac equations~(\ref{Eq:3.2.DiracRHF}) are coupled integro-differential ones
due to the non-local Fock terms \cite{Bouyssy1987_PRC36-380}.
By introducing the effective local potentials $X_G$, $X_F$, $Y_G$, and $Y_F$ defined in Eqs.~(\ref{Eq:3.2.XYGF}),
the integro-differential equations can be formally rewritten as equivalent differential ones as shown in Eq.~(\ref{Eq:3.2.Diracloc}).
Finally, one is able to estimate the Hartree and Fock contributions to the SO splittings in the Dirac negative-energy spectrum by using Eqs.~(\ref{Eq:3.2.V1VPSO}).

The denominator ${\Sigma(r)-E}$ in Eqs.~(\ref{Eq:3.2.V1VPSO}) contains a state-dependent potential $Y_G(r)$. However, the quantity $Y_G(r)$ is around a few MeV and is negligible in comparison with ${\Sigma(r)-E}$ which is of
the order of $1$~GeV.
Similar argument also holds for the time component of the vector potential
$V(r)$ which contains the rearrangement term from Fock channels \cite{Liang2010_EPJA44-119}.

Within the RMF framework, it has been pointed out that the strong
centrifugal barrier and weak spin-orbit potential lead to the
PSS in the single-nucleon spectrum \cite{Meng1998_PRC58-R628}
and the SS in the single-anti-nucleon spectrum \cite{Zhou2003_PRL91-262501}.
In Ref.~\cite{Liang2010_EPJA44-119}, this point has also been examined in the RHF scheme.

\begin{figure}[tbhp]
\begin{center}
\includegraphics[width=12cm]{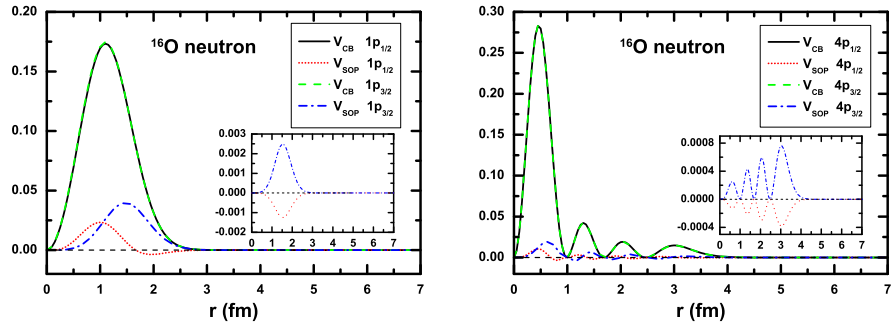}
\end{center}
\caption{(Color online) Centrifugal barriers $V_{\rm CB}$ and
spin-orbit potentials $V_{\rm SO}$ multiplied by the factor $\mp
F^2/(\Sigma-E)$ for the spin doublets $\nu1p$ (the left
panel) and $\nu4p$ (the right panel) in the negative-energy spectrum of $^{16}$O.
The insets show the Hartree contributions of the spin-orbit potentials.
Taken from Ref.~\cite{Liang2010_EPJA44-119} and modified to present notations.
}
\label{fig:cbsop03pD}
\end{figure}

In Fig.~\ref{fig:cbsop03pD} are shown the centrifugal barriers
$V_{\rm CB}$ and the spin-orbit potentials $V_{\rm SO}$ multiplied by
the factor $\mp F^2/(\Sigma-E)$ for the neutron spin doublets $1p$ and $4p$,
and their integrals over $r$ are respectively proportional to their
contributions to the single-particle energy. It is clearly shown
that the contribution of $V_{\rm CB}$ is much
larger than that of $V_{\rm SO}$.
Therefore, similar reasons as in the RMF theory lead
to the SS in the negative-energy spectrum in the RHF theory, and
the SO splitting is due to the different SO
potentials $V_{\rm SO}$ of the spin doublets.

In the insets of Fig.~\ref{fig:cbsop03pD} are given the Hartree
contributions to the SO potentials. It was found that the
contributions from the Fock terms to $V_{\rm SO}$ are one order of
magnitude larger than those from the Hartree terms. Therefore, the
Fock terms must play important roles in the SO splittings.

\begin{figure}[tbhp]
\begin{center}
\includegraphics[width=8cm]{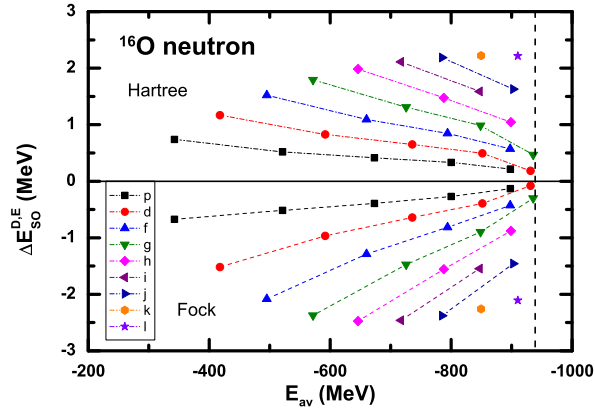}
\end{center}
\caption{(Color online) Hartree and Fock contributions to spin-orbit
splitting
    in the negative-energy spectrum of $^{16}$O
    versus the average energy of the spin doublets.
    The vertical dashed line shows the continuum limit.
Taken from Ref.~\cite{Liang2010_EPJA44-119} and modified to present notations.
}
    \label{fig:splittingHF}
\end{figure}

From Eq.~(\ref{Eq:3.2.V1VPSO}), the contributions to the
single-particle energies $E$ from different channels can be
estimated quantitatively.
To examine the role of the Fock terms, the contributions
from the Hartree and Fock channels to the SO splittings in
the negative-energy spectrum of $^{16}$O versus the average energies
of the spin doublets are shown in Fig.~\ref{fig:splittingHF}.
It is found that the absolute contributions from both Hartree
and Fock parts decrease monotonously with the average energy $E_{\rm
av}$. The contributions from the Hartree terms have an energy
dependence similar to those in the RMF scheme \cite{Zhou2003_PRL91-262501}.
It is also found that the contributions of $V^E_{\rm SO}$ are one order of magnitude larger than $V^D_{\rm SO}$ as shown in Fig.~\ref{fig:cbsop03pD}, but they are substantially cancelled by the other Fock contributions, i.e., the $V^E_1$ and $V_2$ terms \cite{Liang2010_EPJA44-119}.
In total, the
contributions from the Fock terms have an opposite tendency and cancel
with the Hartree ones, thus leading to better spin symmetry. The
competition between the Hartree and Fock terms determines the
sign of the spin-orbit splitting.

In summary, the spin symmetry in the anti-nucleon spectra has been shown to be
very well conserved in nuclei and has also been
tested by examining the wave functions.
The spin symmetry in single-(anti-)particle spectra in various potentials, including
local and non-local potentials, has been investigated.
Nowadays, the studies of the spin symmetry and the pseudospin symmetry are usually combined together
and many interesting topics arise, e.g., the perturbative
nature of these two symmetries and the contribution of Fock terms.
For the spin symmetry in anti-nucleon spectra, the problem concerning the polarization
effects of an anti-nucleon on this symmetry is still open.

\subsection{From nucleon spectra to hyperon spectra}\label{Sect:3.6}

As it has been emphasized in Ref.~\cite{Zhou2003_PRL91-262501},
the annihilation probability of the anti-nucleon
in the nucleus is very large and makes it
very difficult to observe the small spin-orbit splitting of
the anti-nucleon levels experimentally.
Due to the additional strangeness degree
of freedom, it is expected that the annihilation probability of an anti-hyperon
in a normal nucleus is much smaller than that of an anti-nucleon.
Therefore, it might be easier to observe the small spin-orbit splitting of
the anti-hyperon levels, when the spin symmetry is approximately conserved.
In Refs.~\cite{Song2009_CPL26-122102, Song2010_ChinPhysC34-1425,Song2011_CPL28-092101},
Song, Yao, and Meng investigated the spin symmetry in the single-$\bar{\Lambda}$ spectrum, as well as
the corresponding polarization and tensor effects with the relativistic mean-field theory.

In this Section, the spin symmetry in the single-$\bar\Lambda$ spectrum will be introduced, and the polarization and tensor
effects will be discussed.

\subsubsection{SS in single-anti-Lambda spectra}

In the RMF theory, the $\bar{\Lambda}$ hyperon is described as a
Dirac spinor moving in the potentials generated by the meson fields \cite{Lu2003_EPJA17-19,
Lu2011_PRC84-014328},
 \begin{equation}\label{eq:Dirac00}
 \left\{ \boldsymbol{\alpha}\cdot \mathbf{p}
 + V_{\bar{\Lambda}}(\mathbf{r})
 + \beta [M_{\bar{\Lambda}}
 +S_{\bar{\Lambda}}(\mathbf{r})]
 \right\} \psi_{\bar{\Lambda}}(\mathbf{r})
 = \epsilon_{\bar{\Lambda}} \psi_{\bar{\Lambda}}(\mathbf{r})\,,
\end{equation}
where $M_{\bar{\Lambda}}$ is the rest mass of $\bar\Lambda$ and chosen as
$M_{\bar{\Lambda}}=1115.7$~MeV, $\epsilon_{\bar{\Lambda}}$ is the
single-particle energy. As $\bar{\Lambda}$ is charge neutral and
isoscalar, it couples only to the $\sigma$ and $\omega$
mesons. As a consequence, the scalar $S_{\bar{\Lambda}}(\mathbf{r})$ and
vector $V_{\bar{\Lambda}}(\mathbf{r})$ potentials are given by,
\begin{equation}
    S_{\bar{\Lambda}}(\mathbf{r})=g_{\sigma \bar{\Lambda}}\sigma(\mathbf{r})\,,\qquad
    V_{\bar{\Lambda}}(\mathbf{r})=g_{\omega \bar{\Lambda}}\omega(\mathbf{r})\,.
\end{equation}
According to the charge conjugation transformation, the coupling constants for $\bar\Lambda$
are related to those for $\Lambda$ by the following relations,
\begin{equation}
    g_{\sigma \bar{\Lambda}} =  \xi g_{\sigma \Lambda}\,,\qquad
    g_{\omega \bar{\Lambda}} = -\xi g_{\omega \Lambda}\,.
\end{equation}

The Dirac equation for $\bar\Lambda$ (\ref{eq:Dirac00}) can be solved similarly
as those for nucleons.
In Ref.~\cite{Song2009_CPL26-122102}, the SS in the $\bar\Lambda$ spectrum in atomic
nuclei was studied by taking $^{16}$O as an example.
With the mean-field and no-sea approximations, the coupled Dirac
equations for nucleons and $\bar{\Lambda}$ together with the
Klein-Gordon equations for mesons can be self-consistently solved.
The effective interaction PK1~\cite{Long2004_PRC69-034319} is adopted
for the nucleon part, and the coupling constants for $\bar{\Lambda}$
are chosen as $\xi=1$, while $g_{\sigma \bar{\Lambda}}=g_{\sigma \Lambda}={2}/{3}g_{\sigma N}$,
and $g_{\omega \bar{\Lambda}}=-g_{\omega \Lambda}=-{2}/{3}g_{\omega N}$
according to the SU(3) symmetry in naive quark model.

\begin{figure}[tbhp]
\begin{center}
\includegraphics[width=8cm]{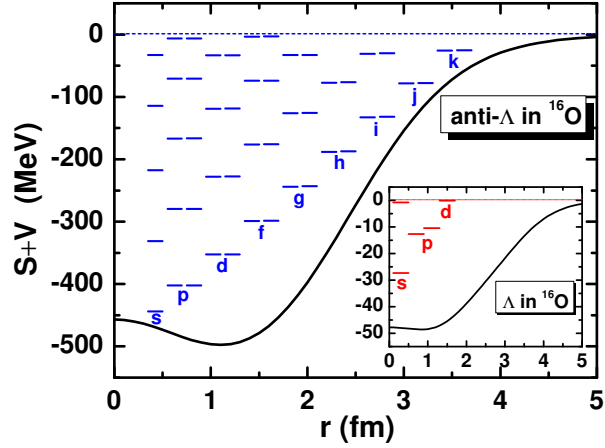}
\end{center}
\caption{(Color online) Potential and spectrum of $\bar{\Lambda}$ in $^{16}$O. For
each pair of spin doublets, the left level is with $\kappa<0$
and the right one with $\kappa>0$. The inset gives the potential
and spectrum of $\Lambda$ in $^{16}$O.
Taken from Ref.~\cite{Song2009_CPL26-122102}.
} \label{fig:spec-o16-AL}
\end{figure}

The potential and single-$\bar{\Lambda}$ spectrum in $^{16}$O are
plotted in Fig.~\ref{fig:spec-o16-AL}, where for each pair of
spin doublets, the left level is with $\kappa<0$ and the right
one with $\kappa>0$.
For comparison, the potential and single-$\Lambda$ spectrum in $^{16}$O
are given as well. As seen in
Fig.~\ref{fig:spec-o16-AL}, the single-$\bar\Lambda$ energies for each
spin doublets are almost identical, and the energy differences
between spin doublets
$\epsilon_{\bar{\Lambda},j_<}-\epsilon_{\bar{\Lambda},j_>}$
in the $\bar\Lambda$ spectrum are around $0.09\sim0.17$~MeV for the $p$ states,
which are much smaller than that in the $\Lambda$ spectrum, $2.26$~MeV.

\begin{figure}[tbhp]
\begin{center}
\includegraphics[width=8cm]{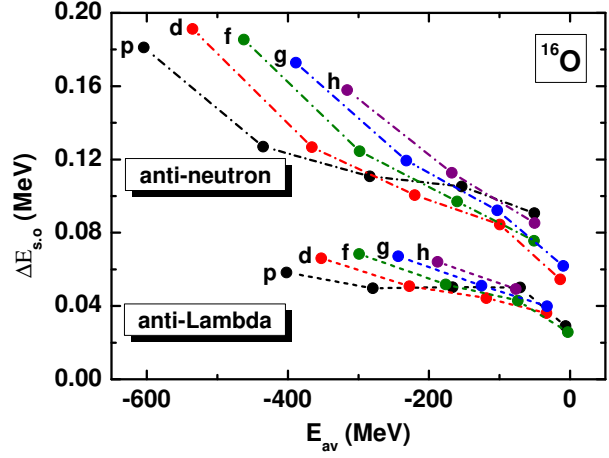}
\end{center}
\caption{(Color online) Reduced spin-orbit splittings $\Delta E_{\rm SO}$ for $\bar{\Lambda}$ and anti-neutrons in
$^{16}$O as a function of the average energy $E_{\rm av}$. For each $l$ orbital, from
left to right read the radial quantum
numbers $n=1,2,\ldots$ Taken from Ref.~\cite{Song2009_CPL26-122102}.
} \label{fig:spli-AL}
\end{figure}

In order to see the splitting and its energy dependence more
clearly, the reduced $\bar{\Lambda}$ SO splittings,
 \begin{equation}
 \label{sosplitting}
 \Delta E_{\rm SO}=(E_{\bar{\Lambda},j_<}-E_{\bar{\Lambda},j_>})/(2l+1)\,,
 \end{equation}
in $^{16}$O are plotted in Fig.~\ref{fig:spli-AL} as a function of the average energies,
 \begin{equation}
 \label{average}
  E_{\rm av}= (E_{\bar{\Lambda},j_<}+E_{\bar{\Lambda},j_>})/2\,,
 \end{equation}
where $E_{\bar{\Lambda}}=\epsilon_{\bar{\Lambda}}-M_{\bar{\Lambda}}$ excluding the rest mass of $\bar\Lambda$.
For comparison, the reduced SO splittings for anti-neutrons are also plotted.
It was found that the $\Delta E_{\rm SO}$ for the $p$ states in the $\bar\Lambda$
spectrum are around $0.03\sim0.06$~MeV,
which is much smaller than those both in the $\Lambda$ spectrum, $0.75$~MeV,
and in the anti-neutron spectrum, $0.09\sim0.18$~MeV.
This indicates that the SS in the $\bar{\Lambda}$ spectrum is even better conserved
than that in the anti-neutron spectrum shown in Ref.~\cite{Zhou2003_PRL91-262501}.
The main reason is that the SO coupling term for $\bar{\Lambda}$ is about
$2/3$ of that for anti-neutrons \cite{Song2009_CPL26-122102}.

\begin{figure}[tbhp]
\begin{center}
\includegraphics[width=8cm]{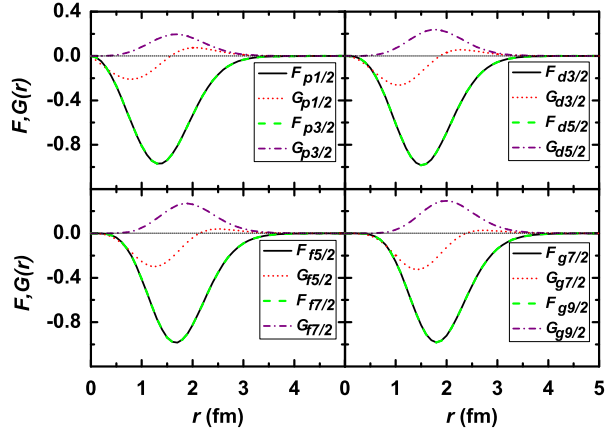}
\end{center}
\caption{Radial wave functions of $\bar{\Lambda}$ spin doublets in $^{16}$O.
Taken from Ref.~\cite{Song2009_CPL26-122102} and modified to present notations.
}
\label{fig:wf-o16-pdfg}
\end{figure}

In Fig.~\ref{fig:wf-o16-pdfg}, the radial wave functions
$F(r)$ and $G(r)$ for several $\bar{\Lambda}$ spin doublets in
$^{16}$O are plotted. Since the SO
splittings in the single-$\bar{\Lambda}$ spectrum are so
small, the dominant components $F(r)$ of the wave functions of
spin doublets are almost identical, while the small components $G(r)$
are quite different.
The nodal relation~(\ref{eq:node2}) is satisfied.

\subsubsection{Polarization effects on SS}

In the above discussions, the polarization
effects of $\bar{\Lambda}$ on the SS was neglected.
For a $\bar{\Lambda}$ really inside $^{16}$O, i.e., the $^{17}_{\bar{\Lambda}}$O system,
the mean fields including the scalar and vector ones are
modified by the $\bar{\Lambda}$ \cite{Burvenich2002_PLB542-261}.

\begin{figure}[tbhp]
\begin{center}
\includegraphics[width=8cm]{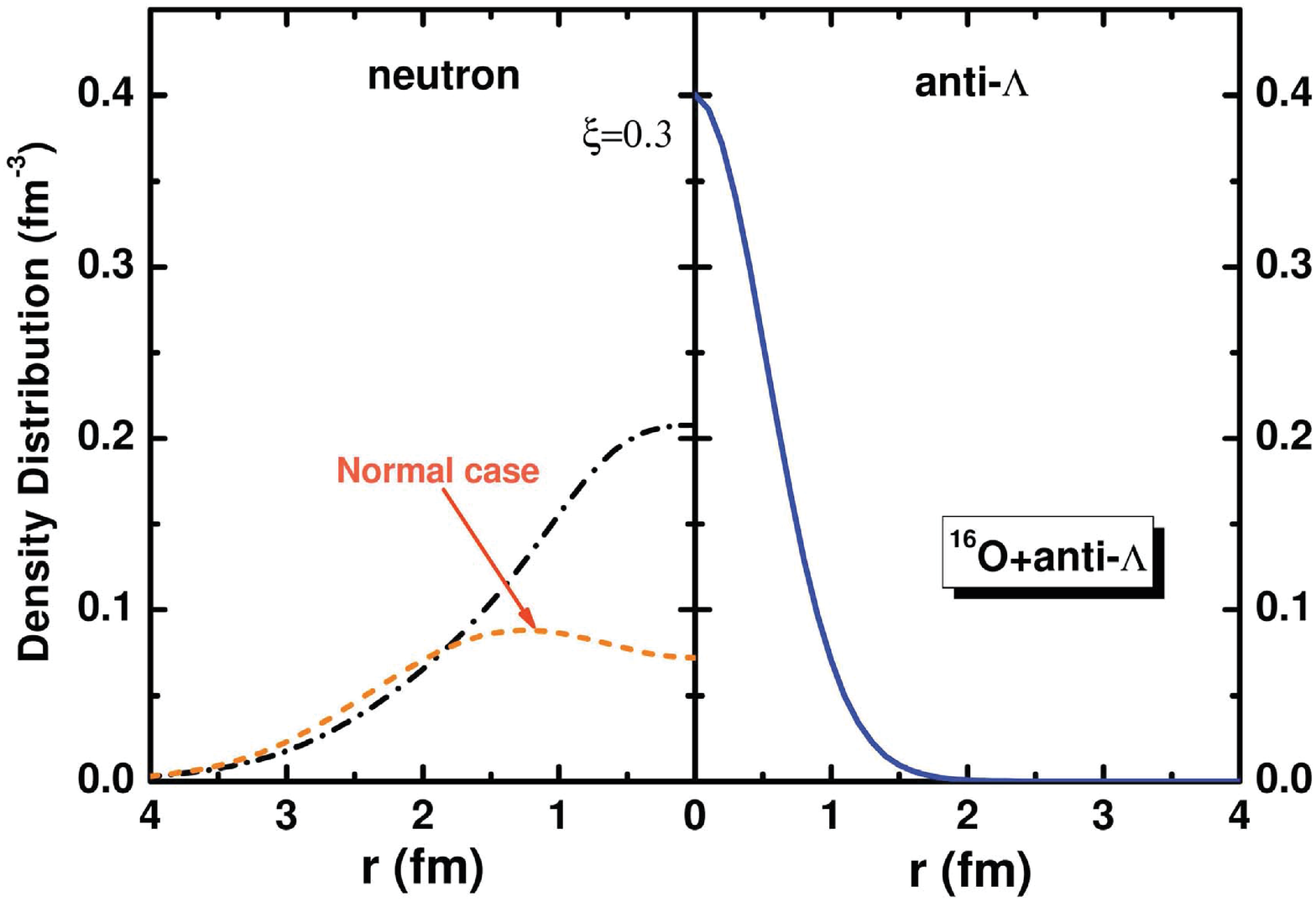}
\end{center}
\caption{
(Color online) Density distributions for $\bar{\Lambda}$ (solid line) and neutrons (dashed-dotted line) in the $^{16}$O+$\bar{\Lambda}$ system.
For comparison, the density distribution for neutrons in $^{16}$O is plotted with the dashed line.
Taken from Ref.~\cite{Song2010_ChinPhysC34-1425}.
}
\label{fig:3.6.Song2010_ChinPhysC34-1425_Fig2}
\end{figure}

In Ref.~\cite{Song2010_ChinPhysC34-1425}, the polarization effect on the SS
due to the $\bar{\Lambda}$ in the $^{16}$O+$\bar{\Lambda}$ system has been studied
with the reduction factor $\xi = 0.3$ for $\bar\Lambda$-meson couplings.
The polarization effect from the valence $\bar\Lambda$ hyperon leads to a highly compression of
nucleus with the central density up to $2\sim3$ times of the normal
saturation density, as seen in Fig.~\ref{fig:3.6.Song2010_ChinPhysC34-1425_Fig2}.
As a result, the energy differences between the spin doublets in the
$\bar\Lambda$ spectrum are around $0.10\sim0.73$~MeV for the $p_{\bar\Lambda}$ states,
which is larger than the results without polarization effect,
$0.09\sim0.17$~MeV,
but still much smaller than the SO splittings in the $\Lambda$ spectrum, $2.26$~MeV.
The dominant components of the Dirac spinor for the $\bar\Lambda$
spin doublets are found to be near identical.

\subsubsection{Tensor effects on SS}

The tensor force has been discussed over decades.
Recently, the tensor force was shown to have a distinct effect on the evolution of the nuclear shell structure and appropriate conservation of pseudospin symmetry, see Section~\ref{Sect:3.3}.
The importance of tensor effects on reducing the spin-orbit splitting of the $\Lambda$ single-particle spectrum has been extensively discussed in the single-$\Lambda$ hypernuclei \cite{Noble1980_PLB89-325,
Jennings1990_PLB246-325, Chiapparini1991_NPA529-589, Yao2008_CPL25-1629}.
Therefore, it is essential to examine further the SS of $\bar{\Lambda}$ in $\bar{\Lambda}$-nucleus system with the presence
of $\bar\Lambda\bar\Lambda\omega$-tensor coupling,
\begin{equation}
  T_{\bar\Lambda} = -\frac{\alpha}{2m_{\bar\Lambda}} i \mathbf{\gamma} \cdot \mathbf{\nabla} V_{\bar\Lambda}\,,
\end{equation}
with $\alpha=f_{\omega\bar\Lambda} / g_{\omega\bar\Lambda}$ in the RMF theory.
This has been done in Ref.~\cite{Song2011_CPL28-092101}.
Note that the tensor coupling discussed here is also of the Lorentz type, as shown with $T(r)$ in Eq.~(\ref{Eq:3.3.Dirac}).

\begin{figure}[tbhp]
\begin{center}
 \includegraphics[width=6cm]{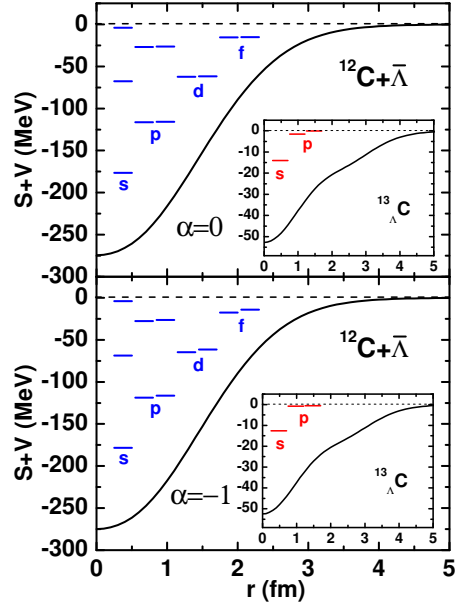}
\end{center}
\caption{(Color online) Single-particle energies for $\bar{\Lambda}$
with ($\alpha=f_{\omega\bar\Lambda}/g_{\omega\bar\Lambda}=-1$) and without ($\alpha=0$) tensor coupling in
$^{12}$C+$\bar{\Lambda}$. For comparison, the insets show the
corresponding results for $\Lambda$ in $^{13}_{\Lambda}$C.
Taken from Ref.~\cite{Song2011_CPL28-092101}.
}
 \label{fig1}
\end{figure}

Taking the $^{12}$C+$\bar{\Lambda}$ system as the first example, the effects of
tensor coupling on the SS in the single-$\bar{\Lambda}$
spectrum were studied in Ref.~\cite{Song2011_CPL28-092101} by using the self-consistent RMF theory with the effective interaction PK1-Y1 \cite{Song2010_IJMPE19-2538, Wang2013_CTP60-479}.
Figure~\ref{fig1} shows the single-particle spectrum for $\bar{\Lambda}$
in $^{12}$C+$\bar{\Lambda}$.
In order to illustrate the tensor effects on the SO
splittings, the single-particle spectrum for $\bar{\Lambda}$
without ($\alpha=0$) tensor coupling is also plotted.
For comparison, the corresponding results for
$\Lambda$ in $^{13}_{\Lambda}$C are given in the insets.
It is shown clearly that the SO splittings of each spin doublets
for $\bar{\Lambda}$ are much smaller than those for $\Lambda$ if the
tensor coupling is not considered.
However, the opposite is true after taking into account the tensor
coupling ($\alpha=-1$), i.e., the SO-splitting size becomes
negligible for $\Lambda$ states as found in
Refs.~\cite{Jennings1990_PLB246-325, Chiapparini1991_NPA529-589,
Cohen1991_PRC44-1181, Mares1994_PRC49-2472, Ma1996_NPA608-305},
but quite noticeable for the $\bar{\Lambda}$ states.

\begin{figure}[tbhp]
\begin{center}
 \includegraphics[width=12cm]{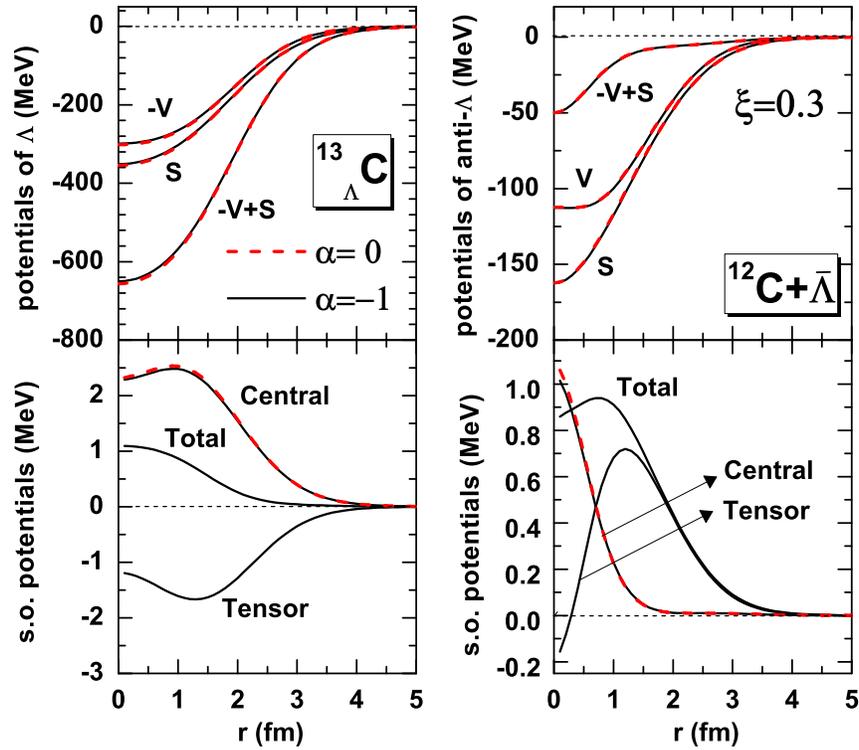}
\end{center}
\caption{(Color online) Comparison of the scalar, vector, and total potentials
(upper panels) and spin-orbit potentials (lower panels) for $\Lambda$ in
$^{13}_{\Lambda}$C and $\bar\Lambda$ in $^{12}$C+$\bar{\Lambda}$
calculated by the self-consistent RMF
theory without ($\alpha=0$, dashed line)
and with ($\alpha=-1$, solid line) the tensor coupling.
Taken from Ref.~\cite{Song2011_CPL28-092101}.
}
 \label{fig2}
\end{figure}

In Fig.~\ref{fig2} are shown the scalar potential $S(r)$, the vector potential
$V(r)$, and $S(r)-V(r)$, as well as the SO potentials for
$\bar{\Lambda}$ in $^{12}$C+$\bar{\Lambda}$ calculated by the self-consistent RMF
theory without and with the
$\bar\Lambda\bar\Lambda\omega$-tensor coupling. For comparison, the
corresponding results for $\Lambda$ in $^{13}_{\Lambda}$C are given
as well. As seen in Fig.~\ref{fig2}, the vector potential of
$\bar{\Lambda}$ changes its sign because of G-parity symmetry. The
derivative of the difference between the vector and scalar
potentials changes dramatically with the radial coordinate only
for $r<1.5$~fm, which leads to the central part of
SO potential decreasing rapidly to zero at $\sim 1.5$~fm.
However, for $\Lambda$ in $^{13}_{\Lambda}$C, it is shown that the
difference between the vector and scalar potentials is quite large.
As the consequence, the corresponding central part of SO
potential is much larger than that for $\bar{\Lambda}$.

Of particular interest is the onset of almost opposite phenomena
after taking into account the tensor effects. The
contribution from tensor coupling (``Tensor'') reduces the SO
potential for $\Lambda$, but enhances that for $\bar\Lambda$.
These effects can be observed from the splittings of SO partner
states, as seen in Fig.~\ref{fig1}.

\begin{table}
\begin{center}
 \caption{Spin-orbit splittings of $\Lambda$ in $^{13}_{\Lambda}$C and
 of $\bar\Lambda$ in $^{12}$C+$\bar{\Lambda}$ calculated by the self-consistent RMF
 theory without ($\alpha=0$) and with ($\alpha=-1$) tensor
 coupling. In the calculations with tensor
 coupling, the expectation values of the SO potentials labelled with ``Central'', ``Tensor'', and
 ``Total'' in Fig.~\ref{fig2} are calculated with the dominant
 components in the Dirac spinors of spin doublets. Their differences
 are shown respectively in column $\Delta E_{\rm SO}$. All energies are in units of MeV.
 The data are taken from Ref.~\cite{Song2011_CPL28-092101}.
}
 \label{Tab1}
\begin{tabular}{@{}ccccccc@{}}
\hline
  & &\multirow{2}{*}{$\Delta E^{\alpha=0}$}& \multicolumn{3}{c}{$\Delta E_{\rm SO}$}&\multirow{2}{*}{$\Delta E^{\alpha=-1}$} \\ \cline{4-6}
  & &                     & Central & Tensor & Total & \\ \hline
  $_{\Lambda}^{13}$C        & $1p$ & $1.51$   &  $1.47$             & $-1.20$   & $0.27$   & $0.26$      \\
 \hline
\multirow{4}{*}{$^{12}$C+$\bar{\Lambda}$}  &$1p$ &  $0.64$ & $0.64$ & $~~1.85$ & $2.49$ &$2.49$ \\
                                           &$2p$ &  $0.33$ & $0.32$ & $~~1.03$ & $1.35$ &$1.37$\\
                                           &$1d$ &  $0.48$ & $0.50$ & $~~2.87$ & $3.37$ &$3.37$\\
                                           &$1f$ &  $0.28$ & $0.30$ & $~~3.18$ & $3.48$ &$3.47$\\
 \hline
\end{tabular}
\end{center}
\end{table}

In Table~\ref{Tab1}, the SO splittings $\Delta E_{\rm SO}=E_{j_<} - E_{j_>}$ are given
for $\Lambda$ in $^{13}_{\Lambda}$C and for $\bar\Lambda$ in $^{12}$C+$\bar{\Lambda}$
calculated by the RMF theory without ($\alpha=0$) and with ($\alpha=-1$) the tensor
coupling.
To show the tensor effects, in Ref.~\cite{Song2011_CPL28-092101}
the expectation values of the
SO potentials with the dominant components in the Dirac spinors
were divided into two parts, ``Central'' and ``Tensor'', respectively.
The difference of the expectation values of total SO potentials
between the spin doublets $\Delta E_{\rm SO}$ gives mainly the observed
SO splittings.

As seen in Table~\ref{Tab1}, the SO splitting for the $p_\Lambda$
states of $^{13}_{\Lambda}$C is $0.26$~MeV, which is in agreement with
the corresponding data $152\pm54\pm36$~keV \cite{Ajimura2001_PRL86-4255}. For
$\bar{\Lambda}$, the SO splittings of the $1p$, $2p$, $1d$, and $1f$
states with the tensor coupling are found to be
$1.37\sim3.47$~MeV, which is an order of magnitude larger than
those without the tensor coupling, $0.28\sim0.64$~MeV.

It is noted that the SO splitting without the tensor
coupling is almost the same as the ``Central'' part of the SO
splitting in the calculations with the tensor coupling.
This indicates that the tensor coupling has negligible contribution to the
``Central'' part of SO potential through the rearrangement of
the mean fields. However, the additional contribution from the tensor
coupling to the SO potential of $\bar\Lambda$, corresponding
to the ``Tensor'' term, dominates the final SO splittings in
the calculations with the tensor coupling. Table~\ref{Tab1} shows
clearly that the ``Tensor'' part of the SO splitting almost
cancels the ``Central'' part for the $\Lambda$ states in
$^{13}_\Lambda$C, but enhances that for the $\bar\Lambda$ states greatly
in $^{12}_\Lambda$C+$\bar\Lambda$.

\begin{figure}[tbhp]
\begin{center}
 \includegraphics[width=12cm]{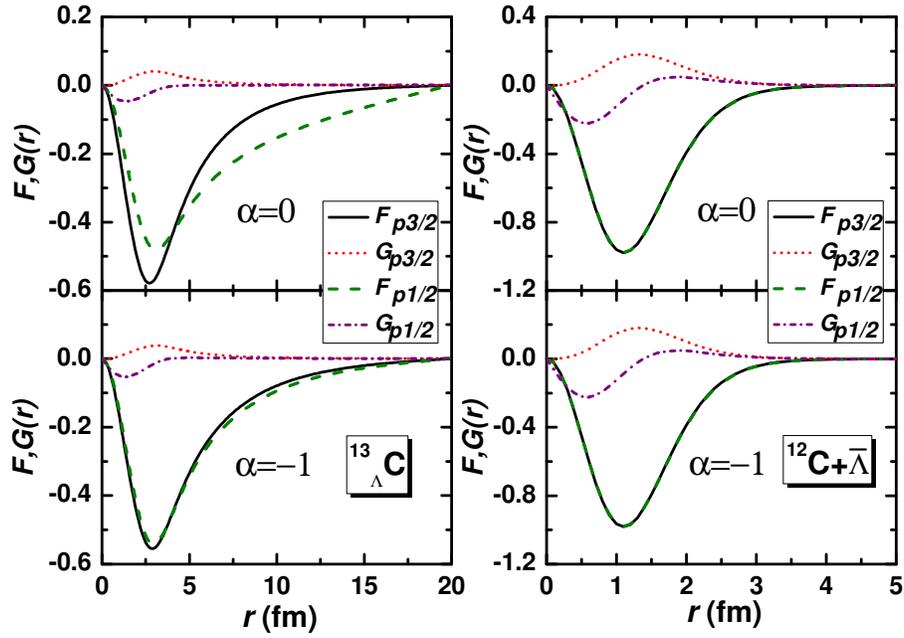}
\end{center}
\caption{(Color online) Radial wave functions for the $p_\Lambda$ states in
$^{13}_{\Lambda}$C (left panels) and $p_{\bar\Lambda}$ states in
$^{12}$C+$\bar{\Lambda}$ (right panels). In each case, the top panel
represents the results without tensor coupling ($\alpha=0$) and the
lower part displays the results with tensor coupling ($\alpha=-1$).
Taken from Ref.~\cite{Song2011_CPL28-092101} and modified to present notations.}
\label{fig3}
\end{figure}

In Fig.~\ref{fig3} are shown the radial wave functions for
the $p_\Lambda$ states in $^{13}_{\Lambda}$C and $p_{\bar\Lambda}$
states in $^{12}$C+$\bar{\Lambda}$ calculated by the RMF theory with
and without the tensor coupling.
It shows clearly that the tensor effect is significant on
the dominant components of Dirac spinors, which recovers the SS on the wave functions of the $p_\Lambda$ spin doublets.
For $\bar\Lambda$, the SS is also well conserved
from the calculations with and without the tensor coupling,
because the SO potential of $\bar\Lambda$ ($\sim1$~MeV) is
much smaller than the corresponding total potential
$V_{\bar\Lambda}+S_{\bar\Lambda}$ ($\sim280$~MeV).
Therefore, the changes of SO potentials due to the tensor coupling has
negligible influence on the final wave functions of the $\bar\Lambda$ states.

\begin{table}
\begin{center}
 \caption{Spin-orbit splittings for different single-particle states of $\bar\Lambda$
 in $^{16}$O+$\bar{\Lambda}$, $^{40}$Ca+$\bar{\Lambda}$, and $^{208}$Pb+$\bar{\Lambda}$
calculated by the
RMF theory without ($\alpha=0$) and with ($\alpha=-1$) tensor
coupling. All energies are in units of MeV.
The data are taken from Ref.~\cite{Song2011_CPL28-092101}.
}
 \label{Tab2}
\begin{tabular}{@{}ccccccc@{}} \hline
  & &\multirow{2}{*}{$\Delta E^{\alpha=0}$}& \multicolumn{3}{c}{$\Delta E_{\rm SO}$}&\multirow{2}{*}{$\Delta E^{\alpha=-1}$} \\ \cline{4-6}
  & &                     & Central& Tensor&Total&\\ \hline
                          &$1p$ & $0.39$ & $0.40$ & $1.48$  & $1.88$ & $1.88$\\
                          &$2p$ & $0.23$ & $0.22$ & $0.89$  & $1.11$ & $1.12$\\
$^{16}$O+$\bar{\Lambda}$  &$1d$ & $0.29$ & $0.30$ & $2.11$  & $2.41$ & $2.41$\\
                          &$2d$ & $0.16$ & $0.16$ & $0.95$  & $1.11$ & $1.12$\\
                          &$1f$ & $0.18$ & $0.19$ & $2.30$  & $2.49$ & $2.48$\\
\hline
                          &$1p$ & $0.26$ & $0.28$ & $1.22$  & $1.50$ & $1.48$\\
                          &$2p$ & $0.23$ & $0.23$ & $0.90$  & $1.13$ & $1.14$\\
$^{40}$Ca+$\bar{\Lambda}$ &$1d$ & $0.08$ & $0.09$ & $0.73$  & $0.82$ & $0.82$\\
                          &$2d$ & $0.17$ & $0.18$ & $1.29$  & $1.47$ & $1.46$\\
                          &$1f$ & $0.05$ & $0.05$ & $0.70$  & $0.75$ & $0.75$\\
\hline
                          &$1p$ & $0.12$ & $0.15$ & $0.64$  & $0.79$ & $0.76$\\
                          &$2p$ & $0.15$ & $0.15$ & $0.60$  & $0.75$ & $0.76$\\
$^{208}$Pb+$\bar{\Lambda}$&$1d$ & $0.00$ & $0.00$ & $0.05$  & $0.05$ & $0.05$\\
                          &$2d$ & $0.03$ & $0.03$ & $0.27$  & $0.30$ & $0.30$\\
                          &$1f$ & $0.00$ & $0.01$ & $0.06$  & $0.07$ & $0.06$\\
 \hline
\end{tabular}
\end{center}
\end{table}

The tensor effects on the SO splittings for
$\bar{\Lambda}$ have been studied systematically for
$\bar{\Lambda}$-nucleus in different mass regions, including
$^{16}$O+$\bar{\Lambda}$, $^{40}$Ca+$\bar{\Lambda}$, and
$^{208}$Pb+$\bar{\Lambda}$ \cite{Song2011_CPL28-092101}.
The corresponding results are given in Table~\ref{Tab2}. The tensor
effects on the SO splittings for these three cases are similar as those for
$^{12}$C+$\bar{\Lambda}$. Specifically, the SO splittings of
$\bar\Lambda$ in the calculations with the tensor coupling are found
to be $1.12\sim2.48$~MeV in $^{16}$O+$\bar{\Lambda}$,
$0.75\sim1.48$~MeV in $^{40}$Ca+$\bar{\Lambda}$, and
$0.05\sim0.76$~MeV in $^{208}$Pb+$\bar{\Lambda}$, which are an
order of magnitude larger than those from the calculations without
the tensor coupling, i.e., $0.16\sim0.39$~MeV in
$^{16}$O+$\bar{\Lambda}$, $0.05\sim0.26$~MeV in
$^{40}$Ca+$\bar{\Lambda}$, and $0\sim0.15$~MeV in
$^{208}$Pb+$\bar{\Lambda}$.
Moreover, it was found that the
SO splittings for $\bar{\Lambda}$ decrease with the mass
number $A$, no matter the tensor coupling is considered or not.

In summary, it has been shown that the spin symmetry in the $\bar\Lambda$ spectrum
is even better conserved than that in the anti-nucleon spectrum.
Even if the polarization and tensor effects, both enlarging the spin-orbit splittings,
are included, the spin symmetry in the $\bar\Lambda$ spectrum remains well conserved.

\subsection{From spherical nuclei to deformed nuclei}\label{Sect:3.7}

After it was proposed in spherical nuclei, the pseudospin symmetry in deformed nuclei was found to be
an important physical concept also in axially deformed nuclei~\cite{Ratna-Raju1973_NPA202-433,
Draayer1984_AP156-41, Troltenier1994_NPA576-351, Troltenier1995_NPA586-53} and
even in triaxially deformed nuclei~\cite{Blokhin1997_NPA612-163,Beuschel1997_NPA619-119}.

The pseudospin symmetry in deformed nuclei has been investigated extensively.
On one hand, a lot of nuclear structure phenomena have been
interpreted in connection with the pseudospin symmetry,
such as nuclear superdeformed configurations \cite{Dudek1987_PRL59-1405, Bahri1992_PRL68-2133},
identical bands \cite{Nazarewicz1990_PRL64-1654, Nazarewicz1990_NPA512-61, Zeng1991_PRC44-R1745},
and pseudospin partner bands \cite{Xu2008_PRC78-064301, Hua2009_PRC80-034303}.
On the other hand,
in early years, much efforts were devoted
to revealing connections between the normal spin-orbit representation
and the ``pseudo'' spin-orbit one and
to exploring the microscopic origin of the pseudospin symmetry
with deformed harmonic oscillator potentials \cite{Bohr1982_PS26-267, Castanos1992_PLB277-238,
Bahri1992_PRL68-2133, Blokhin1995_PRL74-4149, Blokhin1996_JPA29-2039, Blokhin1997_NPA612-163, Beuschel1997_NPA619-119}.
In particular, it was found that the pseudospin symmetry is conserved exactly for
the Nilsson Hamiltonian
with one-body orbit-orbit ($v_{ll}$) and spin-orbit ($v_{ls}$) interaction
strengths satisfying the condition $v_{ls} = 4 v_{ll}$.
Moreover, this condition is found to be consistent with the relativistic mean-field results \cite{Bahri1992_PRL68-2133}.

Following the success of understanding the pseudospin symmetry in spherical nuclei~\cite{Ginocchio1997_PRL78-436,
Meng1998_PRC58-R628},
the study of the pseudospin symmetry within the relativistic framework was
quickly extended to deformed systems~\cite{Lalazissis1998_PRC58-R45,
Sugawara-Tanabe1998_PRC58-R3065, Sugawara-Tanabe2000_PRC62-054307,
Sugawara-Tanabe2002_PRC65-054313, Ginocchio2004_PRC69-034303,
Sugawara-Tanabe2005_RMP55-277}.
First, the pseudospin symmetry limit in deformed nuclei was discussed \cite{Sugawara-Tanabe1998_PRC58-R3065} and
the evolution of the pseudospin symmetry with deformation was revealed \cite{Lalazissis1998_PRC58-R45}.
Later, the pseudospin symmetry in realistic deformed nuclei was tested by examining the single-particle wave functions \cite{Ginocchio2004_PRC69-034303}.
In Ref.~\cite{Sun2012_EPJA48-18}, the influences from different fields of mesons
on the pseudospin symmetry were investigated for deformed nuclei.
Quite recently, the similarity renormalization group was used to study the pseudospin symmetry
in axially deformed Dirac Hamiltonian \cite{Guo2014_PRL112-062502}.

In this Section, we will focus on progress on the pseudospin symmetry in deformed systems
made within the relativistic framework.

\subsubsection{PSS in deformed nuclei}

In Ref.~\cite{Sugawara-Tanabe1998_PRC58-R3065}, two kinds of conditions were derived for the PSS in deformed nuclei.
One is exact,
\begin{equation}\label{Eq:3.7.PSSexact}
  \frac{\partial \Sigma}{\partial r_\bot} = 0\,,\qquad
  \frac{\partial \Sigma}{\partial z} = 0\,,
\end{equation}
and the other is approximate,
\begin{equation}\label{Eq:3.7.PSSapprox}
  \left|\frac{1}{M_-} \frac{\partial M_-}{\partial r_\bot} \frac{\tilde\Omega}{r_\bot}\right|
  \ll \left|\frac{\tilde\Omega^2}{r^2_\bot}\right|\,,\qquad
  \left|\frac{1}{M_-} \frac{\partial M_-}{\partial z} \frac{\tilde\Omega\pm1}{r_\bot}\right|
  \ll \left|\frac{\tilde\Omega^2}{r^2_\bot}\right|\,,
\end{equation}
where $r_\bot = \sqrt{x^2+y^2}$, $M_- = -M-\Sigma+\epsilon$, and $\tilde\Omega=\Omega+1/2$ corresponding to $\tilde l_z$.
The exact conditions correspond to the PSS limit but, similar to the spherical case, they are certainly not satisfied in realistic nuclei, while the approximate ones can be checked in realistic nuclei.
In Refs.~\cite{Sugawara-Tanabe2000_PRC62-054307, Sugawara-Tanabe2002_PRC65-054313},
the axially deformed RMF theory \cite{Gambhir1990_APNY198-132, Ring1997_CPC105-77}
was used to make the numerical check of the approximate conditions for
the PSS in a well-deformed nucleus $^{154}$Sm.
It was found that these conditions are well satisfied.

The contributions from different fields of mesons and photons to
the PSS has been investigated for spherical
nuclei \cite{Guo2010_EPJA45-179}.
In Ref.~\cite{Sun2012_EPJA48-18}, such study was extended to
deformed nuclei and it was found that (i) the $\sigma$- and $\omega$-fields
are dominant in influencing the PSS, and (ii) the PSO splitting
is mainly determined by the cancellation of these two fields.

\subsubsection{Evolution of PSS with deformation}

\begin{figure}[tbhp]
\begin{center}
\includegraphics[width=6cm]{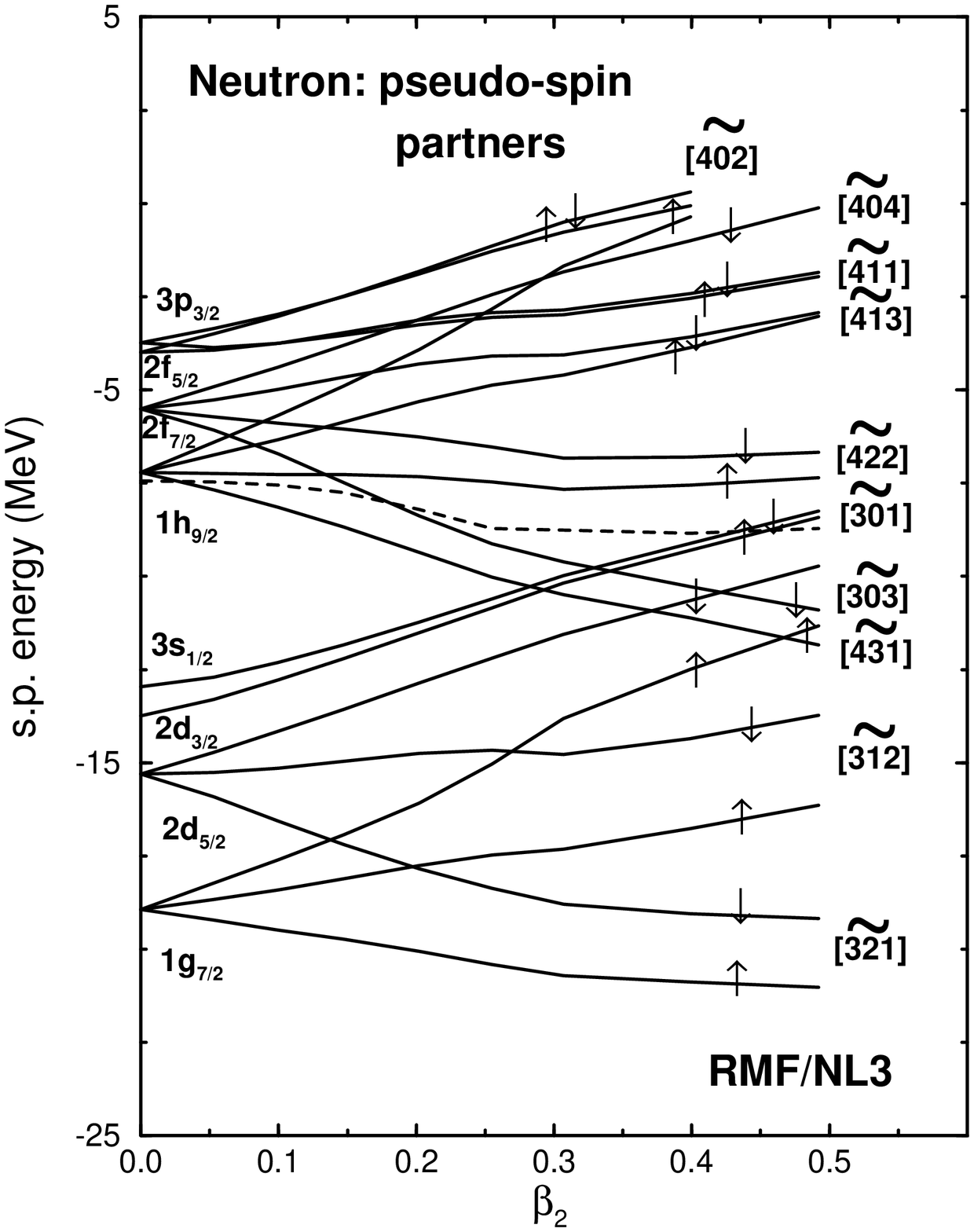}
\includegraphics[width=6cm]{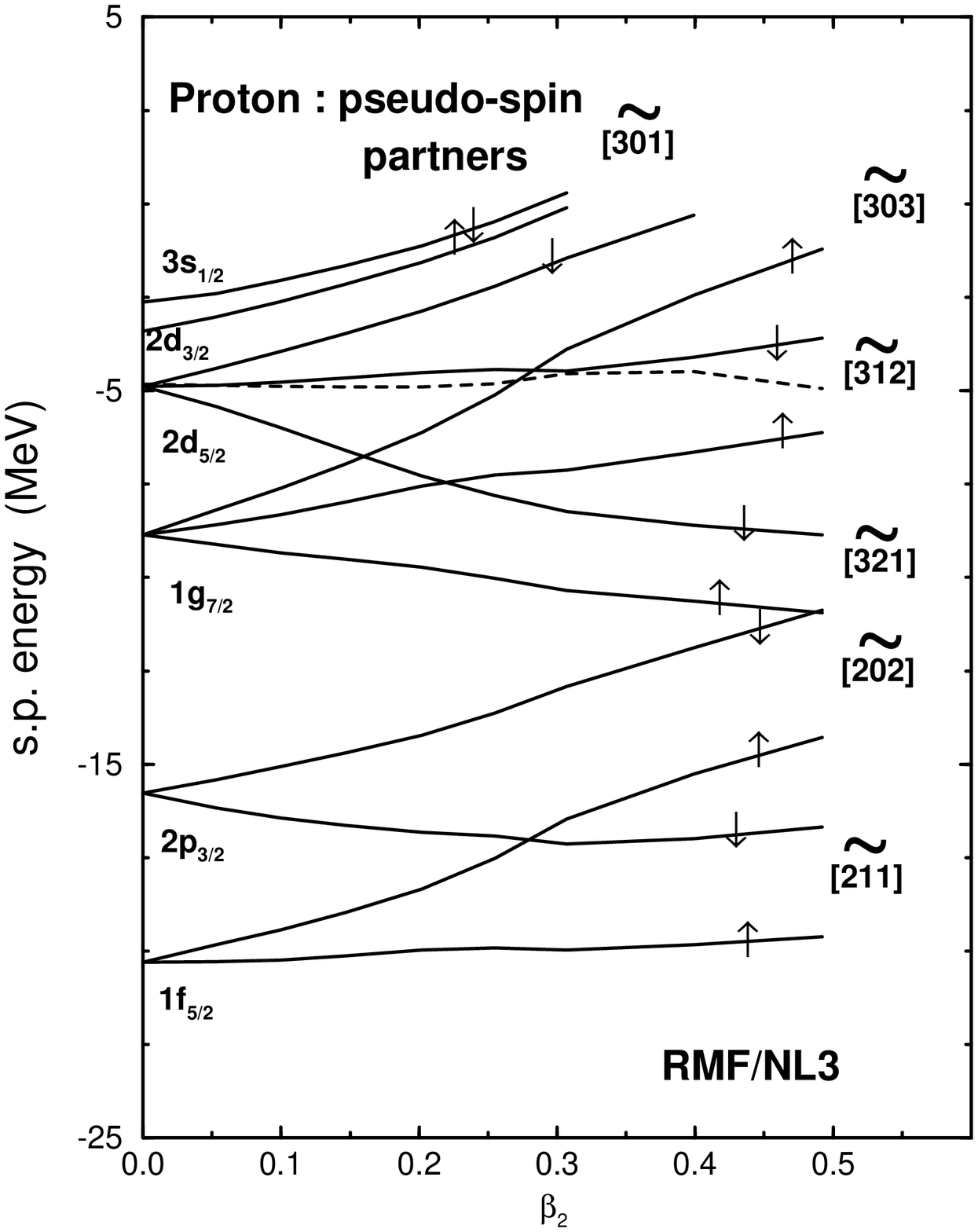}
\end{center}
\caption{Single-particle spectra in the deformed nucleus $^{154}$Dy
as a function of the quadrupole deformation parameter $\beta_2$.
Asymptotic pseudospin quantum numbers are given and the
pseudospin partners are indicated by arrows
$\uparrow $ and $\downarrow$.
Taken from Ref.~\cite{Lalazissis1998_PRC58-R45}.
} \label{fig:Lalazissis1998_PRC58-R45_fig4}
\end{figure}

Since the PSS is observed in both spherical and deformed nuclei,
how the PSS evolves with the shape of nuclei becomes an interesting problem.
In Ref.~\cite{Lalazissis1998_PRC58-R45},
the broken PSS was investigated in axially deformed nuclei
and the evolution of the PSS with deformation was studied by making
the constrained RMF calculations \cite{Gambhir1990_APNY198-132, Ring1997_CPC105-77}
with the effective interaction NL3 \cite{Lalazissis1997_PRC55-540}.
The quasi-degenerate pseudospin doublets were confirmed to exist
near the Fermi surface for deformed nuclei.

In Fig.~\ref{fig:Lalazissis1998_PRC58-R45_fig4},
the single-particle states corresponding to
pseudospin doublets in $^{154}$Dy are plotted against the deformation $\beta_{2}$.
The asymptotic Nilsson quantum numbers $[N,n_{3},\Lambda ,\Omega]$ are good
for large values of the deformation parameter $\beta_{2}$.
The pseudospin doublets $[\tilde{N},\tilde{n}_{3},\tilde{\Lambda},
\tilde{\Omega}=\tilde{\Lambda}\pm 1/2]$ \cite{Bohr1982_PS26-267} are indicated
by $[\tilde{N},\tilde{n}_{3},\tilde{\Lambda}]$ with
$\uparrow $ and $\downarrow $.
For zero deformation ($\beta _{2}=0$) the orbitals are
indicated by the corresponding spherical states.
In Fig.~\ref{fig:Lalazissis1998_PRC58-R45_fig4}, it is found that
the PSO splitting stays almost constant and does not
vary with deformation for $\beta_{2}$ sufficiently large.
Furthermore, the energy difference between the $\downarrow$ and
$\uparrow$ partners always remains positive except for $[\tilde{{{404}}}]$,
where there appears a crossing at $\beta \sim 0.3$. Similar crossing
has been observed in Ref.~\cite{Bohr1982_PS26-267}.

\subsubsection{PSS in deformed single-particle wave functions}

One consequence of the fact that the PSS is a relativistic symmetry of
the Dirac Hamiltonian is that the relativistic wave functions of
the corresponding pseudospin doublets satisfy certain relations.
The relations (\ref{Eq:2.3.FatPSS}) and (\ref{Eq:2.3.GatPSS}) have been tested in spherical nuclei \cite{Ginocchio1998_PRC57-1167,
Ginocchio2002_PRC66-064312, Lalazissis1998_PRC58-R45}.
In Ref.~\cite{Ginocchio2004_PRC69-034303},
the Dirac wave functions in deformed nuclei were examined extensively.
The conditions of the Dirac single-particle wave functions constrained by the PSS
were derived and tested in the self-consistent RMF calculations.

In summary, the pseudospin symmetry is connected to many experimental observations in deformed nuclei.
Theoretically, much effort has been devoted to the study of the conditions of the pseudospin symmetry, the
evolution of the pseudospin symmetry with deformation, and the pseudospin symmetry in single-particle wave functions.
In a recent work \cite{Guo2014_PRL112-062502}, the pseudospin and spin symmetries of the Dirac Hamiltonian with axially deformed
scalar and vector potentials were studied by using the similarity renormalization group theory, see Section~\ref{Sect:4.3.3} for details.
There are still many open problems, e.g., the evolution of the pseudospin symmetry with the triaxial
deformation.

%
\section{Open Issues on PSS and SS}\label{Sect:4}

\subsection{Perturbative or not}\label{Sect:4.1}

Since the pseudospin symmetry was recognized as a relativistic symmetry of the Dirac Hamiltonian \cite{Ginocchio1997_PRL78-436}, the perturbative nature of this symmetry has become a hot topic.
The main concern is there are no bound states at the exact pseudospin symmetry limit, $d\Sigma(r)/dr = 0$ (\ref{Eq:2.1.PSSlimit}), thus the pseudospin symmetry is always broken in realistic nuclei.
The non-perturbative behaviors of pseudospin symmetry have been carefully considered since the quantitative investigations by Marcos \textit{et al.} in 2001 \cite{Marcos2001_PLB513-30}.
It is found that the small splitting of the two pseudospin partners cannot be justified by the smallness of the pseudospin-orbit term (\ref{Eq:2.1.VPSOandVPCB}), but rather by a complicated counter balance between contributions from different terms due to the characteristics of the whole Dirac equation itself.
Following Arima's definition of a dynamical symmetry \cite{Arima1999_RIKEN-AF-NP-276},
Alberto \textit{et al.} \cite{Alberto2001_PRL86-5015,Alberto2002_PRC65-034307} related such non-perturbative behaviors to the dynamical nature of the pseudospin symmetry.

In a sense, such kind of complicated counter balance between contributions from different terms indicates the pseudospin-orbit term should not be considered as the appropriate symmetry-breaking term \cite{Marcos2008_EPJA37-251}.
This triggered recent investigations on the alternative PSS limits, from which a smooth transition to realistic nuclei can be performed \cite{Marcos2008_EPJA37-251,Typel2008_NPA806-156}.
In particular, Typel \cite{Typel2008_NPA806-156} derived the pseudospin symmetry-breaking potential by using the supersymmetric quantum mechanics, and showed that the relativistic harmonic oscillator potential is the simplest case leading to a vanishing pseudospin symmetry-breaking potential.
Therefore, the Dirac Hamiltonian with the relativistic harmonic oscillator potential can be regarded as an alternative pseudospin symmetry limit, based on which one hopes to understand the pseudospin symmetry breaking in realistic nuclei in a perturbative way.

In Ref.~\cite{Liang2011_PRC83-041301R}, Liang \textit{et al.} used the perturbation theory to investigate the spin and pseudospin symmetries of the Dirac Hamiltonian and their breaking in realistic nuclei.
The perturbation corrections to the single-particle energies and wave functions were calculated order by order.
In such a way, the link between the single-particle states in realistic nuclei and their counterparts in the symmetry limits is constructed explicitly.
It is found that the energy splitting of the pseudospin doublets can be regarded as a result of small perturbation around the Dirac Hamiltonian with the relativistic harmonic oscillator potentials.

In this Section, we will focus on these discussions on the non-perturbative or perturbative nature of the pseudospin symmetry.

\subsubsection{Non-perturbative behaviors of PSS}

In Ref.~\cite{Marcos2001_PLB513-30}, the Schr\"odinger-like equation (\ref{Eq:2.1.SchrF}) for the lower component of the Dirac spinor was expressed as
\begin{equation}\label{Eq:4.1.SchrF}
  -F'' + \ls \frac{M'_-}{M_-} \lb\frac{F'}{F} -\frac{\kappa}{r}\rb +\frac{\tilde l(\tilde l+1)}{r^2} +2M(S+V) + 2EV +(S^2-V^2) -E^2\rs F
  = 2MEF\,,
\end{equation}
where $E=\epsilon-M$ excluding the rest mass of nucleon and $M_-(r)=E-V(r)-S(r)$.
It was proven in Ref.~\cite{Marcos2000_PRC62-054309} that the term $({M'_-}/{M_-})({F'}/{F} -{\kappa}/{r})$ is not singular even at the so-called singularity point $r_0$ with $M_-(r_0)=0$, and thus, the integrals, e.g.,
\begin{equation}\label{Eq:4.1.Eeach}
  \int_0^\infty \frac{M'_-}{M_-}\frac{\kappa}{r}F^2dr\,,
\end{equation}
are finite.
Similar to Eq.~(\ref{Eq:3.2.Eeach}), one can evaluate the contributions of the terms containing $F''$, $F'$, $S,V$ and/or $E$, $\tilde l$, and $\kappa$ to the single-particle energy $E$,
\begin{equation}\label{Eq:4.1.decompose}
  E(F'') +E(F') +E(S,V,E) + E(\tilde l) + E(\kappa) = E\,.
\end{equation}

\begin{table}[tbhp]
\begin{center}
\caption{Contributions from different terms in Eq.~(\ref{Eq:4.1.decompose}) to the single-particle energy $E$ in $^{40}$Ca calculated by the RMF theory with the effective interaction NL3 \cite{Lalazissis1997_PRC55-540}.
All units are in MeV.
The data are taken from Ref.~\cite{Marcos2001_PLB513-30}.\label{Tab:4.1.decompose}}
\begin{tabular}{@{}llrrrrrr@{}} \hline
$m_\sigma$ & State & $E(F'')$ &  $E(F')$ & $E(S,V,E)$ & $E(\tilde l)$ & $E(\kappa)$ & $E$~~~ \\ \hline
$508.2$ & $2s_{1/2}$ & $18.53$ & $-0.86$ & $-45.16$ & $12.34$ & $-1.82$ & $-16.96$ \\
        & $\kappa=0$ & $13.14$ &  $3.07$ & $-49.86$ & $11.66$ &  $0.00$ & $-21.98$ \\
        & $1d_{3/2}$ & $14.00$ &  $1.99$ & $-47.58$ & $11.76$ &  $3.65$ & $-16.17$ \\ \hline
$541.0$ & $2s_{1/2}$ & $19.01$ & $-0.82$ & $-46.39$ & $12.79$ & $-1.54$ & $-16.95$ \\
        & $\kappa=0$ & $13.68$ &  $3.29$ & $-51.33$ & $11.97$ &  $0.00$ & $-22.40$ \\
        & $1d_{3/2}$ & $14.93$ &  $2.14$ & $-49.14$ & $12.10$ &  $3.02$ & $-16.95$ \\ \hline
$565.0$ & $2s_{1/2}$ & $19.26$ & $-0.76$ & $-47.10$ & $13.12$ & $-1.28$ & $-16.77$ \\
        & $\kappa=0$ & $14.04$ &  $3.47$ & $-52.22$ & $12.18$ &  $0.00$ & $-22.52$ \\
        & $1d_{3/2}$ & $15.65$ &  $2.24$ & $-50.15$ & $12.31$ &  $2.38$ & $-17.56$ \\ \hline
\end{tabular}
\end{center}
\end{table}

In Table~\ref{Tab:4.1.decompose}, the contributions from different terms in Eq.~(\ref{Eq:4.1.decompose}) to the single-particle energy $E$ are shown by taking the $2s_{1/2}$ and $1d_{3/2}$ pseudospin doublets in $^{40}$Ca as examples.
The results obtained by using three different values of the $\sigma$-meson mass are compared.
In order to illustrate the self-consistent effects caused by the $\kappa$ term, the results appearing in the intermediate line correspond to the solution of Eq.~(\ref{Eq:4.1.decompose}) by taking $\kappa=0$, i.e., neglecting the PSS-breaking term but keeping the PCB.

It is pointed out from these results that \cite{Marcos2001_PLB513-30}:
(i) For a pair of pseudospin doublets, the difference between the contributions of the $\kappa$ term $|E(\kappa_a)-E(\kappa_b)|$ is not negligible in comparison with the contributions of their PCB, $[E(\tilde l_a),E(\tilde l_b)]$.
(ii) In all cases considered, the difference between the contributions to the single-particle energy coming from the different terms $[|E(F''_a)-E(F''_b)|, |E(F'_a)-E(F'_b)|,\ldots]$, in particular of the $\kappa$ term, is larger than the net energy splitting $|E_a-E_b|$ itself, except for the PCB term.
(iii) The $\kappa$ term cannot be considered at all as a perturbation, as it can be inferred from the two previous assertions.

Point (ii) reveals the importance of the self-consistent effects due to the $\kappa$ term, which modify the small
component $F(r)$ of the Dirac spinor.
These effects are not so small as it has been supposed.
As a result, they yield the differences between the contributions to the single-particle energy from most of terms in Eq.~(\ref{Eq:4.1.decompose}) larger than the net splitting $|E_a-E_b|$ itself.

Point (iii) becomes more evident if one compares the results for the $2s_{1/2}$ state ($\kappa=-1$) to the corresponding fictitious case of $\kappa=0$.
Supposing the correction is perturbative, the effect of the $\kappa$ term on the energy of the $2s_{1/2}$ state should be negative.
However, the total energy $E$ of the $2s_{1/2}$ state indeed increases a few MeV when such a $\kappa$ term is included.
That is to say, $E_{\kappa=-1}-E_{\kappa=0}$ has an opposite sign than the case that the $\kappa$ term can be regarded as a small perturbation.

Furthermore, the ordering of the pseudospin doublets strongly depends on the particular model considered, mainly through the value of $m_\sigma$, which has a strong influence on the smoothness of the nuclear surface.
An exact degeneration of the pseudospin doublets can even be achieved with a particular value of $m_\sigma$.
Thus, the accomplishment of the PSS does not necessarily require a small $\kappa$ term.

Therefore, based on the Schr\"odinger-like equation for the lower component of the Dirac spinor and the energy decomposition given in Eq.~(\ref{Eq:4.1.decompose}), the small splitting of the two pseudospin partners cannot be justified by the smallness of the $\kappa$ term, but rather by a complicated counter balance between contributions from different terms due to the characteristics of the whole Dirac equation itself \cite{Marcos2001_PLB513-30}.

\subsubsection{PSS as a dynamical symmetry}

The non-perturbative behaviors of PSS was then related to the dynamical nature of PSS \cite{Alberto2001_PRL86-5015,Alberto2002_PRC65-034307}, in the sense of Arima's definition of a dynamical symmetry \cite{Arima1999_RIKEN-AF-NP-276}: (i) a symmetry of the Hamiltonian which is not geometrical in nature, or (ii) an ordered breaking symmetry from dynamical reasons.

By using the Woods-Saxon potentials in the Dirac equation (\ref{Eq:2.1.DiraceqR}),
\begin{equation}\label{Eq:4.1.WS}
  \Sigma(r) = \frac{\Sigma_0}{1+e^{(r-R)/a}}
  \qquad\mbox{and}\qquad
  \Delta(r) = \frac{\Delta_0}{1+e^{(r-R)/a}}\,,
\end{equation}
the sensitiveness of the PSO splitting with the potential parameters $\Sigma_0$, $\Delta_0$, $R$, and $a$ has been investigated systematically \cite{Alberto2001_PRL86-5015}.
It is found that the degeneracy of pseudospin doublets is very much dependent on the shape of the nuclear mean field potential.
The energy splittings decrease as $a$ increases and $R$ or $|\Sigma_0|$ decreases.
Varying $|\Sigma_0|$ and $R$ simultaneously but keeping their product $|\Sigma_0|R^2$ fixed, the pseudospin doublet splittings remain almost constant.
That is the reason why the PSS is better realized for neutrons than for protons in an isotopic chain towards neutron-rich nuclei.
It is also found that the pseudospin levels can cross each other if $a$ increases or $R$, $|\Sigma_0|$ decreases enough.
This stresses an aspect of the dynamical nature of PSS \cite{Alberto2002_PRC65-034307}.

As a step further, in Ref.~\cite{Alberto2002_PRC65-034307} the contributions to the single-particle energies $E$ are decomposed into
\begin{equation}\label{Eq:4.1.WSdecom}
  \lc \frac{p^2}{2M^*}\rc + \lc V_{\rm PSO}\rc + \lc V_{\rm Darwin}\rc + \lc \Sigma\rc = E\,,
\end{equation}
corresponding to the kinetic, PSO, Darwin, and mean-field potential terms defined in Eq.~(\ref{Eq:3.3.Tdecom}).
This decomposition is quite close to that shown in Eq.~(\ref{Eq:4.1.decompose}), while $\lc p^2/(2M^*)\rc$ stands for the sum of $E(F'')$ and $E(\tilde l)$.
The calculated results can be found in Table~\ref{Tab:3.3.omega} for the cases without tensor potential.
Taking the $1\tilde{h}$ pseudospin doublets as an example, it is clearly seen that the contribution to the PSO splittings of the PSO term is larger than the net splitting itself, and it is at least of the same order of magnitude but has the opposite sign compared to the kinetic and $\lc \Sigma\rc$ terms.

Therefore, it is concluded that the onset of the PSS in nuclei is dynamical, since it results mainly from cancellations of several terms contributing to the single-particle levels, instead of being a consequence of a small PSO coupling  \cite{Alberto2001_PRL86-5015,Alberto2002_PRC65-034307}.

\subsubsection{RHO as a PSS limit}\label{Sect:4.1.3}

In the above studies, either the non-perturbative or the dynamical behaviors of the PSS shows that the approximate PSS in realistic nuclei cannot be justified by the smallness of the PSO potential (\ref{Eq:2.1.VPSOandVPCB}) in the Schr\"odinger-like equations (\ref{Eq:2.1.SchrF}) for the lower component $F(r)$.
That is to say, this term should not be considered as the appropriate symmetry-breaking term \cite{Marcos2008_EPJA37-251}.
An explicit and quantitative connection between the ideal PSS limits and realistic nuclei was still missing.
That triggered recent investigations on the alternative PSS limits, from which a smooth transition to realistic nuclei can be performed \cite{Marcos2008_EPJA37-251,Typel2008_NPA806-156}.

In Ref.~\cite{Marcos2008_EPJA37-251}, Marcos \textit{et al.} aimed at separating the single-particle Hamiltonian $H$ as $H=H_0+W$, where $H_0$ exhibits exact degeneracy of pseudospin doublets, i.e.,
\begin{equation}\label{Eq:4.1.PSSlimit}
  E_a=E_b\,,
\end{equation}
while $W$ represents the PSS breaking term.
Various kinds of potentials have been investigated, including the harmonic oscillator, Woods-Saxon, relativistic Nilsson, and self-consistent RMF potentials.
Among these potentials, the RHO potential exhibits the exact PSS, in which $\Sigma(r)$ is a harmonic oscillator and $\Delta(r)$ is a constant.
Different from Eq.~(\ref{Eq:2.3.FatPSS}), at such a PSS limit, the lower components of a pair of pseudospin doublets are no longer equal to each other, i.e.,
\begin{equation}\label{Eq:4.1.FatPSS}
  F_a(r)\neq F_b(r)\,.
\end{equation}

By comparing their single-particle wave functions, it was pointed out that the transition from the RHO model, satisfying exact PSS, to a more realistic one as the Woods-Saxon or self-consistent RMF potential, with broken PSS, can be considered not far from being perturbative \cite{Marcos2008_EPJA37-251}.
Meanwhile, because of the relation shown in Eq.~(\ref{Eq:4.1.FatPSS}), for a pair of pseudospin doublets, the conditions $E_a\approx E_b$ and $F_a(r)\approx F_b(r)$ are no longer dependent on each other.

In Ref.~\cite{Typel2008_NPA806-156}, Typel derived the PSS breaking potential based on the Schr\"odinger-like equation for the upper component of the Dirac spinor, by using the SUSY quantum mechanics.
The details will be presented in Section~\ref{Sect:4.3}, but one of the essential conclusions is that the RHO potential is the simplest case leading to a vanishing PSS breaking potential in the SUSY framework.

Therefore, the Dirac Hamiltonian with the RHO potential can be regarded as an alternative PSS limit, based on which one hopes for understanding the PSS breaking in realistic nuclei in a perturbative way.

\subsubsection{Perturbative investigation of PSS}

In Ref.~\cite{Liang2011_PRC83-041301R}, the perturbation theory was used to investigate the symmetries of the Dirac Hamiltonian (\ref{Eq:2.1.HDirac}) and their breaking in realistic nuclei.
The perturbation corrections to the single-particle energies and wave functions were calculated order-by-order.
In such a way, the link between the single-particle states in realistic nuclei and their counterparts in the symmetry limits can be constructed explicitly.

First of all, following the idea of Rayleigh-Schr\"odinger perturbation theory \cite{Greiner1994}, the Dirac Hamiltonian $H$ in Eq.~(\ref{Eq:2.1.HDirac}) or (\ref{Eq:2.1.DiraceqR}) can be split as
\begin{equation}\label{Eq:4.1.HH0W}
    H = H_0 + W\,,
\end{equation}
or equivalently
\begin{equation}\label{Eq:4.1.H0HW}
    H_0 = H - W\,,
\end{equation}
where $H_0$ leads to the exact SS or PSS, and $W$ is identified as the corresponding symmetry-breaking potential. The condition
\begin{equation}\label{Eq:4.1.PTcondition}
    \left|\frac{W_{mk}}{E_k-E_m}\right|\ll 1
    \quad\mbox{for}\quad m\neq k\,,
\end{equation}
where $W_{mk}=\left< \psi_m | W | \psi_k \right>$, determines whether $W$ can be treated as a small perturbation and governs the convergence of the perturbation series \cite{Greiner1994}.

For the spin and pseudospin symmetry limits shown in Eqs.~(\ref{Eq:2.1.SSlimit}) and (\ref{Eq:2.1.PSSlimit}), respectively, the Dirac Hamiltonian with exact symmetries reads
\begin{equation}\label{Eq:4.1.H0SU2}
    H_0^{\rm SS}=
    \lb\begin{array}{cc}
        M+\Sigma & -\displaystyle\frac{d}{dr}+\displaystyle\frac{\kappa}{r} \\
        \displaystyle\frac{d}{dr}+\displaystyle\frac{\kappa}{r} & -M+\Delta_0
    \end{array}\rb\,,\qquad
    H_0^{\rm PSS}=
    \lb\begin{array}{cc}
        M+\Sigma_0 & -\displaystyle\frac{d}{dr}+\displaystyle\frac{\kappa}{r} \\
        \displaystyle\frac{d}{dr}+\displaystyle\frac{\kappa}{r} & -M+\Delta
    \end{array}\rb\,,
\end{equation}
whose eigenenergies are denoted as $E_0$, and the corresponding symmetry-breaking potentials are
\begin{equation}\label{Eq:4.1.WSU2}
W^{\rm SS}=
    \lb\begin{array}{cc}
        0 & 0 \\
        0 & \Delta-\Delta_0
    \end{array}\rb\,,\qquad
    W^{\rm PSS}=
    \lb\begin{array}{cc}
        \Sigma-\Sigma_0 & 0 \\
        0 & 0
    \end{array}\rb\,.
\end{equation}

In contrast to adopting the Schr\"{o}dinger-like equations in the above studies \cite{Ginocchio1997_PRL78-436,Marcos2001_PLB513-30,Alberto2001_PRL86-5015,Alberto2002_PRC65-034307}, it is remarkable that all involved operators $H$, $H_0$, and $W$ are Hermitian, and they do not contain any singularity.
This allows us to perform the order-by-order perturbation calculations.
In addition, only the $W$ term corresponds to the symmetry-breaking potential within the present decomposition, thus the ambiguity caused by the strong cancellations among the different terms in the Schr\"{o}dinger-like equations can also be avoided.
Therefore, this method can provide a clear and quantitative way for investigating the perturbative nature of SS and PSS.
For the case that the nature of the symmetry is perturbative, the link between the single-particle states in realistic nuclei and their counterparts in the symmetry limits can be constructed quantitatively.
For the case that the nature of the symmetry is non-perturbative, the divergence of the perturbation series can be found explicitly.

\begin{figure}[tbhp]
\begin{center}
\includegraphics[width=8cm]{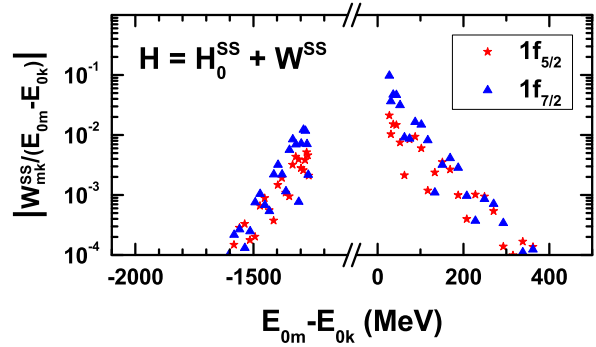}
\end{center}
\caption{(Color online) Values of $\left|W_{mk}/(E_m-E_k)\right|$ versus the energy differences $E_m-E_k$ for the spin doublets $k=1f$.
The unperturbed eigenstates are chosen as those of $H_0^{\rm SS}$, and the single-particle states $m$ include the states in the Dirac sea and Fermi sea.
Taken from Ref.~\cite{Liang2011_PRC83-041301R}.
\label{Fig:4.1.WSU2}}
\end{figure}

The neutrons in $^{132}$Sn will be taken as examples in the following discussions.
The corresponding potentials and single-particle calculated by the self-consistent RMF theory with the effective interaction PK1 \cite{Long2004_PRC69-034319} are shown in Figs.~\ref{Fig:2.3.132potl} and \ref{Fig:2.3.132spectra}, respectively.

For the SS case, the values of $\left|W_{mk}/(E_m-E_k)\right|$ for the spin doublets $k=1f$ are plotted as functions of the energy differences $E_m-E_k$ in Fig.~\ref{Fig:4.1.WSU2}, where the unperturbed eigenstates are chosen as those of $H_0^{\rm SS}$ in Eq.~(\ref{Eq:4.1.H0SU2}), and the constant potential are chosen as $-M+\Delta_0=-350$~MeV.
It can be checked that the convergence of the perturbation calculations is not sensitive to the value of $\Delta_0$.
For the completeness of the single-particle basis, the single-particle states $m$ must include not only the states in the Fermi sea, but also those in the Dirac sea.
It is seen that the values of $\left|W_{mk}/(E_m-E_k)\right|$ decrease as a general tendency when the energy differences $|E_m-E_k|$ increase.
From the mathematical point of view, this property provides natural cut-offs of the single-particle states in the perturbation calculations.
It is critical to find that the largest value of $\left|W_{mk}/(E_m-E_k)\right|$ is $\sim0.1$.
This indicates the criterion in Eq.~(\ref{Eq:4.1.PTcondition}) can be well fulfilled.

\begin{figure}[tbhp]
\begin{center}
\includegraphics[width=8cm]{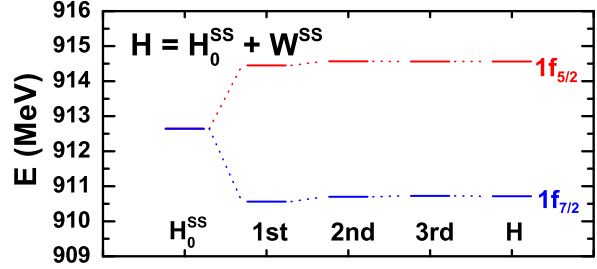}
\end{center}
\caption{(Color online) Single-particle energies of spin doublets $1f$ obtained at the exact SS limit, and by the first-, second-, and third-order perturbation calculations, as well as those by the RMF theory.
The unperturbed eigenstates are chosen as those of $H_0^{\rm SS}$.
Taken from Ref.~\cite{Liang2011_PRC83-041301R}.
\label{Fig:4.1.ESU2}}
\end{figure}

The perturbation corrections to the single-particle energies of the spin doublets $1f$ are then examined.
In Fig.~\ref{Fig:4.1.ESU2}, by choosing the unperturbed eigenstates as those of $H_0^{\rm SS}$, the single-particle energies obtained at the exact SS limit, and their counterparts obtained by the first-, second-, and third-order perturbation calculations, as well as those obtained by the self-consistent RMF theory, are shown from left to right.
It can be clearly seen that the SO splitting is well reproduced by the second-order perturbation calculations.
The perturbation corrections to the single-particle wave functions can be examined in the same way, and the same conclusions hold \cite{Liang2011_PRC83-041301R}.

Similar perturbation calculations have been performed for the PSS case in Ref.~\cite{Liang2011_PRC83-041301R}.
In the PSS case, since there are no bound states in the pseudospin-symmetric Hamiltonian $H_0^{\rm PSS}$, the perturbation calculations are only performed from $H$ to $H_0^{\rm PSS}$, i.e., the unperturbed eigenstates are chosen as those of $H$ and the perturbation is taken as $-W^{\rm PSS}$ in Eq.~(\ref{Eq:4.1.WSU2}).
It is critical to find that the largest value of $\left|W_{mk}/(E_m-E_k)\right|$ is about $0.6$ for the PSS case, compared to $0.1$ for the SS case.
That is because, although the potentials satisfy $\left|\Delta-\Delta_0\right|\gg \left|\Sigma-\Sigma_0\right|$, one should never forget that different components of the Dirac spinor are involved:
\begin{equation}
    W^{\rm SS}_{mk} = \lc F_m\rl (\Delta-\Delta_0)\lr F_k\rc\,,\qquad
    W^{\rm PSS}_{mk} = \lc G_m\rl (\Sigma-\Sigma_0) \lr G_k\rc\,,
\end{equation}
where for the Fermi states the upper component $G(r)\sim O(1)$, and the lower component $F(r)\sim O(1/10)$.

Therefore, from the perturbative point of view, the bridge can be constructed to connect the Dirac Hamiltonian in realistic nuclei with the symmetry limit of constant $\Delta$, but not constant $\Sigma$.
This indicates that the realistic system can be treated as a perturbation of the spin-symmetric Hamiltonian.
This also confirms in an explicit way that the behavior of PSS is non-perturbative, if $d\Sigma/dr=0$ in Eq.~(\ref{Eq:2.1.PSSlimit}) is regarded as the symmetry limit.
These conclusions are in agreement with those given in Ref.~\cite{Marcos2008_EPJA37-251} by observing the single-particle wave-function behaviors, but now they are demonstrated in a quantitative way.

However, it is pointed out in the previous Subsection that the energy splitting of the pseudospin doublets in realistic nuclei can be alternatively considered as the breaking of their degeneracy appearing in the Hamiltonian with the RHO potentials \cite{Marcos2008_EPJA37-251,Typel2008_NPA806-156}.
In order to assess this statement in a perturbative way, the Dirac Hamiltonian $H$ in Eq.~(\ref{Eq:2.1.DiraceqR}) is split as
\begin{equation}\label{Eq:4.1.HU3}
    H = H_0^{\rm RHO} + W^{\rm RHO}\,,
\end{equation}
with
\begin{equation}\label{Eq:4.1.H0U3}
H_0^{\rm RHO}=
    \lb\begin{array}{cc}
        M+\Sigma_{\rm HO} & -\displaystyle\frac{d}{dr}+\displaystyle\frac{\kappa}{r} \\
        \displaystyle\frac{d}{dr}+\displaystyle\frac{\kappa}{r} & -M+\Delta_0
    \end{array}\rb\,,
\end{equation}
and
\begin{equation}\label{Eq:4.1.WU3}
    W^{\rm RHO} =
    \lb\begin{array}{cc}
        \Sigma-\Sigma_{\rm HO} & 0 \\
        0 & \Delta-\Delta_0
    \end{array}\rb\,,
\end{equation}
where $\Sigma_{\rm HO}(r) = c_0 + c_2 r^2$ has the form of a harmonic oscillator.
Here, $H_0^{\rm RHO}$ leads to the energy degeneracy of the whole major shell, and $W^{\rm RHO}$ is identified as the corresponding symmetry-breaking potential.
The constants $-M+\Delta_0=-350$~MeV and $M+c_0=865$~MeV are chosen in $H_0^{\rm RHO}$.
As discussed before, the perturbative properties are not sensitive to these two constants.
Meanwhile, the coefficient $c_2$ is chosen as $1.00$~MeV/fm$^2$ to minimize the perturbations to the $pf$ states.

\begin{figure}[tbhp]
\begin{center}
\includegraphics[width=8cm]{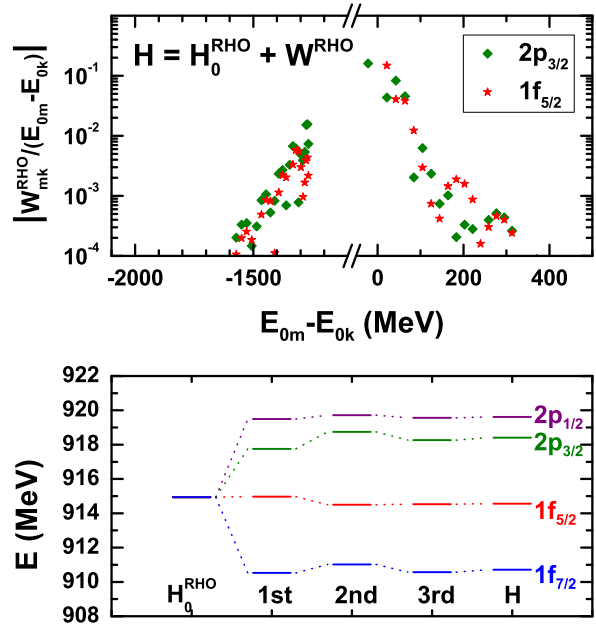}
\end{center}
\caption{(Color online) Upper panel: Same as Fig.~\ref{Fig:4.1.WSU2}, but for the case of the RHO potentials.
Lower panel: Same as Fig.~\ref{Fig:4.1.ESU2}, but for all single-particle states in the $pf$ major shell.
The unperturbed eigenstates are chosen as those of $H_0^{\rm RHO}$.
Taken from Ref.~\cite{Liang2011_PRC83-041301R}.
\label{Fig:4.1.U3}}
\end{figure}

In the upper panel of Fig.~\ref{Fig:4.1.U3}, the values of $\left|W^{\rm RHO}_{mk}/(E_m-E_k)\right|$ for the pseudospin doublets
$k=1\tilde{d}$ are shown as functions of the energy differences $E_m-E_k$.
It is found that the general patterns shown in this panel are the same as those in Fig.~\ref{Fig:4.1.WSU2}, and the largest perturbation correction is $\sim0.16$.
This indicates that the criterion in Eq.~(\ref{Eq:4.1.PTcondition}) is fulfilled, even though not as well as in the SS case.

In the lower panel of Fig.~\ref{Fig:4.1.U3}, the perturbation corrections to the single-particle energies of the states in the $pf$ major shell are shown.
As shown in this panel, both SO and PSO splittings are well reproduced by the third-order perturbation calculations.
Thus, the link between the $pf$ states in realistic nuclei and their counterparts in the symmetry limit with RHO potential can be explicitly established.
Furthermore, it is found that the single-particle wave functions of $H$ can also be reproduced by the second-order perturbation calculations starting from $H_0^{\rm RHO}$.

Therefore, the quantitative connection between the Dirac Hamiltonian in realistic nuclei and that with RHO potentials is constructed by using perturbation theory.
This indicates that the energy splitting of the pseudospin doublets can be regarded as a result of small perturbation around the Dirac Hamiltonian with RHO potentials, where the exact degeneracy of the pseudospin doublets appears \cite{Liang2011_PRC83-041301R}.

In summary, whether or not the pseudospin symmetry breaking behaves perturbatively depends on whether an appropriate symmetry limit is chosen and an appropriate symmetry-breaking term is identified.
In Refs.~\cite{Marcos2001_PLB513-30,Alberto2001_PRL86-5015,Alberto2002_PRC65-034307,
Lisboa2010_PRC81-064324,Ginocchio2011_JPCS267-012037}, the non-perturbative behaviors of PSS were shown from various aspects, if one takes the condition $d\Sigma/dr=0$ in Eq.~(\ref{Eq:2.1.PSSlimit}) as the symmetry limit.
This has also been confirmed with perturbation calculations in Ref.~\cite{Liang2011_PRC83-041301R}.
However, by using the supersymmetric quantum mechanics, it is shown that a Dirac Hamiltonian with the relativistic harmonic oscillator potential is the simplest example that holds the exact pseudospin symmetry \cite{Typel2008_NPA806-156}.
Based on this symmetry limit (\ref{Eq:4.1.H0U3}), the perturbation corrections to the single-particle energies and wave functions have been investigated order by order \cite{Liang2011_PRC83-041301R}.
In such a way, the link between the single-particle states in realistic nuclei and their counterparts in the symmetry limits is constructed explicitly.
It is found that the energy splitting of the pseudospin doublets can be regarded as a result of small perturbation around this symmetry limit.
Further discussions on perturbation calculations with supersymmetric quantum mechanics will be continued in Section~\ref{Sect:4.3}.

\subsection{Intruder states}\label{Sect:4.2}

For the nuclear single-particle spectra in the normal spin-orbit scheme, the $s$ orbitals with $l=0$ correspond to the spin singlets, and all other orbitals with $l>0$ have their spin partners.
In contrast, in the pseudospin-orbit scheme, one observes that not all states with $\tilde l>0$ have their own pseudospin partners.
More precisely, one finds the $\tilde s$ orbitals with $\tilde l=0$ corresponding to the pseudospin singlets, and most of the $\tilde l>0$ orbitals showing up by pairs as pseudospin doublets, but not for the $1s_{1/2}, 1p_{3/2}, 1d_{5/2}, 1f_{7/2}, 1g_{9/2}\ldots$ states.
These states are in general of $j=l+1/2$ and have no internal nodes in neither the upper nor lower component of the Dirac spinor, $n_G=n_F=0$.
Beyond the $1f_{7/2}$, these particular states are usually called the intruder states, which intrude from the the major shell above to the shell below, and form the conventional nuclear magic numbers $28$, $50$, $82$, etc.
For convenience, we call all these states, i.e., $1s_{1/2}$, $1p_{3/2}$, as well as $1d_{5/2}$, $1f_{7/2}$, and $1g_{9/2}$, etc., the pseudospin-unpaired states.

Such a special difference between the spin and pseudospin schemes has been one of the longstanding puzzles in the studies of pseudospin symmetry.
The origin and the meaning of this special feature are still not fully understood.
In this Section, we will discuss this puzzle from several different and interesting aspects.

\subsubsection{Bound states in non-confining potentials}

Leviatan and Ginocchio \cite{Leviatan2001_PLB518-214} have pointed out that, for the single-particle bound states in the Fermi sea, the numbers of internal nodes of the upper and lower components of their Dirac spinor, $n_G$ and $n_F$, obey
\begin{equation}\label{Eq:4.2.GFnodes}
  n_F = n_G
  \quad\mbox{for}\quad
  \kappa<0\,,\qquad
  n_F = n_G + 1
  \quad\mbox{for}\quad
  \kappa>0\,.
\end{equation}
This was proven analytically for the case that the spherical scalar $S(r)$ and vector $V(r)$ potentials are local and $S(r),V(r)\rightarrow0$ as $r\rightarrow\infty$.
In this case, the eigenstates in the Fermi sea are bound by the $\Sigma(r)=V(r)+S(r)$ potential.
The demonstration is shown in detail in Section~\ref{Sect:3.1.1}.

In addition, it is generally believed that, for a pair of pseudospin doublets, the lower components $F(r)$ of their wave functions are very similar to each other \cite{Ginocchio1997_PRL78-436}, as illustrated in Fig.~\ref{Fig:2.3.wf}.
Even in the alternative PSS limit, the Dirac Hamiltonian with RHO potentials discussed in Section~\ref{Sect:4.1.3}, it is also true that the numbers of internal nodes of their $F(r)$ components are equal.
Namely, for any pair of pseudospin doublets with $\kappa_a<0$ and $\kappa_b=1-\kappa_a>0$, there holds
\begin{equation}\label{Eq:4.2.nF}
  n_F(\kappa_a) = n_F(\kappa_b)\,.
\end{equation}

Combining Eqs.~(\ref{Eq:4.2.GFnodes}) and (\ref{Eq:4.2.nF}), one naturally obtains
\begin{equation}\label{Eq:4.2.nG}
  n_G(\kappa_a) = n_G(\kappa_b)+ 1\,.
\end{equation}
This explains the reason why the $nj_>$ orbitals pair with the $(n-1)j_<$ ones in the pseudospin scheme.
In addition, the eigenstates $1j_>$ with $n_G=0$ cannot find their pseudospin partners simply because $n_G$ cannot be $-1$.

In such a way, the special feature concerning the intruder states can be understood.
However, many interesting aspects about this puzzle are still open, which will be discussed in the following parts.

\subsubsection{Bound states in confining potentials}

All states with $\tilde l>0$ have their own pseudospin partners, if the single-particle eigenstates in the Fermi sea are bound by the $\Delta(r)=V(r)-S(r)$ potential instead of $\Sigma(r)$ \cite{Chen2003_CPL20-358,Alberto2013_PRC87-031301R}.
It is the case when $\lim_{r\rightarrow\infty}\Delta(r)>2M+\Sigma(r)$ shown in Eq.~(\ref{Eq:3.1.boundatPSS}) is satisfied \cite{Chen2003_CPL20-358}.

On one hand, Alberto, de Castro, and Malheiro \cite{Alberto2013_PRC87-031301R} have proven that the number of internal nodes of the upper and lower components, $n_G$ and $n_F$, now obeys
\begin{equation}\label{Eq:4.2.GFnodescon}
  n_F = n_G - 1
  \quad\mbox{for}\quad
  \kappa<0\,,\qquad
  n_F = n_G
  \quad\mbox{for}\quad
  \kappa>0\,,
\end{equation}
for the local and spherical scalar $S(r)$ and vector $V(r)$ potentials.
On the other hand, the relation~(\ref{Eq:4.2.nF}) is held as pseudospin doublets have similar lower components $F(r)$.
As a result, one still holds $n_G(\kappa_a) = n_G(\kappa_b)+ 1$ in Eq.~(\ref{Eq:4.2.nG}) for all pairs of pseudospin doublets with $\kappa_a<0$ and $\kappa_b=1-\kappa_a>0$.
However, it is interesting and important to note that now there exist no eigenstates with quantum numbers $n_G=0$ and $j=l+1/2$, because $n_F$ cannot be $-1$ in the first expression of Eq.~(\ref{Eq:4.2.GFnodescon}).

In such a way, all states with $\tilde l>0$ have their own pseudospin partners.
The corresponding typical examples are shown in Fig.~\ref{Fig:3.1.HOE}.
However, one can comment that in the present case not all states with $l>0$ have their own spin partners, i.e., there are some spin-unpaired states which do not have spin partners.
This may be even more serious because the nuclear shell structure and magic numbers are substantially changed.

\subsubsection{Zeros of Jost function: Bound and resonant states}

In Section~\ref{Sect:3.4}, it has been shown that the PSS in the single-particle
resonant states can be investigated by studying the asymptotic behavior of the nucleon
Dirac wave functions \cite{Lu2012_PRL109-072501}.
In particular, by examining the zeros of the Jost function in Eq.~(\ref{eq:Jost_function})
corresponding to the small component of the radial Dirac wave function,
one can investigate the PSS not only for bound states but also
for resonant states.
Here, the problem concerning the intruder states will be discussed.

\begin{figure}[tbhp]
\begin{center}
\includegraphics[width=8cm]{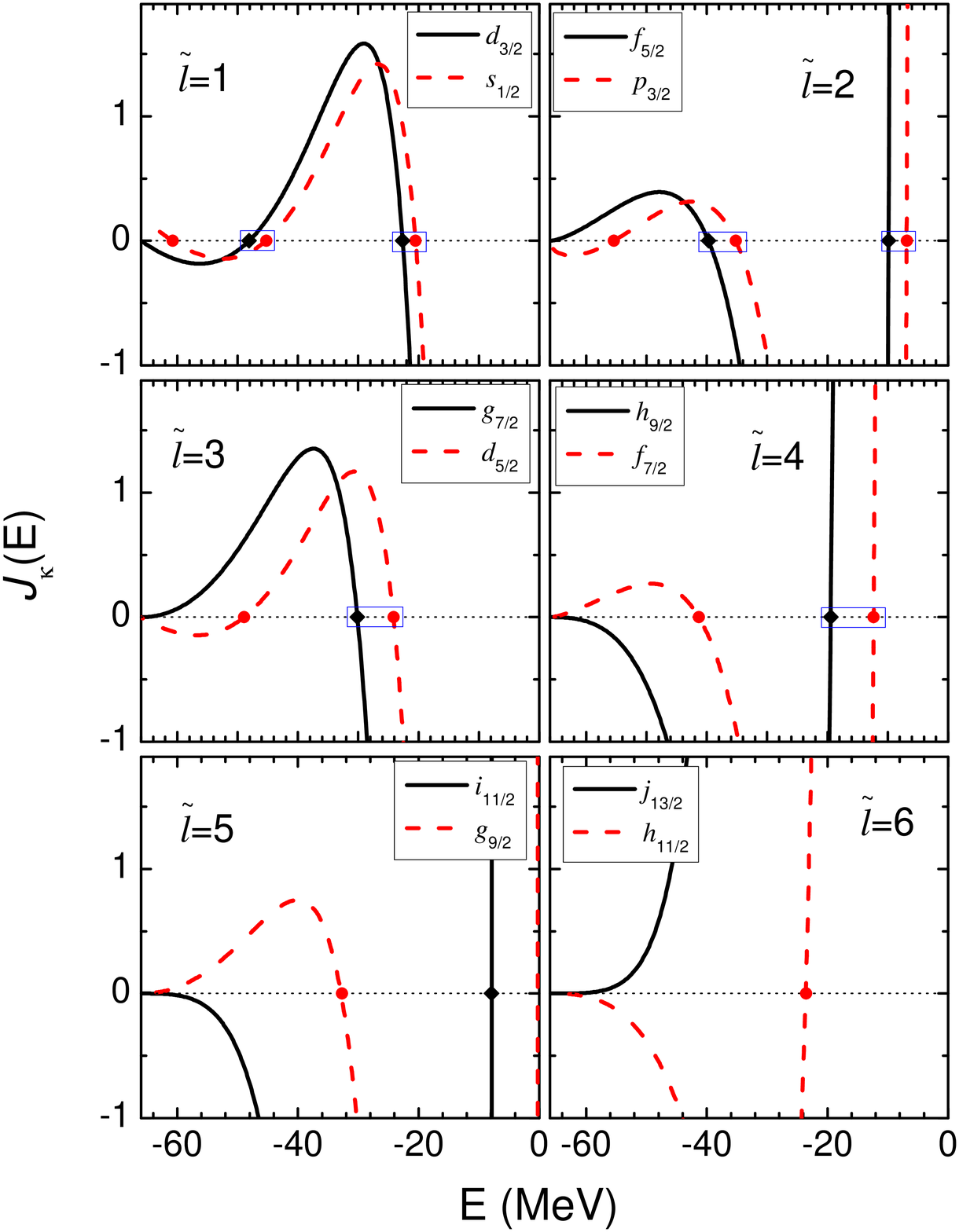}
\end{center}
\caption{(Color online)
Jost functions $\mathcal{J}_\kappa(E)$ (in arbitrary unit)
on the real $E$ axis for several pairs of pseudospin partners
in square-well potentials~(\ref{eq:S.W.}) with $C=-66$~MeV and $D=650$~MeV.
The results for the $j=\tilde{l}\pm 1/2$ orbitals
are denoted as solid and dashed curves, and their zero points representing the bound states
are denoted as black and red dots, respectively.
Taken from Ref.~\cite{Lu2013_PRC88-024323}.
\label{fig:comparison_of_Jost_functions}
}
\end{figure}

In Ref.~\cite{Lu2013_PRC88-024323},
the Jost functions~(\ref{eq:Jost_function}) corresponding to
the small component of the radial Dirac wave function in square-well potentials~(\ref{eq:S.W.})
with $C=-66$~MeV and $D=650$~MeV were investigated in detail.
The Jost functions are plotted as a function of the binding energy
$E\equiv\epsilon-M$ for several bound pairs of pseudospin partners
in Fig.~\ref{fig:comparison_of_Jost_functions}.
For each pseudo-orbital angular momentum $\tilde{l}$, there exist two
$\kappa$'s, one with a positive value $\kappa = \tilde{l}+1$ and
the other with a negative value $\kappa = -\tilde{l}$, respectively.
From Fig.~\ref{fig:comparison_of_Jost_functions}, one can clearly find that
the number of zeros for Jost functions with negative $\kappa$ is
always one more than that with positive $\kappa$.
For example, for $\tilde{l}=1$ there are two zeros for the $d_{3/2}$ states
but three for the $s_{1/2}$ states.
This means that there are two bound states for $d_{3/2}$ orbitals and three for $s_{1/2}$ orbitals.
Therefore, there is always one bound state with $\kappa<0$ which does not
have a partner.
This state is simply an intruder state.

\begin{figure}[tbhp]
\begin{center}
\includegraphics[width=6cm]{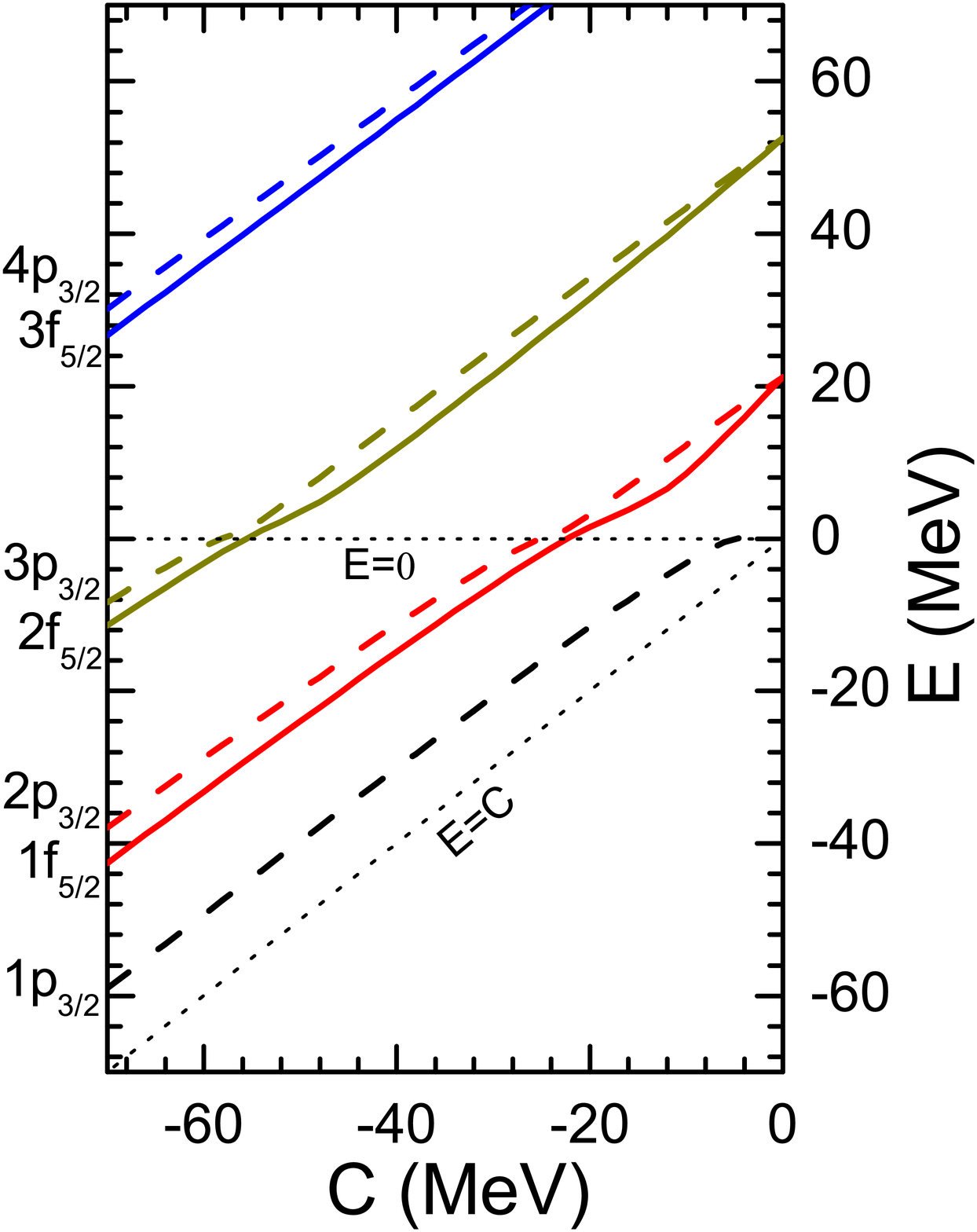}
\end{center}
\caption{(Color online)
Energies of bound and resonant states for $p_{3/2}$ and $f_{5/2}$ with $\tilde{l}=2$
in square-well potentials~(\ref{eq:S.W.}) as functions of the potential depth $C$.
The results for the $j=\tilde l \pm 1/2$ orbitals
are denoted as solid and dashed curves, respectively.
All the levels are paired off except the lowest one.
The bottom of the potential $E=C$ and the bound state threshold $E=0$
are shown as dotted lines.
Taken from Ref.~\cite{Lu2013_PRC88-024323}.
}\label{fig:Efunc}
\end{figure}

In Fig.~\ref{fig:Efunc} are shown the calculated energies of bound and resonant
states for the $\tilde{l}=2$ pseudospin doublets in the square-well potentials as functions of the potential
depth $C$.
As the potential depth varies from $0$ to $-70$~MeV, the energies are always paired
off except the lowest one which is an intruder states.

This is an alternative way to show the origin of the appearance of intruder states:
The lowest zero of the Jost functions~(\ref{eq:Jost_function}) with negative $\kappa$
is always isolated while the others are paired off with those of the Jost functions
with positive $\kappa$ \cite{Lu2013_PRC88-024323}.

\subsubsection{A continuous transformation between SS and PSS}

Recently, Desplanques and Marcos \cite{Desplanques2010_EPJA43-369} re-examined the puzzle of the intruder states in a different way, by setting up two single-particle spectra satisfying the SS and PSS, and between these two limits constructing a continuous and unitary transformation that commutes with the kinetic-energy operator.
This was performed in the non-relativistic framework but involving non-local potentials.

The starting point is the potentials
\begin{equation}\label{Eq:4.2.V0V1}
  V_1 = V_1(r)\,,\qquad
  \tilde V_2 = (\boldsymbol{\sigma}\cdot\hat{\mathbf{p}}) V_2(r) (\boldsymbol{\sigma}\cdot\hat{\mathbf{p}})\,,
\end{equation}
where $V_1(r)$ and $V_2(r)$ are local and spin independent.
The operator $(\boldsymbol{\sigma}\cdot\hat{\mathbf{p}})$ is unitary, but it makes the potential $\tilde V_2$ non-local.
It is clear that $[V_1, \boldsymbol{\sigma}] = 0$, and consequently the potential $V_1$ satisfies the spin symmetry.
Analogously, $[\tilde V_2, (\boldsymbol{\sigma}\cdot\hat{\mathbf{p}}) \boldsymbol{\sigma} (\boldsymbol{\sigma}\cdot\hat{\mathbf{p}})] = 0$, since $(\boldsymbol{\sigma}\cdot\hat{\mathbf{p}}) (\boldsymbol{\sigma}\cdot\hat{\mathbf{p}}) = 1$ and $[V_2, \boldsymbol{\sigma}] = 0$, and consequently $\tilde V_2$ satisfies the pseudospin symmetry.
Therefore, the Hamiltonian $\tilde H_2$ can be written as
\begin{equation}\label{Eq:4.2.Htilde}
  \tilde H_2 = \frac{p^2}{2M} + \tilde V_2
  = (\boldsymbol{\sigma}\cdot\hat{\mathbf{p}}) \ls \frac{p^2}{2M} + V_2(r)\rs (\boldsymbol{\sigma}\cdot\hat{\mathbf{p}})\,,
\end{equation}
implying that the spectra for the potentials $\tilde V_2$ and $V_2(r)$ are the same.
This indicates that the pseudospin-symmetric Hamiltonian $\tilde H_2$ has the same spectra as those of the spin-symmetric Hamiltonian $H_2=p^2/(2M) + V_2(r)$.

In order to construct a continuous and unitary transformation from the SS to PSS limits, a superposition of the above two potentials is made as \cite{Desplanques2010_EPJA43-369}
\begin{equation}\label{Eq:4.2.Vx}
  V = (1-x)V_1 + x\tilde V_2 = (1-x)V_1(r) + x (\boldsymbol{\sigma}\cdot\hat{\mathbf{p}}) V_2(r) (\boldsymbol{\sigma}\cdot\hat{\mathbf{p}})\,,
\end{equation}
where $x$ is supposed to vary from $0$ to $1$.

\begin{figure}[tbhp]
\begin{center}
\includegraphics[width=8cm]{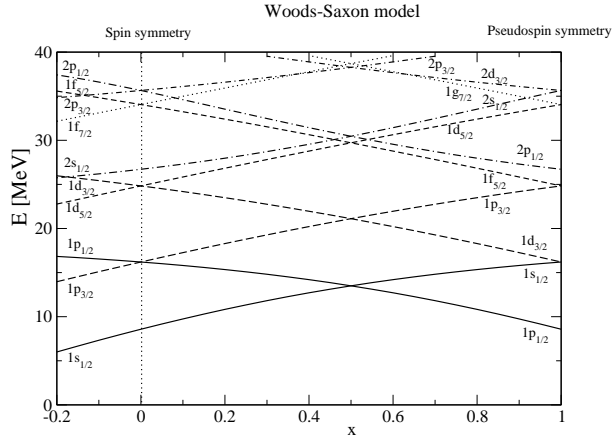}
\end{center}
\caption{(Color online) Single-particle spectra at the SS and PSS limits and the continuous transition in between.
The potentials are chosen as the Woods-Saxon type with parameters appropriate for a medium-size nucleus like $^{40}$Ca.
Taken from Ref.~\cite{Desplanques2010_EPJA43-369}.
\label{Fig:4.2.UtranE}}
\end{figure}

In Fig.~\ref{Fig:4.2.UtranE}, the single-particle spectra at the SS and PSS limits as well as the continuous transition in between are shown.
In these results, the potentials are chosen as the Woods-Saxon form,
\begin{equation}\label{Eq:4.2.VWS}
  V_1(r) = V_2(r) = -\frac{V_0}{1+e^{(r-R)/a}}
\end{equation}
with $V_0=40$~MeV, $R=5$~fm, $a=0.65$~fm, and $x$ is varied from $-0.2$ to $1$.

First of all, it is more or less as expected but still very important that a full continuity between the spectra at the SS and PSS limits is obtained.
Focusing on the single-particle spectrum around $x = -0.2$, it agrees with the expectation for a medium-size nucleus like $^{40}$Ca.
In particular, the quasi-degeneracies between the standard pseudospin doublets can be seen, such as $(2s_{1/2},1d_{3/2})$ and $(2p_{3/2},1f_{5/2})$, but these combinations do not correspond to those at the hypothetical PSS limit with $x=1$ discussed here.
Looking at how the spectrum changes when going from $x = -0.2$ to $1$, a striking feature is observed.
The energies of the standard pseudospin doublets, $(2s_{1/2},1d_{3/2})$ or $(2p_{3/2},1f_{5/2})$, tend to get apart from each other, whereas those for states $(1s_{1/2},1d_{3/2})$ and $(1p_{3/2},1f_{5/2})$ get closer and closer and eventually coincide at $x=1$.
These results suggest a pattern quite different from the one generally proposed.
In such a pattern, not only all states with $l>0$ have their own spin partners at the SS limit, but also all states with $\tilde l>0$ have their own pseudospin partners at the PSS limit \cite{Desplanques2010_EPJA43-369}.

\begin{figure}[tbhp]
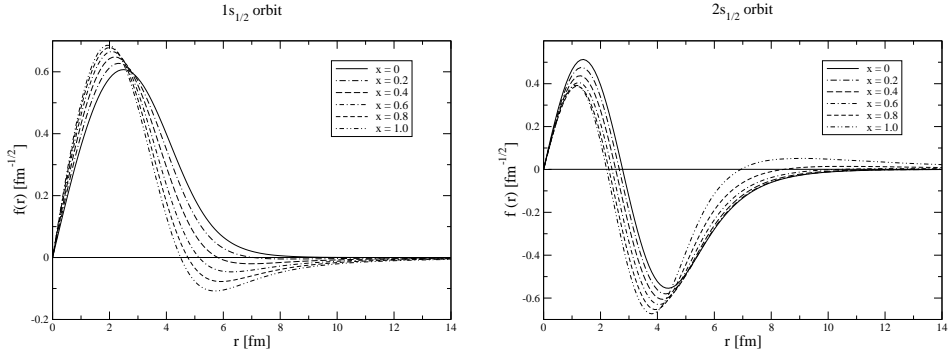

\begin{center}
\includegraphics[width=6cm]{figs/4.2.UtranWF-a.eps}~~~~
\includegraphics[width=6cm]{figs/4.2.UtranWF-b.eps}
\end{center}
\caption{Wave functions $R(r)$ [$f(r)$ in the original figures] of the $1s_{1/2}$ (the left panel) and $2s_{1/2}$ (the right panel) states from the SS to the PSS limits given by the Woods-Saxon model.
Taken from Ref.~\cite{Desplanques2010_EPJA43-369}.
\label{Fig:4.2.UtranWF}}
\end{figure}

As a step further, the corresponding single-particle wave functions are studied.
The wave functions $R(r)$ of the $1s_{1/2}$ and $2s_{1/2}$ states are shown in Fig.~\ref{Fig:4.2.UtranWF} from the SS to PSS limit.
While it confirms that the transition shown in the wave functions is continuous, it shows an interesting phenomenon that an extra node progressively appears in their wave functions when $x$ varies from $0$ to $1$.
This is indeed the case for all the $j=l+1/2$ orbitals, whereas no extra node appears for all the $j=l-1/2$ orbitals \cite{Desplanques2010_EPJA43-369}.
Unfortunately, the analytical demonstrations about the nodal structure shown in Section~\ref{Sect:3.1} cannot be straightforwardly applied, since the non-local potentials are involved here.
Therefore, in a sense, it is not clear whether these states should be labelled as $1s_{1/2},2s_{1/2}$ or as $2s_{1/2},3s_{1/2}$ at the PSS limit, according to the number of internal nodes.
In Ref.~\cite{Desplanques2010_EPJA43-369}, it is argued that these extra nodes are not the usual ones, because they occur outside the range of the potential.
The contribution to the norm beyond the extra node amounts to about $3\%$ in the largest case, compared with the usual contribution which amounts up to several $10\%$.

\begin{figure}[tbhp]
\begin{center}
\includegraphics[width=6cm]{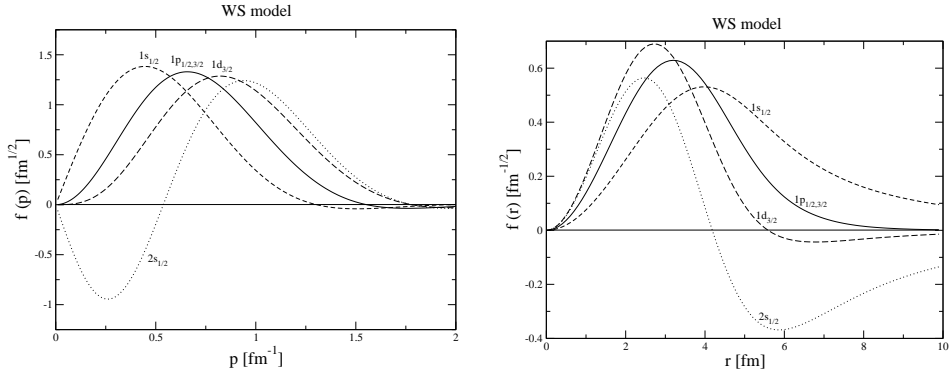}~~~~
\includegraphics[width=6cm]{figs/4.2.UtranWFp-b.eps}
\end{center}
\caption{Left panel: Wave functions of the $1s_{1/2}$, $2s_{1/2}$, $1p_{1/2}$, $1p_{3/2}$, and $1d_{3/2}$ states in the momentum space $R(p)$ [$f(p)$ in the original figure] calculated at the SS limit $x=0$.
Right panel: The corresponding wave functions in the coordinate space $\mathscr R(r)$ [$f(r)$ in the original figure] calculated by Eq.~(\ref{Eq:4.2.Bessel}).
Taken from Ref.~\cite{Desplanques2010_EPJA43-369}.
\label{Fig:4.2.UtranWFp}}
\end{figure}

To investigate the relations in the wave functions of pseudospin doublets, one gets benefit from the facts that the wave functions of spin doublets are the same at the SS limit and the operator $(\boldsymbol{\sigma}\cdot\hat{\mathbf{p}})$ for the continuous and unitary transformation is local in the momentum space.
Thus, the effect of this operator will not affect the angular-momentum-independent part of the wave function denoted as $R(p)$.

In the left panel of Fig.~\ref{Fig:4.2.UtranWFp}, the wave functions of the $1s_{1/2}$, $2s_{1/2}$, $1p_{1/2}$, $1p_{3/2}$, and $1d_{3/2}$ states in momentum space $R(p)$ are compared.
These results are calculated at the SS limit $x=0$.
Since $V_1(r)=V_2(r)$ is set, the wave functions of the $1s_{1/2}$ and $1d_{3/2}$ states at the PSS limit $x=1$ are nothing but those of the $1p$ states at the SS limit.
In addition, to make a fair comparison between these functions in the coordinate space, the same spherical Bessel function is used for the transformation from the $p$- to $r$-space, i.e., \cite{Desplanques2010_EPJA43-369}
\begin{equation}\label{Eq:4.2.Bessel}
  \mathscr R(r) = \sqrt{\frac{2}{\pi}}r\int dp R(p) j_1(pr)\,.
\end{equation}
The corresponding results are shown in the right panel of Fig.~\ref{Fig:4.2.UtranWFp}.
Note that they are different from $R(r)$ shown in Fig.~\ref{Fig:4.2.UtranWF}.

From these results, it was pointed out that \cite{Desplanques2010_EPJA43-369} the wave functions of $1s_{1/2}$ and $1d_{3/2}$ do not show much qualitative difference.
Furthermore, it is interesting that they indeed evolve towards each other when $x$ goes from $0$ to $1$.
In contrast, the wave functions of the standard pseudospin doublets $2s_{1/2}$ and $1d_{3/2}$ evidence a striking difference for the whole range of $x$.

In short, a new pattern of the SS and PSS limits was set up in Ref.~\cite{Desplanques2010_EPJA43-369}.
There exists a continuous and unitary transformation between these two symmetry limits, while the non-local potentials are involved.
In this pattern, all states with $l>0$ have their own spin partners at the SS limit and all states with $\tilde l>0$ have their own pseudospin partners at the PSS limit.
By observing the evolutions in the single-particle energies and wave functions, it is tempting to assign the pseudospin doublets as, e.g., $(1s_{1/2},1d_{3/2})$, instead of the standard ones, e.g., $(2s_{1/2},1d_{3/2})$.
It represents a departure from the standard PSS.
To clarify its validity, analytical investigations similar to those shown in Section~\ref{Sect:3.1} are highly demanded.

\subsubsection{Supersymmetric transformation}

Motivated by the special feature of the intruder states in the pseudospin-orbit scheme, the SUSY quantum mechanics was used to investigate the spin and pseudospin symmetries \cite{Leviatan2004_PRL92-202501, Typel2008_NPA806-156, Leviatan2009_PRL103-042502, Liang2013_PRC87-014334, Shen2013_PRC88-024311}.
It is found that by employing both exact and broken SUSY, the phenomenon that all states with $\tilde{l}>0$ have their own pseudospin partners except for the intruder states can be interpreted within a unified scheme.
Furthermore, the ``striking'' difference between the wave functions of standard pseudospin doublets mentioned above can be understood, as they are indeed almost identical by performing the SUSY transformation \cite{Typel2008_NPA806-156, Liang2013_PRC87-014334} as shown in Figs.~\ref{Fig:4.3.SUSY1WF} and \ref{Fig:4.3.SUSY2WF} below.
It will be the main task in the next Section to illustrate the key ideas of the related investigations.

In summary, the spin-unpaired (pseudospin-unpaired) states, i.e., the states with $l>0$ ($\tilde l>0$) which do not have spin (pseudospin) partners, are listed in Table~\ref{Tab:4.2.intruder} for the five different schemes discussed in this Section.

\begin{table}[tbhp]
\begin{center}
\caption{Spin-unpaired (S-unpaired) and pseudospin-unpaired (PS-unpaired) states within different schemes. The symbol ``---'' means that there is no S-unpaired or PS-unpaired states.
\label{Tab:4.2.intruder}}
\begin{tabular}{@{}lcc@{}} \hline
        & S-unpaired states & PS-unpaired states \\ \hline
Non-confining potentials \cite{Leviatan2001_PLB518-214} & --- & $n=1$ \& $j=l+1/2$ \\
Confining potentials \cite{Chen2003_CPL20-358, Alberto2013_PRC87-031301R}    & $n=1$ \& $j=l-1/2$ & --- \\
Jost functions \cite{Lu2012_PRL109-072501} & --- & $n=1$ \& $j=l+1/2$ \\
Non-local unitary transformation \cite{Desplanques2010_EPJA43-369} & --- & --- \\
SUSY transformation \cite{Typel2008_NPA806-156, Liang2013_PRC87-014334} & --- & $n=1$ \& $j=l+1/2$ \\ \hline
\end{tabular}
\end{center}
\end{table}

\subsection{SUSY and SRG}\label{Sect:4.3}

In the above Sections, we have discussed many special features of pseudospin symmetry, for example, the singularities in the pseudospin-orbit potential (\ref{Eq:2.1.VPSOandVPCB}), the different nodal structures for the pseudospin doublets, the perturbative nature, the puzzle of intruder states, and so on.
These special features strongly motivated recent studies on the pseudospin symmetry \cite{Leviatan2004_PRL92-202501,Typel2008_NPA806-156,Leviatan2009_PRL103-042502,Liang2013_PRC87-014334,Shen2013_PRC88-024311} by using the supersymmetric quantum mechanics \cite{Cooper1995_PR251-267,Cooper2001}, instead of the Schr\"odinger-like equations for the lower component of the Dirac spinor.
By employing both exact and broken supersymmetries, the phenomenon that all states with $\tilde{l}>0$ have their own pseudospin partners except for the intruder states can be interpreted within a unified scheme.
In addition, the wave functions of each pair of pseudospin doublets become almost identical by performing the supersymmetric transformation.

Typel \cite{Typel2008_NPA806-156} investigated the pseudospin symmetry by using the supersymmetric quantum mechanics, based on the Schr\"odinger-like equation~(\ref{Eq:2.1.SchrG}) for the upper component of the Dirac spinor.
Different from the Schr\"odinger-like equation~(\ref{Eq:2.1.SchrF}) for the lower component, the equation for the upper component has no singularities.
As a result, a regular pseudospin symmetry-breaking term can be obtained.
Unfortunately, the effective Hamiltonian involved in this scheme is not Hermitian, which prevents us from being able to perform quantitative perturbation calculations.

Recent works by Guo and coauthors \cite{Guo2012_PRC85-021302R,Li2013_PRC87-044311,Guo2014_PRL112-062502} bridged the perturbation calculations and the supersymmetric descriptions by using the similarity renormalization group.
With the similarity renormalization group, the Dirac Hamiltonian is transformed into a diagonal form, expanding in a series of $1/M$.
The effective Hamiltonian in the Schr\"odinger-like equation thus obtained is Hermitian, which makes the perturbation calculations possible.

By combining the similarity renormalization group, supersymmetric quantum mechanics, and perturbation theory, Liang and coauthors \cite{Liang2013_PRC87-014334,Shen2013_PRC88-024311} pointed out that it is now promising to understand the pseudospin symmetry and its breaking mechanism in a fully quantitative way.

Alternatively, Leviatan \cite{Leviatan2004_PRL92-202501} established the supersymmetric scheme directly based on the first-order differential Dirac Hamiltonian by using the intertwining relation.
As the supersymmetric scheme is established directly on the Dirac Hamiltonian, the higher-order terms in the effective Hamiltonian transformed by the similarity renormalization group can be avoided.
However, it is still an open problem how to identify the perturbative pseudospin symmetry limit and the corresponding symmetry-breaking term within such a scheme.

In this Section, we will start with a brief introduction of supersymmetric quantum mechanics and its application to the Schr\"odinger and Schr\"odinger-like equations.
We will then introduce the similarity renormalization group.
Detailed discussions will be mainly focused on the perturbation calculations with similarity renormalization group and supersymmetric quantum mechanics.
Finally, the application of supersymmetric quantum mechanics to the Dirac equations will be illustrated.

\subsubsection{Supersymmetric quantum mechanics}

It has been shown that every second-order differential Hamiltonian can be factorized in a product of two Hermitian conjugate first-order differential operators, i.e., \cite{Infeld1951_RMP23-21}
\begin{equation}\label{Eq:4.3.SUSYH1}
    H_1=B^+B^-\,,
\end{equation}
with $B^-=[B^+]^\dag$.
Its SUSY partner Hamiltonian can thus be constructed by \cite{Cooper1995_PR251-267,Cooper2001}
\begin{equation}\label{Eq:4.3.SUSYH2}
    H_2=B^-B^+\,.
\end{equation}

The Hermitian operators
\begin{equation}\label{Eq:4.3.SUSYQ}
  Q_1=
  \lb\begin{array}{cc}
    0 & B^+ \\ B^- & 0
  \end{array}\rb\,,\qquad
  Q_2=
  \lb\begin{array}{cc}
    0 & -iB^+ \\ iB^- & 0
  \end{array}\rb\,,
\end{equation}
are the so-called supercharges with respect to the involution
\begin{equation}\label{Eq:4.3.SUSYtau}
  \tau=\tau^\dag=
  \lb\begin{array}{cc}
    1 & 0 \\ 0 & -1
  \end{array}\rb\,,
\end{equation}
because $\{Q_1,\tau\}=\{Q_2,\tau\}=0$.
The extended SUSY Hamiltonian $H_S$ is the square of these Hermitian supercharges,
\begin{equation}\label{Eq:4.3.SUSYHS}
    H_S=Q^2_1=Q^2_2=
    \lb\begin{array}{cc}
        H_1 & 0 \\ 0 & H_2
    \end{array}\rb\,.
\end{equation}
The supercharges $Q_1,Q_2$ and the extended Hamiltonian $H_S$, together with the commutators,
\begin{equation}\label{Eq:4.3.SUSYcommute}
  [H_S,Q_1]=[H_S,Q_2]=0\,,
\end{equation}
and the anti-commutator,
\begin{equation}\label{Eq:4.3.SUSYanticom}
  \{Q_1,Q_2\}=0\,,
\end{equation}
form the most simple example of a supersymmetric algebra.

Since the extended Hamiltonian $H_S$ is the square of the supercharges, all eigenvalues $E_S(n)$ of the eigenvalue equation
\begin{equation}\label{Eq:4.3.SUSYeq}
    H_S\Psi_S(n)=E_S(n)\Psi_S(n)
\end{equation}
are non-negative, and the two-component wave functions read
\begin{equation}\label{Eq:4.3.SUSYPsiS}
    \Psi_S(n)=
    \lb\begin{array}{c}
        \psi_1(n) \\ \psi_2(n)
    \end{array}\rb\,,
\end{equation}
where $\psi_1(n)$ and $\psi_2(n)$ are the eigenfunctions of $H_1$ and $H_2$, respectively.
For each eigenstate with $E_S(n)>0$, it is an eigenstate for both $H_1$ and $H_2$, and the corresponding eigenfunctions satisfy
\begin{equation}\label{Eq:4.3.SUSYWFtran}
    \psi_2(n)=\frac{B^-}{\sqrt{E_S(n)}}\psi_1(n)\,,\qquad
    \psi_1(n)=\frac{B^+}{\sqrt{E_S(n)}}\psi_2(n)\,,
\end{equation}
with the normalization factor $1/\sqrt{E_S(n)}$.

\begin{figure}[tbhp]
\begin{center}
\includegraphics[width=8cm]{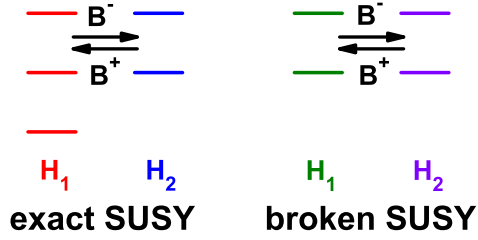}
\end{center}
\caption{(Color online) Schematic patterns of the exact and broken supersymmetries.
Taken from Ref.~\cite{Liang2013_PRC87-014334}.}
\label{Fig:4.3.SUSY}
\end{figure}

The property of SUSY can be either exact (also called unbroken) or broken \cite{Cooper1995_PR251-267}.
The SUSY is exact when the eigenvalue equation~(\ref{Eq:4.3.SUSYeq}) has a zero energy eigenstate $E_S(0)=0$.
In this case, as a usual convention, the Hamiltonian $H_1$ has an additional eigenstate at zero energy that does not appear in its partner Hamiltonian $H_2$, because
\begin{equation}\label{Eq:4.3.SUSYpsi0}
    B^-\psi_1(0)=0\,, \qquad \psi_2(0)=0\,,
\end{equation}
i.e., the trivial eigenfunction of $H_2$ identically equals zero.
The SUSY is broken when the eigenvalue equation~(\ref{Eq:4.3.SUSYeq}) does not have any zero energy eigenstate.
In this case, the partner Hamiltonians $H_1$ and $H_2$ have identical spectra.
The schematic patterns of both cases are illustrated in Fig.~\ref{Fig:4.3.SUSY}.

For detailed derivations and various examples of SUSY quantum mechanics, the readers are referred to Ref.~\cite{Cooper1995_PR251-267} for a review and Ref.~\cite{Cooper2001} for a textbook.

\subsubsection{SUSY for Schr\"odinger-like equations}\label{Sect:4.3.2}

A second-order differential Hamiltonian (\ref{Eq:4.3.SUSYH1}) is the starting pointing of the SUSY quantum mechanics presented above, so this scheme is straightforwardly applicable to the Schr\"odinger or Schr\"odinger-like equations.
Based on the Schr\"odinger-like equation~(\ref{Eq:2.1.SchrG}) for the upper component of the Dirac spinor, Typel \cite{Typel2008_NPA806-156} investigated the PSS and derived a regular symmetry-breaking term.

In Ref.~\cite{Typel2008_NPA806-156}, the effects of tensor interactions have also been taken into account.
The corresponding radial Dirac equations read [cf. Eq.~(\ref{Eq:3.3.Dirac})]
\begin{align}\label{Eq:4.3.Dirac}
  \lb\frac{d}{dr}+\frac{\kappa}{r}-T(r)\rb G_\kappa(r) &= M_+(r)F_\kappa(r)\,,\nonumber\\
  \lb-\frac{d}{dr}+\frac{\kappa}{r}-T(r)\rb F_\kappa(r)&= M_-(r)G_\kappa(r)\,,
\end{align}
where $T(r)$ is the tensor potential and $M_\pm(r)=\epsilon_{n\kappa}\pm M-V(r)\pm S(r)$.
In the Schr\"odinger-like equation for the upper component of the Dirac spinor,
\begin{equation}\label{Eq:4.3.SchrG}
  H_G(\kappa) G_{n\kappa} = \epsilon_{n\kappa} G_{n\kappa}\,,
\end{equation}
the effective Hamiltonian reads
\begin{equation}\label{Eq:4.3.SchrHG}
  H_G(\kappa)
  =\frac{1}{M_+} \ls -\frac{d^2}{dr^2} +\frac{\kappa(\kappa+1)}{r^2} -2T\frac{\kappa}{r} +T^2 +\frac{M'_+}{M_+} \lb\frac{d}{dr} +\frac{\kappa}{r} -T\rb\rs + (M+\Sigma)\,.
\end{equation}

The main task in the SUSY description is to construct the operators $B^+_\kappa$ and $B^-_\kappa$.
The particular form of the Hamiltonian in Eq.~(\ref{Eq:4.3.SchrHG}) suggests the ansatz
\begin{equation}\label{Eq:4.3.SUSY1B+B-}
    B^+_\kappa = \ls Q_\kappa(r)-\frac{d}{dr}\rs\frac{1}{\sqrt{M_+(r)}}\,,
    \qquad
    B^-_\kappa = \frac{1}{\sqrt{M_+(r)}}\ls Q_\kappa(r)+\frac{d}{dr}\rs\,,
\end{equation}
where the superpotentials $Q_\kappa(r)$ are the functions of $r$ to be determined.
Then, the SUSY partner Hamiltonians read
\begin{align}\label{Eq:4.3.SUSY1HQ}
  H_1(\kappa) &= B^+_\kappa B^-_\kappa
    = \frac{1}{M_+}\ls Q^2_\kappa - Q'_\kappa -\frac{d^2}{dr^2} +\frac{M'_+}{M_+}\lb Q_\kappa +\frac{d}{dr}\rb\rs\,,\nonumber\\
  H_2(\kappa) &= B^-_\kappa B^+_\kappa
    = \frac{1}{M_+}\ls Q^2_\kappa + Q'_\kappa -\frac{d^2}{dr^2} +\frac{M'_+}{M_+}\frac{d}{dr} +\frac{M''_+}{2M_+} -\frac{3(M'_+)^2}{4M_+^2}\rs\,.
\end{align}
In order to identify explicitly the $\kappa(\kappa+1)$ structure and the tensor $T(r)$ terms shown in Eq.~(\ref{Eq:4.3.SchrHG}), the reduced superpotentials $q_\kappa(r)$ are introduced as \cite{Typel2008_NPA806-156}
\begin{equation}\label{Eq:4.3.SUSY1qQ}
    q_\kappa(r) = Q_\kappa(r) - \frac{\kappa}{r} + T(r)\,.
\end{equation}
The Hamiltonians $H_1$ and $H_2$ can be further expressed as
\begin{align}\label{Eq:4.3.SUSY1Hq}
  H_1(\kappa) =
    \frac{1}{M_+}\bigg[& -\frac{d^2}{dr^2} +\frac{\kappa(\kappa+1)}{r^2} +q^2_\kappa +2q_\kappa\frac{\kappa}{r} -q'_\kappa -2q_\kappa T +T^2\nonumber\\
    &-2T\frac{\kappa}{r} +T' +\frac{M'_+}{M_+}\lb q_\kappa +\frac{d}{dr} +\frac{\kappa}{r}-T\rb\bigg]\,,\nonumber\\
  H_2(\kappa)
    = \frac{1}{M_+}\bigg[& -\frac{d^2}{dr^2} +\frac{\kappa(\kappa-1)}{r^2} +q^2_\kappa +2q_\kappa\frac{\kappa}{r} +q'_\kappa -2q_\kappa T +T^2\nonumber\\
    &-2T\frac{\kappa}{r} +T' +\frac{M'_+}{M_+}\frac{d}{dr} +\frac{M''_+}{2M_+} -\frac{3(M'_+)^2}{4M_+^2}\bigg]\,.
\end{align}

In general, the effective Hamiltonian $H_G$ in the Schr\"odinger-like equation (\ref{Eq:4.3.SchrHG}) differs from the SUSY Hamiltonian $H_1$ in Eq.~(\ref{Eq:4.3.SUSY1Hq}) by a constant, i.e.,
\begin{equation}\label{Eq:4.3.Eshift}
    H_G(\kappa) = H_1(\kappa)+e(\kappa)\,,
\end{equation}
where $e(\kappa)$ is the so-called energy shift \cite{Cooper1995_PR251-267}.
Thus, the reduced superpotentials $q_\kappa(r)$ satisfy the first-order differential equation,
\begin{equation}\label{Eq:4.3.SUSY1q}
  q^2_\kappa +\lb 2\frac{\kappa}{r} -2T +\frac{M'_+(\kappa)}{M_+(\kappa)}\rb q_\kappa -q'_\kappa = -M_+(\kappa) N(\kappa) - T'\,.
\end{equation}
Note that $N(\kappa)=e(\kappa)-M-\Sigma(r)$ depends on the energy shift but $M_+(\kappa)=\epsilon_{n\kappa}+M-\Delta(r)$ depends on the single-particle energy.
For regular $S(r)$, $V(r)$, and $T(r)$ potentials, the boundary condition for the reduced superpotentials reads
\begin{equation}\label{Eq:4.3.SUSY1q0}
  q_\kappa(0) = 0\,.
\end{equation}
At small radius, $q_\kappa(r)$ behaves asymptotically as a linear function of $r$,
\begin{equation}\label{Eq:4.3.SUSY1qsmall}
  \lim_{r\rightarrow0} q_\kappa(r)= \frac{M_+(\kappa) N(\kappa)+T'}{1-2\kappa} r\,,
\end{equation}
and at large radius, $q_\kappa(r)$ becomes a constant,
\begin{equation}\label{Eq:4.3.SUSY1qlarge}
  \lim_{r\rightarrow\infty} q_\kappa(r)= \sqrt{(M+\epsilon_{n\kappa})(M-e(\kappa))}\,,
\end{equation}
if the $S$ ,$V$, and $T$ potentials vanish there.

It is very important to examine the asymptotic behaviors of the full superpotentials $Q_\kappa(r)$, because they determine the type of SUSY \cite{Cooper1995_PR251-267}.
If there is a change of sign in $Q_\kappa(r)$ when comparing the limits $r\rightarrow0$ with $r\rightarrow\infty$, the exact SUSY follows, and thus there exists a single non-degenerate state at zero energy.
In contrast, if the $Q_\kappa(r)$ does not change the sign between these limits, the SUSY is broken, and thus all eigenstates are doubly degenerate with positive energy.

In the present cases, $Q_\kappa(r)$ are always positive at $r\rightarrow\infty$.
In contrast, $Q_\kappa(r)$ for $r\rightarrow0$ is determined by the angular-momentum term $\kappa/r$.
One finds $\lim_{r\rightarrow0} Q_\kappa(r)=-\infty$ for $\kappa<0$, and $\lim_{r\rightarrow0} Q_\kappa(r)=+\infty$ for $\kappa>0$.
In other words, the SUSY is exact for all cases of $\kappa<0$, whereas SUSY is broken for all cases of $\kappa>0$.
This is crucial to understand the puzzle of intruder states discussed in Section~\ref{Sect:4.2}.

The $\kappa$-dependent energy shifts $e(\kappa)$ can be determined in the following ways:
(i) For the case of $\kappa_a<0$, the SUSY is exact, and it requires
\begin{equation}\label{Eq:4.3.SUSY1e1}
    e(\kappa_a) = \epsilon_{1\kappa_a}\,.
\end{equation}
(ii) For the case of $\kappa_b>0$, the SUSY is broken, and thus the corresponding energy shift can be, in principle, any number which makes the whole set of $H_1$ eigenstates positive.
In practice, the energy shifts are determined by assuming that the PSO potentials vanish as $r\rightarrow0$.
This vanishing behavior is similar to that of the usual surface-peaked SO potentials.
Therefore, $\lim_{r\rightarrow0}q_{\kappa_b}(r) = \lim_{r\rightarrow0}q_{\kappa_a}(r)$ with $\kappa_a+\kappa_b=1$ is satisfied for pseudospin doublets.
Considering $M_+(\kappa_a)$ and $M_+(\kappa_b)$ are almost identical as $\epsilon_a\approx \epsilon_b$, and neglecting the contribution of $T'(0)$, the energy shifts read \cite{Typel2008_NPA806-156}
\begin{equation}\label{Eq:4.3.SUSY1e2}
    e(\kappa_b) = 2\left.(M+\Sigma)\right|_{r=0} - e(\kappa_a)\,.
\end{equation}

Finally, one can evaluate the corresponding PSS-breaking terms.
The Hamiltonians $H_2(\kappa_a) + e(\kappa_a)$ and $H_2(\kappa_b) + e(\kappa_b)$ for the pseudospin doublets are almost identical, and their difference is given by the potential \cite{Typel2008_NPA806-156}
\begin{equation}\label{Eq:4.3.SUSY1PSO}
  \tilde W^{\rm PSS} = [H_2(\kappa_a) + e(\kappa_a)] - [H_2(\kappa_b) + e(\kappa_b)]
  = \frac{2}{\sqrt{M_+}}\frac{d}{dr}\frac{q_{\kappa_a}-q_{\kappa_b}}{\sqrt{M_+}} - 2T\frac{\kappa_a-\kappa_b}{r M_+}\,,
\end{equation}
where the difference between $M_+(\kappa_a)$ and $M_+(\kappa_b)$ is neglected.
It is essential that this symmetry-breaking potential is a regular function for all $r$, in contrast to that shown in Eq.~(\ref{Eq:2.1.VPSOandVPCB}).
One of the simplest cases that such a symmetry-breaking potential vanishes is nothing but the relativistic harmonic oscillator potentials as
\begin{equation}\label{Eq:4.3.SUSY1RHO}
  S(r)=V(r)=\frac{M}{4}\omega^2 r^2\qquad\mbox{and}\qquad
  T(r)=0\,.
\end{equation}

\begin{figure}[tbhp]
\begin{center}
  \includegraphics[width=8cm]{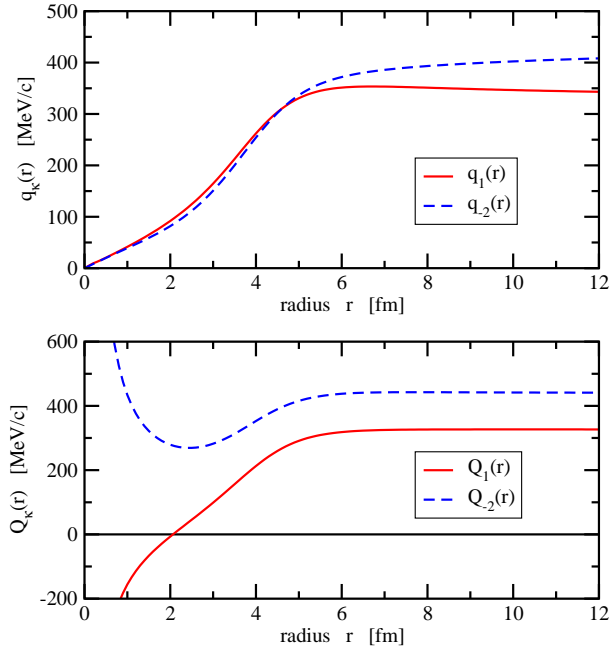}
\end{center}
\caption{(Color online) Reduced superpotentials $q_\kappa(r)$ (the upper panel) and full superpotentials $Q_\kappa(r)$ (the lower panel) for the $s_{1/2}$ and $d_{3/2}$ orbitals.
Taken from Ref.~\cite{Typel2008_NPA806-156}, and the values of $\kappa$ for $s_{1/2}$ and $d_{3/2}$ orbitals were shown with $1$ and $-2$ in the original figures, respectively.}
\label{Fig:4.3.SUSY1q}
\end{figure}

In Fig.~\ref{Fig:4.3.SUSY1q}, a typical example of the reduced $q_\kappa(r)$ and full $Q_\kappa(r)$ superpotentials is shown for the $\tilde p$ pseudospin doublets, in which the tensor potential is neglected and scalar and vector potentials are chosen as the Woods-Saxon type, $S(r)=S_0/[1+\exp((r-R)/a)]$ and $V(r)=V_0/[1+\exp((r-R)/a)]$ with $S_0 = -450$~MeV, $V_0 = 370$~MeV, $R = 3.8$~fm, and $a = 0.65$~fm.
The single-particle energies of $1\tilde p$ pseudospin doublets thus obtain are $E_{2s_{1/2}}=-15.604$~MeV and $E_{1d_{3/2}}=-15.424$~MeV, and the corresponding energy shifts are $e(-1)-M=-56.849$~MeV and $e(2)-M=-102.680$~MeV.
It is shown that the reduced superpotentials $q_\kappa(r)$ are almost identical for radii below $5$~fm.
At large $r$ they become constant and approach different values for $r\rightarrow\infty$ as predicted by Eq.~(\ref{Eq:4.3.SUSY1qlarge}).
The full superpotentials $Q(r)$ contain the angular-momentum contribution and, hence, they diverge for $r\rightarrow0$.
The change of sign for $Q_{-1}(r)$ indicates the exact SUSY with a single state at zero energy for its $H_1$.
In contrast, no change of sign for $Q_{2}(r)$ corresponds to the broken SUSY.

\begin{figure}[tbhp]
\begin{center}
  \includegraphics[width=8cm]{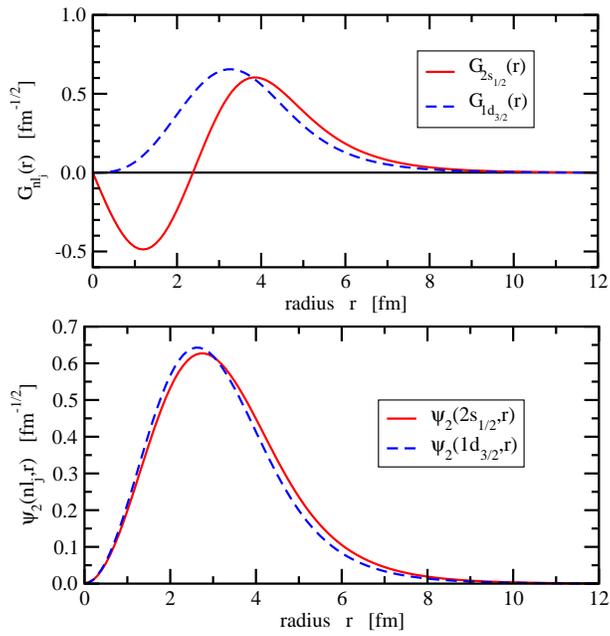}\\
  \includegraphics[width=8cm]{figs/4.3.SUSY1WF-b.eps}
\end{center}
\caption{(Color online) Upper panel: Single-particle wave functions of the original Hamiltonian for the $2s_{1/2}$ and $1d_{3/2}$ pseudospin doublets.
Lower panel: Those of the supersymmetric partner Hamiltonian.
Taken from Ref.~\cite{Typel2008_NPA806-156} and modified to present notations.}
\label{Fig:4.3.SUSY1WF}
\end{figure}

In Fig.~\ref{Fig:4.3.SUSY1WF}, the single-particle wave functions for the $1\tilde p$ pseudospin doublets are shown.
The upper panel shows the wave functions of the original Hamiltonian $H_G$ in Eq.~(\ref{Eq:4.3.SchrHG}).
No doubt they have different numbers of nodes and look quite different.
However, as shown in the lower panel, the eigenfunctions of the SUSY partner Hamiltonian in Eq.~(\ref{Eq:4.3.SUSY1Hq}) are remarkably similar to each other.
They are nothing but the original wave functions with the transformation shown in Eq.~(\ref{Eq:4.3.SUSYWFtran}).

Furthermore, it is proven in the SUSY quantum mechanics that the number of nodes in the radial wave functions $\psi_2(r)$ is one less than those in their counterparts $\psi_1(r)$ when the SUSY is exact, while the number of nodes in $\psi_2(r)$ is the same as those in their counterparts $\psi_1(r)$ when the SUSY is broken \cite{Cooper1995_PR251-267}.
This indicates that the nodal relation between the pseudospin doublets \cite{Leviatan2001_PLB518-214} discussed in Section~\ref{Sect:3.1.1},
\begin{equation}\label{Eq:4.3.nodes}
    \tilde{n}=n-1\quad\mbox{for}\quad\kappa<0\,,\qquad
    \tilde{n}=n\quad\mbox{for}\quad\kappa>0\,,
\end{equation}
is one of the intrinsic properties of the SUSY quantum mechanics \cite{Liang2013_PRC87-014334}.

The quasi-degeneracy in the single-particle energies, the similarity in the wave functions, and the smallness of the symmetry-breaking term (\ref{Eq:4.3.SUSY1PSO}), all of these facts imply that one should be able to understand the PSS breaking in a quantitative and perturbative way.
Unfortunately, both the effective Hamiltonian $H_G$~(\ref{Eq:4.3.SchrHG}) and its SUSY partner~(\ref{Eq:4.3.SUSY1Hq}) are not Hermitian, since the upper component wave functions, as the solutions of the Schr\"{o}dinger-like equation, are not orthogonal to each other.
This prevents us from being able to perform quantitative perturbation calculations.

\subsubsection{Similarity renormalization group}\label{Sect:4.3.3}

Recent works in Ref.~\cite{Guo2012_PRC85-021302R,Li2013_PRC87-044311,Guo2014_PRL112-062502} bridged the perturbation calculations and the SUSY descriptions by using the SRG.

The idea of SRG \cite{Glazek1993_PRD48-5863, Glazek1994_PRD49-4214, Wegner1994_AP506-77, Bylev1998_PLB428-329, Wegner2001_PR348-77} is to drive the Hamiltonian toward a band-diagonal form via the flow equation and unitary transformations that suppress off-diagonal matrix elements.
In recent years, the SRG is also widely used in nuclear effective field theory for decoupling the low-energy physics from the high-energy physics \cite{Bogner2007_PLB649-488, Bogner2007_PRC75-061001, Anderson2008_PRC77-037001}.
The SRG-evolved two-nucleon and three-nucleon interactions have been used for the \textit{ab initio} calculations of nuclear ground states \cite{Jurgenson2009_PRL103-082501, Roth2012_PRL109-052501}, low-lying spectra \cite{Roth2011_PRL107-072501}, nucleon scatterings \cite{Navratil2010_PRC82-034609}, and so on.
The in-medium SRG has been developed for the medium-mass open-shell nuclei \cite{Tsukiyama2011_PRL106-222502, Tsukiyama2012_PRC85-061304, Hergert2013_PRC87-034307, Hergert2013_PRL110-242501}, which decouples the physics of valence nucleons from the full Hilbert space, enabling exact diagonalizations in the valence space that are impossible in the full problem where all nucleons are active.
Recent reviews on the related topics can be found in, e.g., Refs.~\cite{Bogner2010_PPNP65-94, Hammer2013_RMP85-197}.

For the Dirac Hamiltonian shown in Eq.~(\ref{Eq:2.1.HDirac}), it can be transformed with the SRG into a diagonal form and expanded in a series of $1/M$.
It is crucial that the effective Hamiltonian in the Schr\"odinger-like equation thus obtained is Hermitian.
This makes the perturbation calculations possible.

Fist of all, according to the commutation and anti-commutation relations with respect to the $\beta$ matrix, the Dirac Hamiltonian in Eq.~(\ref{Eq:2.1.HDirac}) is separated into the diagonal $\varepsilon$ and off-diagonal $o$ parts, $H=\varepsilon+o$, which satisfy
\begin{equation}\label{Eq:4.3.SRGcommute}
  [\varepsilon,\beta] = 0 \qquad\mbox{and}\qquad
  \{o,\beta\} = 0\,.
\end{equation}
In order to obtain the equivalent Schr\"odinger-like equation for nucleons, the main task is to decouple the eigenvalue equations for the upper and lower components of the Dirac spinor.
A possible way is to make the off-diagonal part of the Dirac Hamiltonian vanish with a proper unitary transformation.

According to the SRG \cite{Glazek1993_PRD48-5863, Glazek1994_PRD49-4214, Wegner1994_AP506-77, Bylev1998_PLB428-329, Wegner2001_PR348-77}, the Hamiltonian $H$ is transformed by a unitary operator $U(l)$ as
\begin{equation}\label{Eq:4.3.SRGHl}
    H(l) = U(l)HU^\dag(l)\,,
\end{equation}
with $H(l)=\varepsilon(l)+o(l)$, $H(0) = H$, and a flow parameter $l$.
Then, the so-called Hamiltonian flow equation can be obtained by taking the differential of the above equation,
\begin{equation}\label{Eq:4.3.SRGdHl}
    \frac{d}{dl}H(l) = [\eta(l),H(l)]\,,
\end{equation}
with the anti-Hermitian generator
\begin{equation}\label{Eq:4.3.SRGeta}
  \eta(l) = \frac{dU(l)}{dl}U^\dag(l)\,.
\end{equation}
As pointed out in Ref.~\cite{Bylev1998_PLB428-329}, one of the proper choices of $\eta(l)$ for letting the off-diagonal part $o(l)=0$ as $l\rightarrow\infty$ reads
\begin{equation}\label{Eq:4.3.SRGetaD}
  \eta(l) = [\beta M,H(l)]\,.
\end{equation}
Then, the diagonal part of the Dirac Hamiltonian $\varepsilon(l)$ at the $l\rightarrow\infty$ limit can be derived analytically in a series of $1/M$ \cite{Guo2012_PRC85-021302R},
\begin{align}\label{Eq:4.3.SRGSchr}
    \varepsilon(\infty) =&~ M\varepsilon_0(\infty) +\varepsilon_1(\infty) +\frac{\varepsilon_2(\infty)}{M} +\frac{\varepsilon_3(\infty)}{M^2} +\frac{\varepsilon_4(\infty)}{M^3} +\cdots\nonumber\\
    =&~
    \beta M+(\beta S+V)+\frac{1}{2M}\beta (\boldsymbol{\alpha}\cdot\mathbf{p})^2
    +\frac{1}{8M^2} \ls\ls \boldsymbol{\alpha}\cdot\mathbf{p},(\beta S+V)\rs, \boldsymbol{\alpha}\cdot\mathbf{p}\rs \nonumber\\
    &~+\frac{1}{32M^3}\beta\left( -4(\boldsymbol{\alpha}\cdot\mathbf{p})^4 +\Lb\boldsymbol{\alpha}\cdot\mathbf{p}, \ls\ls \boldsymbol{\alpha}\cdot\mathbf{p}, (\beta S+V)\rs, (\beta S+V)\rs\Rb 
    \right.
    \nonumber \\
    &~\left.\phantom{~+\frac{1}{32M^3}\beta}
     -2\ls \boldsymbol{\alpha}\cdot\mathbf{p}, (\beta S+V)\rs^2\right) +\cdots
\end{align}
In such a way, the eigenvalue equations for the upper and lower components of the Dirac spinor are decoupled.
The equivalent Schr\"odinger-like equations for nucleons with Hermitian effective Hamiltonians can be obtained.
The corresponding details can be found in Refs.~\cite{Guo2012_PRC85-021302R,Li2013_PRC87-044311} for the spherical case and Ref.~\cite{Guo2014_PRL112-062502} for the axially deformed case.

For example, for the spherical case, the effective Hamiltonian for the nucleons in the Fermi sea up to the $(1/M^3)$-th order reads \cite{Guo2012_PRC85-021302R}
\begin{align}\label{Eq:4.3.SRGHFermi}
  H_F =&~ M +\Sigma +\frac{p_F^2}{2M} -\frac{1}{2M^2}\lb Sp_F^2 -S'\frac{d}{dr}\rb -\frac{\kappa}{r}\frac{\Delta'}{4M^2} +\frac{\Sigma''}{8M^2}\nonumber\\
  &~+\frac{S}{2M^3}\lb Sp_F^2 -2S'\frac{d}{dr}\rb +\frac{\kappa}{r}\frac{S\Delta'}{2M^3} -\frac{(\Sigma')^2 -2\Sigma'\Delta' +4S\Sigma''}{16M^3} -\frac{p_F^4}{8M^3}
\end{align}
with the operator $p_F^2 = -d^2/(dr^2) + \kappa(\kappa+1)/r^2$.
This Hamiltonian can be decomposed into five Hermitian components:
the non-relativistic term $H_n$, the spin-orbit term $H_c$, the dynamical term $H_d$,
the relativistic modification of kinetic energy $H_k$, and the Darwin term $H_w$ as
\begin{align}
  H_n &= M +\Sigma +\frac{p_F^2}{2M}\,,\nonumber\\
  H_c &= -\frac{\kappa}{r}\frac{\Delta'}{4M^2}+\frac{\kappa}{r}\frac{S\Delta'}{2M^3}\,,\nonumber\\
  H_d &= -\frac{1}{2M^2}\lb Sp_F^2 -S'\frac{d}{dr}\rb+\frac{S}{2M^3}\lb Sp_F^2 -2S'\frac{d}{dr}\rb\,,\nonumber\\
  H_k &= -\frac{p_F^4}{8M^3}\,,\nonumber\\
  H_w &= +\frac{\Sigma''}{8M^2}-\frac{(\Sigma')^2 -2\Sigma'\Delta' +4S\Sigma''}{16M^3}\,.
\end{align}
Since all these terms are Hermitian, one can calculate the contribution of each
term to the single-particle energies, which is very helpful to
disclose the origin of relativistic symmetries.

In Ref.~\cite{Guo2014_PRL112-062502}, Guo \textit{et al.} have
made a new exploration of the PSS in deformed nuclei
by carrying out such a decomposition for axially deformed Dirac Hamiltonian with
Woods-Saxon-like vector and scalar potentials.
The single-particle states for $^{154}$Dy were obtained and
the SO and PSO splittings were analyzed in detail.

\begin{figure}[tbhp]
\begin{center}
\includegraphics[width=6cm]{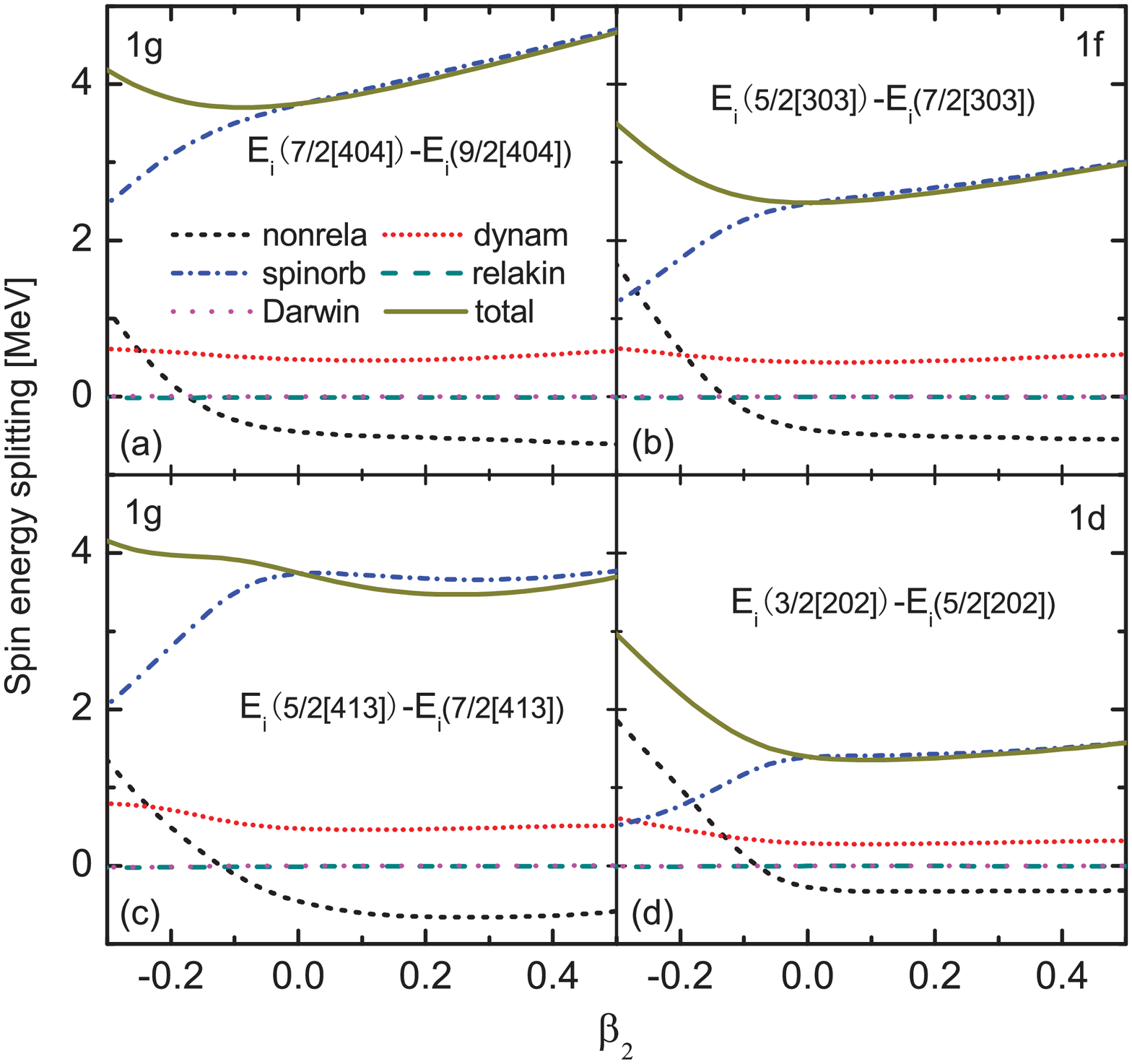}
\includegraphics[width=6cm]{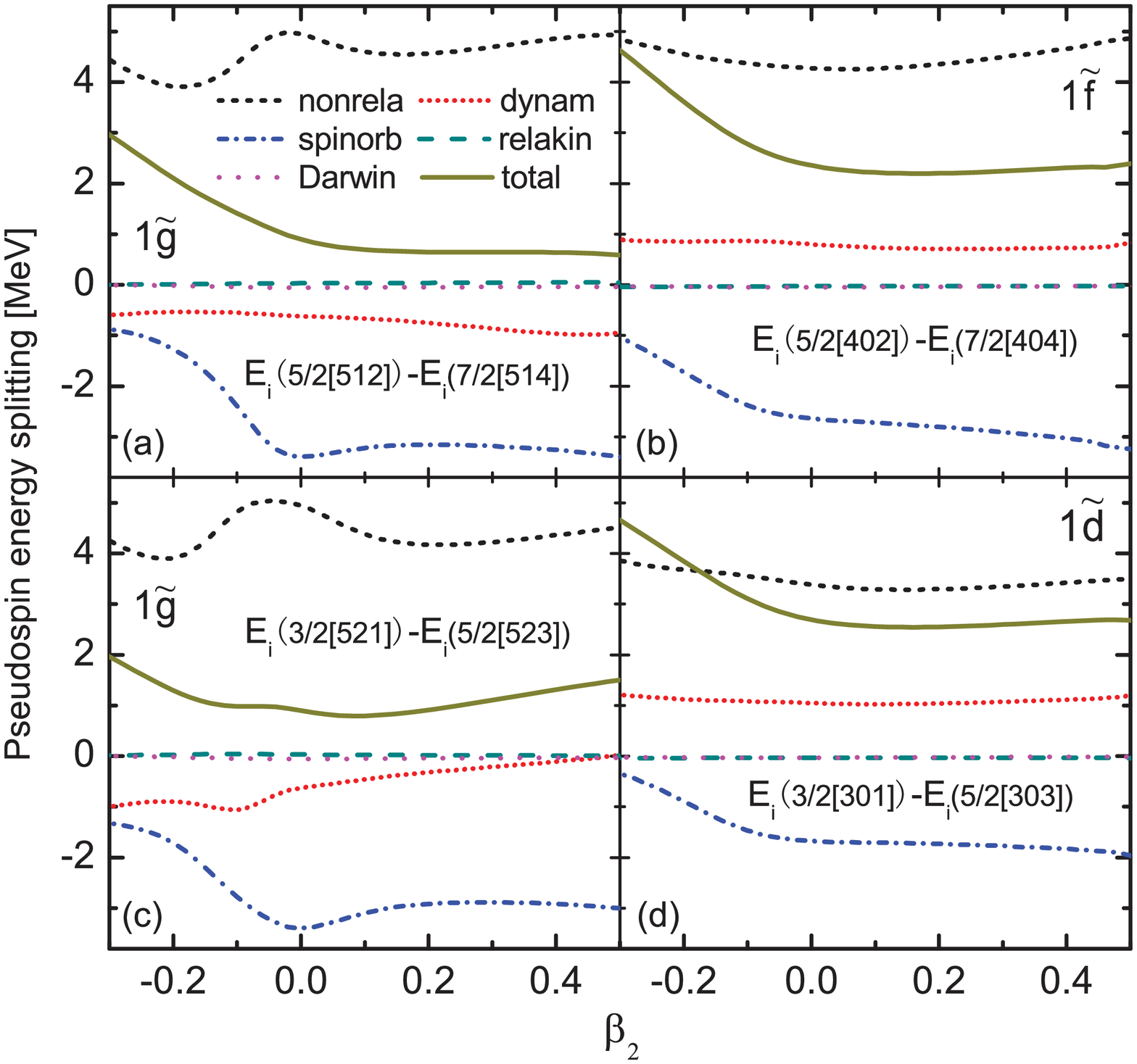}
\end{center}
\caption{(Color online)
Comparison of the contributions of all
the terms in $H_P$ to the spin (left panels) and pseudospin (right panels)
energy splitting and their correlations
with the deformation parameter $\beta_2$ for four pairs of spin (pseudospin) doublets.
Here ``nonrela'', ``dynam'', ``spinorb'', ``relakin'', and ``Darwin''
denote the non-relativistic part, the dynamical term, the spin-orbit
term, the relativistic modification of kinetic energy, and the spin (pseudospin)
Darwin term, respectively. As a guide to the eyes, the total
energy splittings are marked as ``total''.
Taken from Ref.~\cite{Guo2014_PRL112-062502}.
} \label{fig:Guo2014_PRL112-062502}
\end{figure}

In Fig.~\ref{fig:Guo2014_PRL112-062502}, the variation of the energy splitting
between the spin (pseudospin) doublets with the quadrupole deformation
parameter $\beta_2$ is shown. It can be seen that the SO splitting
is mainly from the contribution of the SO term $H_c$ when the system
is prolate deformed.
However, on the oblate side, the situation becomes more complicated:
The non-relativistic, the SO coupling, and the dynamical terms all
influence very much the SO splitting.
The different behaviors between the SO energy splitting on the prolate and
oblate sides are quite interesting.
This difference was attributed to the presence of configuration
mixing on the oblate side.
This should be examined in detail in the future.

Compared with the SO splitting, the change of the PSO splitting
with the deformation is more complicated, as one can see in the right panel
of Fig.~\ref{fig:Guo2014_PRL112-062502}.
The energy difference between a pair of pseudospin partners is mainly determined
by three parts,
the non-relativistic term $H_n$, the spin-orbit term $H_c$, and the dynamical term $H_d$.
Four pairs of pseudospin partners are shown here
and their energies are in the range of $0\sim-25$~MeV.
In this energy range and in the whole shown range of $\beta_2$,
the contribution of $H_n$ to the PSO splitting
is always positive, which means its contribution to
the pseudospin anti-aligned state is larger. The opposite is that of $H_c$, i.e.,
its contribution to the pseudospin aligned state is larger.
Therefore, the SO term always plays a role in favor of the PSS.
The contribution of the dynamical term $H_d$ to the PSO
splitting varies with the deformation and changes its sign
when the pseudospin doublets approach the threshold.

\subsubsection{Perturbation with SRG and SUSY}

Gathering all pieces presented above, it is promising to understand the PSS and its breaking mechanism in a fully quantitative way by combining the SRG, SUSY quantum mechanics, and perturbation theory \cite{Liang2013_PRC87-014334,Shen2013_PRC88-024311}.

The Dirac equation can be transformed into a diagonal form in a series of $1/M$ by using the SRG as shown in Eq.~(\ref{Eq:4.3.SRGHFermi}), and its lowest-order approximation corresponds to a Schr\"odinger equation.
In Ref.~\cite{Liang2013_PRC87-014334}, taking the lowest-order approximation as an example, the idea for applying the SUSY quantum mechanics to trace the origin of the PSS was illustrated and the PSS breaking mechanism was explored quantitatively by the perturbation theory.

By assuming the spherical symmetry, the radial Schr\"odinger equation is cast in the form of
\begin{equation}\label{Eq:4.3.SUSY2Schr}
    H(\kappa) R(r) = E R(r)
\end{equation}
with the single-particle Hamiltonian
\begin{equation}\label{Eq:4.3.SUSY2HSchr}
    H(\kappa) = -\frac{d^2}{2Mdr^2} + \frac{\kappa(\kappa+1)}{2Mr^2} + V(r)\,,
\end{equation}
and the single-particle wave functions
\begin{equation}\label{Eq:4.3.SUSY2psi}
    \psi_\alpha(\mathbf{r}) = \frac{R_a(r)}{r}\mathscr Y_{j_a m_a}^{l_a}(\hat{\mathbf{r}})\,.
\end{equation}
Here $V(r)$ is the non-relativistic central potential standing for the sum of the scalar and vector potentials, $\Sigma(r)$, in Eq.~(\ref{Eq:4.3.SRGHFermi}).

It is clear that $H$ conserves the explicit SS for the spin doublets $a$ and $b$ with $\kappa_a+\kappa_b=-1$, which leads to the same centrifugal barrier $\kappa(\kappa+1)/(2Mr^2)$.
Similarly, in order to investigate the origin of the PSS and its breaking mechanism, it is crucial to identify the corresponding term proportional to $\tilde{l}(\tilde{l}+1)=\kappa(\kappa-1)$, which leads to the same $\kappa(\kappa-1)$ values for the pseudospin doublets $a$ and $b$ with $\kappa_a+\kappa_b=1$.
The SUSY quantum mechanics is one of promising approaches for identifying such $\kappa(\kappa-1)$ structure.

Following the similar procedures shown in Section~\ref{Sect:4.3.2}, one can start with a couple of Hermitian conjugate first-order operators
\begin{equation}\label{Eq:4.3.SUSY2B+B-}
    B^+_\kappa = \ls Q_\kappa(r)-\frac{d}{dr}\rs\frac{1}{\sqrt{2M}}\,,
    \qquad
    B^-_\kappa = \frac{1}{\sqrt{2M}}\ls Q_\kappa(r)+\frac{d}{dr}\rs\,,
\end{equation}
and the reduced superpotentials
\begin{equation}\label{Eq:4.3.SUSY2q}
    q_\kappa(r) = Q_\kappa(r) - \frac{\kappa}{r}\,,
\end{equation}
and end up with the SUSY partner Hamiltonians
\begin{align}\label{Eq:4.3.SUSY2Hq}
    H_1(\kappa) &= B^+_\kappa B^-_\kappa = \frac{1}{2M} \ls-\frac{d^2}{dr^2} +\frac{\kappa(\kappa+1)}{r^2} +q_\kappa^2 +\frac{2\kappa}{r}q_\kappa -q'_\kappa\rs\,,\nonumber\\
    H_2(\kappa) &= B^-_\kappa B^+_\kappa = \frac{1}{2M} \ls-\frac{d^2}{dr^2} +\frac{\kappa(\kappa-1)}{r^2} +q_\kappa^2 +\frac{2\kappa}{r}q_\kappa +q'_\kappa\rs\,.
\end{align}
It is important to note that these Hamiltonians are Hermitian, but those in Eqs.~(\ref{Eq:4.3.SUSY1Hq}) are not \cite{Liang2013_PRC87-014334}.

The reduced superpotentials $q_\kappa(r)$ satisfy the first-order differential equation \cite{Liang2013_PRC87-014334},
\begin{equation}\label{Eq:4.3.SUSY2qeq}
    \frac{1}{2M}\ls q_\kappa^2(r)+\frac{2\kappa}{r}q_\kappa(r)-q'_\kappa(r)\rs + e(\kappa) = V(r)\,,
\end{equation}
with the asymptotic behaviors
\begin{equation}\label{Eq:4.3.SUSY2qasy}
    \lim_{r\rightarrow\infty}q_\kappa(r) = \sqrt{-2Me(\kappa)}
    \qquad\mbox{and}\qquad
    \lim_{r\rightarrow0}q_{\kappa}(r)=\frac{2M(e(\kappa)-V)}{(1-2\kappa)}r\,.
\end{equation}
The energy shifts are determined in the same way as that shown in Eqs.~(\ref{Eq:4.3.SUSY1e1}) and (\ref{Eq:4.3.SUSY1e2}), i.e,
\begin{equation}\label{Eq:4.3.SUSY2e}
    e(\kappa_a) = E_{1\kappa_a}
    \qquad\mbox{and}\qquad
    e(\kappa_b) = 2\left.V\right|_{r=0} - e(\kappa_a)\,,
\end{equation}
for the cases of $\kappa_a<0$ and $\kappa_b>0$, respectively.

Before the numerical calculations, it is interesting to seek a possible exact PSS limit analytically.
The SUSY partner Hamiltonian reads
\begin{equation}\label{Eq:4.3.SUSY2Htil}
    \tilde{H}(\kappa)=H_2(\kappa) + e(\kappa)
    = -\frac{d^2}{2Mdr^2} + \frac{\kappa(\kappa-1)}{2Mr^2} + \tilde{V}_\kappa(r)\,,
\end{equation}
with
\begin{equation}\label{Eq:4.3.SUSY2Vtil}
    \tilde{V}_\kappa(r)=V(r)+q'_\kappa(r)/M\,.
\end{equation}
By definition the exact PSS limit holds $E_{n\kappa_a} = E_{(n-1) \kappa_b}$ with $\kappa_a<0$ and $\kappa_a+\kappa_b=1$,
which indicates $H_2(\kappa_a) + e(\kappa_a) = H_2(\kappa_b) + e(\kappa_b)$.
By combining Eqs.~(\ref{Eq:4.3.SUSY2Hq}) and (\ref{Eq:4.3.SUSY2qeq}), as well as the boundary condition $q_\kappa(0)=0$, one can readily have
\begin{equation}\label{Eq:4.3.SUSY2PSSq}
    q_{\kappa_a}(r) = q_{\kappa_b}(r) = M\omega_{\kappa} r\,,
\end{equation}
with a known constant $\omega_{\kappa} \equiv (e(\kappa_a)-e(\kappa_b))/(\kappa_b-\kappa_a)$.
As the reduced superpotentials $q_\kappa(r)$ are linear functions of $r$, the central potential $V(r)$ in $H$ has the form [cf. Eq.~(\ref{Eq:4.3.SUSY1RHO})]
\begin{equation}\label{Eq:4.3.SUSY2VHO}
    V_{\rm HO}(r) = \frac{M}{2}\omega_{\kappa}^2 r^2+V(0)\,.
\end{equation}
The corresponding PSS limit is nothing but the well known case with harmonic oscillator potentials, which leads to the energy degeneracy of the whole major shell.

In the following calculations, the mass of nucleon is taken as $M=939.0$~MeV, and the central potential $V(r)$ is chosen as a Woods-Saxon form
\begin{equation}\label{Eq:4.3.SUSY2WS}
    V(r) = \frac{V_0}{1+e^{(r-R)/a}}\,,
\end{equation}
with the parameters $V_0=-63.297$~MeV, $R=6.278$~fm, and $a=0.615$~fm, which corresponds to the neutron potential provided in Ref.~\cite{Koepf1991_ZPA339-81} by taking $N=82$ and $Z=50$.
This potential is illustrated as the solid line in Fig.~\ref{Fig:4.3.SUSY2V} below.
In this Section, we use a tilde to denote the operators, potentials, and wave functions belonging to $\tilde{H}$.

\begin{figure}[tbhp]
\begin{center}
  \includegraphics[width=8cm]{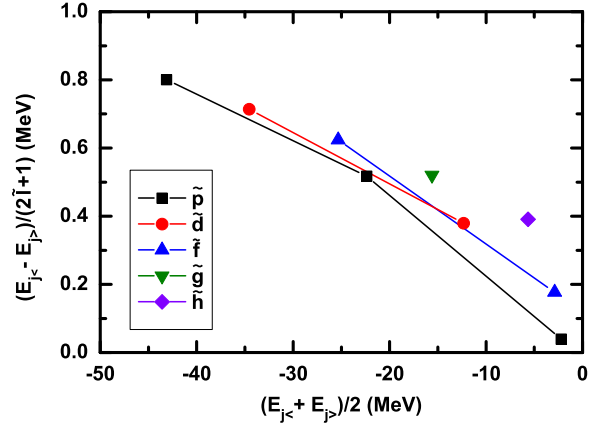}
\end{center}
\caption{(Color online) Reduced PSO splittings $(E_{j_<}-E_{j_>})/(2\tilde{l}+1)$ versus their average single-particle energies $(E_{j_<}+E_{j_>})/2$.
Taken from Ref.~\cite{Liang2013_PRC87-014334}.}
\label{Fig:4.3.SUSY2dE}
\end{figure}

In Fig.~\ref{Fig:4.3.SUSY2dE}, the reduced PSO splittings $(E_{j_<}-E_{j_>})/(2\tilde{l}+1)$ versus their average  single-particle energies $E_{\rm av}=(E_{j_<}+E_{j_>})/2$ are plotted, where $j_<$ ($j_>$) denotes the states with $j=\tilde{l}-1/2$ ($j=\tilde{l}+1/2$).
It is seen that the amplitudes of the reduced PSO splittings are less than 1~MeV.
Moreover, as a general tendency, the splittings become smaller with the increasing single-particle energies, which is in agreement with the results shown in Figs.~\ref{Fig:2.3.132split} and \ref{Fig:3.0.Sn} by the self-consistent relativistic mean-field and relativistic continuum Hartree-Bogoliubov theories.
It is very interesting to investigate the physical mechanism for such energy-dependent behavior.
This also helps to figure out whether or not the PSS is an accidental symmetry \cite{Marcos2008_EPJA37-251}.

\begin{figure}[tbhp]
\begin{center}
  \includegraphics[width=6cm]{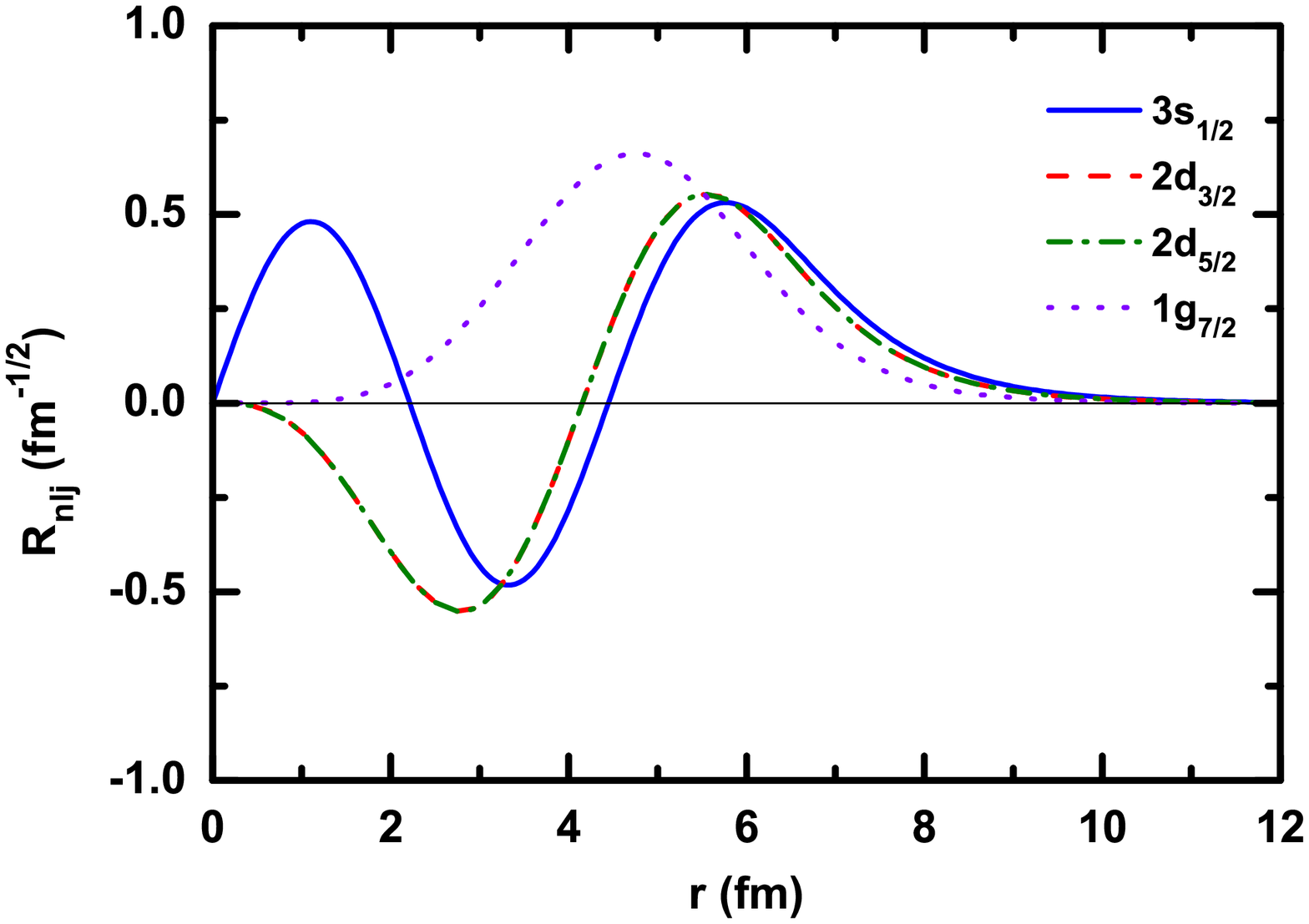}
  \includegraphics[width=6cm]{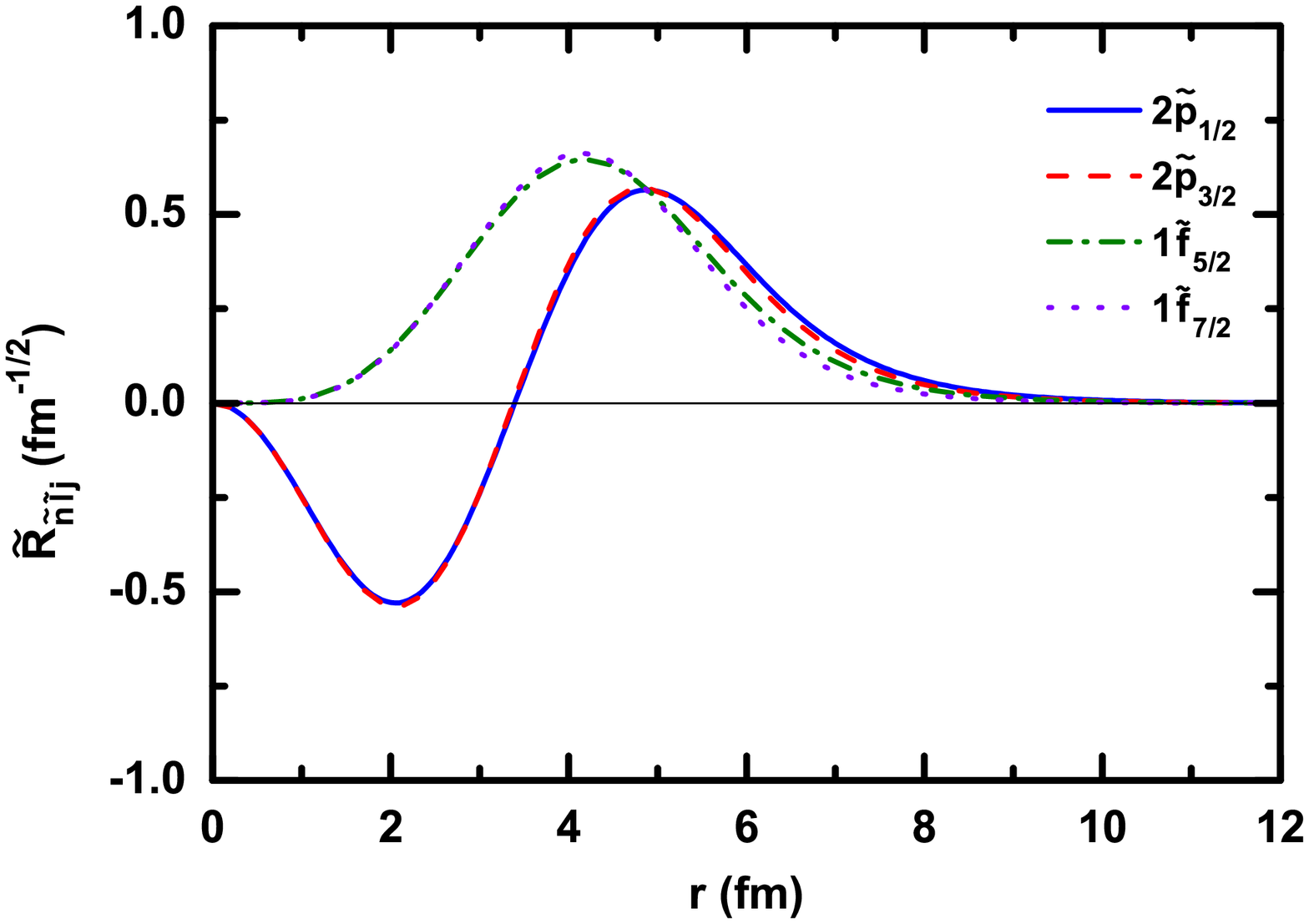}
\end{center}
\caption{(Color online) Single-particle wave functions $R_{nlj}(r)$ of $H$ (the left panel) and $\tilde{R}_{\tilde{n}\tilde{l}j}(r)$ of $\tilde{H}$ (the right panel) for the $3s_{1/2}$, $2d_{3/2}$, $2d_{5/2}$, and $1g_{7/2}$ states.
Taken from Ref.~\cite{Liang2013_PRC87-014334}.}
\label{Fig:4.3.SUSY2WF}
\end{figure}

In the left panel of Fig.~\ref{Fig:4.3.SUSY2WF}, the single-particle radial wave functions $R_{nlj}(r)$ of $H$ are shown by taking the pseudospin doublets $2\tilde{p}$ and $1\tilde{f}$ as examples.
It is clear that the wave functions of the spin doublets are identical since there is no spin-orbit term in $H$.
In contrast, the wave functions of the pseudospin doublets are very different from each other.
This leads to difficulties in analyzing the origin of the PSS and its breaking.

\begin{table}
\begin{center}
\caption{Contributions from the kinetic term (kin), centrifugal barrier (CB), and central potential (cen) to the single-particle energies $E$ and the corresponding PSO splittings $\Delta E_{\rm PSO}$ for the pseudospin doublets $2\tilde{p}$ and $1\tilde{f}$.
All units are in MeV.
The data are taken from Ref.~\cite{Liang2013_PRC87-014334}.
\label{Tab:4.3.SUSY2dE1}}
\begin{tabular}{@{}lrrrr@{}} \hline
State & \multicolumn{1}{c}{$E_{\rm kin}$} & \multicolumn{1}{c}{$E_{\rm CB}$} & \multicolumn{1}{c}{$E_{\rm cen}$} & \multicolumn{1}{c}{$E$} \\ \hline
            $3s_{1/2}$ & $28.953$ &   $0.000$ & $-50.545$ & $-21.591$ \\
            $2d_{3/2}$ & $16.845$ &  $11.758$ & $-51.746$ & $-23.143$ \\
  $\Delta E_{\rm PSO}$ & $12.109$ & $-11.758$ &   $1.201$ &   $1.552$ \\ \hline
            $2d_{5/2}$ & $16.845$ &  $11.758$ & $-51.746$ & $-23.143$ \\
            $1g_{7/2}$ &  $6.197$ &  $20.483$ & $-54.188$ & $-27.508$ \\
  $\Delta E_{\rm PSO}$ & $10.648$ &  $-8.725$ &   $2.442$ &   $4.365$ \\ \hline
\end{tabular}
\end{center}
\end{table}

Prior to the quantitative analysis by using the perturbation theory in Ref.~\cite{Liang2011_PRC83-041301R}, the investigation of PSO splittings was usually done by decomposing the contributions by terms, where each contribution is calculated with the corresponding operator $\hat O_i$ by
\begin{equation}\label{Eq:4.3.SUSY2Ei}
    E_i = \int R^*(r)\hat O_i R(r) dr\,.
\end{equation}
Within the representation of $H$ shown in Eq.~(\ref{Eq:4.3.SUSY2HSchr}), the operators of the kinetic term, centrifugal barrier, and central potential read $-d^2/(2Mdr^2)$, $\kappa(\kappa+1)/(2Mr^2)$, and $V(r)$, respectively.
In Table~\ref{Tab:4.3.SUSY2dE1}, the contributions from these terms to the single-particle energies $E$ as well as the corresponding PSO splittings $\Delta E_{\rm PSO}$ are shown for the pseudospin doublets $2\tilde{p}$ and $1\tilde{f}$.
It is not surprising that, within this representation, the contributions to $\Delta E_{\rm PSO}$ come from all channels, while they substantially cancel to each other in a sophisticated way.

As discussed in Section~\ref{Sect:4.1}, the phenomenon of such strong cancellations among different terms was usually associated with the dynamical \cite{Alberto2001_PRL86-5015,Alberto2002_PRC65-034307} and even the non-perturbative~\cite{Marcos2001_PLB513-30,Lisboa2010_PRC81-064324,Ginocchio2011_JPCS267-012037} nature of PSS.
However, such connection is mystified and sometimes even misleading.
Indeed, by using the perturbation calculations based on the PSS limit shown in Eq.~(\ref{Eq:4.3.SUSY2VHO}), we demonstrated that the nature of PSS discussing here is perturbative \cite{Liang2013_PRC87-014334}, the corresponding calculated results are quite similar to those shown in Fig.~\ref{Fig:4.1.U3}.

Much more important is now the origin of the PSS and its breaking mechanism can be studied in an explicit way within the representation of the SUSY partner Hamiltonian $\tilde{H}$ \cite{Liang2013_PRC87-014334}.

\begin{figure}[tbhp]
\begin{center}
  \includegraphics[width=8cm]{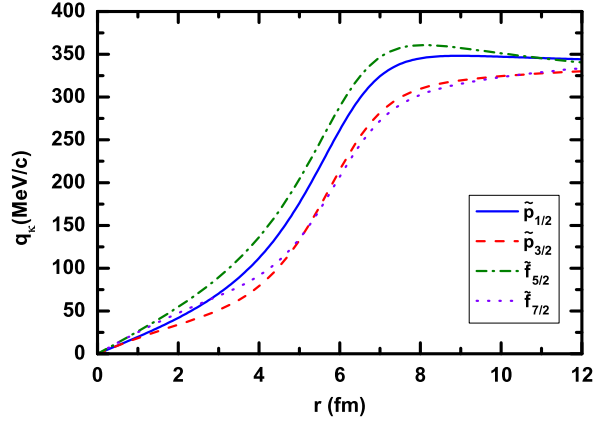}
\end{center}
\caption{(Color online) Reduced superpotentials $q_\kappa(r)$ for the $\tilde{p}$ and $\tilde{f}$ orbitals.
Taken from Ref.~\cite{Liang2013_PRC87-014334}.}
\label{Fig:4.3.SUSY2q}
\end{figure}

To obtain the SUSY partner Hamiltonian $\tilde{H}$ in Eq.~(\ref{Eq:4.3.SUSY2Htil}), one solves the first-order differential equation~(\ref{Eq:4.3.SUSY2qeq}) for the reduced superpotentials $q_\kappa(r)$ with the boundary condition $q_\kappa(0)=0$.
The corresponding $q_\kappa(r)$ are shown in Fig.~\ref{Fig:4.3.SUSY2q} in units of MeV/$c$.
The $q_\kappa(r)$ are $\kappa$-dependent since the left-hand side of Eq.~(\ref{Eq:4.3.SUSY2qeq}) contains a $\kappa$-dependent term.
However, it should be emphasized that $q_\kappa(r)$ does not depend on the radial quantum number $n$ for a given $\kappa$.
One will discover that such an $n$-independent property is essential for understanding the general pattern of $\Delta E_{\rm PSO}$ versus $E_{\rm av}$ shown in Fig.~\ref{Fig:4.3.SUSY2dE}.

\begin{figure}[tbhp]
\begin{center}
  \includegraphics[width=6cm]{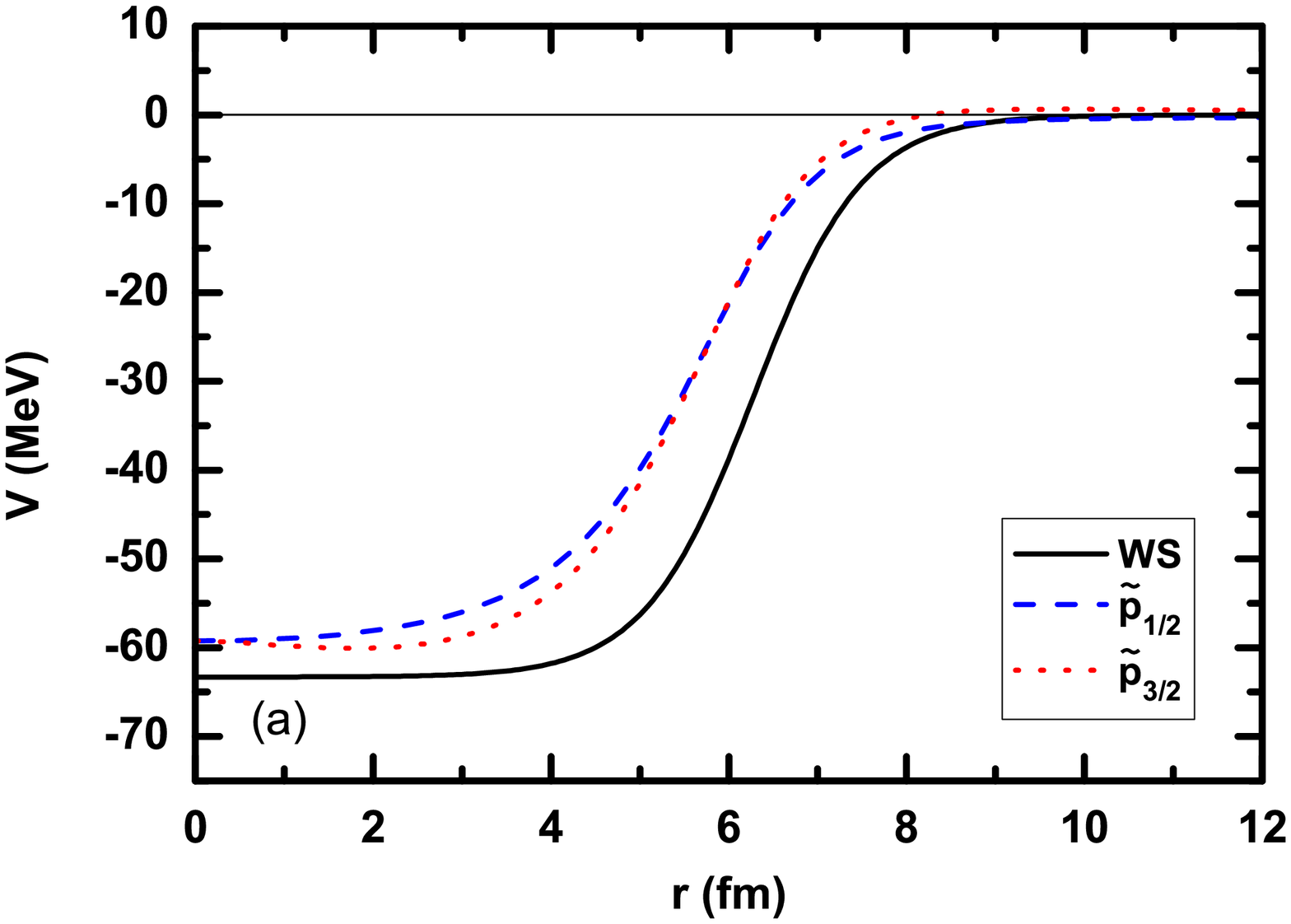}
  \includegraphics[width=6cm]{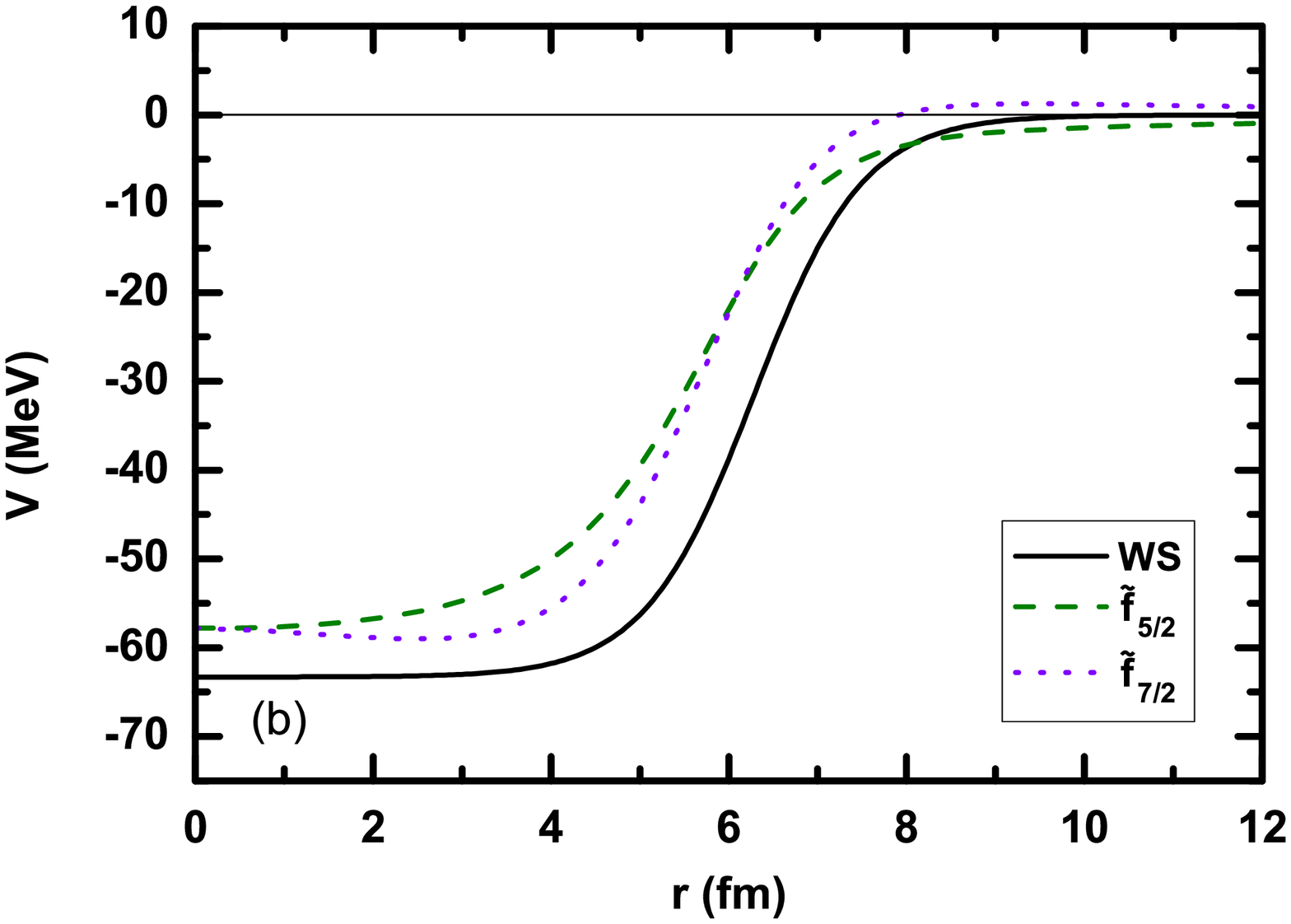}\\
  \includegraphics[width=6cm]{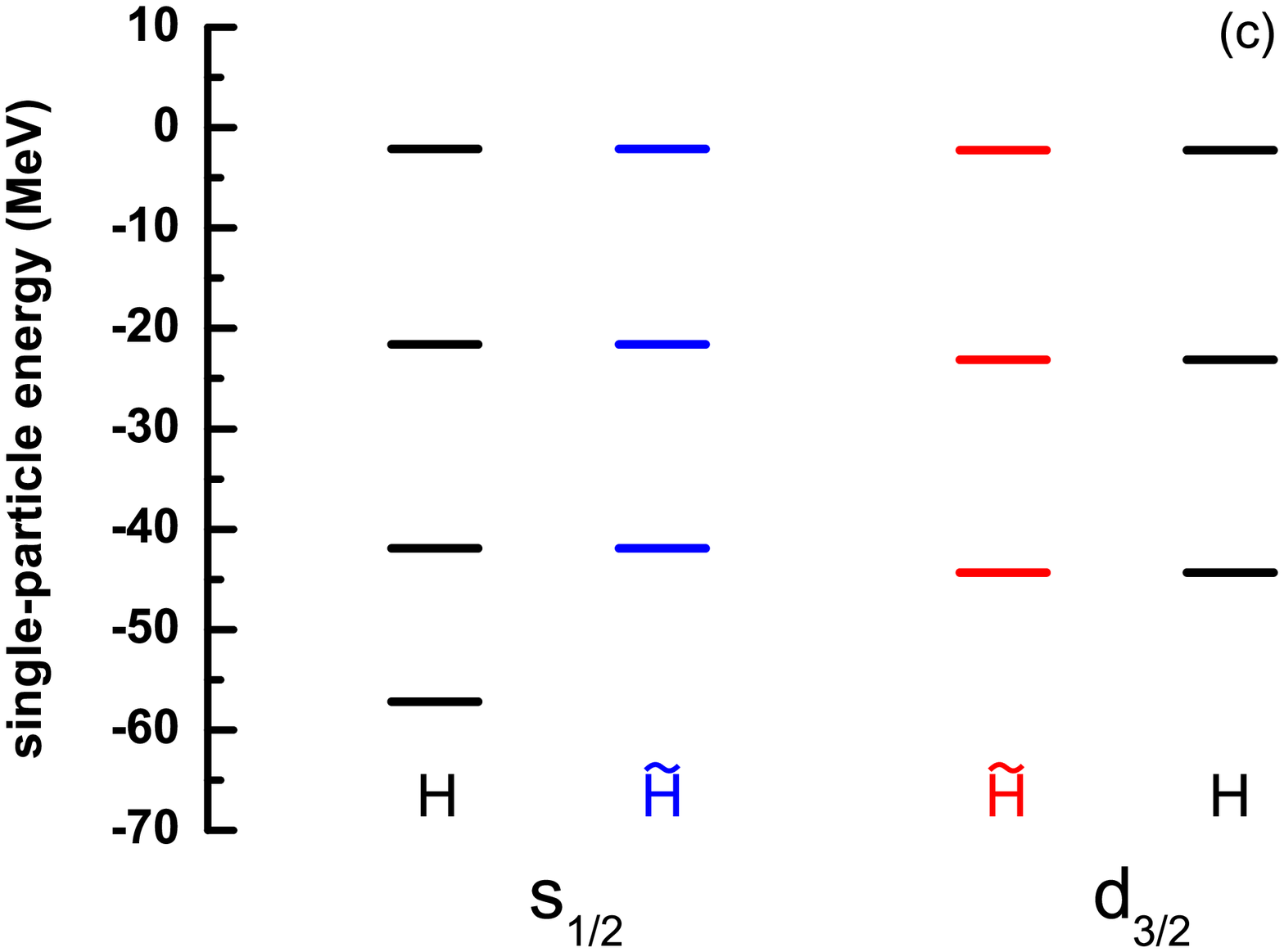}
  \includegraphics[width=6cm]{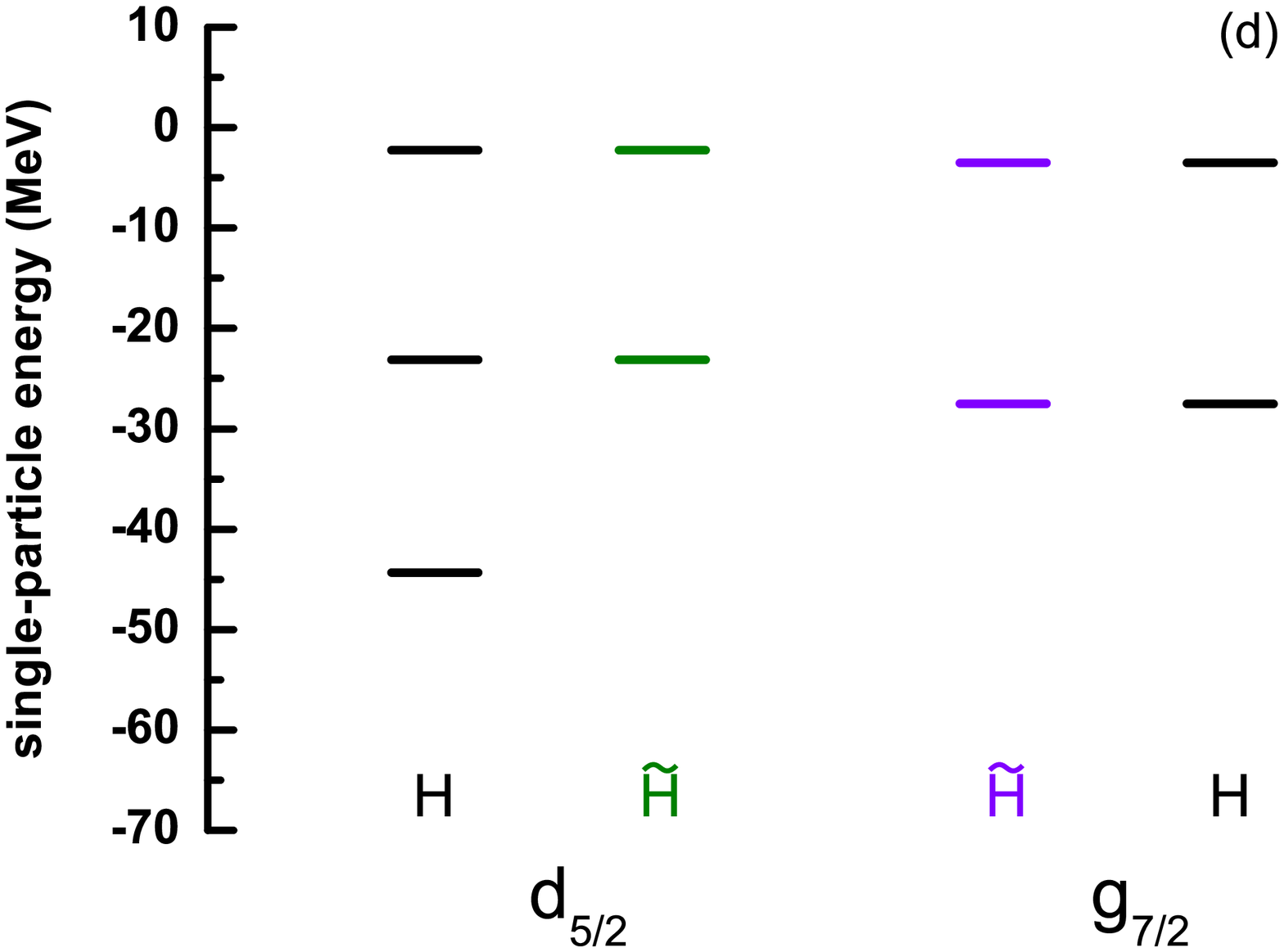}
\end{center}
\caption{(Color online) Upper panels: $\kappa$-dependent central potentials $\tilde{V}_\kappa(r)$ in $\tilde{H}$ as a function of $r$ for the (a) $\tilde{p}$ and (b) $\tilde{f}$ orbitals, while the Woods-Saxon potential in $H$ is shown for comparison.
Lower panels: The corresponding single-particle energies obtained with $H$ and $\tilde{H}$.
Taken from Ref.~\cite{Liang2013_PRC87-014334}.}
\label{Fig:4.3.SUSY2V}
\end{figure}

Then the $\kappa$-dependent central potentials $\tilde V_\kappa(r)$ in $\tilde{H}$ can be obtained, and their asymptotic behaviors read
\begin{equation}\label{Eq:4.3.SUSY2Vasym}
    \lim_{r\rightarrow0}\tilde{V}_{\kappa}(r)=V+\frac{2(e(\kappa)-V)}{(1-2\kappa)}
    \qquad\mbox{and}\qquad
    \lim_{r\rightarrow\infty}\tilde{V}_{\kappa}(r) = 0\,.
\end{equation}
It is important that these potentials are regular and converge at both $r\rightarrow0$ and $r\rightarrow\infty$.
In the upper panels of Fig.~\ref{Fig:4.3.SUSY2V}, these central potentials $\tilde V_\kappa(r)$ are shown, while the Woods-Saxon potential $V(r)$ in $H$ is also shown for comparison.
For all $\kappa$, the potentials $\tilde V_\kappa(r)$ approximately remain a Woods-Saxon shape, and they become shallower than the original potential $V(r)$.
By comparing two upper panels, it is seen that the amplitude of the difference between $\tilde V_\kappa(r)$ for a pair of pseudospin partners increases with the difference of their quantum numbers $|\kappa_a-\kappa_b|$.

After getting the central potentials $\tilde V_\kappa(r)$, one is ready to calculate the single-particle energies and wave functions of the SUSY partner Hamiltonians $\tilde{H}(\kappa)$.
In the lower panels of Fig.~\ref{Fig:4.3.SUSY2V}, the discrete single-particle energies obtained with $\tilde{H}$ are compared with those obtained with $H$.
It is clear that the eigenstates of Hamiltonians $H$ and $\tilde{H}$ are identical, except for the lowest eigenstates with $\kappa<0$ in $H$, which are the so-called intruder states.
In other words, the fact that the intruder states have no pseudospin partners can be interpreted as a natural result of the exact SUSY for $\kappa<0$ and broken SUSY for $\kappa>0$.
By holding the one-to-one mapping relation in the two sets of spectra, the origin of PSS, which is deeply hidden in $H$, can be now traced by employing its SUSY partner Hamiltonian $\tilde{H}$.

The single-particle radial wave functions $\tilde{R}_{\tilde{n}\tilde{l}j}(r)$ of $\tilde{H}$ for the $2\tilde{p}$ and $1\tilde{f}$ pseudospin doublets are shown in the right panel of Fig.~\ref{Fig:4.3.SUSY2WF}.
The nodal relation in Eq.~(\ref{Eq:4.3.nodes}) can be seen by comparing the wave functions shown in the left and right panels.
In fact, not only are the numbers of nodes equal, but also the wave functions of pseudospin doublets are almost identical to each other.
Therefore, within this representation, the quasi-degeneracy of pseudospin doublets is closely related to the similarity of their wave functions, and vice versa \cite{Liang2013_PRC87-014334}.

\begin{table}
\begin{center}
\caption{Contributions from kinetic term (kin), pseudo-centrifugal barrier (PCB), and central potential (cen) to the single-particle energies $E$ and the corresponding PSO splittings $\Delta E_{\rm PSO}$ for the pseudospin doublets $2\tilde{p}$ and $1\tilde{f}$.
All units are in MeV.
The data are taken from Ref.~\cite{Liang2013_PRC87-014334}.
\label{Tab:4.3.SUSY2dE2}}
\begin{tabular}{@{}lrrrr@{}} \hline
State & \multicolumn{1}{c}{$E_{\rm kin}$} & \multicolumn{1}{c}{$E_{\rm PCB}$} & \multicolumn{1}{c}{$E_{\rm cen}$} & \multicolumn{1}{c}{$E$} \\ \hline
    $2\tilde{p}_{1/2}$ & $16.602$ &  $6.723$ & $-44.916$ & $-21.591$ \\
    $2\tilde{p}_{3/2}$ & $17.331$ &  $6.857$ & $-47.332$ & $-23.143$ \\
  $\Delta E_{\rm PSO}$ & $-0.729$ & $-0.134$ &   $2.415$ &   $1.552$ \\ \hline
    $1\tilde{f}_{5/2}$ &  $5.710$ & $16.286$ & $-45.139$ & $-23.143$ \\
    $1\tilde{f}_{7/2}$ &  $6.293$ & $16.591$ & $-50.392$ & $-27.508$ \\
  $\Delta E_{\rm PSO}$ & $-0.584$ & $-0.305$ &   $5.253$ &   $4.365$ \\ \hline
\end{tabular}
\end{center}
\end{table}

The same strategy as done in Table~\ref{Tab:4.3.SUSY2dE1} is then adopted to investigate the PSO splittings, but now within the representation of $\tilde{H}$ shown in Eq.~(\ref{Eq:4.3.SUSY2Htil}) instead.
The corresponding operators include the kinetic term $-d^2/(2Mdr^2)$, the PCB $\kappa(\kappa-1)/(2Mr^2)$, and the central potential $\tilde{V}_\kappa(r)$.
The corresponding results for the pseudospin doublets $2\tilde{p}$ and $1\tilde{f}$ are listed in Table~\ref{Tab:4.3.SUSY2dE2}.
It can be seen that for each pair of pseudospin doublets the energy contributions from the PSS-conserving terms, i.e., the kinetic and PCB, are very similar.
The PSO splittings $\Delta E_{\rm PSO}$ are mainly contributed by the difference in the central potentials $\Delta E_{\rm cen}$, which is due to the slight $\kappa$-dependence of $\tilde{V}_\kappa(r)$ as shown in Fig.~\ref{Fig:4.3.SUSY2V}.
In other words, the sophisticated cancellations among different terms in $H$ can be clearly understood by using a proper decomposition with the help of the SUSY quantum mechanics \cite{Liang2013_PRC87-014334}.

In order to perform the perturbation calculations, the Hamiltonian $\tilde{H}$ is split as
\begin{equation}\label{Eq:4.3.SUSY2HH0W}
  \tilde{H} = \tilde{H}^{\rm PSS}_0 + \tilde{W}^{\rm PSS}\,,
\end{equation}
where $\tilde{H}^{\rm PSS}_0$ and $\tilde{W}^{\rm PSS}$ are the corresponding PSS-conserving and PSS-breaking terms, respectively.
By requiring that $\tilde{W}^{\rm PSS}$ should be proportional to $\kappa$, which is similar to the case of the SO term in the conventional scheme, one has
\begin{align}\label{Eq:4.3.SUSY2H0W}
    \tilde{H}^{\rm PSS}_0&=\frac{1}{2M}\ls-\frac{d^2}{dr^2}+\frac{\kappa(\kappa-1)}{r^2}\rs + \tilde{V}_{\rm PSS}(r)\,,\nonumber\\
    \tilde{W}^{\rm PSS}&=\kappa \tilde{V}_{\rm PSO}(r)\,.
\end{align}
In such a way, the PSS-conserving $\tilde{V}_{\rm PSS}(r)$ and PSS-breaking $\tilde{V}_{\rm PSO}(r)$ potentials can be uniquely determined as \cite{Liang2013_PRC87-014334}
\begin{equation}\label{Eq:4.3.SUSY2VPSO}
  \tilde{V}_{\rm PSS}(r) = \frac{\kappa_b\tilde{V}_{\kappa_a}(r) - \kappa_a\tilde{V}_{\kappa_b}(r)}{\kappa_a-\kappa_b}
  \qquad\mbox{and}\qquad
  \tilde{V}_{\rm PSO}(r) =\frac{1}{M}\frac{q'_{\kappa_a}(r) - q'_{\kappa_b}(r)}{\kappa_a-\kappa_b}\,.
\end{equation}

\begin{figure}[tbhp]
\begin{center}
  \includegraphics[width=6cm]{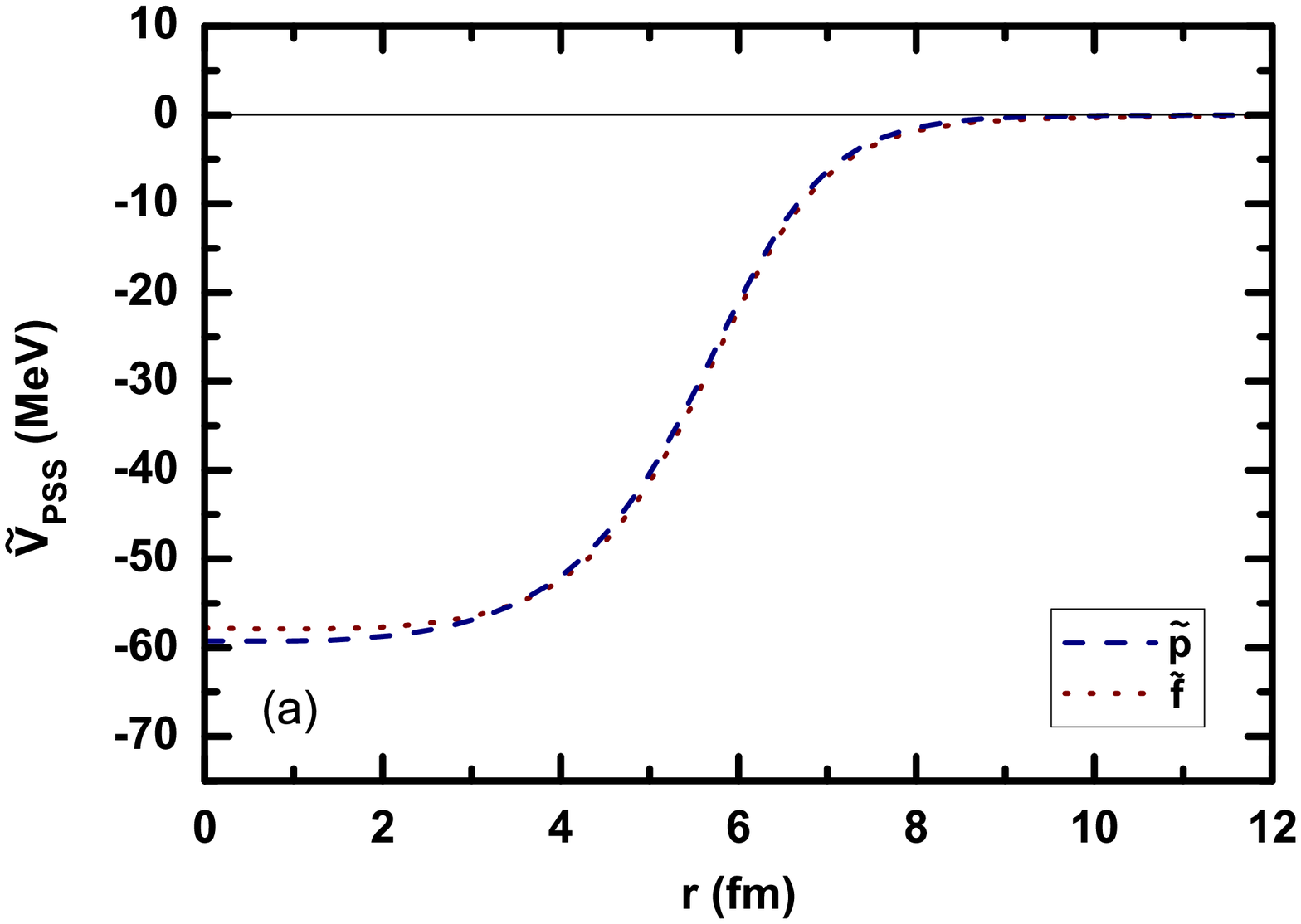}
  \includegraphics[width=6cm]{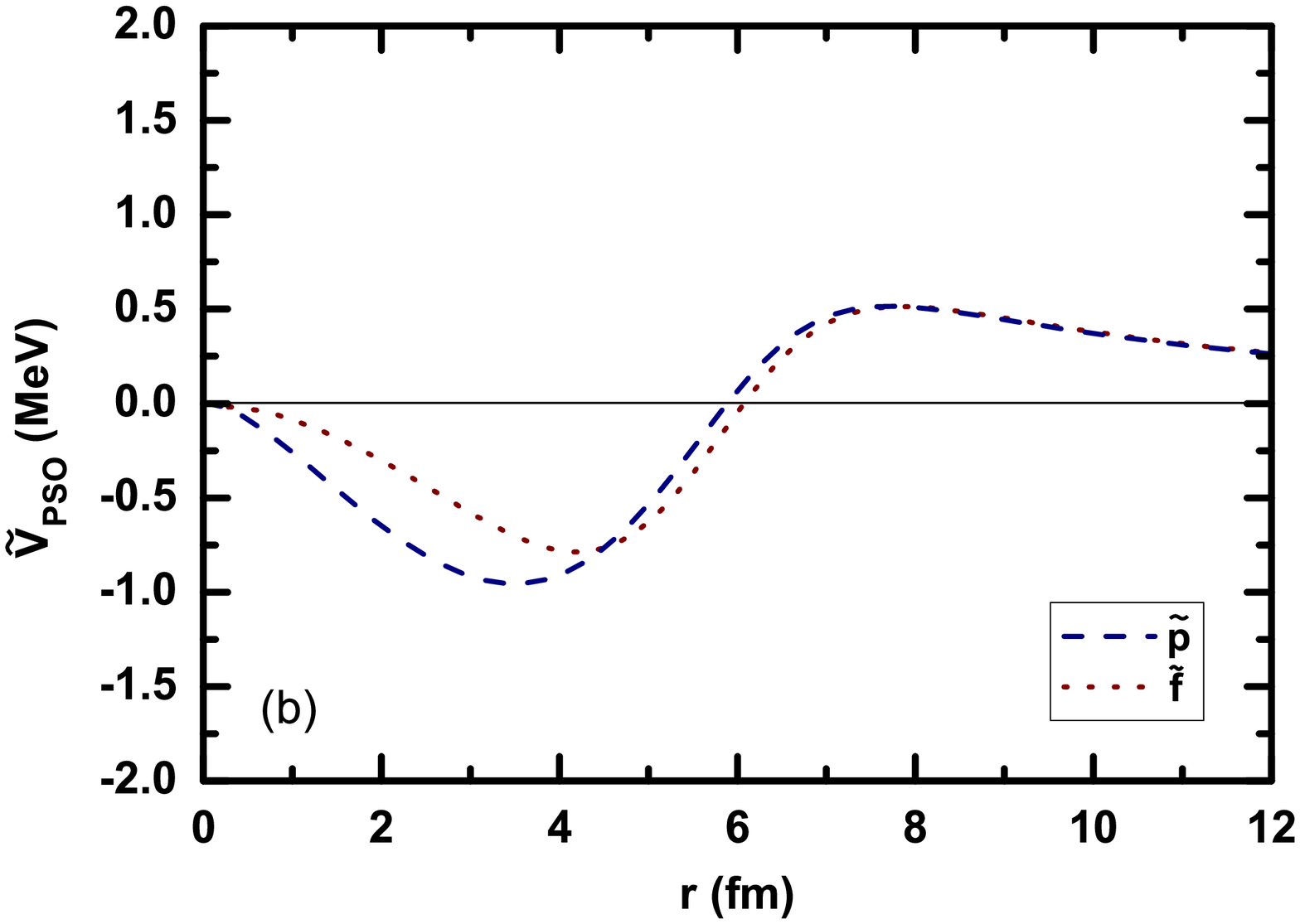}
\end{center}
\caption{(Color online) PSS-conserving potentials $\tilde{V}_{\rm PSS}(r)$ (the left panel) and PSS-breaking potentials $\tilde{V}_{\rm PSO}(r)$ (the right panel) for the $\tilde{p}$ and $\tilde{f}$ orbitals.
Taken from Ref.~\cite{Liang2013_PRC87-014334}.}
\label{Fig:4.3.SUSY2H0W}
\end{figure}

In Fig.~\ref{Fig:4.3.SUSY2H0W}, the $\tilde{V}_{\rm PSS}(r)$ and $\tilde{V}_{\rm PSO}(r)$ potentials are illustrated by taking the $\tilde{p}$ and $\tilde{f}$ orbitals as examples.
On one hand, it can be seen that the PSS-conserving potentials $\tilde{V}_{\rm PSS}(r)$ remain an approximate Woods-Saxon shape, and they are $\kappa$-dependent to a small extent.
On the other hand, the PSS-breaking potentials $\tilde{V}_{\rm PSO}(r)$ show several special features \cite{Liang2013_PRC87-014334}:
(i) These PSS-breaking potentials are regular functions of $r$, in particular, they vanish at $r\rightarrow\infty$.
(ii) It can be seen that the amplitudes of $\tilde{V}_{\rm PSO}$ are around 1~MeV, which directly lead to the reduced PSO splittings $\Delta E_{\rm PSO}\lesssim1$~MeV as shown in Fig.~\ref{Fig:4.3.SUSY2dE}.
(iii) More importantly, the PSO potentials $\tilde{V}_{\rm PSO}(r)$ are negative at small radius but positive at large radius with a node at the surface region, which is totally different from the usual SO potentials with a surface-peaked shape.
The particular shape of the PSO potentials $\tilde{V}_{\rm PSO}(r)$ can explain well the variations of the PSO splitting with the single-particle energy.

More details for point (iii) are as follows:
First of all, $\tilde{V}_{\rm PSO}(r)$ do not depend on the radial quantum number $n$.
Meanwhile, the single-particle wave functions $\tilde{R}(r)$ extend to larger distances with higher energies.
Thus, the matrix element $\langle\tilde{R}|\tilde{V}_{\rm PSO}|\tilde{R}\rangle$ is negative when the wave function is centralized in the inner part.
As the wave function becomes more extended, the positive part of $\tilde{V}_{\rm PSO}(r)$ compensates for the negative value of the matrix element.
In this way, the PSO splittings $\Delta E_{\rm PSO}$ decrease while the radial quantum numbers $\tilde{n}$ increase.
The splittings can even reverse for the resonance states, where the outer part of the PSO potentials plays the major role.

\begin{figure}[tbhp]
\begin{center}
  \includegraphics[width=8cm]{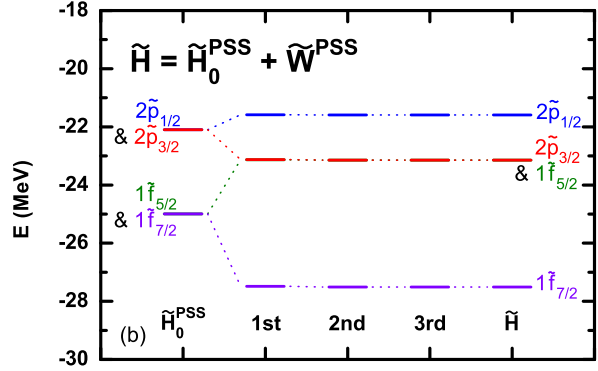}
\end{center}
\caption{(Color online) Single-particle energies obtained at the exact PSS limit $\tilde H^{\rm PSS}_0$, and their counterparts obtained by the first-, second-, and third-order perturbation calculations with $\tilde W^{\rm PSS}$, as well as those obtained with $\tilde H$.
Taken from Ref.~\cite{Liang2013_PRC87-014334}.}
\label{Fig:4.3.SUSY2PT}
\end{figure}

Finally, the perturbation calculations are performed based on the pseudospin symmetric Hamiltonian $\tilde{H}^{\rm PSS}_0$ with the perturbation $\tilde{W}^{\rm PSS}$.
In Fig.~\ref{Fig:4.3.SUSY2PT}, the single-particle energies obtained at the PSS limit $\tilde{H}^{\rm PSS}_0$, and their counterparts obtained by the first-, second-, and third-order perturbation calculations, as well as those obtained with $\tilde{H}$ are shown from left to right.
It can be seen that the pseudospin doublets are exactly degenerate at the PSS limit $\tilde{H}^{\rm PSS}_0$.
For the present decomposition, the largest perturbation correction $|\tilde{W}^{\rm PSS}_{mk}/(E_m-E_k)|$ is less than $0.03$ \cite{Liang2013_PRC87-014334}.
This indicates that the criterion in Eq.~(\ref{Eq:4.1.PTcondition}) is satisfied very well.
As shown in Fig.~\ref{Fig:4.3.SUSY2PT}, not only the PSO splittings but also the energy degeneracy of the spin doublets are excellently reproduced by the first-order perturbation calculations.

In such an explicit and quantitative way, the PSO splittings $\Delta E_{\rm PSO}$ can be directly understood by the PSS-breaking term $\tilde{W}^{\rm PSS}$ within the representation of the SUSY partner Hamiltonian $\tilde{H}$.
Furthermore, this symmetry-breaking term can be treated as a very small perturbation on the exact PSS limit $\tilde{H}^{\rm PSS}_0$.
Therefore, the PSS discussing here is of pertubative nature \cite{Liang2013_PRC87-014334}.

Recently, it has been demonstrated in Ref.~\cite{Shen2013_PRC88-024311} that the perturbative nature of PSS maintains even when a substantial SO potential is included.
The SO term shows both indirect and direct effects on the PSS-breaking potentials $\tilde{V}_{\rm PSO}(r)$.
The indirect effect due to the changes of the reduced superpotentials $q_\kappa(r)$ is rather small.
In contrast, the direct effect, i.e., the SO potential itself appearing in both $H$ and $\tilde H$,
reduces the PSO splittings $\Delta E_{\rm PSO}$ systematically and substantially.

\subsubsection{SUSY for Dirac equations}

The above discussions are based on the second-order differential Hamiltonian, i.e., the factorizable Hamiltonian.
Alternatively, Leviatan established the SUSY schemes directly based on the first-order differential Dirac Hamiltonian by using the intertwining relation \cite{Leviatan2004_PRL92-202501}.

In usual applications of SUSY quantum mechanics, one starts from a factorizable Hamiltonian $H_1$, then identifies the pair of Hermitian conjugate operators $B^+$ and $B^-$ in Eq.~(\ref{Eq:4.3.SUSYH1}), and eventually obtains the SUSY partner Hamiltonian $H_2$ in Eq.~(\ref{Eq:4.3.SUSYH2}).

This procedure can be regarded in a different way.
Assuming one holds
\begin{equation}\label{Eq:4.3.SUSYDitw}
  B^-H_1 = H_2 B^-\,,
\end{equation}
which is the so-called intertwining relation between $H_1$ and $H_2$ \cite{Nieto2003_AP305-151},
this intertwining relation ensures that, if $\psi_1(n)$ is an eigenstate of $H_1$, then also $\psi_2(n)\propto B^-\psi_1(n)$ (\ref{Eq:4.3.SUSYWFtran}) is an eigenstate of $H_2$ with the same energy $E_S(n)$, unless $B^-\psi_1(n)$ vanishes (\ref{Eq:4.3.SUSYpsi0}) or produces an unphysical state, e.g., non-normalizable.
In other words, the SUSY schemes shown in Fig.~\ref{Fig:4.3.SUSY} can be set up as long as the intertwining relation is satisfied, but Hamiltonians $H_1$ and $H_2$ are not necessarily factorizable.

For example, one can insist that both SUSY partner Hamiltonians $H(\kappa_a)$ and $H(\kappa_b)$ be the Dirac Hamiltonian of the form prescribed in Eq.~(\ref{Eq:2.1.DiraceqR}), and looks for possible solutions of $B^-_\kappa$ that satisfy
\begin{equation}\label{Eq:4.3.SUSYDHB}
  B^-_\kappa
  \lb\begin{array}{cc}
    M+\Sigma(r) & -\displaystyle\frac{d}{dr}+\displaystyle\frac{\kappa_a}{r} \\
    \displaystyle\frac{d}{dr}+\displaystyle\frac{\kappa_a}{r} & -M+\Delta(r)
  \end{array}\rb
  =
  \lb\begin{array}{cc}
    M+\Sigma(r) & -\displaystyle\frac{d}{dr}+\displaystyle\frac{\kappa_b}{r} \\
    \displaystyle\frac{d}{dr}+\displaystyle\frac{\kappa_b}{r} & -M+\Delta(r)
  \end{array}\rb
  B^-_\kappa\,.
\end{equation}

\begin{figure}[tbhp]
\begin{center}
  \includegraphics[width=8cm]{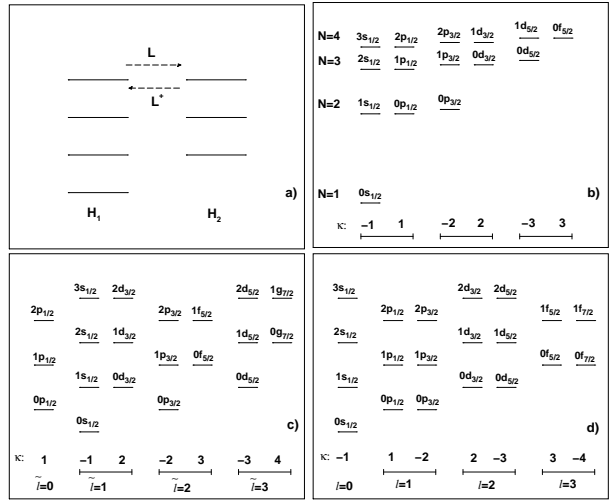}
\end{center}
\caption{Schematic patterns in (a) SUSY quantum mechanics and in the (b) Coulomb, (c) pseudospin, and
(d) spin symmetry limits of the Dirac Hamiltonian.
Taken from Ref.~\cite{Leviatan2004_PRL92-202501}, and the notation of the single-particle spectra starts from $n=0$.}
\label{Fig:4.3.SUSYD}
\end{figure}

By considering a matrical Darboux transformation operator,
\begin{equation}\label{Eq:4.3.SUSYDB}
  B^-_\kappa = P_\kappa(r)\frac{d}{dr}+Q_\kappa(r)\,,
\end{equation}
where $P_\kappa$ and $Q_\kappa$ are $2\times2$ matrices, and assuming certain forms in the functions $[P_\kappa(r)]_{ij}$ and $[Q_\kappa(r)]_{ij}$, Leviatan found three different kinds of solutions in Ref.~\cite{Leviatan2004_PRL92-202501}.
They correspond to three different kinds of symmetry limits: (i) Coulomb symmetry limit, (ii) spin symmetry limit, and (iii) pseudospin symmetry limit.
The schematic patterns in these symmetry limits are illustrated in Fig.~\ref{Fig:4.3.SUSYD}.

In the Coulomb symmetry limit, the orbitals with $\kappa_a+\kappa_b=0$ form the degenerate doublets, e.g., $(2s_{1/2},1p_{1/2})$, $(2p_{3/2},1d_{3/2})$, and the scalar and vector potentials are in the shape of
\begin{equation}\label{Eq:4.3.SUSYDCV}
  S(r) = \alpha_S/r
  \qquad\mbox{and}\qquad
  V(r) = \alpha_V/r\,,
\end{equation}
respectively.
The corresponding transformation operator reads \cite{Leviatan2004_PRL92-202501}
\begin{equation}\label{Eq:4.3.SUSYDCB}
  B^-_\kappa =
  \lb\begin{array}{cc}
    \displaystyle\frac{d}{dr}+\displaystyle\frac{\varepsilon_+}{r}+\displaystyle\frac{M\alpha_+}{\kappa_a} & \quad-\displaystyle\frac{\alpha_S}{\kappa_a}\displaystyle\frac{d}{dr}+\displaystyle\frac{\alpha_V}{r} \\
    \displaystyle\frac{\alpha_S}{\kappa_a}\displaystyle\frac{d}{dr}-\displaystyle\frac{\alpha_V}{r} & \quad\displaystyle\frac{d}{dr}-\displaystyle\frac{\varepsilon_-}{r}-\displaystyle\frac{M\alpha_-}{\kappa_a}
  \end{array}\rb\,,
\end{equation}
where $\varepsilon_\pm=\kappa_a+\alpha_S\alpha_\pm/\kappa_a$ and $\alpha_\pm=\alpha_S\pm\alpha_V$.

The SS limit corresponds to that shown in Eq.~(\ref{Eq:2.1.SSlimit}), i.e., $V(r)-S(r)=\Delta_0$, and the transformation operator reads \cite{Leviatan2004_PRL92-202501}
\begin{equation}\label{Eq:4.3.SUSYDSSB}
  B^-_\kappa =
  \lb\begin{array}{cc}
    2M+\Sigma(r)-\Delta_0 & \quad-\displaystyle\frac{d}{dr}+\displaystyle\frac{\kappa_a}{r} \\
    \displaystyle\frac{d}{dr}+\displaystyle\frac{\kappa_b}{r} & \quad0
  \end{array}\rb\,,
\end{equation}
with $\kappa_a+\kappa_b=-1$.

The PSS limit corresponds to that shown in Eq.~(\ref{Eq:2.1.PSSlimit}), i.e., $V(r)+S(r)=\Sigma_0$, and the transformation operator reads \cite{Leviatan2004_PRL92-202501}
\begin{equation}\label{Eq:4.3.SUSYDPSSB}
  B^-_\kappa =
  \lb\begin{array}{cc}
    0 & \quad-\displaystyle\frac{d}{dr}+\displaystyle\frac{\kappa_b}{r} \\
    \displaystyle\frac{d}{dr}+\displaystyle\frac{\kappa_a}{r} & \quad-2M+\Delta(r)-\Sigma_0
  \end{array}\rb\,,
\end{equation}
with $\kappa_a+\kappa_b=1$.

The higher-order terms in the effective Hamiltonian (\ref{Eq:4.3.SRGHFermi}) transformed by SRG can be avoided, if the SUSY scheme is established directly on the Dirac Hamiltonian by using the intertwining relation (\ref{Eq:4.3.SUSYDitw}).
However, the present studies have not included the perturbative PSS limit shown in Eq.~(\ref{Eq:4.1.H0U3}).
Furthermore, it is still an open problem about how to identify the corresponding PSS-breaking term $\tilde W^{\rm PSS}$ in such a scheme.

In summary, it was pointed out in Refs.~\cite{Liang2013_PRC87-014334,Shen2013_PRC88-024311} that it is promising to understand the pseudospin symmetry and its breaking mechanism in a fully quantitative way, by combining the similarity renormalization group \cite{Guo2012_PRC85-021302R,Li2013_PRC87-044311,Guo2014_PRL112-062502}, the supersymmetric quantum mechanics \cite{Leviatan2004_PRL92-202501,Typel2008_NPA806-156,Leviatan2009_PRL103-042502}, and the perturbation theory \cite{Liang2011_PRC83-041301R,Li2011_ChinPhysC35-825}.
It is shown that while the spin-symmetry-conserving term appears in the single-particle Hamiltonian $H$, the pseudospin-symmetry-conserving term appears naturally in its supersymmetric partner Hamiltonian $\tilde H$.
The eigenstates of Hamiltonians $H$ and $\tilde H$ are exactly one-to-one identical except for the so-called intruder states.
In such a way, the origin of pseudospin symmetry deeply hidden in $H$ can be traced in its supersymmetric partner Hamiltonian $\tilde H$.

By using the similarity renormalization group, the Dirac Hamiltonian can be transformed into a diagonal form and expanded in a series of $1/M$.
So far, the cases corresponding to the lowest-order approximation and the lowest-order spin-orbit potential have been investigated in Refs.~\cite{Liang2013_PRC87-014334, Shen2013_PRC88-024311}, respectively.
The perturbative nature of pseudospin symmetry has been demonstrated by the perturbation calculations for both cases without and with a substantial spin-orbit potential.
A general tendency that the pseudospin-orbit splittings become
smaller with increasing single-particle energies can also be interpreted in an explicit way.
The corresponding studies with the effective mass term and higher-order terms are expected in the near future.

Alternatively, the supersymmetric scheme can be directly established based on the first-order differential Dirac Hamiltonian by using the intertwining relation \cite{Leviatan2004_PRL92-202501}.
In such a way, the higher-order terms in the effective Hamiltonian transformed by the similarity renormalization group can be avoided.
Meanwhile, it is still an open problem how to identify in such a scheme the perturbative pseudospin symmetry limit, e.g., a Dirac Hamiltonian with relativistic harmonic oscillator potential, as well as the corresponding symmetry-breaking term.

\section{Summary and Perspectives}\label{Sect:5}


After the independent observation of the near degeneracy between pairs of single-particle states $(n, l, j = l + 1/2)$ and $(n-1, l + 2, j = l + 3/2)$, i.e., the so-called pseudospin symmetry, by Hecht and Adler \cite{Hecht1969_NPA137-129} and Arima, Harvey, and Shimizu \cite{Arima1969_PLB30-517} in 1969, it raised a fascinating question whether such near degeneracy is accidental (a degeneracy not explained by an obvious symmetry) or due to symmetry breaking (more descriptively hidden symmetry).
The pioneering work by examining the ratio between the strengths of the spin-orbit and orbit-orbit potentials by Bohr, Hamamoto, and Mottelson in 1982 \cite{Bohr1982_PS26-267} opened the door to explore the hidden symmetry in understanding the origin of pseudospin symmetry.

With the success of the relativistic mean-field theory \cite{Ring1996_PPNP37-193,Vretenar2005_PR409-101,Meng2006_PPNP57-470,Niksic2011_PPNP66-519,Meng2011_PP31-199,Meng2013_FPC8-55},  it is quite encouraging to find that the special ratio between the strengths of the spin-orbit and orbit-orbit interactions can be reproduced by the relativistic mean-field theory \cite{Bahri1992_PRL68-2133}.
The reveal of the pseudo-orbital angular momentum as the orbital angular momentum of the lower component of the Dirac spinor by Ginocchio in 1997 \cite{Ginocchio1997_PRL78-436} proved to be an unexpected success of the relativistic mean-field theory.
It is then followed by lots of exciting discoveries.
For examples, the vanishing of the derivative for the sum of the scalar and vector potentials, i.e., $d\Sigma(r)/dr=0$, can lead to the exact pseudospin symmetry \cite{Meng1998_PRC58-R628}, which means that the pseudospin symmetry becomes much better for exotic nuclei with a highly diffused potential \cite{Meng1999_PRC59-154}.
While developing the relativistic mean-field theory in a Dirac Woods-Saxon basis \cite{Zhou2003_PRC68-034323}, Zhou, Meng, and Ring \cite{Zhou2003_PRL91-262501} discovered the spin symmetry in the anti-nucleon spectra.
A rigorous verification of the pseudospin symmetry in the single-particle resonant states is given by Lu, Zhao, and Zhou \cite{Lu2012_PRL109-072501} by examining the Jost functions corresponding to small components of the radial Dirac wave functions.
By using the similarity renormalization group, Guo and coauthors \cite{Guo2012_PRC85-021302R,Li2013_PRC87-044311} have made a new exploration of the pseudospin symmetry, including the axially deformed systems \cite{Guo2014_PRL112-062502}.
Combining the supersymmetric quantum mechanics, perturbation theory, and similarity renormalization group, Liang and coauthors \cite{Liang2013_PRC87-014334,Shen2013_PRC88-024311} have provided a promising way to understand the origin of pseudospin symmetry and its breaking fully quantitatively.


All the pioneering works and exciting discoveries have triggered tremendous enthusiasms in understanding the pseudospin symmetry in various physical systems as well as exploring their origin.
Although the present paper is intended to provide a comprehensive overview of the related works done recently in the exploration of the pseudospin and spin symmetries, it is unavoidable to have overlooked some of them.


The historical development and recent progress have been summarized in the introduction by trying to exhaust the existing literatures.
In Section~\ref{Sect:2}, the general features for the Dirac equation and its corresponding Schr\"odinger-like equations were discussed.
The pseudospin symmetry and spin symmetry in various systems and potentials have been reviewed in Section~\ref{Sect:3}.
The discussions ranged from stable to exotic nuclei, from non-confining to confining potentials, from local to non-local potentials, from central to tensor potentials, from bound to resonant states, from nucleon to anti-nucleon spectra, from nucleon to hyperon spectra, and from spherical to deformed nuclei.
Extensive discussions have been devoted in Section~\ref{Sect:4} to the open issues on the pseudospin and spin symmetries as well, including the perturbative nature, the puzzle of intruder states, and the connection with the supersymmetric quantum mechanics and similarity renormalization group.



The current Review is focused on the theoretical exploration of the hidden pseudospin and spin symmetries and their origins in atomic nuclei.
There might be questions on the unique experimental signal of these symmetries.
While introducing some themes in the study of very deformed rotating nuclei,
Mottelson \cite{Mottelson1991_NPA522-1c} preluded the link between the pseudospin symmetry and newly observed experiments. In particular,  a lot of phenomena in nuclear structure have been successfully interpreted directly or implicitly by the pseudospin symmetry, including nuclear superdeformed configurations \cite{Dudek1987_PRL59-1405,Bahri1992_PRL68-2133}, identical bands \cite{Nazarewicz1990_PRL64-1654,Nazarewicz1990_NPA512-61,Zeng1991_PRC44-R1745}, quantized alignment \cite{Stephens1990_PRL65-301}, and pseudospin partner bands \cite{Xu2008_PRC78-064301,Hua2009_PRC80-034303}.
In addition, the relevance of pseudospin symmetry in the structure of halo nuclei \cite{Long2010_PRC81-031302R} and superheavy nuclei \cite{Jolos2007_PAN70-812,Li2014_PLB732-169} was pointed out. The fingerprint for pseudospin doublet bands and their difference with the chiral doublet bands \cite{Frauendorf1997Nucl.Phys.A131} have been briefly listed in Ref.~\cite{Meng2010_JPG37-064025}.

The experimental verification of the pseudospin symmetry can also be done from the single-particle energies.
The observed single-particle energies obtained by pickup or knockout reactions contain polarization effects, except for the cases where the spectroscopic factors of the single-particle states are close to one \cite{Cottle2010_Nature465-430,Liang2011_PRC83-011302R}.
Therefore, one should take into account the particle-vibration coupling and polarization effects \cite{Litvinova2011_PRC84-014305,Niu2012_PRC85-034314}.
Litvinova and Afanasjev \cite{Litvinova2011_PRC84-014305} have studied the impact of particle-vibration coupling and polarization effects due to deformation and time-odd mean fields on the single-particle spectra
systematically in doubly magic nuclei from $^{56}$Ni up to superheavy ones.
It has been shown that the particle-vibration coupling substantially improves the description of splitting energies in the pseudospin doublets.
However, there are still cases where the observed pseudospin-orbit splittings are poorly reproduced.


Finally, the exploration for the deep reasons on the approximate pseudospin symmetry observed in realistic nuclei is also a challenging problem. There are already some discussions on the physics behind
the near equality of the vector and scalar potentials in the
Dirac Hamiltonian \cite{Cohen1995_PPNP35-221, Furnstahl2000_NPA673-298, Ke2010_IJMPA25-1123} and more investigations are expected. In particular, it is highly necessary to search for the links with more fundamental models on strong interactions or realistic interactions between nucleons.

\section*{Acknowledgements}

We would like to express our gratitude to all the collaborators and colleagues who contributed to the investigations presented here, in particular to A. Arima, T. S. Chen, L. S. Geng, J. N. Ginocchio, J. Y. Guo, X. T. He, R. V. Jolos, A. Leviatan, F. Q. Li, W. H. Long, B. N. Lu, H. F. L\"u, P. Ring, H. Sagawa, W. Scheid, S. H. Shen, C. Y. Song, K. Sugawara-Tanabe, H. Toki, N. Van Giai, S. Yamaji, S. C. Yang, J. M. Yao, S. Q. Zhang, Y. Zhang, E. G. Zhao, and P. W. Zhao.
We acknowledge the fruitful discussions with K. Arita, I. Hamamoto, A. Ikeda, K. Matsuyanagi, T. Nakatsukasa, K. Sato, T. T. Sun, T. Suzuki, and Y. Zhang during the iTHES workshop: Exploration of hidden symmetries in atomic nuclei at RIKEN.
We appreciate C. L. Bai, J. Y. Guo, W. H. Long, B. N. Lu, Z. M. Niu, S. H. Shen, N. Wang, S. Q. Zhang, and Y. Zhang for the careful reading of the manuscript and the valuable suggestions.
We also thank S. H. Shen and B. Zhao for redrawing some of the figures used in the paper.
This work was supported in part by the Major State 973 Program of China (Grant No.~2013CB834400),
the Natural Science Foundation of China (Grants No.~10975008, No.~11175002, No.~11105006, and No.~11335002),
the Research Fund for the Doctoral Program of Higher Education (RFDP) (Grant No.~20110001110087),
the Grant-in-Aid for JSPS Fellows under Grant No. 24-02201,
the RIKEN iTHES Project,
and the RIKEN Foreign Postdoctoral Researcher Program.




\appendix
\section{}

During the preparation of the present Review, we found the symbols and notations used in different papers are indeed quite similar.
However, subtle differences, like a sign, a factor, etc., are sometimes crucial for understanding the investigations.
Therefore, some of the symbols and notations have been changed from the original papers and unified through the present Review by choosing the common conventions.
We also recommend these unified conventions for the future studies.

In this Appendix, we list the key symbols and notations as well as the abbreviations used in this Review in an alphabetic way.

\subsection*{List of key symbols and notations}

\begin{longtable}[l]{@{~~~~~~~~~~~~~~~~}lcl@{}}
$A$ & & atomic mass number \\
$B^+,B^-$ & & pair of Hermitian conjugate operators in SUSY \\
$E$ & & single-particle energy excluding rest mass, $E = \epsilon-M$ \\
$E_{\bar{\Lambda}}$ & & single-$\bar{\Lambda}$ energy excluding rest mass, $E_{\bar{\Lambda}}=\epsilon_{\bar{\Lambda}}-M_{\bar{\Lambda}}$ \\
$E_{\rm av}$ & & average single-particle energy, $E_{\rm av}=(E_{j_<} + E_{j_>})/2$ \\
$E_i$ & & contribution from each potential to \\
      & &                                    the total single-particle energy $E$ \\
$F(r)$ & & lower component of Dirac spinor in spherical case \\
$\mathcal{F}_\kappa(r)$ & & $\mathcal{F}_\kappa(r)=r^{-\kappa} F_\kappa(r)$ \\
$G(r)$ & & upper component of Dirac spinor in spherical case \\
$\mathcal{G}_\kappa(r)$ & & $\mathcal{G}_\kappa(r)=r^\kappa G_\kappa(r)$ \\
$H, \tilde H$ & & single-particle Hamiltonian and its SUSY partner \\
$H_0$ & & symmetry-conserving Hamiltonian \\
$\mathscr H$ & & system Hamiltonian density \\
$\mathcal{J}_\kappa$ & & Jost function \\
$\mathscr L$ & & Lagrangian density \\
$M$ & & mass of nucleon \\
$M_+$ & & effective mass, $M_+=M-\Delta+\epsilon$\\
$M_-$ & & effective mass, $M_-=-M-\Sigma+\epsilon$\\
$N$ & & neutron number \\
$[N,n_{3},\Lambda ,\Omega]$ & & asymptotic Nilsson quantum numbers \\
$Q_\kappa$ & & SUSY superpotential \\
$R(r)$ & & non-relativistic single-particle radial wave function \\
       & &                                                          in the spherical case \\
$S(\mathbf{r})$ & & scaler potential, negative value means attractive \\
$\mathbf{T}(\mathbf{r})$ & & tensor potential \\
$V(\mathbf{r})$ & & vector potential, positive value means repulsive \\
$V_{\rm (P)CB}$ & & (pseudo-)centrifugal barrier \\
$V_{\rm (P)SO}$ & & (pseudo)spin-orbit potential \\
$W$ & & symmetry-breaking term \\
$\tilde W^{\rm PSS}$ & & PSS-breaking term in SUSY scheme \\
$X,Y$ & & non-local Fock potentials\\
$X_G, X_F, Y_G, Y_F$ & & localized Fock potentials\\
$\mathscr Y^l_{jm}$ & & spherical harmonics spinor \\
$Z$ & & proton number \\
\\
$a,b$ & & labels of single-particle orbitals \\
$c^\dag, c$ & & nucleon creation and annihilation operators \\
$e(\kappa)$ & & SUSY energy shift \\
$f_{\rho,\pi}$ & & meson tensor coupling strengths \\
$g_{\sigma,\omega,\rho}$ & & meson coupling strengths \\
$i,j$ & & run over $1,2,3$ \\
$j$ & & total angular momentum of single-particle states \\
$j_<, j_>$ & & $j=l\mp1/2$ for spin doublets or \\
           & &                                     $j=\tilde l\mp1/2$ for pseudospin doublets \\
$l,\tilde l$ & & angular momenta of upper component $G(r)$ and \\
             & &                                                  lower component $F(r)$ \\
$m_{\sigma,\omega,\rho,\pi}$ & & mass of mesons \\
$n$ & & main quantum number of single-particle states, \\
    & &                                                   starting from $n=1$ \\
$n_G,n_F$ & & number of internal nodes of $G(r)$ and $F(r)$ \\
$\mathbf{p}$ & & momentum operator \\
$q_\kappa$ & & SUSY reduced superpotential \\
$r, \hat{\mathbf{r}}$ & & radial and angular coordinates in spherical coordinate system \\
$r_\bot, z$ & & radial and height coordinates in cylindrical coordinate system \\
$\mathbf{s}, \tilde{\mathbf{s}}$ & & spin operator $\mathbf{s}=\boldsymbol{\sigma}/2$ and pseudospin operator \\
                                 & &                                                                          $\tilde{\mathbf{s}} = (\boldsymbol{\sigma}\cdot\hat{\mathbf{p}}) \mathbf{s} (\boldsymbol{\sigma}\cdot\hat{\mathbf{p}})$ \\
\\
$\Gamma$ & & width of a resonant state \\
$\Delta(r)$ & & $\Delta(r)=V(r)-S(r)$, positive value means repulsive\\
$\Delta_0$ & & for the case of constant $V-S$ \\
$\Delta E_{\rm SO}$ & & reduced spin-orbit splitting, $\Delta E_{\rm SO}=(E_{j_<} - E_{j_>})/(2l+1)$ \\
$\Delta E_{\rm PSO}$ & & reduced pseudospin-orbit splitting, $\Delta E_{\rm PSO}=(E_{j_<} - E_{j_>})/(2\tilde l+1)$ \\
$\Sigma(r)$ & & $\Sigma(r)=S(r)+V(r)$, negative value means attrative\\
$\Sigma_0$ & & for the case of constant $S+V$ \\
$\Sigma_{\rm HO}$ & & $\Sigma(r)$ of the form of a harmonic oscillator\\
$\lr\Phi_0\rc$ & & system ground-state wave function \\
$\lr-\rc$ & & physical vacuum \\
\\
$\alpha,\beta,\gamma^\mu, \gamma^5, \sigma^{\mu\nu}$ & & Dirac matrices \\
$\alpha$ & & set of quantum numbers, $\alpha=(a, m_a)=(n_a, l_a, j_a, m_a)$ \\
$\epsilon$ & & single-particle energy including rest mass $M$\\
$\epsilon_{\rm A},\epsilon_{\bar{\Lambda}}$ & & single-anti-nucleon and single-$\bar{\Lambda}$ energies including rest mass\\
$\kappa$ & & quantum number, $\kappa=\mp(j+1/2)$ for $j=l\pm1/2$ states \\
$\mu,\nu$ & & run over $0,1,2,3$ \\
$\rho_C, \rho_S, \rho_{V}, \rho^{(3)}_{V}$ & &  charge and scalar densities, isoscalar and isovector\\
                                           & &                                                       baryonic densities \\
$\boldsymbol{\sigma}$ & & Pauli matrices \\
$\vec\tau,\tau_3$ & & isospin operator and its third component \\
$\psi(\mathbf{r})$ & & single-particle wave function \\
$\psi_{\rm A}(\mathbf{r}),\psi_{\bar{\Lambda}}(\mathbf{r})$ & & single-anti-nucleon and single-$\bar{\Lambda}$ wave functions \\
\end{longtable}

\subsection*{List of abbreviations}

\begin{longtable}[l]{@{~~~~~~~~~~~~~~~~}lcl@{}}
ACCC & & analytical continuation in coupling constant \\
CB & & centrifugal barrier \\
CDFT & & covariant density functional theory\\
CSM & & complex scaling method \\
HO & & harmonic oscillator \\
NU & & Nikiforov-Uvarov \\
PCB & & pseudo-centrifugal barrier \\
PSO & & pseudospin-orbit \\
PSS & & pseudospin symmetry \\
RCHB & & relativistic continuum Hartree-Bogoliubov \\
RHF & & relativistic Hartree-Fock \\
RHO & & relativistic harmonic oscillator \\
RMF & & relativistic mean-field \\
SO & & spin-orbit \\
SRG & & similarity renormalization group \\
SS & & spin symmetry \\
SUSY & & supersymmetry or supersymmetric \\
\end{longtable}

\section*{References}
\bibliographystyle{elsarticle-num}
\bibliography{ref}


%
%
%
\end{document}